\documentclass[11pt]{article}
\usepackage{verbatim,amsmath,amssymb}
\usepackage{epsfig,float,color}
\usepackage{epstopdf}
\usepackage{geometry}
\usepackage{setspace}
\usepackage{wrapfig}
\usepackage{hyperref}
\usepackage[utf8]{inputenc}
\usepackage{graphicx}
\usepackage{flushend}
\usepackage[linesnumbered,ruled]{algorithm2e}

\usepackage{subcaption}
\usepackage[font={footnotesize}]{caption}
\usepackage[font={footnotesize}]{subcaption}
\usepackage{footnote}
\usepackage{multicol}
\usepackage{multirow}
\usepackage{enumitem}   
\usepackage{soul}
\usepackage{framed}
\usepackage[titletoc,title]{appendix}
\usepackage{bm}
\usepackage{cite}
\usepackage[makeroom]{cancel} 
\definecolor{darkblue}{rgb}{0,0,1}
\hypersetup{pdftex=true, colorlinks=true, breaklinks=true, linkcolor=darkblue, menucolor=darkblue, pagecolor=darkblue, citecolor=darkblue, urlcolor=darkblue}

\usepackage[dvipsnames]{xcolor}
\begin{document} 

	\title{On Topology Optimization with Elliptical Masks and Honeycomb Tessellation with Explicit Length Scale Constraints}
\author{Nikhil Singh \\Indian Institute of Technology Kanpur INDIA 208016\\ \\  Prabhat Kumar\\ Technical University of Denmark, Lyngby Denmark\\ \\
	Anupam Saxena\\ Indian Institute of Technology Kanpur INDIA 208016\\\\
Published\footnote{This pdf is the personal version of an article whose final publication is available at \href{https://link.springer.com/article/10.1007/s00158-020-02548-w}{https://www.springer.com/journal/158}}\,\,\,in \textit{Structural and Multidisciplinary Optimization}, 
\href{https://doi.org/10.1007/s00158-020-02548-w}{DOI:10.1007/s00158-020-02548-w} \\
Submitted on 20.~August 2019, Revised on 03. February 2020, Accepted on 13. February 2020}

\date{}

\maketitle
\begin{abstract}

Topology optimization using gradient search with negative and positive elliptical masks and honeycomb tessellation is presented. Through a novel skeletonization algorithm for topologies defined using filled and void hexagonal cells/elements, explicit minimum and maximum length scales are imposed on solid states in the solutions.  An analytical example is presented suggesting that for a skeletonized topology, optimal solutions may not always exist for any specified volume fraction, minimum and maximum length scales, and that there may exist implicit interdependence between them.  A Sequence for Length Scale (SLS) methodology is  proposed wherein solutions are sought by specifying only the minimum and maximum length scales with volume fraction getting determined within a specified range systematically. Through four benchmark problems in small deformation topology optimization, it is demonstrated that solutions by-and-large satisfy the length scale constraints though the latter may get violated at certain local sites. The proposed approach seems promising, noting especially that solutions, if rendered perfectly {\it black and white} with minimum length scale explicitly imposed and boundaries smoothened, are quite close in performance compared to the parent topologies. Attaining {\it volume distributed} topologies, wherein members are more or less of the same thickness, may also be possible with the proposed approach. 
\end{abstract}


\section{Introduction and Background}
\label{background}
\indent  Topology optimization formulations, which entail finding optimal continua for given sets of objectives and constraints, in 2D, are fairly well-developed \cite{Eschenauer2003,Bendsoe2005,Guo2010, sigmund_maute_review_2013}. These include density based \cite{Martin2003topology}, phase field \cite{wang2004phase,wang2004synthesis}, level set \cite{sethian2000structural,wang2005design,luo2008new} evolutionary \cite{yang1999bidirectional,huang2010evolutionary
} and other approaches with rectangular \cite{Martin2003topology}, regular hexagonal \cite{saxena_and_saxena_2003a,saxena_and_saxena_2007, langelaar2007use,talischi2009honeycomb,saxena2008material,saxena2011topology}, and in general, irregular hexagonal and polygonal \cite{talischi2012polytop,kumar2015topology} discretization of the design domain. Topology optimization methods are generalized to cater to a wide range of problems in mechanics, heat transfer, electrothermal, electrostatic, and other fields \cite{Martin2003topology}. Expectations from a topology optimization formulation are singularity-free (e.g., free of checkerboard and point-connection patterns), almost perfectly binary, mesh independent designs that could be attained as computationally efficiently as possible. Formulations that employ rectangular cells with their densities as design variables often employ filtering (e.g., density based \cite{bourdin_2001_dens_filtera, bruns_tortorelli_2001_dens_filterb}, and/or sensitivity based \cite{sigmund_1997_sens_filter})  to primarily suppress checkerboard patterns and point connections. Filtering also offers indirect control on minimum thickness \cite{wang_lazarov_sigmund_2011} but with regions of gray transitions making it difficult to properly discern contour boundaries. 

\indent To minimize such transitions, numerous projection schemes \cite{guest_P_B_2004_proj1, sigmund_2007_morph_proj2, Kawamoto_et_al_2010_proj3, Xu_et_al_proj4} are proposed to attain close to black and white solutions.  Having length scale control in seeking optimal topologies may also be mandatory so that solutions can be readily fabricated.  As with filtering methods, projection techniques also  impose length scales on solutions.  Guest et al. \cite{guest_P_B_2004_proj1} employ nodal density values as design variables, and use them to compute element densities within a specified circular region through projection, to control the minimum length scale. Guest \cite{Guest2009a} imposes maximum length scale on solutions via a metric corresponding to the radius of a circular test region. Other methods that impose control on length scales implicitly/explicitly include (i) the slope-constrained formulation by Petersson and Sigmund \cite{petersson1998slope} that prevents rapid variation of density, (ii) MOLE method by Poulsen \cite{Poulsen2003} who demonstrates existence of solutions and uses a global functional formulated to capture the monotonicity of densities along specific directions, (iii) methods employing level sets that involve use of strain energy \cite{LS_energy_method_2012}, quadratic energy functional \cite{LS_shape_feature_control_2008, LS_hinge_free_cms}, and feature control \cite{Guo2014explicit} wherein medial surface and signed distance are used for length scale definitions, (iv) those that involve wavelets \cite{multi_resolution_scale_wavelets_2000, poulsen_wavelet_2002} and (v) robust topology optimization formulations \cite{wang_lazarov_sigmund_2011, robust_boundary_uncertainities_2013, manuf_tol_to_2009}. Global stress based constraint is proposed in \cite{le2010stress, zhang2017stress, zhang2016geometry} which implicitly leads to the length scale control. The method avoids stress concentration points but does not provide explicit length scale control pertaining to the manufacturing constraints. Also, imposing a maximum length scale constraint may not be possible, unless a lower bound on stress measure is employed, determining which may not be trivial.  With machinability as focus, Mei et al. \cite{mei2008feature} propose feature based topology optimization by using concepts from constructive solid geometry, topological derivatives and a morphing approach. The structure is gradually constructed to be composed of a finite set of geometric primitives. Decision making in the construction process is performed via topological derivatives which suggest where to subtract the geometric primitive (material) from, and the morphing approach which suggests choosing one from a given set of primitives.

\indent Zhang et. al \cite{expl_ls_simp_2014} highlight some drawbacks with existing length scale control methods in topology optimization. Sensitivity filter based, slope constrained based and projection schemes leave gray cells at boundaries between the solid and void states. Many of these methods are designed to impose only the minimum length scale on the design. Guest’s approach \cite{Guest2009a} for maximum length scale control involves a large number of nonlinear constraints. Approach by Chen et al. \cite{LS_shape_feature_control_2008} offers difficulty in numerical implementation, and also, length scale control is implicit. Method by Guo et al. \cite{Guo2014explicit} is implemented only in the level set setting. Appreciating the need for implementation of explicit length scale control \cite{ sigmund_maute_review_2013} in a SIMP based formulation, Zhang et. al \cite{expl_ls_simp_2014} skeletonize intermediate topologies using the algorithm by Aichholzer et al. \cite{Aichholzer1995}. Gray topologies are first converted into black and white ones using Otsu's method \cite{Otsu_1979}.  Thereafter, a single cell skeleton is obtained in a manner that the original topology remains intact. Using each cell in the skeleton, explicit 
minimum and maximum length scale measures are formulated as sums of the quadratic terms. \textcolor{black}{Zhou et. al \cite{zhou2015minimum} propose a similar approach that does not require explicit determination of the skeleton of a topology. Lazarov et al. \cite{lazarov2016length} review recent advances in manufacturable, topology-optimized designs with focus on methods that intend to restrict the length scales on features from above and below. They note, per \cite{allaire:hal-00985000}, that a perfect formulation for minimum length scale imposition is still being sought. Lazarov and Wang \cite{lazarov2017maximum} remark that sensitivities related to changes in the skeleton, as the continuum topology changes, are neglected in \cite{Guo2014explicit,expl_ls_simp_2014}}.

\indent Methods on topology optimization that determine cell densities in groups, for instance by using a set of masks, also exist, e.g., \cite{saxena2008material, saxena2011topology, kumar2015topology, morphable_bars_2017, guo2014doing, zhang2017structural,wang2012high}. Densities are determined based on whether the cell centroids are inside, or outside the masks. Number of design variables are significantly lower, allowing optimization algorithms to deliver optimal solutions faster. In \cite{saxena2008material, saxena2011topology, kumar2015topology}, negative circular masks are used while in \cite{morphable_bars_2017}, morphable bars, or, positive bar like masks are employed. A bar used as a positive mask therein, is a union of a rectangle and semicircles at the two ends.  Gradient based searches are performed in \cite{saxena2011topology} and \cite{morphable_bars_2017}. Hoang and Jang \cite{morphable_bars_2017} implement both, minimum and maximum thickness constraints in an explicit manner.  Minimum thickness is achieved by directly setting the lower bound on thickness of the bars. Maximum thickness of each bar is addressed by limiting volume of the void around the bar, within a test region.  Using thickness control, joint connection and perimeter constraints,  Hoang and Jang demonstrate not only dimensional control but also, they achieve solutions with uniform thickness. Number of constraints are quite large however, and proportional to the number of bars used. It is noted that use of masks (or morphable bars) as proposed herein and Heaviside/Inverse Heaviside projections are similar, as such projections work similar to circular masks while locally ensuring length-scale on solid/void states. The difference is in the number of masks used, that they are decoupled from the mesh nodes, and whether the latter are of constant size/shape or varying, stationary or mobile.

\section{Aim, Motivation and Organization}
\label{aim_prob_form}

\begin{figure}[H]
    \centering
         \captionsetup{font=scriptsize}
    \begin{subfigure}[b]{.45\textwidth}
      \centering
       \captionsetup{font=scriptsize}
      \includegraphics[scale=0.45]{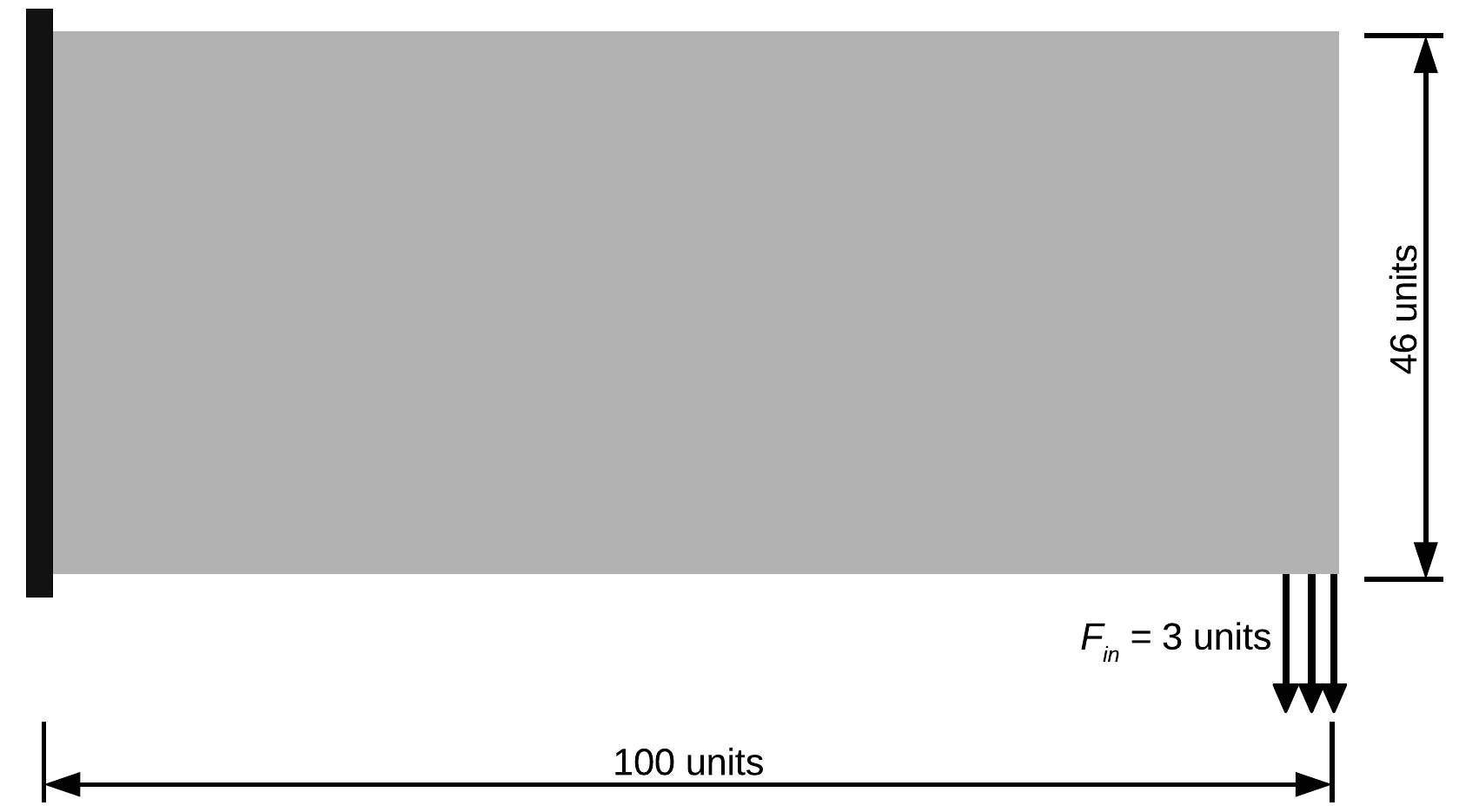}   
      \caption{Beam I: Stiffest continuum sought}
      \label{Fig:2:Eg1}
    \end{subfigure}
    \begin{subfigure}[b]{.45\textwidth}
      \centering
       \captionsetup{font=scriptsize}
      \includegraphics[scale=0.45]{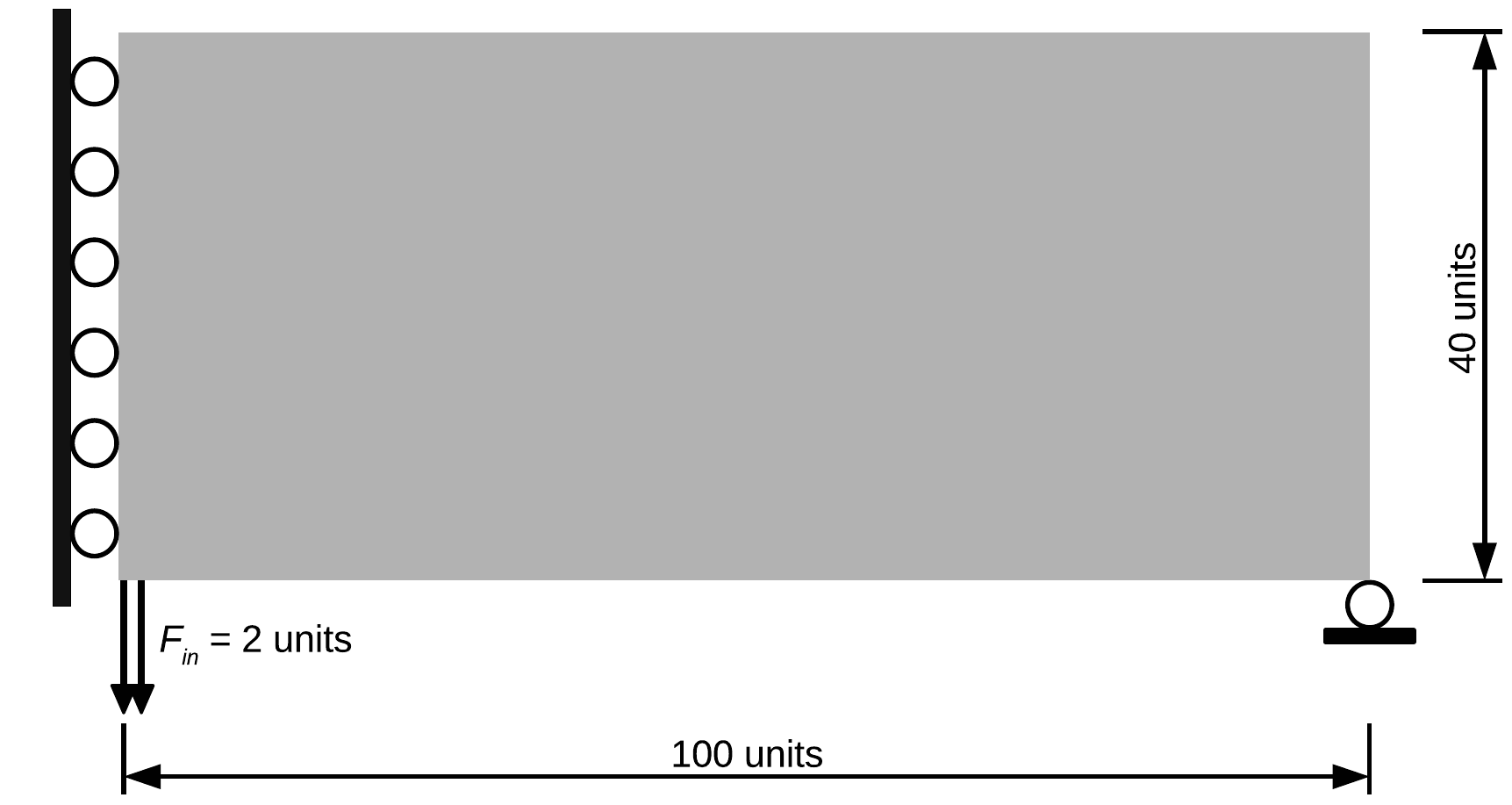}    
      \caption{Beam II: Stiffest continuum sought}
      \label{Fig:2:Eg2}
    \end{subfigure}  \\
 \centering
    \begin{subfigure}[b]{.45\textwidth}
      \centering
       \captionsetup{font=scriptsize}
      \includegraphics[scale=0.45]{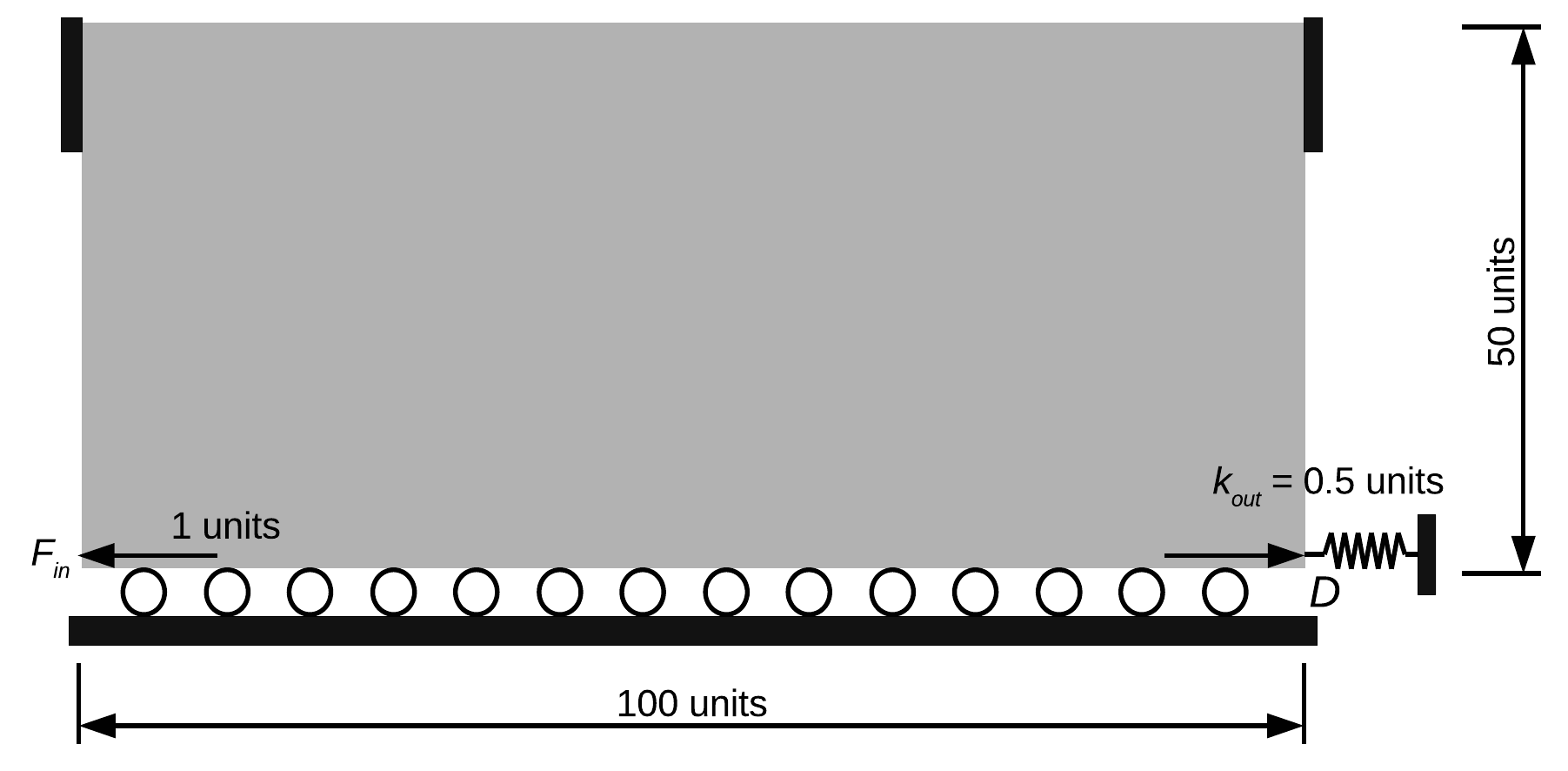}
      \caption{Small Displacement Inverter, displacement $D$ along the direction shown maximized.}
      \label{Fig:2:Eg3}
    \end{subfigure}
    \begin{subfigure}[b]{.45\textwidth}
      \centering
       \captionsetup{font=scriptsize}
      \includegraphics[scale=0.45]{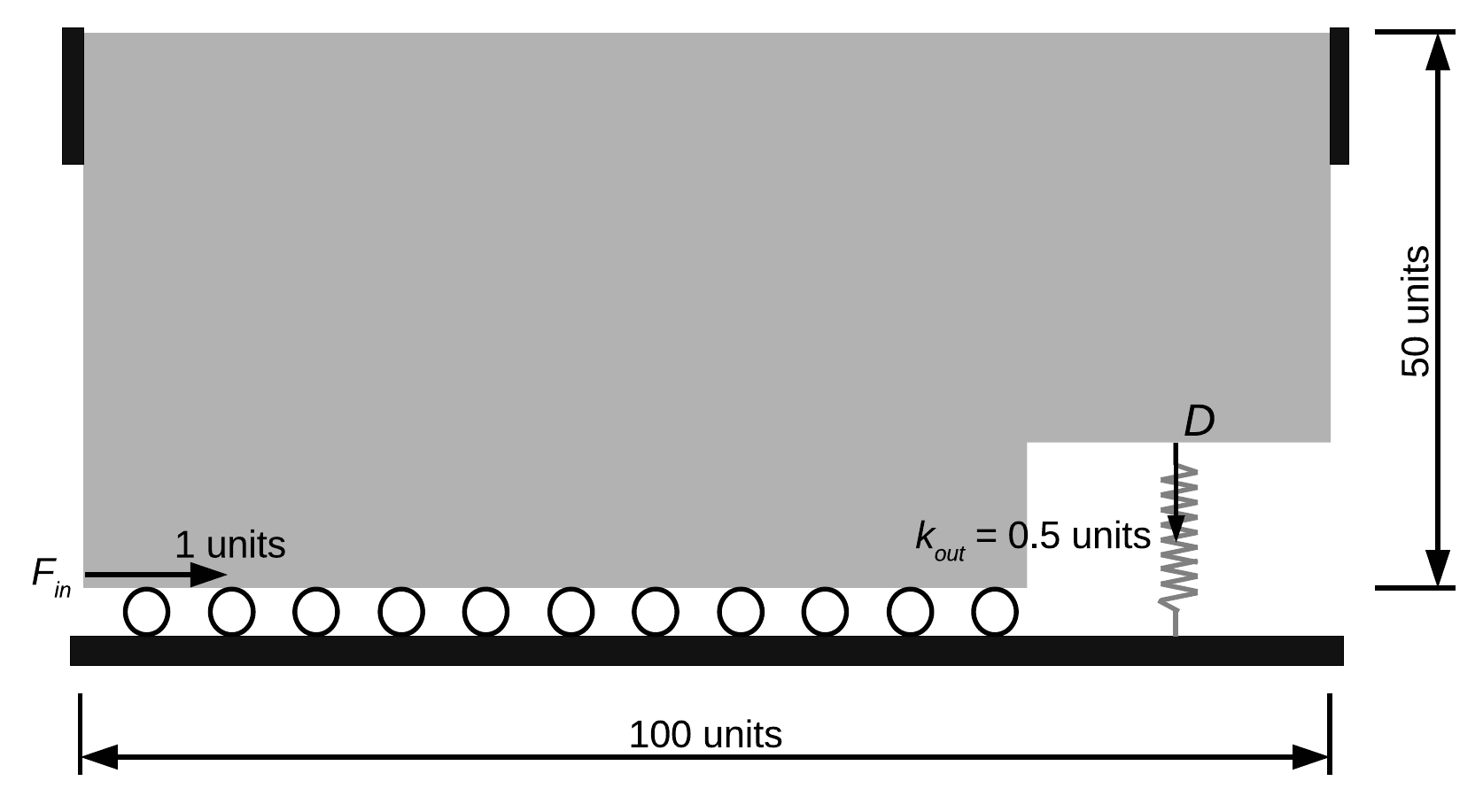}
      \caption{Small Displacement Crimper, displacement $D$ maximized.}
      \label{Fig:2:Eg4}
    \end{subfigure}  \\
\caption{Four benchmark examples to demonstrate topology optimization with honeycomb tessellation using negative and positive elliptical masks. }
     \label{Fig:2}
    \end{figure}

\noindent The intent in this paper is to illustrate how negative and positive elliptical masks can be employed to attain explicit minimum ($min_{ls}$) and/or maximum ($max_{ls}$) length scales induced over honeycomb meshes. Elliptical masks offer more versatality in shape control compared to circular masks (e.g., \cite{ saxena2011topology}), are less involved, and easier to implement compared to say, morphable bars \cite{morphable_bars_2017} wherein density is determined by considering three separate cases. While the formulation presented is extendable to the use of supershapes \cite{super_shapes_2018} or Gielis curves (generalization of superellipses) which are closed contours exhibiting variable symmetry and assymmetry, and can be described using a single relation, focus herein is primarily on elliptical masks. Topology optimization is illustrated via four benchmark examples (Fig. \ref{Fig:2}), two pertaining to minimization of the mean compliance (maximization of stiffness) and the others pertaining to small deformation compliant mechanisms wherein the intended deformation, $D$, at the output port is maximized. The optimization problems solved, are formulated as

\begin{eqnarray}
\label{formulation}
\underset{\{x_j, y_j, a_j, b_j, \theta_j\}, j = 1, ..., M}{\mbox{minimize: }  }  \Phi = \;\;\;\;\;\; \frac{1}{2}\mathbf{f} ^{T}\mathbf{u}  \mbox{  (stiff continua)    or   }  - S \frac{D}{\frac{1}{2}\mathbf{f}^{T}\mathbf{u}}    \mbox{  (compliant mechanisms) }\\
\begin{split}
\mbox{subject to:} & \nonumber \\
\mbox{(i)} \;\;\;\; & \mathbf{K} \mathbf{u} = \mathbf{f} \nonumber \\
\mbox{(ii)} \;\;\;\; & g_1 \equiv v - V^{*} = \sum_{i=0}^{Ne} \rho_{i} - V^{*} \leq 0 \nonumber \\
\mbox{(iii)} \;\;\;\; & g_{min}(\bm{\rho})  \leq \varepsilon_{1}   \nonumber \\
\mbox{(iv)} \;\;\;\; & g_{max}(\bm{\rho})  \leq \varepsilon_{2}  \nonumber \\
\mbox{(v)} \;\;\;\; & \bm{\psi}_{min}  \leq \bm{\psi} \leq \bm{\psi}_{max}, \mbox{ with } \bm{\psi} = \{x_j, y_j, a_j, b_j, \theta_j\},  j = 1, ..., M, \\
\end{split} 
\end{eqnarray}

\noindent where the displacement vector $\mathbf{u}$ is the response to $\mathbf{K} \mathbf{u} = \mathbf{f}$ with $\mathbf{K}$ as the global linear stiffness matrix assembled using element stiffness matrices $\mathbf{K}_{i} = [\rho_{min} + \rho_{i}(\alpha, \eta)(1 -\rho_{min} )]\mathbf{K}_{0}$. $\rho_{i}(\alpha, \eta)$ is the density of the $i^{(\mathrm{th})}$ cell (section \ref{mat_model}), $\mathbf{K}_{0}$ is the stiffness matrix of a solid cell, $\rho_{min}$ is the minimum (specified very small and positive) density a (void) cell can attain, $\mathbf{f}$ is a global force vector of applied loads, $S$ is a scale factor used to adjust the objective (primarily to adjust magnitudes of sensitivities of the objective),  $v$, the summation of cell densities, is the continuum volume bounded from above by $V^{*} = v_{f} V_{max}$ where $v_{f}$ is the volume fraction and $V_{max}$ is the maximum attainable volume\footnote{$V_{max}$ corresponds to all cells in the domain attaining the solid state.}. $g_{min}(\bm{\rho})$ and $g_{max}(\bm{\rho})$ are explicit, global minimum and maximum length scale measures dependent on densities and bounded by relaxation parameters $\varepsilon_{1}$ and $\varepsilon_{2}$. The last set of inequalities represent bounds on  positions, sizes and orientations of the elliptical masks. \\

\noindent Formulation of explicit length scale measures $g_{min}(\bm{\rho}) $ and $g_{max}(\bm{\rho}) $ is relatively straightforward if an intermediate topology can be converted into its skeletonized form. Let $\Omega = \bigcup \Omega_{H}$ be the design domain composed of hexagonal cells $\Omega_{H}$. Let $\Omega_{S}$ be a filled hexagonal cell that is a part of the skeleton (see Section \ref{skeletonization}) of an intermediate topology. With minimum and maximum length scales as $min_{ls}$ and $max_{ls}$ respectively, let two circles $C_{min}^{S}$ of radius $min_{ls}$ and $C_{max}^{S}$ of radius $max_{ls}$ respectively be drawn with center as the centroid of $\Omega_{S}$. Let regions $\mathbb{R}_{min}$ and $\mathbb{R}_{max}$ be such that $\mathbb{R}_{min} = \bigcup{C_{min}^{S}}$ and $\mathbb{R}_{max} = \Omega - \bigcup{C_{max}^{S}}$. Then, as suggested in \cite{expl_ls_simp_2014}, $g_{min}(\bm{\rho}) $ and $g_{max}(\bm{\rho}) $, slightly adapted, are formulated as

\begin{flalign}
\label{min_max_ls}
&& g_{min}(\bm{\rho}) &= \sum_{\mathbb{R}_{min}} \left[ 1 - \rho_{i}(\alpha, \eta) \right]^{p},&  \nonumber \\
\mbox{and} && g_{max}(\bm{\rho}) &= \sum_{\mathbb{R}_{max}} \left[ \rho_{i}(\alpha, \eta) - \rho_{min} \right]^{p},  & 
\end{flalign} 

\noindent where $p > 0$ is a chosen exponent. In \cite{expl_ls_simp_2014}, this exponent is 2. \\

Length scale measures in Eq. \ref{min_max_ls} are effective only when skeletons of the intermediate continuum solutions do not undergo topological alterations \cite{allaire:hal-00985000}. In this regard, these measures are more restrictive as opposed to the point-wise measures based on signed distances, proposed in \cite{allaire:hal-00985000}. In Eq. \ref{min_max_ls}, one notes that irrespective of the value/parity of $p$, each term within the summation, in $g_{min}(\bm{\rho})$ and $g_{max}(\bm{\rho})$ is non-negative as $\rho_{min} \leq  \rho_{i}(\alpha, \eta) \leq 1$. Thus, $g_{min}(\bm{\rho})$ and $g_{max}(\bm{\rho})$ in constraints (iii) and (iv) in Eq. \ref{formulation}  can never be negative. If $\varepsilon_{1}$ and $\varepsilon_{2}$ are chosen as zero, an optimal solution of Eq. \ref{formulation} will lie on the constraint boundaries $g_{min}(\bm{\rho}) = 0$ and $g_{max}(\bm{\rho}) = 0$ implying that all cells within $\mathbb{R}_{min}$ must precisely attain their solid states and those within $\mathbb{R}_{max}$, precisely the void states. As perfectly binary solutions are unlikely with gradient based optimization, $\varepsilon_{1}$ and $\varepsilon_{2}$ must be strictly positive. da Silva et al. \cite{da2019stress, da2019topology}, in their work on a robust formulation for optimal design of small deformation compliant topologies addressing stress constraints and manufacturing uncertainty,  opine that in case of ‘near perfect’ 0-1 solutions, extracting a smooth topology is difficult, and that one gets undesirable stress distribution along the boundary(ies). They suggest that a thin grey sliver should always be present between solid and void regions. $\varepsilon_{1}$ and $\varepsilon_{2}$ must therefore not be very close to zero. However, these relaxation parameters must be adequately small so that $g_{min}(\bm{\rho})$ and $g_{max}(\bm{\rho})$, the otherwise global length scale measures, are effective locally as well. Choosing $\varepsilon_{1}$ and $\varepsilon_{2}$ a priori may not be straightforward, as they may also depend on other parameters in Eq. \ref{formulation}, e.g., the upper bound on volume $V^{*}$ (or $vf$), minimum and maximum length scales $min_{ls}$ and $max_{ls}$ respectively. Moreover, the latter three parameters may themselves be interdependent and influenced by the skeleton which evolves continuously in topology optimization. While this interrelation may be apparent and explicable/quantifiable in case of simple examples (as shown later), the three parameters are usually specified independently/arbitrarily in most previous works on topology optimization with specified length scales.  To our knowledge, situations wherein optimal topologies are not attainable for a given set of these three (or five) parameters have not been addressed yet. \\

In what follows, material model with positive and negative elliptical masks is discussed in section \ref{mat_model}. To compute explicit, global length scale measures, a new skeletonization algorithm for intermediate topologies resulting from hexagonal meshes is developed and presented (Section \ref{skeletonization}). The method is similar to the approach in \cite{Arcelli_Baja_1978}, but implemented with hexagonal cells and therefore is confined to 2-dimensional cases. Through an analytical example (Section \ref{analytical_example}), one observes that, given a skeleton, with regard to Eqs. (\ref{formulation}), arbitrarily and independently  specified upper bound on volume ($vf$), minimum ($min_{ls}$) and maximum ($max_{ls}$) length scales, may not always yield a solution. In other words, the three parameters could be interrelated, whether there are changes in the skeleton or otherwise.  We show in section \ref{prelim_results} that if the formulation in Eq. \ref{formulation} is employed directly, obtained solutions are of inferior quality. We attribute this to altering skeletons corresponding to intermediate topologies and the associated length scale measures in Eq. \ref{min_max_ls}. Realising that these measures  are effective only when a well-defined skeleton exists, and that the upper bound on the volume constraint, minimum and maximum length scale measures are interrelated, a methodology is proposed in section \ref{method_TO} to attain topological solutions by specifying only the minimum and maximum length measures, and the initial volume fraction. The final volume fraction between the specified limits $vf_{min}$ and $vf_{max}$, and tolerances on $g_{min}(\bm{\rho}) $ and $g_{max}(\bm{\rho})$ get computed systematically. Examples are presented and discussed in Sections \ref{examples} and \ref{discussion}, and finally conclusions are drawn.

\section{Material Model and Sensitivities}
\label{mat_model}
\begin{figure}[H]
    \centering
      \includegraphics[scale=1.5]{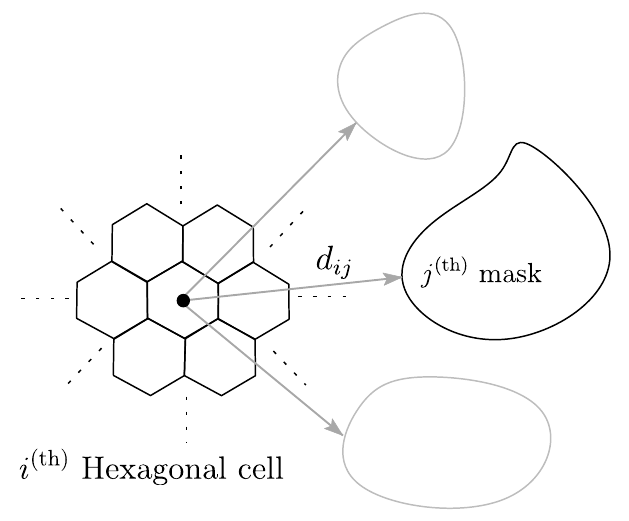}        
\caption{Masks of arbitrary shapes interacting with the {\it i}th hexagonal cell to decide its material status as {\it solid} or {\it void}}
     \label{Fig:1}
    \end{figure}

\indent As conventional, gradient-based topology optimization problems are formulated \cite{Eschenauer2003,Bendsoe2005,Guo2010, sigmund_maute_review_2013}, consider a design region  (Fig. \ref{Fig:1}) modeled with $Ncells$ regular hexagonal cells\footnote{Hexagonal cell is the same as a hexagonal finite element.} wherein, say, the $i^{(\mathrm{th})}$ cell has {\it density} $\rho_{i}$ such that if $\rho_{i} = 1$, the cell is regarded {\it solid} whereas if $\rho_{i} = 0$, the cell is considered {\it void}. Let a set of masks, those represented by simple (non self-intersecting), closed curves in Fig. \ref{Fig:1}, be laid over the domain. Influence of the $j^{(\mathrm{th})}$ mask on density of the $i^{(\mathrm{th})}$ cell  is modeled per the logistic approximation of the Heaviside function as

\begin{equation}
\rho_{ij}(\alpha_j) = \left[ \frac{1}{1 + \exp(-\alpha_j d_{ij})} \right],
\label{dens}
\end{equation}

\noindent
where $\alpha_j > 0$ is a mask specific parameter, and $d_{ij}$ is a measure that determines if the $i^{(\mathrm{th})}$ cell, represented by its centroid (Fig. \ref{Fig:1}), is enclosed within the $j^{(\mathrm{th})}$  mask or is on its boundary in which case $d_{ij} \leq 0$, or otherwise. If $\alpha_j d_{ij}$ is negative, and of large magnitude, $\rho_{ij}(\alpha_j)$ approaches $0$. If  $\alpha_j d_{ij}$  is positive and large, $\rho_{ij}(\alpha_j)$ approaches $1$. This makes the $j^{(\mathrm{th})}$ mask a {\it negative} mask as it {\it extracts} material off the group of hexagonal cell(s) it is laid over. If $\eta_{j}$ masks of identical shape and size are overlaid precisely, contribution may be written in product form as 

\begin{equation}
\rho_{ij}(\alpha_j, \eta_j) = \left[ \frac{1}{1 + \exp(-\alpha_j d_{ij})} \right]^{\eta_{j}}.
\label{dens1}
\end{equation}

\noindent For $M_n$ unique and non-overlapping negative masks over and/or around the domain, the overall density $\rho_{i}$ of the $i^{(\mathrm{th})}$ cell can be computed as 

\begin{equation}
\label{dens2} \rho_{i}(\alpha_j, \eta_j) = \prod_{j = 1}^{M_n} \left[ \frac{1}{1 + \exp(-\alpha_j d_{ij})} \right]^{\eta_j}.
\end{equation}

\noindent Indeed, if the $i^{(\mathrm{th})}$ cell (or its centroid) is not enclosed within any mask and if all masks are far away from it, the cell is {\it solid} ($\rho_{i}(\alpha_j, \eta_j) \approx 1$). If any mask encloses the $i^{(\mathrm{th})}$ cell, $\rho_{i}(\alpha_j, \eta_j) \approx 0$ and the cell is {\it void}.  The above notion could be {\it flipped} for {\it positive} masks which, when laid over the domain, deliver material to the cells beneath them. In that case, either $\alpha_j$ could be chosen negative, or, for $M_{p}$ number of unique, non-overlapping positive masks, Eq. (\ref{dens2}) could be modified as

\begin{equation}
\label{dens3} \rho_{i}(\alpha_j, \eta_j) = \left[1 -  \prod_{j = 1}^{M_p}  \frac{1}{1 + \exp(-\alpha_j d_{ij})} \right]^{\eta_j}.
\end{equation}

\begin{figure}[H]
    \centering
         \captionsetup{font=scriptsize}
    \begin{subfigure}[b]{.45\textwidth}
      \centering
       \captionsetup{font=scriptsize}
      \includegraphics[scale=0.5]{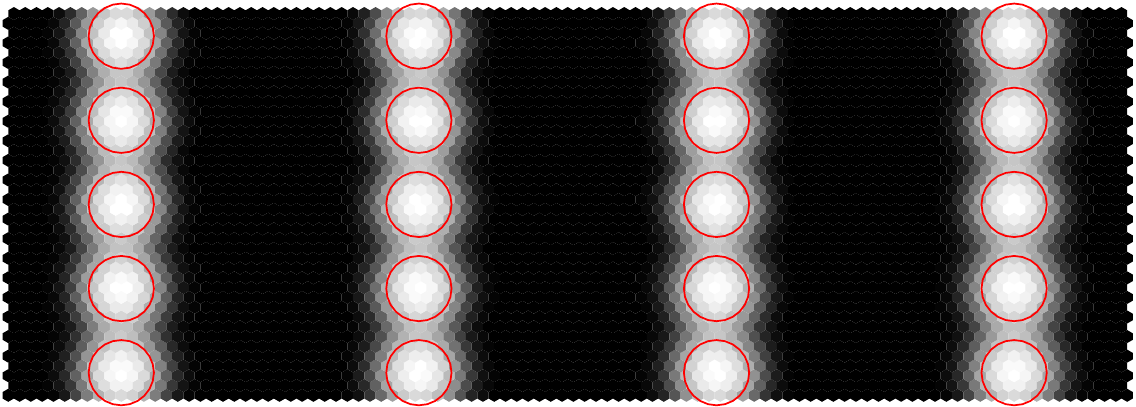}   
      \caption{$\alpha = 1; \eta = 1$ for all masks (default across subfigures)}
      \label{fig:R1}
    \end{subfigure}
    \begin{subfigure}[b]{.45\textwidth}
      \centering
       \captionsetup{font=scriptsize}
      \includegraphics[scale=0.5]{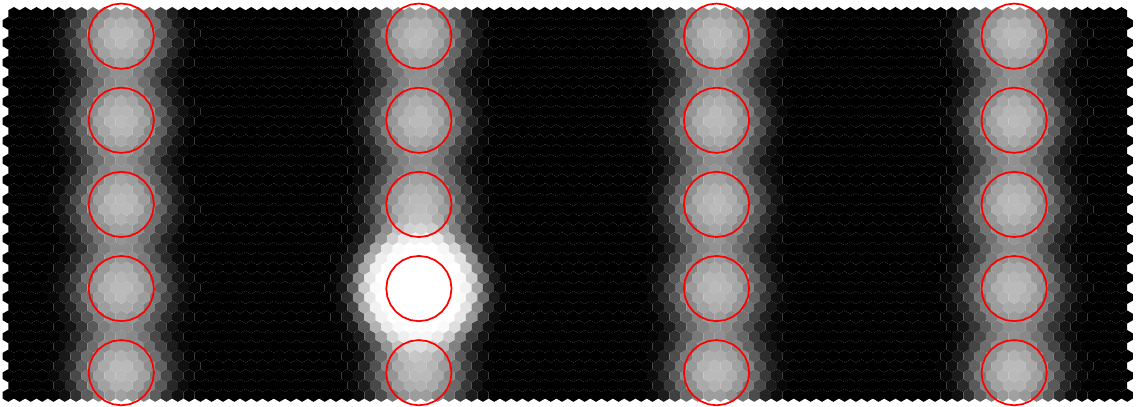}    
      \caption{$\eta = 10$ for mask in row 4, column 2}
      \label{fig:R2}
    \end{subfigure}
    \\
    \centering
    \begin{subfigure}[b]{.45\textwidth}
      \centering
       \captionsetup{font=scriptsize}
      \includegraphics[scale=0.5]{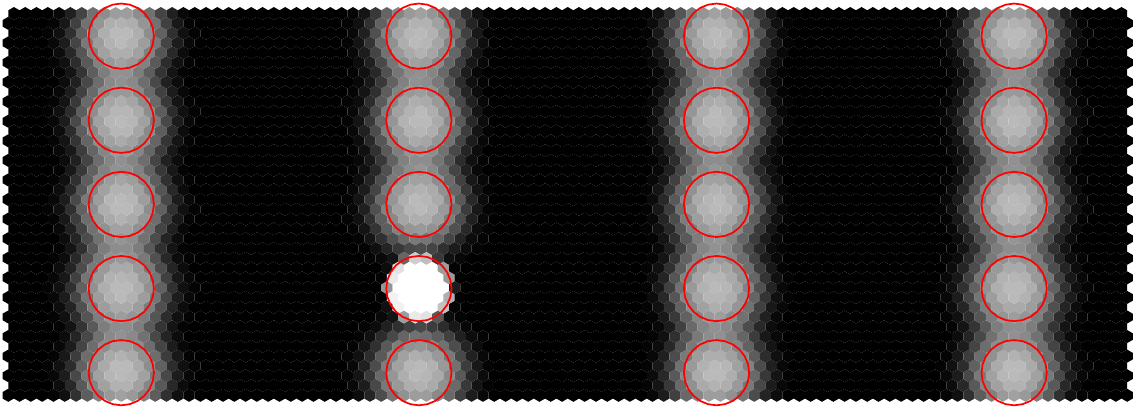}
      \caption{$\alpha = 10$ for mask in row 4, column 2}
      \label{fig:R3}
    \end{subfigure}%
    \begin{subfigure}[b]{.45\textwidth}
      \centering
       \captionsetup{font=scriptsize}
      \includegraphics[scale=0.5]{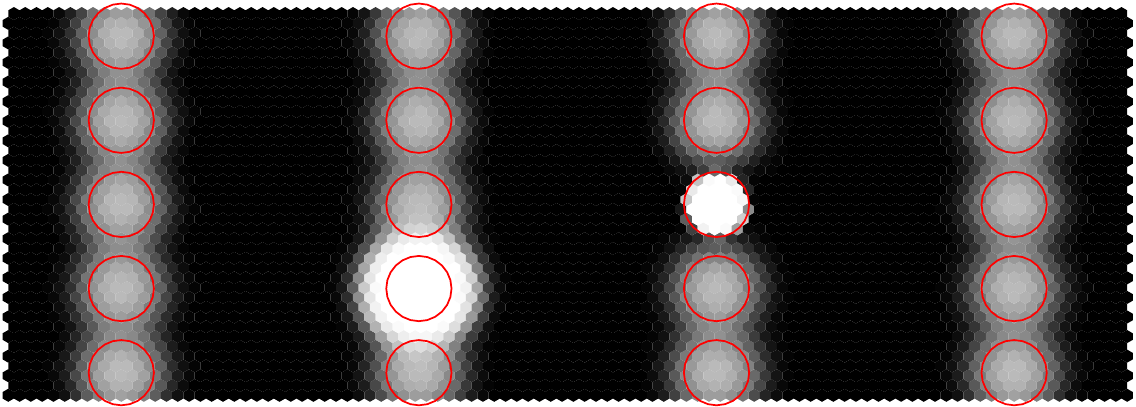}
      \caption{$\alpha = 10$ for mask in row 3, column 3; $\eta = 10$ for mask in row 4, column 2}
      \label{fig:R4}
    \end{subfigure}
    \\
    \centering
    \begin{subfigure}[b]{.45\textwidth}
      \centering
       \captionsetup{font=scriptsize}
      \includegraphics[scale=0.5]{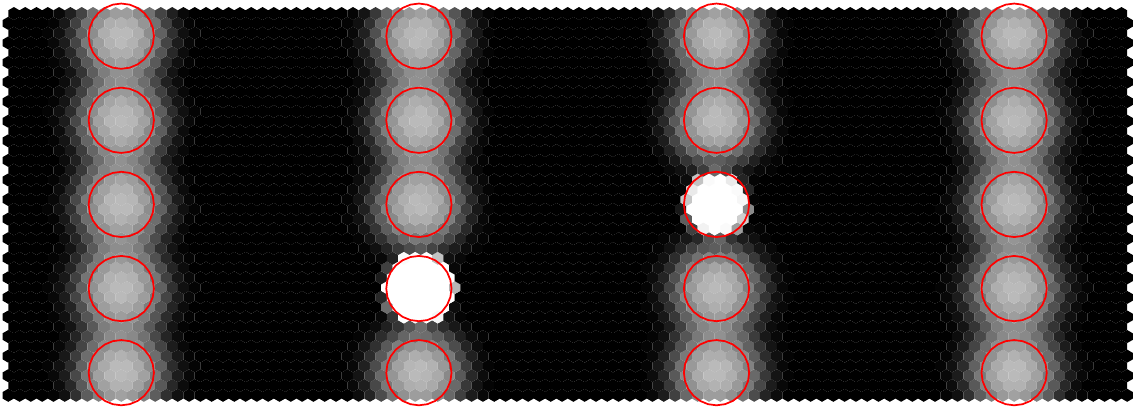}
      \caption{$\alpha = 10, \eta = 10$ for mask in row 4, column 2; \\ $\alpha = 10$ for mask in row 3, column 3}
      \label{fig:R5}
    \end{subfigure}%
   \begin{subfigure}[b]{.45\textwidth}
      \centering
       \captionsetup{font=scriptsize}
      \includegraphics[scale=0.5]{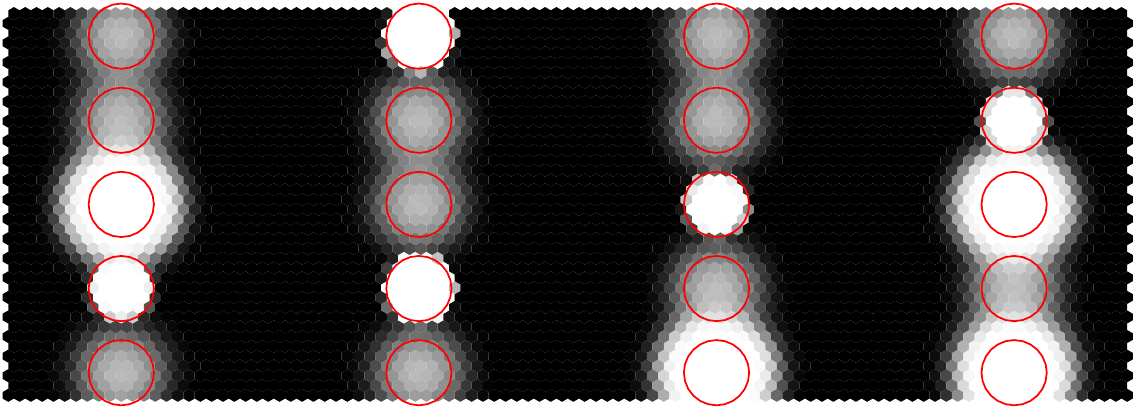}
      \caption{$\alpha$ and $\eta$ made to vary randomly to values of $1$ and $10$}
      \label{fig:R6}
    \end{subfigure}%
    \caption{Effect on cell densities when  mask parameters $\alpha_j$ and $\eta_j$ are varied individually}
     \label{fig:RD11}
    \end{figure}

\noindent One may consider $\alpha_j$ and $\eta_j$ to be parameters specific to the $j^{(\mathrm{th})}$ mask. Effect of variation in $\alpha_j$ and $\eta_j$ is illustrated for negative masks in Fig. \ref{fig:RD11}. With $\eta_j$ increased (e.g., Figs. \ref{fig:R2}, \ref{fig:R4}), local effect is that of {\it density erosion}, similar to that when erosion filter \cite{sigmund_2007_morph_proj2} is used. With $\alpha_j$ increased (e.g., Figs. \ref{fig:R3}, \ref{fig:R4}), cell densities around and outside the respective masks are close to 1. 

Alternatively, $\alpha_j$ and $\eta_j$ may be replaced by two global parameters $\alpha$ and $\eta$. Each separate mask is then $\eta$ identical masks overlaid precisely. With $\rho_{i}(\alpha_j) = \prod_{j = 1}^{M_n} \left[ \frac{1}{1 + \exp(-\alpha_j d_{ij})} \right]$ (Eq. \ref{dens2}) or $\rho_{i}(\alpha_j) = \left[1 -  \prod_{j = 1}^{M_p}  \frac{1}{1 + \exp(-\alpha_j d_{ij})} \right]$ (Eq. \ref{dens3}) as definitions of cell densities, exponent $\eta$ also acts similar to the penalty parameter employed in the SIMP model \cite{ Bendsoe2005 } of topology optimization. Pertaining to Eq. (\ref{dens2}), consider the $j^{(\mathrm{th})}$ mask to be a negative elliptical mask with

\begin{flalign}\nonumber
&& d_{ij} &= \left( \frac{X_{ij}}{a_j}\right)^2 + \left( \frac{Y_{ij}}{b_j}\right)^2 -1, &\\ \nonumber
&\mbox{where} & X_{ij} &= (x_i - x_j)\cos \theta_j + (y_i - y_j) \sin \theta_j, & \\ &\mbox{and} & Y_{ij} &= - (x_i - x_j) \sin \theta_j + (y_i - y_j) \cos \theta_j, &
\label{neg_elliptical_mask}
\end{flalign}

\noindent where $(x_i, y_i)$ are coordinates of centroid of the $i^{(\mathrm{th})}$ hexagonal cell, $(x_j, y_j)$ are center coordinates of the elliptical mask, $a_j$ and $b_j$ are its semi-major and semi-minor axes lengths, and $\theta_j$ is orientation of the mask in relation to the horizontal. If $\{x_j, y_j, a_j, b_j, \theta_j\}$ are modeled as topology design variables with $\psi_j$ representing any one of them generically, sensitivities, as required by a gradient search, can be computed as

\begin{equation}
\label{sens1a} \frac{\partial \rho_{i} (\alpha, \eta) }{\partial \psi_{j}} = \eta \alpha \rho_{i} (\alpha, \eta)  \left[  1 -  \frac{1}{1 + \exp(-\alpha d_{ij})}   \right]  \left[\frac{\partial d_{ij}}{\partial \psi_{j}}  \right], 
\end{equation}

\noindent where 
\begin{flalign}
&&\frac{\partial d_{ij}}{\partial x_{j}} &= - 2 \left( \frac{X}{a_j}\right)\left( \frac{\cos \theta_j}{a_j}\right) + 2 \left( \frac{Y}{b_j}\right)\left( \frac{\sin \theta_j}{b_j}\right), &\nonumber \\
&&\frac{\partial d_{ij}}{\partial y_{j}} &= - 2 \left( \frac{X}{a_j}\right)\left( \frac{\sin \theta_j}{a_j}\right) - 2 \left( \frac{Y}{b_j}\right)\left( \frac{\cos \theta_j}{b_j}\right), &\nonumber \\
&&\frac{\partial d_{ij}}{\partial a_{j}} &= - 2 \frac{X^2}{a_j^3}; \;\;\;\; \frac{\partial d_{ij}}{\partial b_{j}} = - 2 \frac{Y^2}{b_j^3}, & \nonumber \\
\mbox{and} && \frac{\partial d_{ij}}{\partial \theta_{j}} &= 2 \frac{XY}{a_j^2} - 2 \frac{XY}{b_j^2}. &
\label{ell_sens2}
\end{flalign}

\noindent Similar expressions can be obtained for positive elliptical, or circular masks. In case masks are circular, $\theta_j = 0$ in Eq. \ref{ell_sens2}, and $\frac{\partial d_{ij}}{\partial \theta_{j}}$ is not required. Further, as $a_j = b_j$, $\frac{\partial d_{ij}}{\partial a_{j}} = - 2 \left( \frac{X^2 + Y^2}{a_j^3} \right)$.\\

\section{An analytical example}
\label{analytical_example}

\begin{figure}[H]
    \centering
      \includegraphics[scale=1.0]{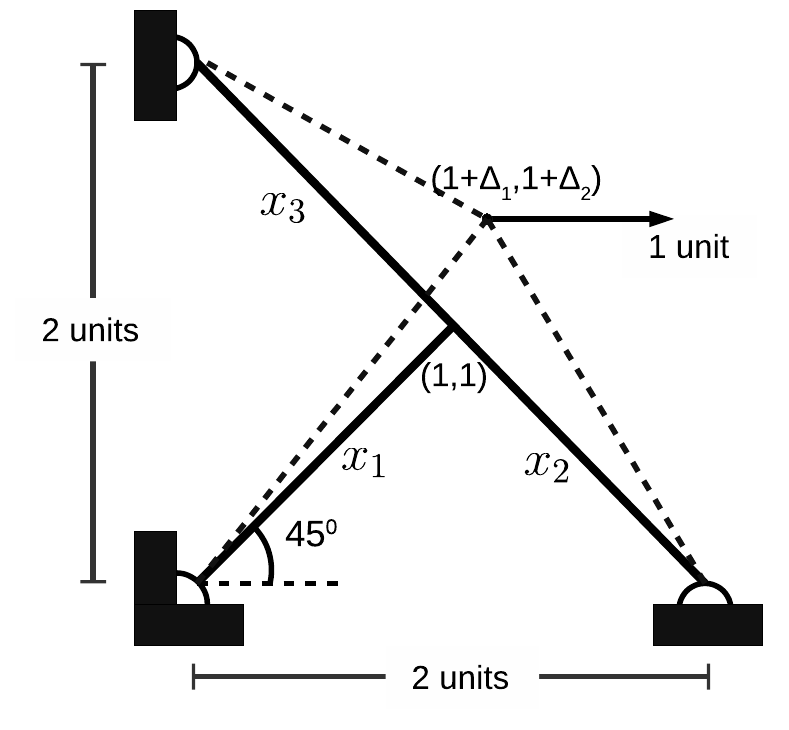}        
\caption{An analytical example with three trusses, all of unit out of plane thickness and elastic modulus}
     \label{Fig:anal_ex}
\end{figure}

\indent As one of the motivations for the methodology in section \ref{method_TO}, it is shown that given a skeleton, upper bound on the volume, $V^{*}$, minimum (and/or maximum) length scale measure(s) and even the associated relaxation parameter(s) may be related in that specifying all of these independently may not always yield a (desirable) solution.  Furthermore, there may exist multiple solutions. Consider Fig. \ref{Fig:anal_ex}  showing an assemblage of three trusses, all of unit elastic modulus, unit out of plane thicknesses and lengths $l_i = \sqrt{2}, i = 1, 2, 3$. Let their in-plane widths be $x_1$, $x_2$ and $x_3$ respectively. Let $V^{*}$ be the upper bound on the summation of $x_i$ and $x_m$ be the minimum length scale imposed on them. For a unit force applied as shown, expression for the strain energy can be obtained, using finite element analysis, or otherwise\footnote{the problem being statically indeterminate, one could solve by assuming horizontal ($\Delta_1$) and vertical ($\Delta_2$) displacements at node $(1, 1)$, compute strains (linearized) and stresses in the three members, strain energy, and then compute $\Delta_1$ and $\Delta_2$ by minimizing the total potential.}, as $SE = C \left( \frac{x_1 + x_2 + x_3}{x_1 x_2 + x_1 x_3} \right)$ where $C = \frac{1}{2\sqrt{2}}$.

\subsection{$x_1, x_2, x_3$ retained in the skeleton}
\label{formulation_analytical_all_members_retained}
\indent We solve the following optimization problem using the KKT (Karush Kuhn Tucker) stationarity conditions. 

\begin{eqnarray}
\label{formulation_analytical_all_retained}
\begin{split}
\underset{x_1, x_2, x_3}{\mbox{minimize: }  }  &   SE  = C \left( \frac{x_1 + x_2 + x_3}{x_1 x_2 + x_1 x_3} \right) \\
\mbox{subject to:} &  \\
\mbox{(i)} \;\;\;\; &  g_1 \equiv x_1 + x_2 + x_3 - V^{*} \leq 0 \\
\mbox{(ii)} \;\;\;\; & g_2 \equiv (x_m - x_1)^p +  (x_m - x_2)^p + (x_m - x_3)^p - \varepsilon_1 \leq 0 \\
\end{split} 
\end{eqnarray}

where $p$ is a natural number. Nature of constraint $g_2$ depends on $p$. For an odd $p, g_2$ acts as a minimum length scale constraint while an even $p$ leads to a fixed length scale constraint in this example. In case of the topology optimization formulation in Eq. \ref{formulation},  $g_{min}(\bm{\rho}) - \varepsilon_1 \leq 0$ and $g_{max}(\bm{\rho}) - \varepsilon_2 \leq 0$ are more strict compared to $g_2$ in Eq. (\ref{formulation_analytical_all_retained}). $g_2$ could become negative, even for the relaxation parameter as $0$, with odd $p$ and some (or all) $x_i > x_m$, $i = 1, 2, 3$. 

With the Lagrangian as $\phi = C \left( \frac{x_1 + x_2 + x_3}{x_1 x_2 + x_1 x_3} \right) + \lambda_1 g_1 + \lambda_2 g_2 $,  where $\lambda_1$ and $\lambda_2$ are Lagrange multipliers, stationarity conditions for $p=1 \mbox{ and } p=2$ are discussed and solved below. For $p=1$ we get
\begin{eqnarray}
\label{anal_stationarity_conditions_1}
&&\frac{\partial \phi }{\partial x_1} = - \frac{C}{x_{1}^{2}} + \lambda_1 - \lambda_2 = 0 \nonumber \\
&&\frac{\partial \phi }{\partial x_2} = - \frac{C}{(x_{2} + x_{3}) ^{2}} + \lambda_1 - \lambda_2 = 0 \nonumber \\
&&\frac{\partial \phi }{\partial x_3} = - \frac{C}{(x_{2} + x_{3}) ^{2}} + \lambda_1 - \lambda_2 = 0 \nonumber \\
&&\lambda_1 g_1 = 0;  \;\;\;\;\;\; \lambda_2 g_2 = 0. 
\end{eqnarray}

\noindent The conditions obtained from $\frac{\partial \phi }{\partial x_2}=0$ and $\frac{\partial \phi }{\partial x_3} = 0$ are identical. We analyse the following cases:

Case I: $\lambda_1 \neq 0, \lambda_2 = 0$: \\
From Eqs. \ref{anal_stationarity_conditions_1}, we have,

\begin{flalign} \nonumber
&& x_1 &= \sqrt{\frac{C}{\lambda_{1}}}~;\;\;\; x_2 + x_3 = \sqrt{\frac{C}{\lambda_{1}}};&\\ \nonumber
&& g_1 &= 0 \implies 2 \sqrt{\frac{C}{\lambda_1}} = V^{*}, \mbox{ or}, \frac{C}{\lambda_1} = \frac{ { V^{*}} ^{2} }{4}&\\ \mbox{so that}&& x_{1} &= \frac{V^{*}}{2} \mbox{ and } x_2 + x_3 = \frac{V^{*}}{2}.& \nonumber
\end{flalign}
For a feasible solution, $g_2 = 3x_m - x_1 - x_2 - x_3 \leq \varepsilon_{1}$ must be true. One can conclude that
\begin{eqnarray}
\label{case1_anal_1}
3x_m \leq V^* + \varepsilon_{1}.
\end{eqnarray}
The above suggests a rather intutive relation between $V^*,x_m~ \mbox{and}~ \varepsilon_{1}$. Here, $\varepsilon_{1}$ acts as a relaxation parameter between $V^*$ and $x_m$. A solution for this case is infeasible if inequality (\ref{case1_anal_1}) is violated.

Case II: $\lambda_1 = 0, \lambda_2 \neq 0$: \\
Eqs. \ref{anal_stationarity_conditions_1}, yield
\begin{eqnarray}
\label{case2_anal_1}
\lambda_2 = \frac{-C}{x_1^2}= \frac{-C}{(x_2 + x_3)^2}.
\end{eqnarray}
As $\lambda_2$ is negative, this case does not provide a solution. 

Case III: $\lambda_1 \neq 0, \lambda_2 \neq 0$: \\
From Eqs. \ref{anal_stationarity_conditions_1}, we have,
\begin{eqnarray}\nonumber
&&\lambda_1 - \lambda_2 = \frac{C}{x_1^2};\;\;\; \lambda_1 - \lambda_2 = \frac{C}{(x_2 + x_3)^2};\\ \nonumber
&&g_1 = 0 \implies x_1 + x_2 + x_3 = V^* ;\\&& g_2 = 0 \implies 3x_m - x_1 - x_2 - x_3 = \varepsilon_{1}. \nonumber
\end{eqnarray}
From the above one concludes that
\begin{flalign}\nonumber
\label{case3_anal_1}
&&x_1 &= \frac{V^{*}}{2} = \frac{3x_m - \varepsilon_{1}}{2} = x_2 + x_3&\\ \mbox{and}&& V^*&= 3x_m - \varepsilon_{1}.&
\end{flalign}
$\varepsilon_{1}$ works as relaxation parameter for the length scales as can be seen from the relation between $x_1, x_m \mbox{ and } \varepsilon_{1}$. Eqs. \ref{case1_anal_1} and \ref{case3_anal_1} suggest that for $p=1$, $V^{*}, x_m \mbox{ and } \varepsilon_{1}$ are interdependent and hence one may not achieve a solution for an independent choice of these parameters, specifically when $V^{*}$ is chosen less than $3x_m - \varepsilon_1$. A similar analysis, for $p=2$, is given below. Stationary conditions are

\begin{eqnarray}
\label{anal_stationarity_conditions}
&&\frac{\partial \phi }{\partial x_1} = - \frac{C}{x_{1}^{2}} + \lambda_1 + 2 \lambda_2 (x_1 - x_m) = 0 \nonumber \\
&&\frac{\partial \phi }{\partial x_2} = - \frac{C}{(x_{2} + x_{3}) ^{2}} + \lambda_1 + 2 \lambda_2 (x_2 - x_m) = 0 \nonumber \\
&&\frac{\partial \phi }{\partial x_3} = - \frac{C}{(x_{2} + x_{3}) ^{2}} + \lambda_1 + 2 \lambda_2 (x_3 - x_m) = 0 \nonumber \\
&&\lambda_1 g_1 = 0;  \;\;\;\;\;\; \lambda_2 g_2 = 0. 
\end{eqnarray}
  
\noindent From $\frac{\partial \phi }{\partial x_2} = \frac{\partial \phi }{\partial x_3} = 0$, one realizes that $\lambda_2(x_2- x_3)=0$. Further, $\lambda_1$ and $\lambda_2$ cannot both be $0$ since $C \neq 0$. We consider the following cases:

Case I: $\lambda_1 \neq 0, \lambda_2 = 0$: \\
$\lambda_2(x_2-x_3)=0$ is satisfied. From Eqs. \ref{anal_stationarity_conditions}, we have,
\begin{flalign}
\label{case1_anal}
&&x_1 &= \sqrt{\frac{C}{\lambda_1}}; \;\;\;\;\;\; x_2 + x_3 = \sqrt{\frac{C}{\lambda_1}};& \nonumber \\
&&g_1 &= 0 \implies 2 \sqrt{\frac{C}{\lambda_1}} = V^{*}, \mbox{ or}, \frac{C}{\lambda_1} = \frac{ { V^{*}} ^{2} }{4}& \nonumber \\ \mbox{so that} && x_{1} &= \frac{V^{*}}{2} \mbox{ and } x_2 + x_3 = \frac{V^{*}}{2};& \nonumber \\&&
\lambda_{1} &= \frac{4C}{ { V^{*}} ^{2}  } > 0.& \nonumber 
\end{flalign}

So that the solution is feasible, 
$g_2 = \left(x_m - \frac{V^{*}}{2}\right)^2 + \left(x_m - x_2\right)^2 + \left(x_m -\frac{V^{*}}{2} + x_2\right)^2 -\varepsilon_{1} \leq 0$ must hold, or,

\begin{flalign}\nonumber
&& &\frac{V^*}{4} - \frac{\sqrt{D}}{4} \leq x_2 \leq \frac{V^*}{4} + \frac{\sqrt{D}}{4} &\\ \mbox{where} && &D = -3V^{*2} + 16x_m V^* -24x_m^2 +8 \varepsilon_{1}. 
\end{flalign}

\noindent For realistic bounds on $x_2$, $D$ must be $\geq 0$. Thus, 
\begin{flalign}
\frac{8x_m}{3} - \frac{2}{3}\sqrt{6\varepsilon_{1} - 2x_m^2} \leq V^* \leq \frac{8x_m}{3} + \frac{2}{3}\sqrt{6\varepsilon_{1} - 2x_m^2}.
\end{flalign}

The above suggests, rather intricate, dependence between $x_m$, $V^{*}$ and $\varepsilon_1$. Given $x_m$, $\varepsilon_1$ depends on it in that  $\varepsilon_1 > \frac{{x_m}^2}{3}$ must hold if $V^{*}$ is to have realistic bounds. Specifically, if $\varepsilon_1 = 0$, no solution exists for this case. Otherwise, $V^{*}$ must be such that it is bounded from both sides by limits depending on $x_m$ and $\varepsilon_1$. \\

Case II: $\lambda_1 = 0, \lambda_2 \neq 0$: \\
As $\lambda_2 \neq 0$, $x_2 = x_3$ must hold. Eqs. \ref{anal_stationarity_conditions}, yield
\begin{flalign}
&& x_1& - x_m = \frac{C}{2 \lambda_2 x_1^2}; \;\;\;\; x_2 - x_m = \frac{C}{8 \lambda_2 x_2^2}; \;\;\;\; x_2 = x_3&  \nonumber \\ \mbox{so that}
&& x_1 &= x_m + \frac{C}{2 \lambda_2 x_1^2} = x_m + \delta_1; \;\;\;\; x_2 = x_3 = x_m + \frac{C}{8 \lambda_2 x_2^2} = x_m + \delta_2; & \nonumber \\ 
&& g_2 &= 0 \implies \frac{C^2}{4 \lambda_2^2 x_1^4} + \frac{C^2}{32 \lambda_2^2 x_2^4} - \varepsilon_1 = 0& \nonumber\\&& \implies \lambda_2 &= \pm \frac{C}{2\sqrt{\varepsilon_1}}\sqrt{ \left( \frac{1}{x_1^4} + \frac{1}{8x_2^4} \right) }.&
\end{flalign}

\noindent Here, $x_1, x_2$ and $x_3$ are all larger than $x_m$ as $\delta_1 > 0 \mbox{ and }  \delta_2 > 0$ for positive $\lambda_2$. One notes that $\varepsilon_1$ must be strictly positive. Thereafter, a free choice of $\varepsilon_1$, howsoever small, can control the magnitude of $\lambda_2$ and thus those of $\delta_1 \mbox{ and }  \delta_2$. Further, that $g_1 < 0$ must be satisfied, $V^{*} > 3x_m + \delta_1 + 2\delta_2$ must hold, suggesting again, dependence between $V^{*}$ and $x_m$. \\

Case III: $\lambda_1 \neq 0, \lambda_2 \neq 0$: \\
Again, $x_2 = x_3$ must hold and further, $g_1 = g_2 = 0$ implies $x_1 + 2x_2 = V^{*} \mbox{ and } (x_1 - x_m)^2  + 2(x_2 - x_m)^2 - \varepsilon_1 = 0$ solving which yields
\begin{flalign}
&&  &6x_2^2 - 4V^{*}x_2 + (3x_m^2 - 2V^{*}x_m + {V^{*}}^{2} - \varepsilon_1) = 0 & \nonumber \\ \mbox{so that}
&& &x_2 = x_3 = \frac{V^{*}}{3} \pm \frac{1}{12}\sqrt{ 16 {V^{*}}^{2} - 24(3 x_m^2 - 2 V^{*} x_m + {V^{*}}^{2} - \varepsilon_1 ) }  = \frac{V^{*}}{3} +  \delta_3. &
\end{flalign}

Discreminant in the above relation must be non-negative which yields
\begin{eqnarray}
\label{case3_anal2}
3x_m - \sqrt{3 \varepsilon_1} \leq V^{*} \leq 3x_m + \sqrt{3 \varepsilon_1}.
\end{eqnarray}

One can solve for $\lambda_1 $ and $\lambda_2$ in Eqs. \ref{anal_stationarity_conditions} to get
\begin{flalign}\nonumber
\label{lambdas_caseIII}
&&\lambda_1 &= \frac{C}{4 (\frac{V^{*}}{3} + \delta_3)^2} - 2 \lambda_2 (\frac{V^{*}}{3} + \delta_3 - x_m) &\\ \mbox{and}
&&\lambda_2 &= - 9 C \frac{V^{*}(V^{*} + 12 \delta_3) } { 8 \delta_3(V^{*} + 3 \delta_3)^2 (V^{*} - 6 \delta_3)^2  }.& 
\end{flalign}

$\delta_3$, which could either be positive or negative, depends on $V^{*}, x_m$ and $\varepsilon_1$. Proper choices, though not independent of each other, of the latter three may yield both Lagrange multipliers positive. Eq. \ref{case3_anal2} however suggests that $x_m$ and $V^{*}$ are related, given $\varepsilon_1$. 

Intuitively, one observes through Eqs. \ref{formulation_analytical_all_retained} that  for $x_1, x_2$ and $x_3$ to concur to the minimum length scale, $V^{*}$ must be larger than $3x_m$. Further,  as $SE = C/(x_2 + x_3) + C/x_1$, the three in-plane widths may be as large as possible until a maximum length scale or a resource constraint is imposed, the latter yielding an upper bound on $V^{*}$. Imposing both, the maximum length scale and resource constraints makes one of the two redundant.

\subsection{$x_2 = 0$, $x_1, x_3$ retained in the skeleton}
\label{formulation_analytical_two_members_retained}

\indent One notices that $x_1$ cannot be zero as then the inclined members $x_2$ and $x_3$ cannot take the transverse load. One now considers $x_2 = 0$ so that the skeleton is composed of $x_1$ and $x_3$. Expression for the strain energy is $SE = C/x_1 + C/x_3$. Per the Karush Kuhn Tucker conditions, 
\begin{eqnarray}
\label{anal_stationarity_conditions2}
&&\frac{\partial \phi }{\partial x_1} = - \frac{C}{x_{1}^{2}} + \lambda_1 + 2 \lambda_2 (x_1 - x_m) = 0 \nonumber \\
&&\frac{\partial \phi }{\partial x_3} = - \frac{C}{x_{3} ^{2}} + \lambda_1 + 2 \lambda_2 (x_3 - x_m) = 0 \nonumber \\
&&\lambda_1 g_1 = 0;  \;\;\; \lambda_2 g_2 = 0. 
\end{eqnarray}

\indent $\lambda_1$ and $\lambda_2$ cannot both be zero as $C$ is non-zero. Considering $\lambda_1 \neq 0$ and $\lambda_2 = 0$, from Eqs. \ref{anal_stationarity_conditions2},
\begin{flalign}\nonumber
\label{case1_anal1}
&&x_1 &= x_3 = \sqrt{\frac{C}{\lambda_1}}; &\\
&&g_1 &= 0 \implies 2 \sqrt{\frac{C}{\lambda_1}} = V^{*}, \mbox{ or}, \frac{C}{\lambda_1} = \frac{ { V^{*}} ^{2} }{4} \nonumber& \\ \mbox{so that} && x_{1} &= x_3 = \frac{V^{*}}{2} \mbox{ and } \lambda_{1} = \frac{4C}{ { V^{*}} ^{2}  } > 0.&  
\end{flalign}

For feasible solution, $g_2 = 2 \left( \frac{ V^{*}}{2} - x_m \right)^2  - \varepsilon_1 \leq 0$ must hold, or,
\begin{eqnarray}
2\left( x_m - \sqrt{\frac{\varepsilon_1}{2}} \right) \leq V^{*} \leq 2\left(x_m + \sqrt{\frac{\varepsilon_1}{2}} \right),
\end{eqnarray}

\noindent suggesting dependence of $V^{*}$ on $x_m$ and $\varepsilon_1$ which, must be non-negative. If $\lambda_1 = 0$ and $\lambda_2 \neq 0$, from Eqs. \ref{anal_stationarity_conditions2},

\begin{flalign}
&&\frac{\partial \phi }{\partial x_1} &= - \frac{C}{x_{1}^{2}} +  2 \lambda_2 (x_1 - x_m) = 0 \implies x_1 - x_m = \frac{C}{2 \lambda_2 x_1^2}  = \delta_1 &\nonumber \\
&&\frac{\partial \phi }{\partial x_3} &= - \frac{C}{x_{3} ^{2}} + 2 \lambda_2 (x_3 - x_m) = 0  \implies x_3 - x_m = \frac{C}{2 \lambda_2 x_3^2}  = \delta_2 & \nonumber \\
&&g_2 &= 0 \implies \delta_1^2 + \delta_2^2 = \varepsilon_1, \mbox{ or }, \varepsilon_1 = \frac{C^2}{4 x_1^4 \lambda_2^2} +  \frac{C^2}{4 x_3^4 \lambda_2^2} & \nonumber \\
\mbox{or,} &&\lambda_2 &= \pm \frac{C}{2\sqrt{\varepsilon_1}}\sqrt{ \left( \frac{1}{x_1^4} + \frac{1}{x_3^4} \right)  }& 
\end{flalign}

\noindent and hence $\varepsilon_1$ must be strictly positive. Further, $g_1 \leq 0 $ implies $2x_m + \delta_1 + \delta_2 \leq V^{*}$ suggesting again, an interdependence between $V^{*}, x_m$ and $\varepsilon_1$.  If $\lambda_1 \neq 0$ and $\lambda_2 \neq 0$, from Eqs. \ref{anal_stationarity_conditions2}, $g_1$ and $g_2$ must both be zero. Thus,

\begin{flalign}
\label{case3_anal1}
&&x_1 &= V^{*} - x_3; &\nonumber \\
&&(x_1& - x_m)^2 +  (x_3 - x_m)^2 =  2 x_3^2  - 2 V^{*} x_3 +  (V^{*})^2 + 2 x_m^2 - 2 V^{*} x_m -  \varepsilon_1 = 0 &\nonumber \\
\mbox{or}, &&x_3 &= \frac{V^{*}}{2} \pm \frac{\sqrt{ -(V^{*} - 2x_m)^2 + 2\varepsilon_1 }}{2}. &
\end{flalign}

\noindent So that $x_3$ has a solution, $(V^{*} - 2x_m)^2 - 2\varepsilon_1 \leq 0$ must hold, or, $2 x_m - \sqrt{ 2\varepsilon_1} \leq V^{*} \leq 2 x_m + \sqrt{ 2\varepsilon_1}$, suggesting that $V^{*}$ is bounded by limits that depend on $x_m$ and $\varepsilon_1 $. Note that, $\varepsilon_1 \geq 0$ must hold. For $\varepsilon_1 = 0$, $V^{*}$ must precisely be $2x_m$. The case wherein $x_3 = 0$, $x_1, x_2$ are retained in the skeleton, is identical.

\section{Sequence of Length Scales (SLS) Methodology}
\label{method_TO}

\noindent It may be possible to comprehend the interplay between the resource, length scale constraints and the corresponding tolerance for simpler examples (sections \ref{formulation_analytical_all_members_retained} and \ref{formulation_analytical_two_members_retained}) but not for more involved problems, as topology/skeleton evolves continuously as optimization progresses. When seeking optimal continuum topologies, a designer may not always have an intuitive notion on upper bound on the continuum volume though the intent would be to keep it as low as possible. One may, however, prefer to specify the minimum and maximum length scales, from failure and/or manufacturing viewpoints, more readily.  The analytical example above, suggests strong corelation between upper bound on the volume, length scales, tolerance specified on the corresponding constraints, and also that there could exist multiple solutions, or possibly none.

\begin{figure}[H]
    \centering
      \includegraphics[scale=0.4]{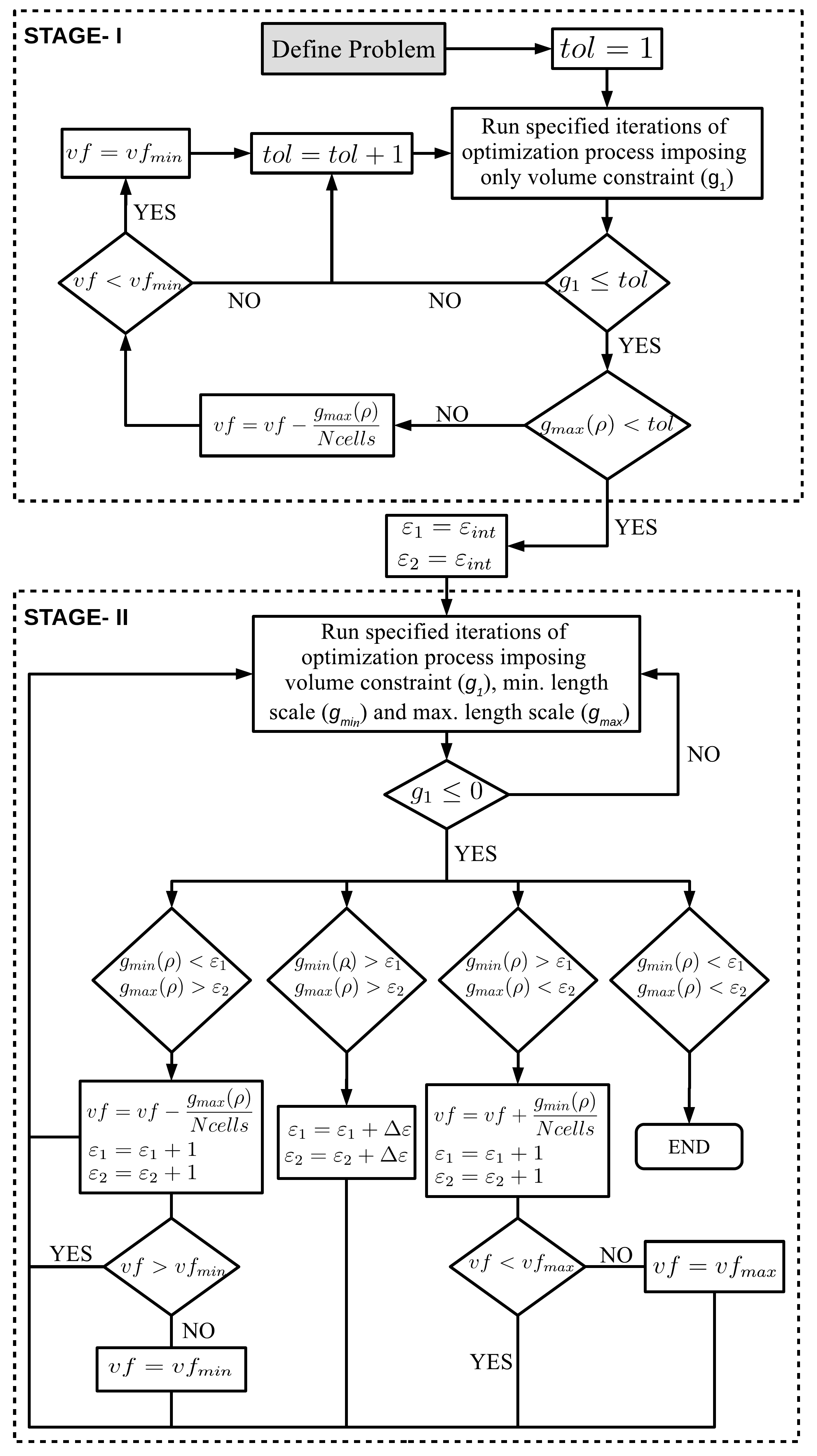}        
\caption{Flow chart for the methodology in Section \ref{method_TO}. $vf_{min}$ and $vf_{max}$ are specified limits on  upper bound of the volume fraction (chosen as $0.1$ and $0.5$ in this paper). $\Delta \varepsilon$ is the specified increment on relaxation parameters (chosen as $10$ for Figures \ref{fig:Eg1_Eg2_Eg3_Eg4_new_logic_NEM}, \ref{fig:Eg1_Eg2_Eg3_Eg4_new_logic_PEM}, and $1$ for Figures \ref{fig:Eg1_Eg2_Eg3_Eg4_post_review_NEM}). $\varepsilon_{int}$ is chosen as $tol$ for the examples. }
     \label{Fig: methodology_flow}
\end{figure}

\noindent The methodology proposed (for $p = 1$ in Eq. \ref{min_max_ls}), rather heuristic,  which uses minimum and maximum length scale measures as design parameters is delineated.  Given the design specifications, initial guess on elliptical masks, a low volume fraction (say $vf = 0.2$), and low relaxation tolerances (say $tol = 1$) in Stage I, one first seeks a topology (skeleton) that optimizes the objective in Eq. \ref{formulation} with only the volume constraint imposed. One checks whether the solution satisfies the maximum length scale constraint. If the latter is violated, the volume fraction $vf$ is reduced as $vf = vf - \frac{g_{max}(\bm{\rho})}{Ncells}$ (step 3 below) ensuring $vf$ is always larger than its lower limit, $vf _{min}$. The intent  in Stage I is to attain a solution with $vf$ low enough so that the maximum length scale is satisfied. We assume such a solution exists, as demonstrated by Rehmatallah and Swan \cite{Rehmatallah_Swan_2005}. We also expect a topology to be available whose skeleton does not change (significantly, though we allow for it) so that the length scale measures $g_{min}(\bm{\rho})$ and $g_{max}(\bm{\rho})$ in Eq. \ref{min_max_ls} are effective (see \cite{allaire:hal-00985000}). With Stage I solution as the initial guess, which one reckons is better than that wherein elliptical masks of uniform sizes are evenly placed since the skeleton is relatively well formed,  one now imposes all constraints, those on volume, minimum and maximum length scales and seeks the optimal solution in Stage II.  Following possibilities exist for an intermediate solution in Stage II. 

\begin{enumerate}
\item If the volume constraint is satisfied,
\begin{enumerate}
\item If $g_{min}(\bm{\rho}) < \varepsilon_1$ and $g_{max}(\bm{\rho}) < \varepsilon_2$, the optimization process is ceased and the Stage II solution is accepted.
\item If $g_{min}(\bm{\rho}) > \varepsilon_1$ and $g_{max}(\bm{\rho}) < \varepsilon_2$, it is reckoned that the volume fraction is not aqequate. Accordingly, $vf$ is readjusted as $vf = vf + \frac{g_{min}(\bm{\rho})}{Ncells}$. Both, $\varepsilon_1$ and  $\varepsilon_2$ are incremented marginally ($\varepsilon_i = \varepsilon_i + 1, i = 1, 2$) and an optimal solution is sought again.
\item If $g_{min}(\bm{\rho}) < \varepsilon_1$ and $g_{max}(\bm{\rho}) > \varepsilon_2$, the volume fraction can be lowered, as, further reduction of $g_{max}(\bm{\rho}) $ will only reduce the overall continuum volume. Accordingly, $vf$ is readjusted as $vf = vf - \frac{g_{max}(\bm{\rho})}{Ncells}$, $\varepsilon_1$ and  $\varepsilon_2$ are incremented ($\varepsilon_i = \varepsilon_i + 1, i = 1, 2$), and the optimization process is commenced again.
\item If $g_{min}(\bm{\rho}) > \varepsilon_1$ and $g_{max}(\bm{\rho}) > \varepsilon_2$, one chooses to keep $vf$ unaltered, and rather, increments both  $\varepsilon_1$ and  $\varepsilon_2$ ($\varepsilon_i = \varepsilon_i + \Delta \varepsilon, i = 1, 2$), envisaging that any of the above three cases will be met within subsequent stage(s) in optimization.  
\end{enumerate}
\item If the volume constraint is not satisfied in stage II, $vf$ is increased marginally and an optimal solution is sought again. 
\end{enumerate}

The overall notion is that in stage I, an optimal topology with well formed skeleton is available satisfying the maximum length scale constraint whereas in Stage II, length scale measures in Eq. \ref{min_max_ls} are used more effectively while also addressing the underlying yet non-apparant interdependence between the design parameters $vf, min_{ls}, max_{ls}$ and relaxation parameters.  In case positive elliptical masks are employed, after each optimization step, one checks for existence of connectivity singularities, i.e., dangling appendages and/or islands. Cells with negligible strain energy densities, and masks enclosing them are identified, and such masks are removed.  A flow chart is depicted in Figure \ref{Fig: methodology_flow}. One must note that $g_{min}(\bm{\rho})$ and $g_{max}(\bm{\rho})$ are global length scale measures in that while it may be possible for minimum and maximum length scale constraints in Eqs. \ref{formulation} to be satisfied with reference to the obtained relaxation parameters after stage II, locally, these constraints may still get violated, as observed in some solutions (section \ref{examples}).

\begin{figure}[H]
	\begin{subfigure}[b]{.3\textwidth}
		\centering
		\captionsetup{font=scriptsize}
\includegraphics[trim={3.5cm 2.5cm 2cm 2cm}, clip, scale = 0.5]{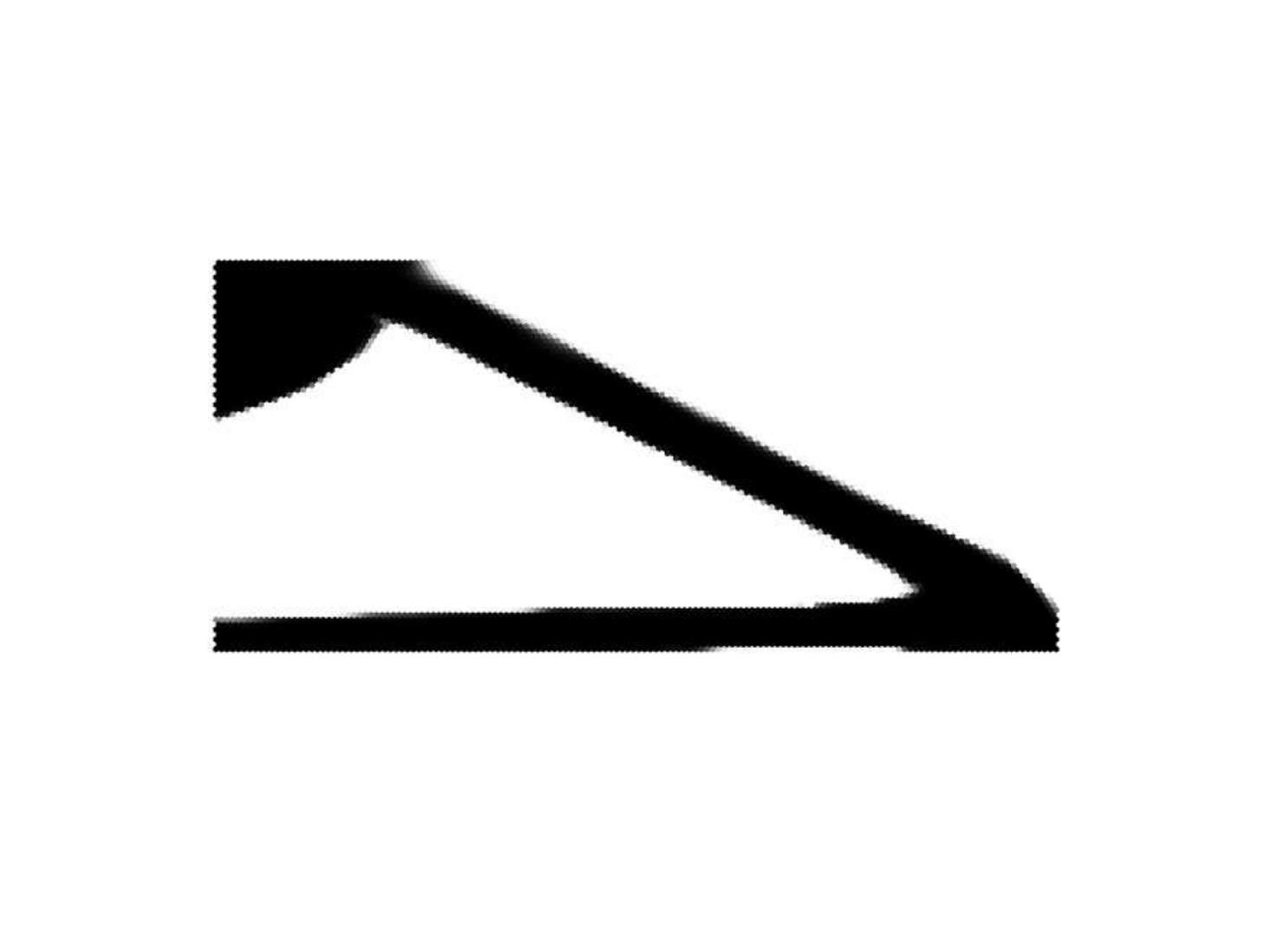}   
		\caption{$\alpha = 4$, $V_{f} =0.3$, $BWI = 0.05$}
		\label{fig:Eg1_test_minimum_ls_a}
	\end{subfigure}~~~~
	\begin{subfigure}[b]{.3\textwidth}
		\centering
		\captionsetup{font=scriptsize}
\includegraphics[trim={3.5cm 2.5cm 2cm 2cm}, clip, scale = 0.5]{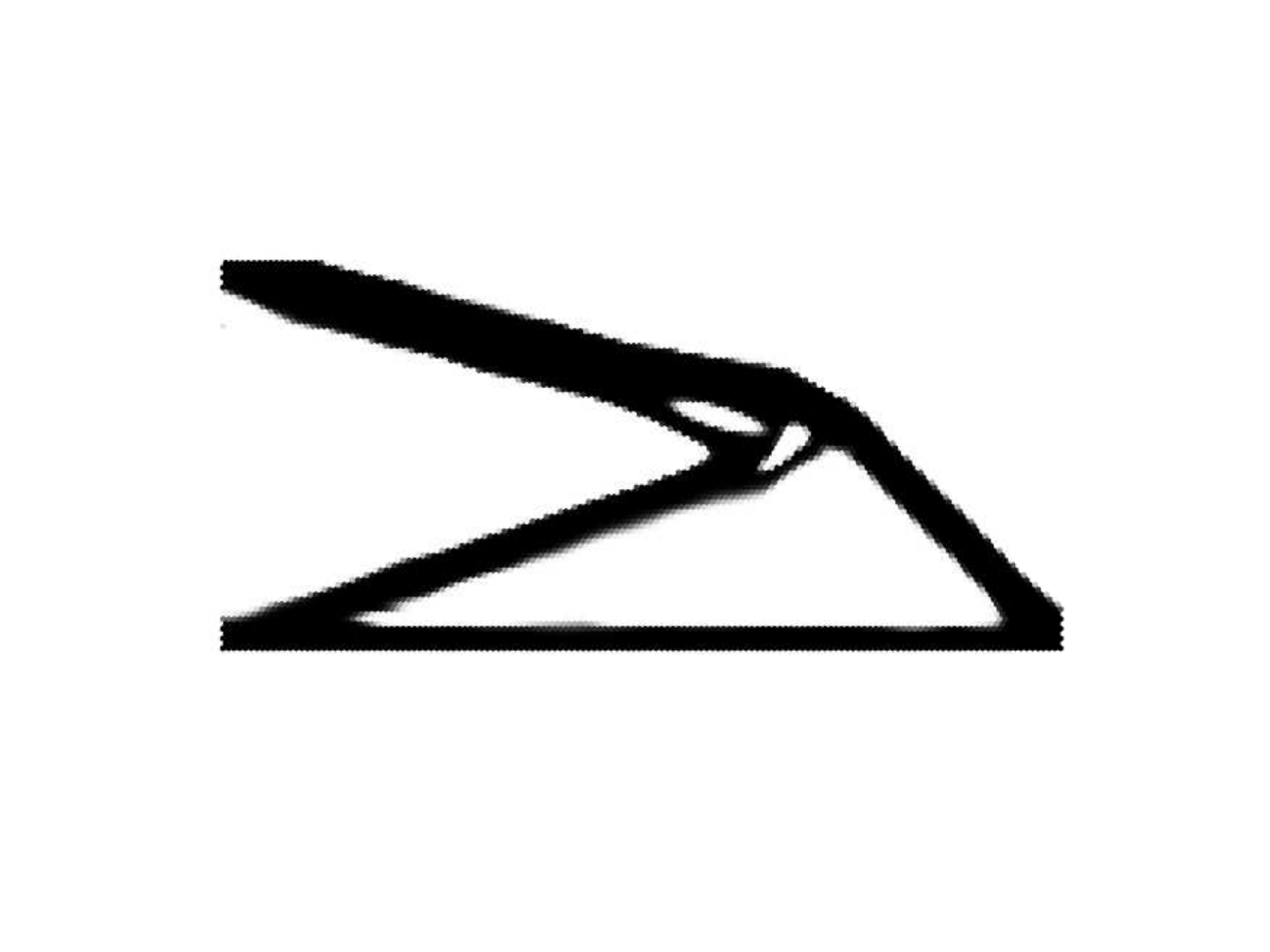}   
		\caption{$\alpha = 6$, $V_{f} =0.3$, $BWI = 0.04$}
		\label{fig:Eg1_test_minimum_ls_b}
	\end{subfigure}~~~~
	\begin{subfigure}[b]{.3\textwidth}
		\centering
		\captionsetup{font=scriptsize}
\includegraphics[trim={3.5cm 2.5cm 2cm 2cm}, clip, scale = 0.5]{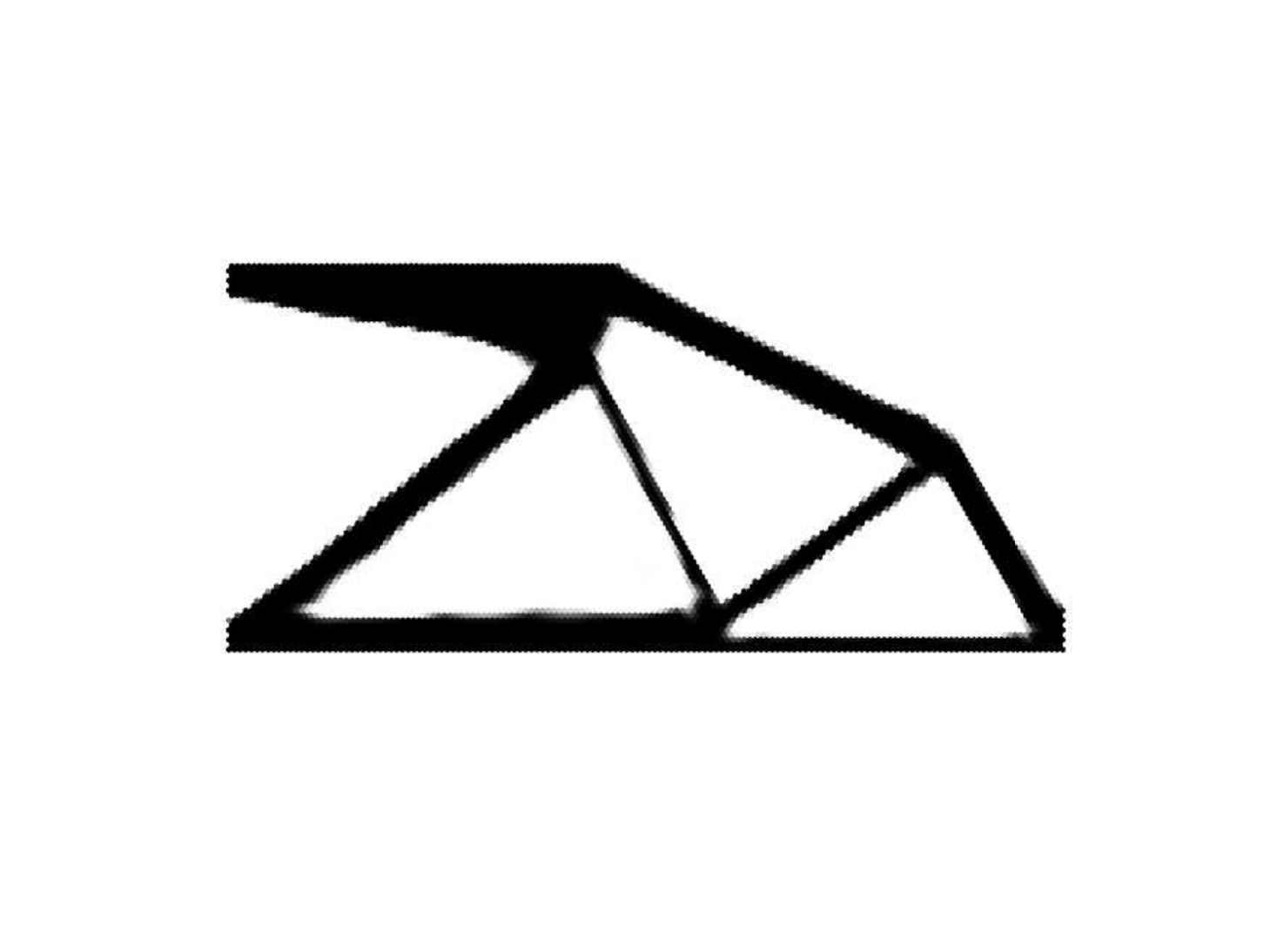}  
		\caption{$\alpha = 8$, $V_{f} = 0.3$, $BWI = 0.04$}
		\label{fig:Eg1_test_minimum_ls_c}
	\end{subfigure}
    \begin{subfigure}[b]{.3\textwidth}
      \centering
       \captionsetup{font=scriptsize}
      \includegraphics[trim={3.5cm 2.5cm 2cm 2cm}, clip, scale = 0.5]{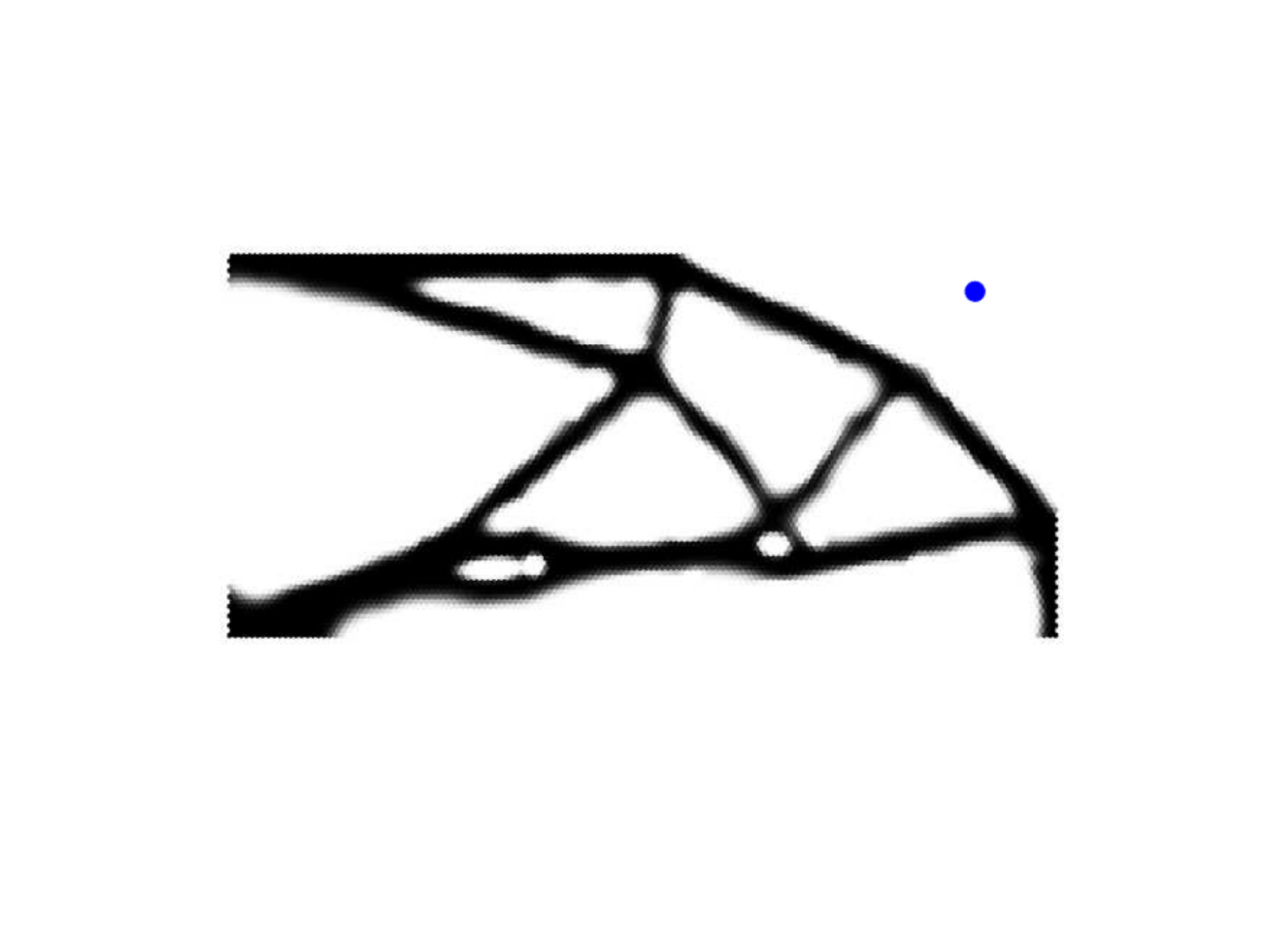}   
      \caption{$\alpha = 4$, $V_{f} =0.3$, $BWI = 0.11$ \\ $g_1 = -24.4$, $g_{min}(\bm{\rho})  = 33.3$, $min_{ls} = 1.52$}
      \label{fig:Eg1_test_minimum_ls_d}
    \end{subfigure}~~~~
    \begin{subfigure}[b]{.3\textwidth}
      \centering
       \captionsetup{font=scriptsize}
      \includegraphics[trim={3.5cm 2.5cm 2cm 2cm}, clip, scale = 0.5]{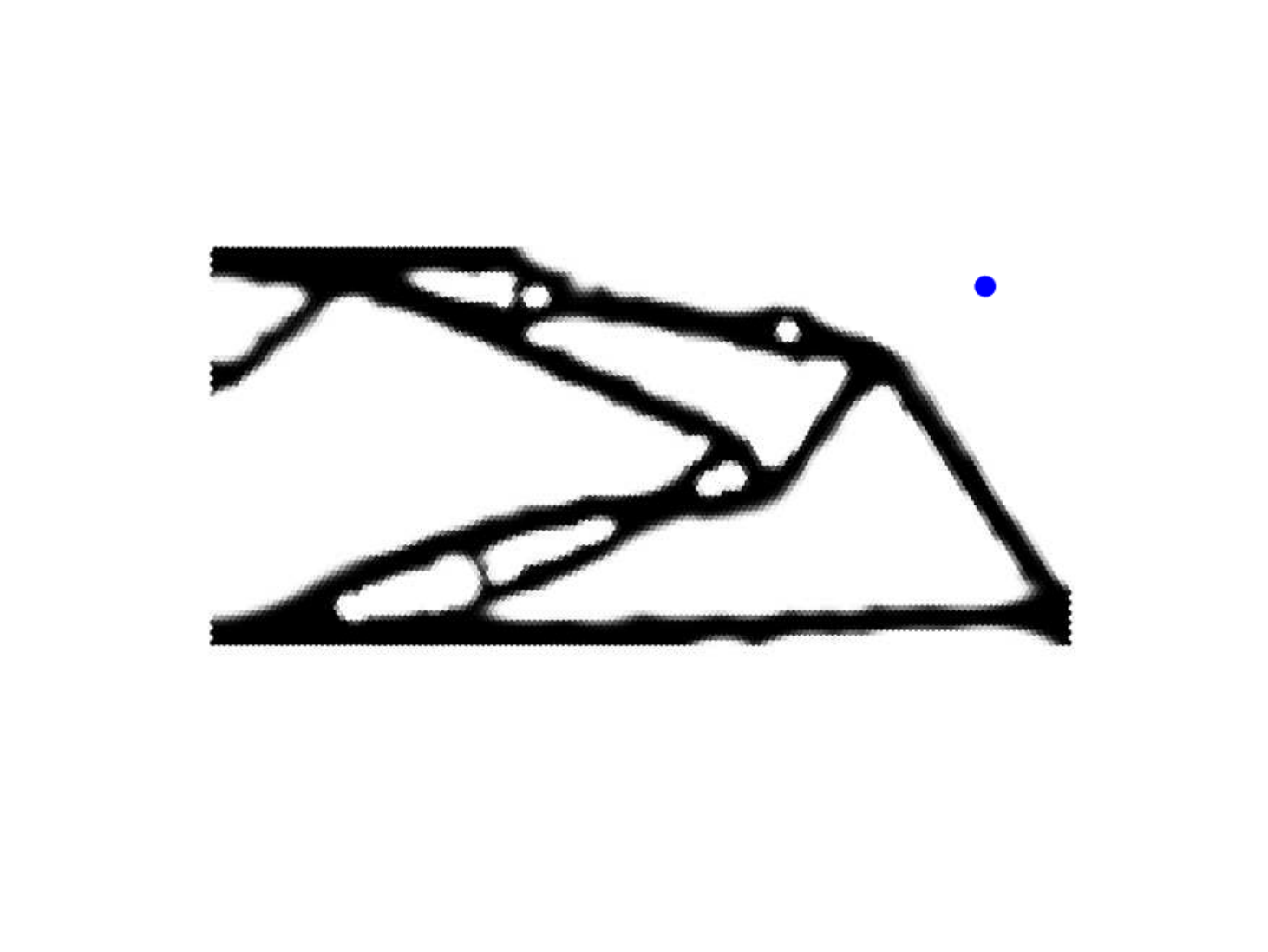}    
      \caption{$\alpha = 6$, $V_{f} =0.3$, $BWI = 0.09$ \\ $g_1 = -11.0$, $g_{min}(\bm{\rho})  = 87.1$, $min_{ls} = 1.52$}
      \label{fig:Eg1_test_minimum_ls_e}
    \end{subfigure}~~~~
    \begin{subfigure}[b]{.3\textwidth}
      \centering
       \captionsetup{font=scriptsize}
      \includegraphics[trim={3.5cm 2.5cm 2cm 2cm}, clip, scale = 0.5]{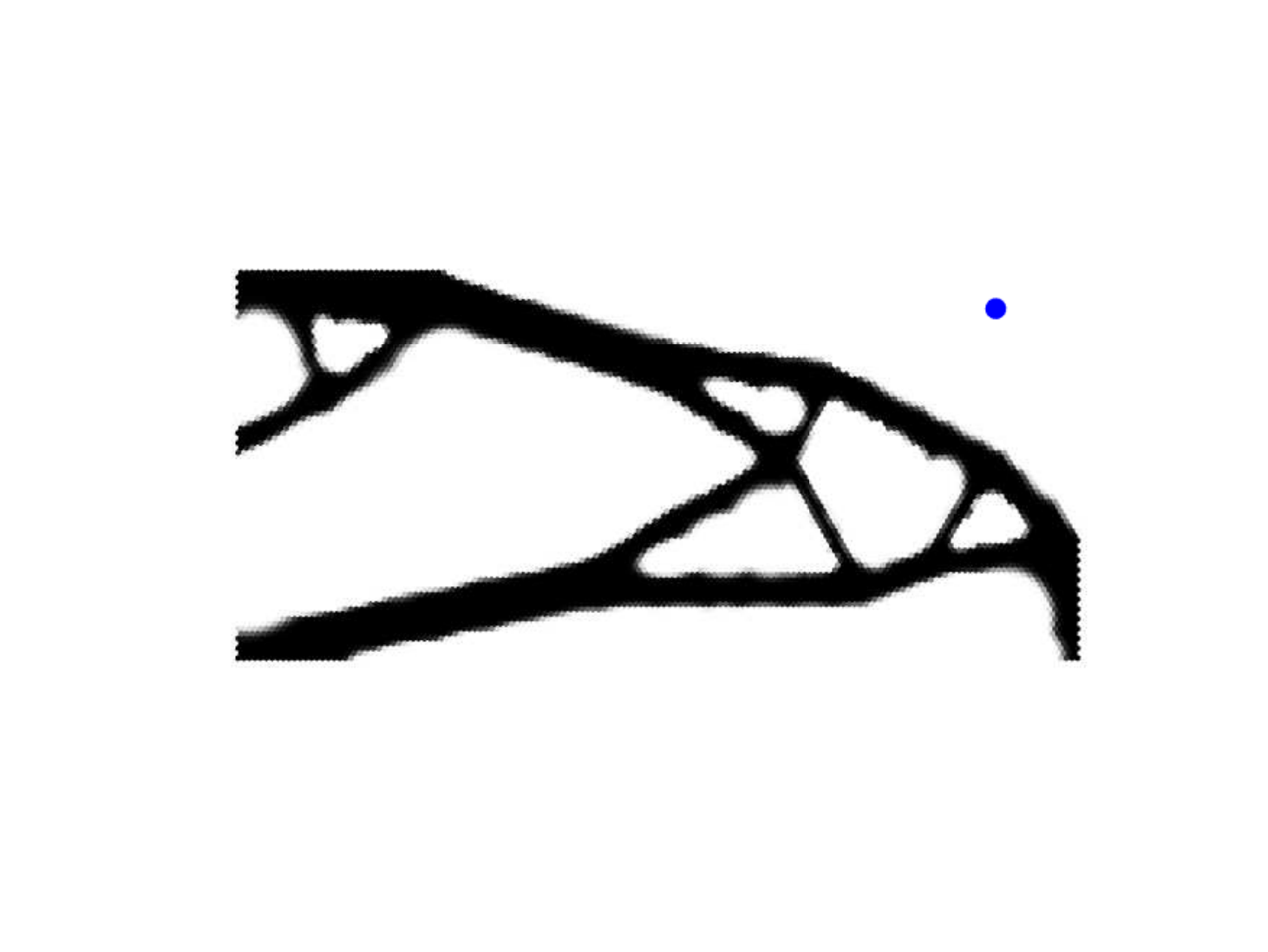}    
      \caption{$\alpha = 8$, $V_{f} = 0.3$, $BWI = 0.06$ \\ $g_1 = -15.0 $, $g_{min}(\bm{\rho})  = 4.5$, $min_{ls} = 1.52$}
      \label{fig:Eg1_test_minimum_ls_f}
    \end{subfigure}
    \begin{subfigure}[b]{.3\textwidth}
      \centering
       \captionsetup{font=scriptsize}
      \includegraphics[trim={3.5cm 2.5cm 2cm 2cm}, clip, scale = 0.5]{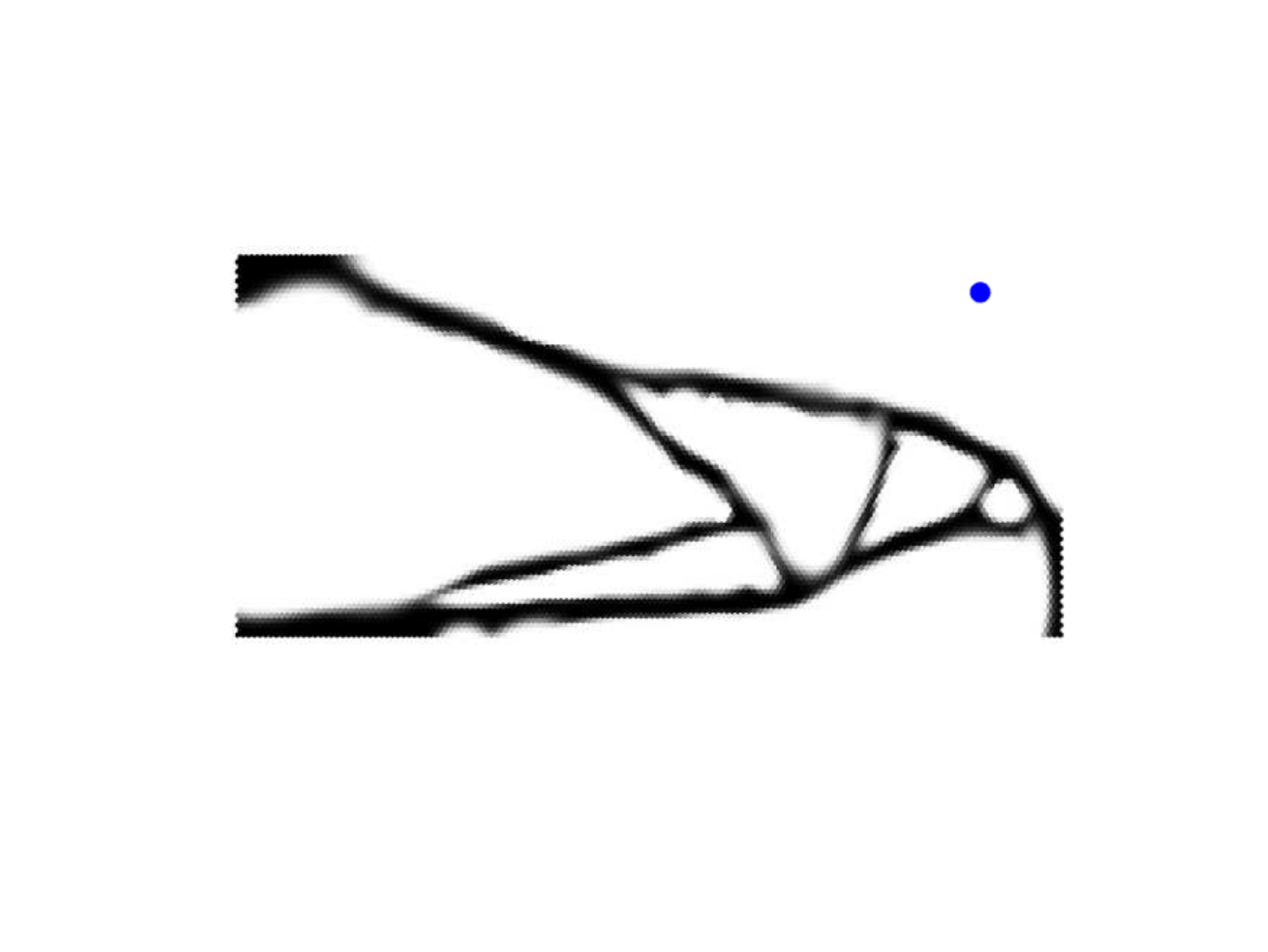}   
      \caption{$\alpha = 4$, $V_{f} =0.2$, $BWI = 0.11$ \\ $g_1 =  98.1$, $g_{min}(\bm{\rho})  =  214.2$, $min_{ls} = 1.52$}
      \label{fig:Eg1_test_minimum_ls_g}
    \end{subfigure}~~~~
    \begin{subfigure}[b]{.3\textwidth}
      \centering
       \captionsetup{font=scriptsize}
      \includegraphics[trim={3.5cm 2.5cm 2cm 2cm}, clip, scale = 0.5]{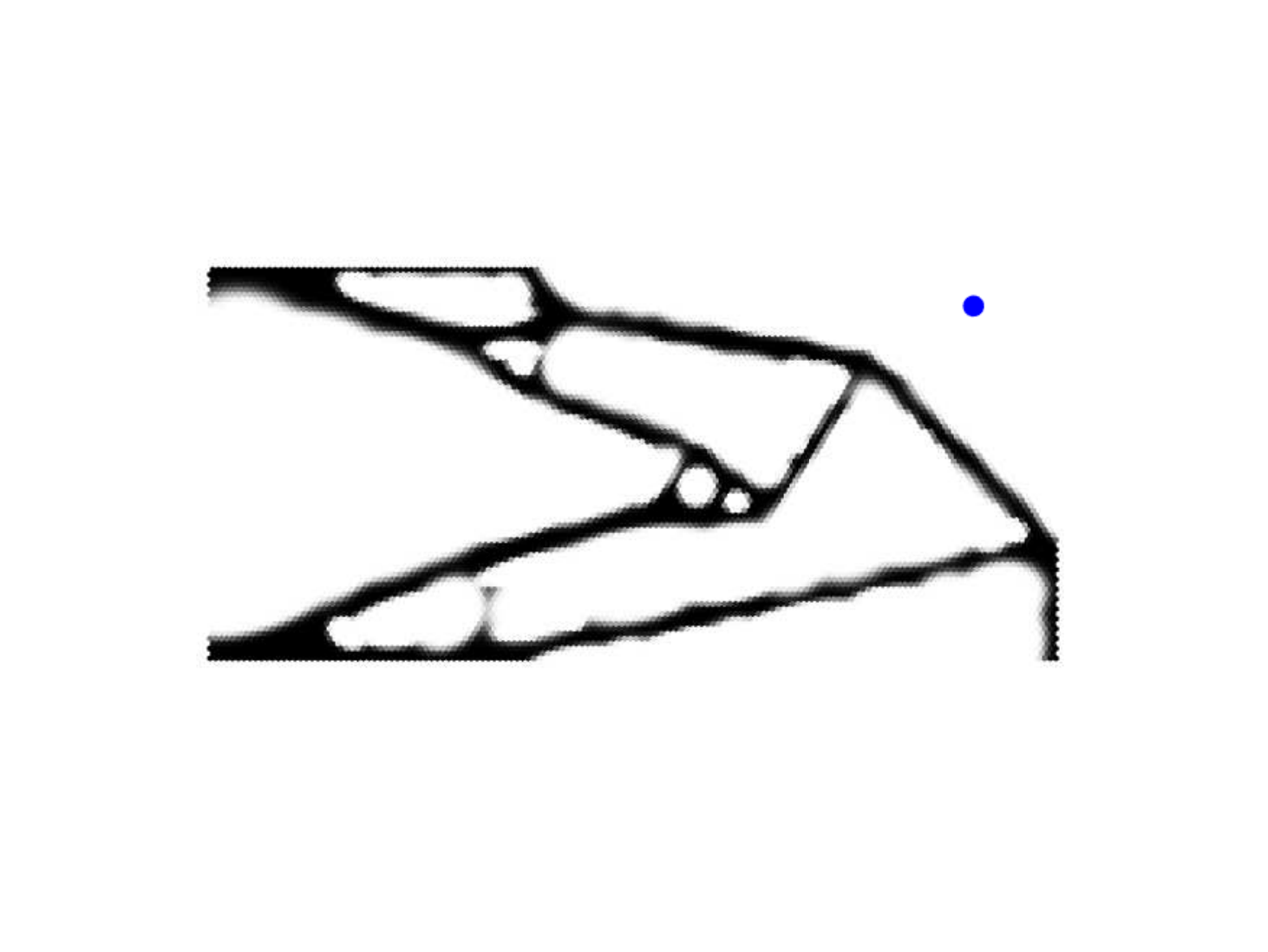}   
      \caption{$\alpha = 6$, $V_{f} =0.2$, $BWI = 0.1$ \\ $g_1 =  215.3$, $g_{min}(\bm{\rho})  =  375.8$, $min_{ls} = 1.52$}
      \label{fig:Eg1_test_minimum_ls_h}
    \end{subfigure}~~~~
    \begin{subfigure}[b]{.3\textwidth}
      \centering
       \captionsetup{font=scriptsize}
      \includegraphics[trim={3.5cm 2.5cm 2cm 2cm}, clip, scale = 0.5]{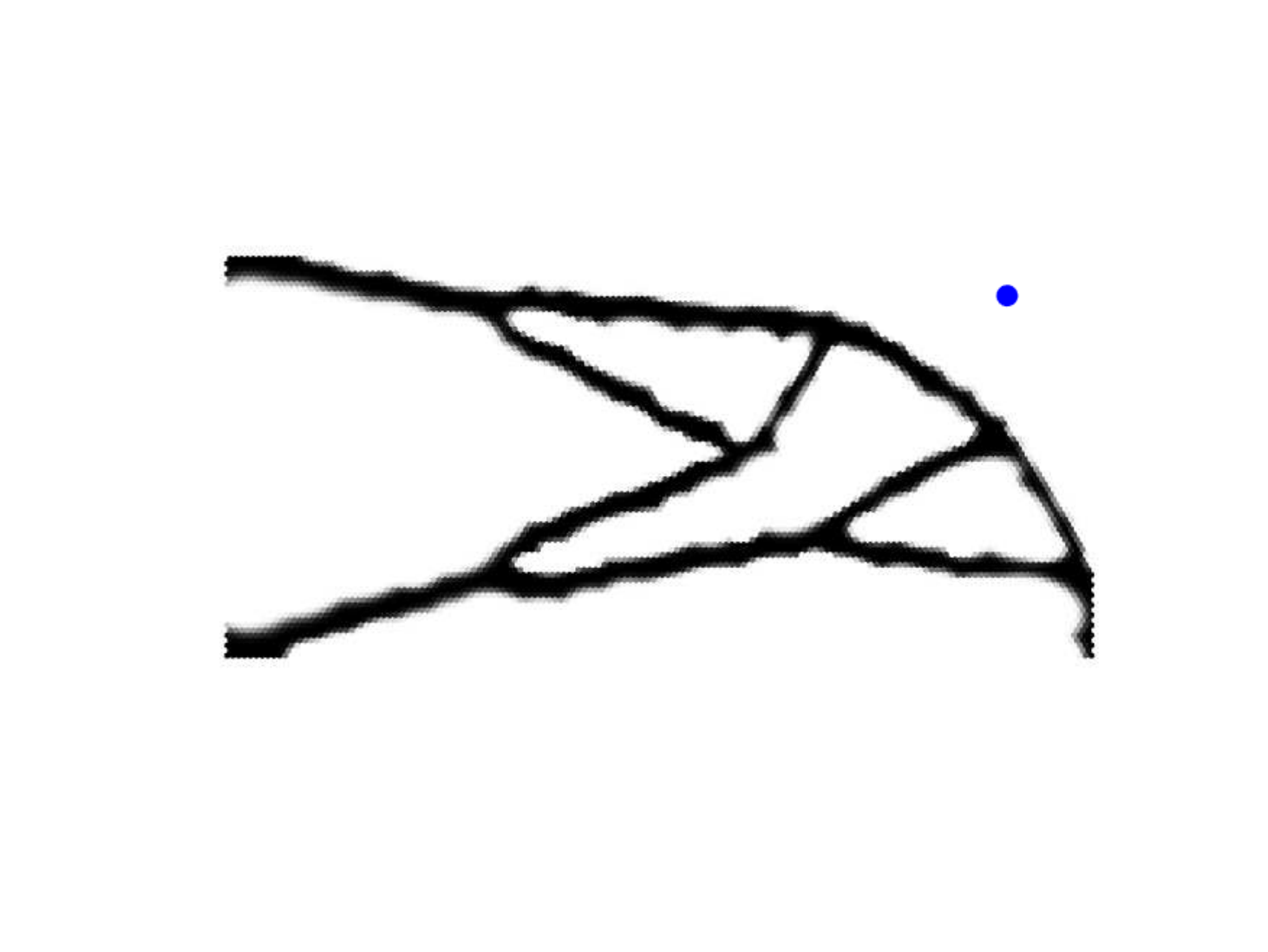}   
      \caption{$\alpha = 8$, $V_{f} =0.2$, $BWI = 0.08$ \\ $g_1 = 48.6$, $g_{min}(\bm{\rho})  = 144.2$, $min_{ls} = 1.52$}
      \label{fig:Eg1_test_minimum_ls_i}
    \end{subfigure}
    \begin{subfigure}[b]{.3\textwidth}
      \centering
       \captionsetup{font=scriptsize}
      \includegraphics[trim={3.5cm 2.5cm 2cm 2cm}, clip, scale = 0.5]{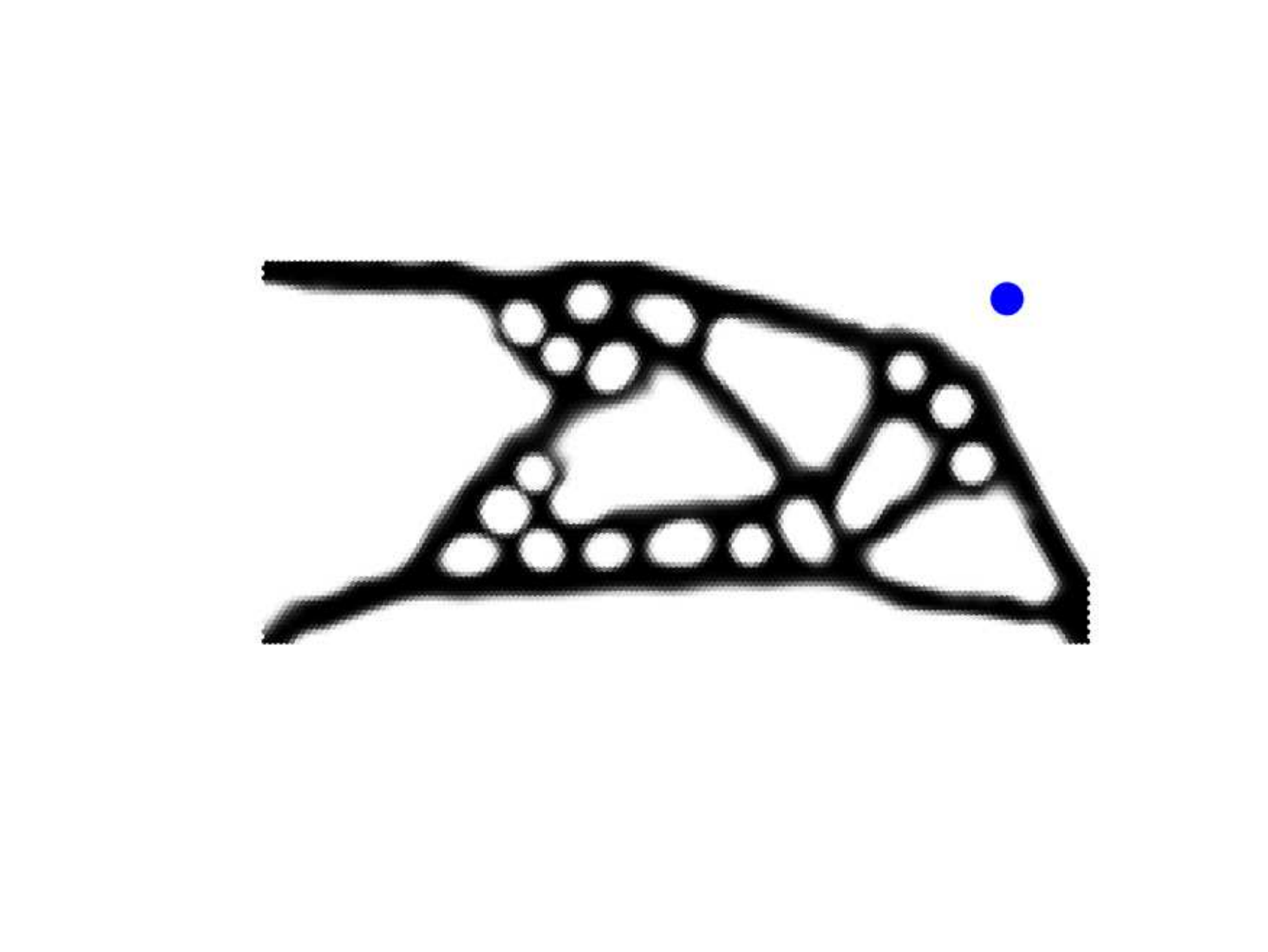}   
      \caption{$\alpha = 4$, $V_{f} =0.3$, $BWI = 0.15$ \\ $g_1 = 612.9$, $g_{min}(\bm{\rho})  = 804.7$, $min_{ls} = 1.9$}
      \label{fig:Eg1_test_minimum_ls_j}
    \end{subfigure}~~~~
    \begin{subfigure}[b]{.3\textwidth}
      \centering
       \captionsetup{font=scriptsize}
      \includegraphics[trim={3.5cm 2.5cm 2cm 2cm}, clip, scale = 0.5]{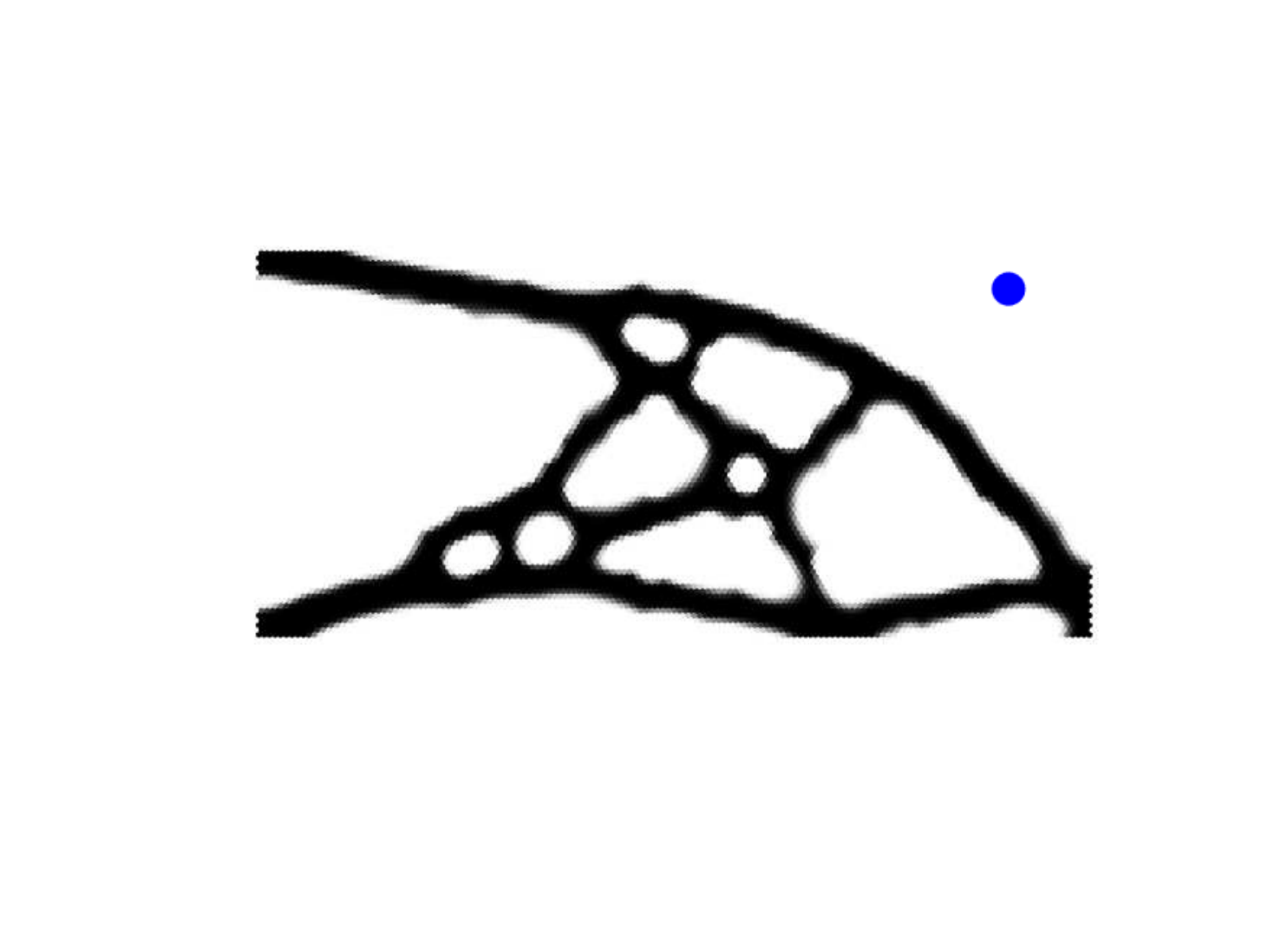}   
      \caption{$\alpha = 6$, $V_{f} =0.3$, $BWI = 0.09$ \\ $g_1 = 142.6 $, $g_{min}(\bm{\rho})  =  272.4$, $min_{ls} = 1.9$}
      \label{fig:Eg1_test_minimum_ls_k}
    \end{subfigure}~~~~
    \begin{subfigure}[b]{.3\textwidth}
      \centering
       \captionsetup{font=scriptsize}
      \includegraphics[trim={3.5cm 2.5cm 2cm 2cm}, clip, scale = 0.5]{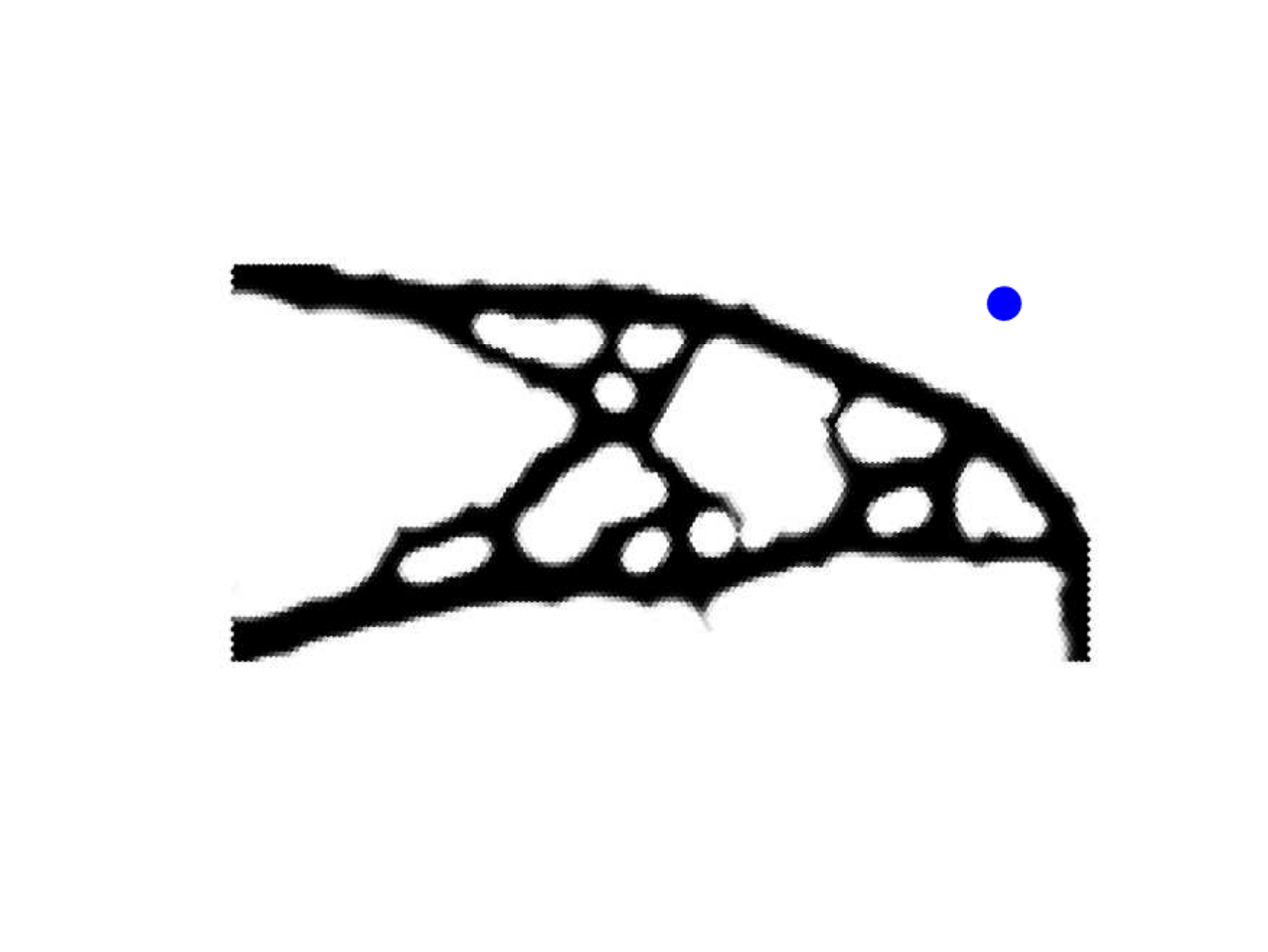}   
      \caption{$\alpha = 8$, $V_{f} =0.3$, $BWI =  0.07$ \\ $g_1 =  229.5$, $g_{min}(\bm{\rho})  =  480.3$, $min_{ls} = 1.9$}
      \label{fig:Eg1_test_minimum_ls_l}
    \end{subfigure}
\begin{subfigure}[b]{.3\textwidth}
      \centering
       \captionsetup{font=scriptsize}
      \includegraphics[trim={3.5cm 2.5cm 2cm 2cm}, clip, scale = 0.5]{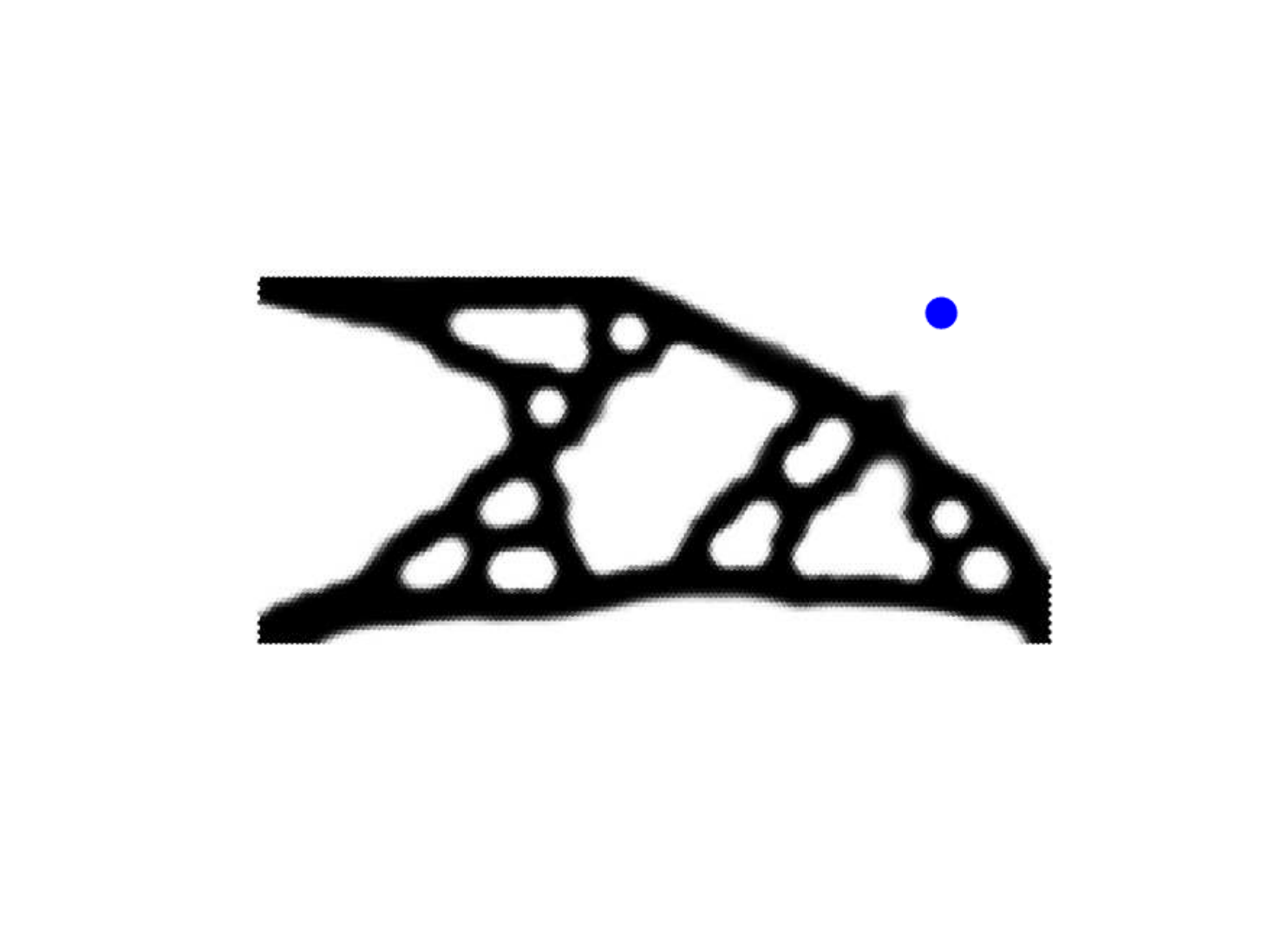}   
      \caption{$\alpha =4$, $V_{f} =0.4$, $BWI = 0.10$ \\ $g_1 =42.8$, $g_{min}(\bm{\rho})  = 179.3$, $min_{ls} = 1.9$}
      \label{fig:Eg1_test_minimum_ls_m}
    \end{subfigure}~~~~
\begin{subfigure}[b]{.3\textwidth}
      \centering
       \captionsetup{font=scriptsize}
      \includegraphics[trim={3.5cm 2.5cm 2cm 2cm}, clip, scale = 0.5]{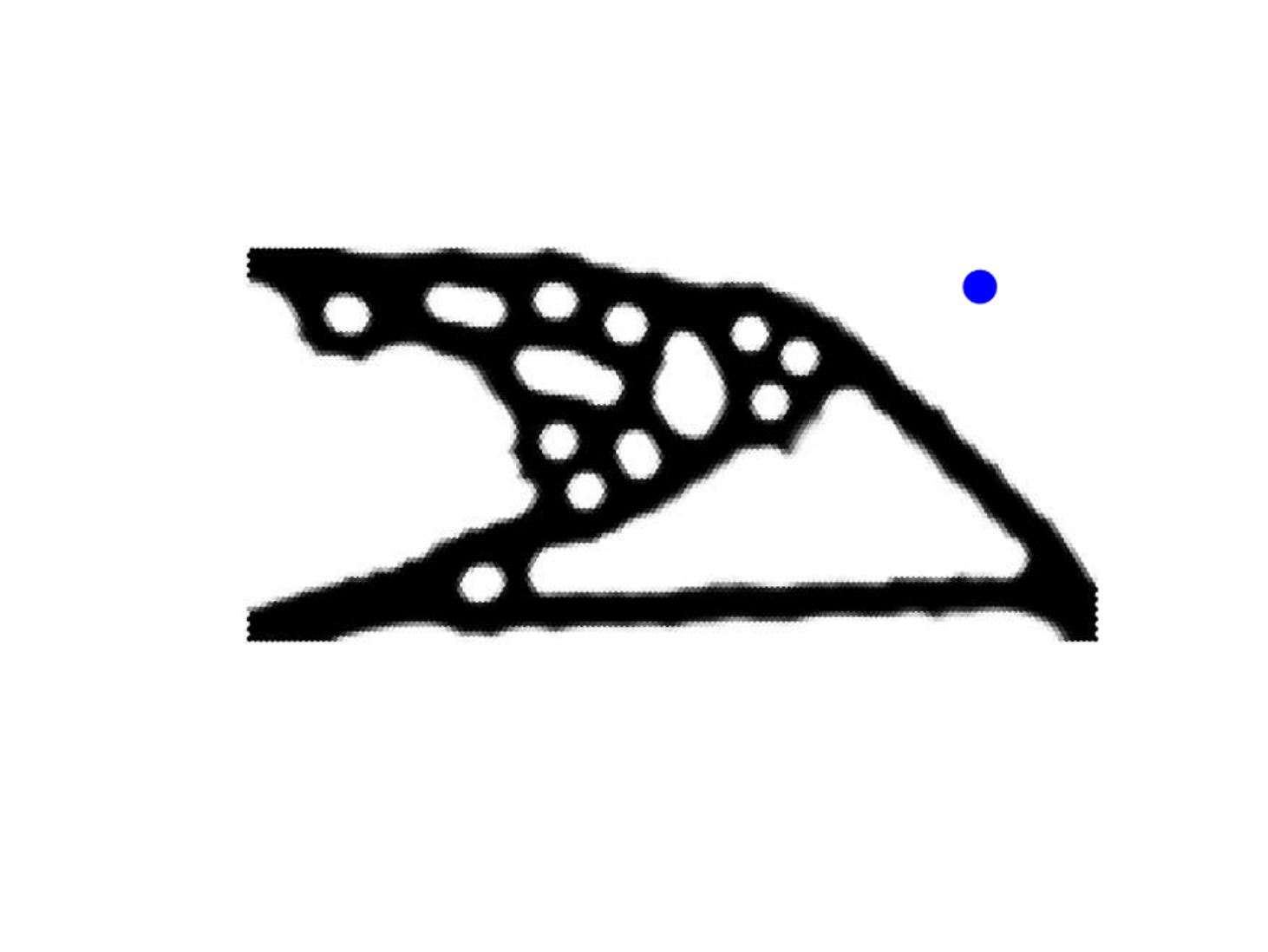}   
      \caption{$\alpha = 6$, $V_{f} =0.4$, $BWI = 0.08$ \\ $g_1 = 21.8$, $g_{min}(\bm{\rho})  = 179.9$, $min_{ls} = 1.9$}
      \label{fig:Eg1_test_minimum_ls_n}
    \end{subfigure}~~~~
\begin{subfigure}[b]{.3\textwidth}
      \centering
       \captionsetup{font=scriptsize}
      \includegraphics[trim={3.5cm 2.5cm 2cm 2cm}, clip, scale = 0.5]{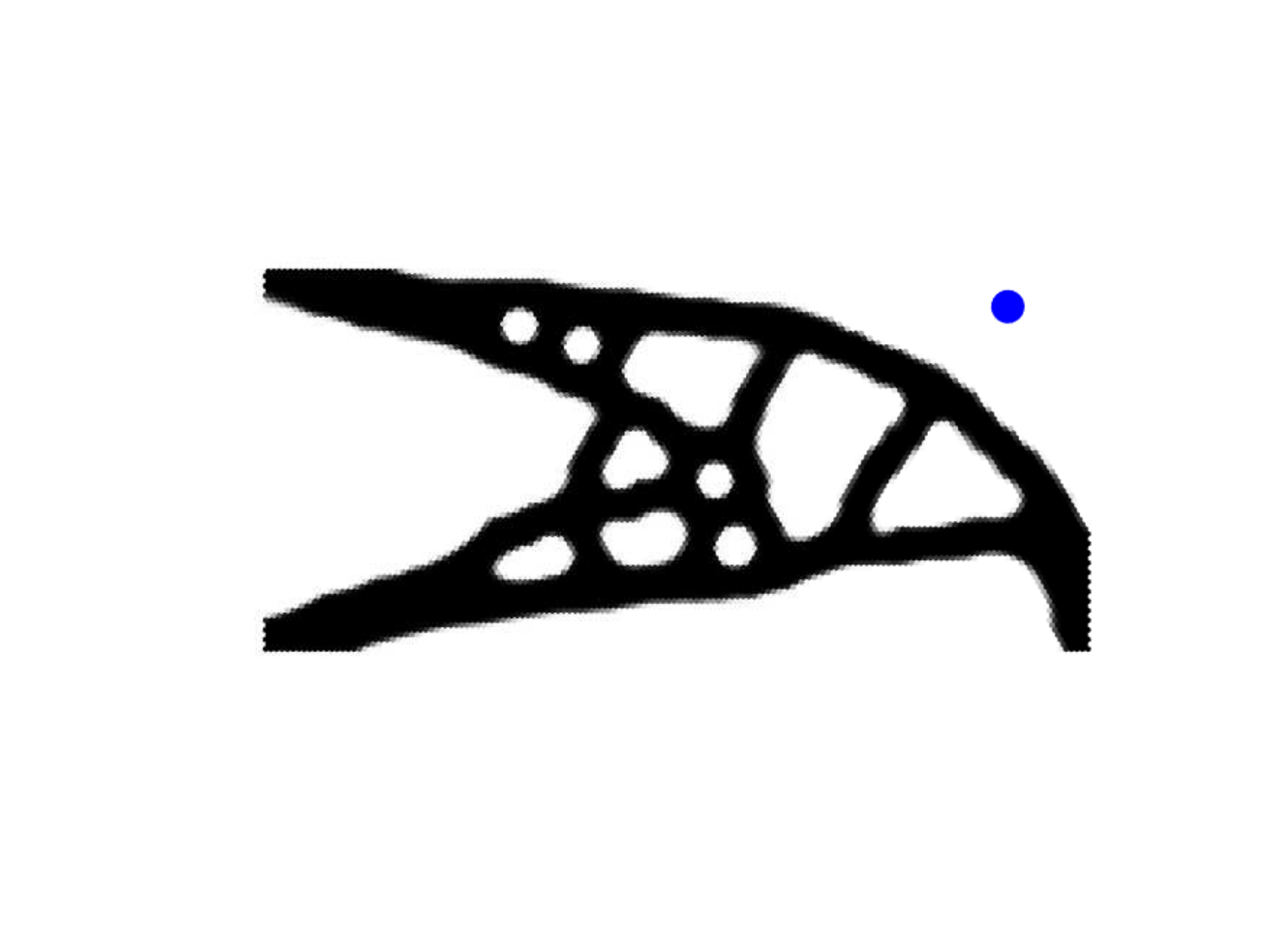}   
      \caption{$\alpha = 8$, $V_{f} =0.4$, $BWI = 0.06$ \\ $g_1 = -9.9$, $g_{min}(\bm{\rho})  =12.1$, $min_{ls} = 1.9$}
      \label{fig:Eg1_test_minimum_ls_o}
    \end{subfigure}

\caption{Topological Solutions for Example I: Solutions are obtained using $20 \times 10$ negative elliptical masks as design variables. Domain of size $100 \times 46 \mbox{ unit}^2$ is discretized via 150 by 80 regular honeycomb mesh. Cell size (radius of the circumcircle) is 0.38 mm. A minimum length scale (blue circles as insets) of $2 min_{ls} \mbox{ mm}$ is imposed. For all solutions, $\eta = 3$, and maximum number of function evaluations is 400. Solutions are obtained for different $\alpha$ and upper bounds on the volume constraint. Values of the volume constraint $g_1$ and minimum length scale constraint $g_{min}(\bm{\rho}) $ are depicted below each solution, wherever imposed. 
Exponent for the constraints, $p = 2$ (Eqs. \ref{min_max_ls}). No maximum length scale is imposed on any problem. }
\label{fig:Eg1_test_minimum_ls}
\end{figure}

\section{Examples}
\label{examples}
\noindent We solve the first two examples (Fig. \ref{Fig:2}) by imposing minimum length scale, and then both length scales respectively. Thereafter, we solve all four examples using the SLS methodology with both, negative and positive elliptical masks. 

\subsection{Results without the SLS methodology}
\label{prelim_results} 

\indent It is known that the design space for stiffness maximization problems for $\eta > 1$ is non-convex \cite{Stolpe2001}. With the flexibility-stiffness multi-criteria formulations,  optimal topologies of compliant mechanisms also depend on the initial guess \cite{Saxena2000,deepak2009comparative}. With many parameters associated with the Material Mask Overlay Strategy \cite{saxena2011topology}, final solutions are expected to be influenced by these. For the examples presented, we standardize the parameters as follows. Maximum dimension of a (rectangular) region is chosen as 100 units. Number of elliptical (circular) masks along each axis is the rounded off value of the length along that axis over $5$ (or $3$), the number of design variables per mask.  $\eta$ is chosen as $3$. No filtering is performed for all examples generated. Before commencing optimization for which the $fmincon$ routine of MATLAB$^{TM}$ is employed, all masks are distributed evenly as in Fig. \ref{fig:RD11}. Maximum possible dimension of the semi-major ($a_j$) or minor ($b_j$) axis is $mR = 10$ units. For negative masks, minimum dimensions correspond to the minimum length scale, $min_{ls}$ while for positive masks, minimum dimensions are chosen close to 0. For negative masks, this is equivalent to imposing a minimum length scale on the void state. Initial values of $a_j$ and $b_j$ are set to $\frac{mR}{4}$ units, and initial orientation $\theta_j$ to $0$ degrees.  With the above standardization, parameter $\alpha$ still remains a free choice along with upper bound on the continuum volume, minimum and maximum length scales. One realizes \cite{saxena2011topology} that optimal topologies may depend on $\alpha$ and the way it is chosen to vary during optimization, e.g., as in continuation methods. One also notes that a high $\alpha$ yields solutions close to the ideal $0-1$ topologies. We employ a $BWI$ (black and white) index \cite{sigmund_2007_morph_proj2, wang_lazarov_sigmund_2011} to evaluate how far solutions are from the originally intended 0-1 topologies. A lower $BWI$ indicates that a topology is closer to the $0-1$ solution. $BWI$ is given as

\begin{equation}
\label{bwi_index}
BWI = \frac{4 \sum_{i = 1}^{Ncells} \rho_{i}(\alpha, \eta) [1 - \rho_{i}(\alpha, \eta) ]}{Ncells}
\end{equation}

We solve the first two examples (Fig. \ref{Fig:2:Eg1} and \ref{Fig:2:Eg2}) without the SLS methodology in section \ref{method_TO} to illustrate that  if the volume fraction $vf$,  minimum ($min_{ls}$) and maximum ($max_{ls}$) length scales are specified arbitrarily and independently, a (desirable) solution may not always be possible. The first example is of a compliance minimization problem for a cantilever system fixed at the left vertical boundary. A force of $3$ units along the downward vertical direction is applied at the right bottom corner (Fig. \ref{Fig:2:Eg1}). Solutions for Example I in Figs. \ref{fig:Eg1_test_minimum_ls} are obtained for a domain size of $100\times 46$ unit$^2$ using 150$\times$80 cells in a honeycomb mesh with 20$\times$10 negative elliptical masks spread evenly over the domain as the initial guess. Solutions for different $\alpha$, volume fraction and minimum length scale constraints are presented. Maximum length scale constraint is not imposed on any solution for this example. Generated topologies are arranged such that along a row, they have the same volume and minimum length scale constraints, while those along the column correspond to the same value of $\alpha$. 

Figs. \ref{fig:Eg1_test_minimum_ls_a}-\ref{fig:Eg1_test_minimum_ls_c} are topologies generated for a volume fraction $vf = 0.3$ with no minimum length scale constraint imposed. As $\alpha$ increases, the black and white index, $BWI$ or the gray scale indicator decreases. Increasing $\alpha$ results in closer to the ideal, black and white solutions, as expected. Topologies also change.  Members are relatively well formed and straight. However, members have uneven thicknesses and hence, imposition of only the volume constraint may result in some constituents having undesirable dimensions that are not manufacturable and/or are prone to failure.

Figs. \ref{fig:Eg1_test_minimum_ls_d}-\ref{fig:Eg1_test_minimum_ls_f} are solutions corresponding to the same specifications as in Figs. \ref{fig:Eg1_test_minimum_ls_a}-\ref{fig:Eg1_test_minimum_ls_c} respectively but with an additional, minimum length scale constraint, $min_{ls}$, of $1.52$ units. In these, both, the volume ($g_1$) and minimum length scale ($g_{min}(\bm{\rho}) $) constraints are considered satisfied as $g_1$ and $g_{min}(\bm{\rho}) $ are either small positive or negative values. For the solution in Fig. \ref{fig:Eg1_test_minimum_ls_e}, $g_{min}(\bm{\rho}) $ is higher. Decrease in $BWI$ with increase in $\alpha$ is consistent. Solutions in Figs. \ref{fig:Eg1_test_minimum_ls_d}-\ref{fig:Eg1_test_minimum_ls_f} have more members (topologies have relatively more holes) in comparison to their counterparts in Figs. \ref{fig:Eg1_test_minimum_ls_a}-\ref{fig:Eg1_test_minimum_ls_c}. Members are almost straight with some possessing certain, although small, curvature and also, some undulations along their boundaries. Some of these solutions could be sub-optimal, perhaps, due to imposition of the explicit minimum length scale via the structural skeleton, which changes continuously as optimization progresses. 

\begin{figure}[H]
	\begin{subfigure}[b]{.3\textwidth}
		\centering
		\captionsetup{font=scriptsize}
\includegraphics[trim={2cm 2.5cm 2cm 2cm}, clip, scale = 0.5]{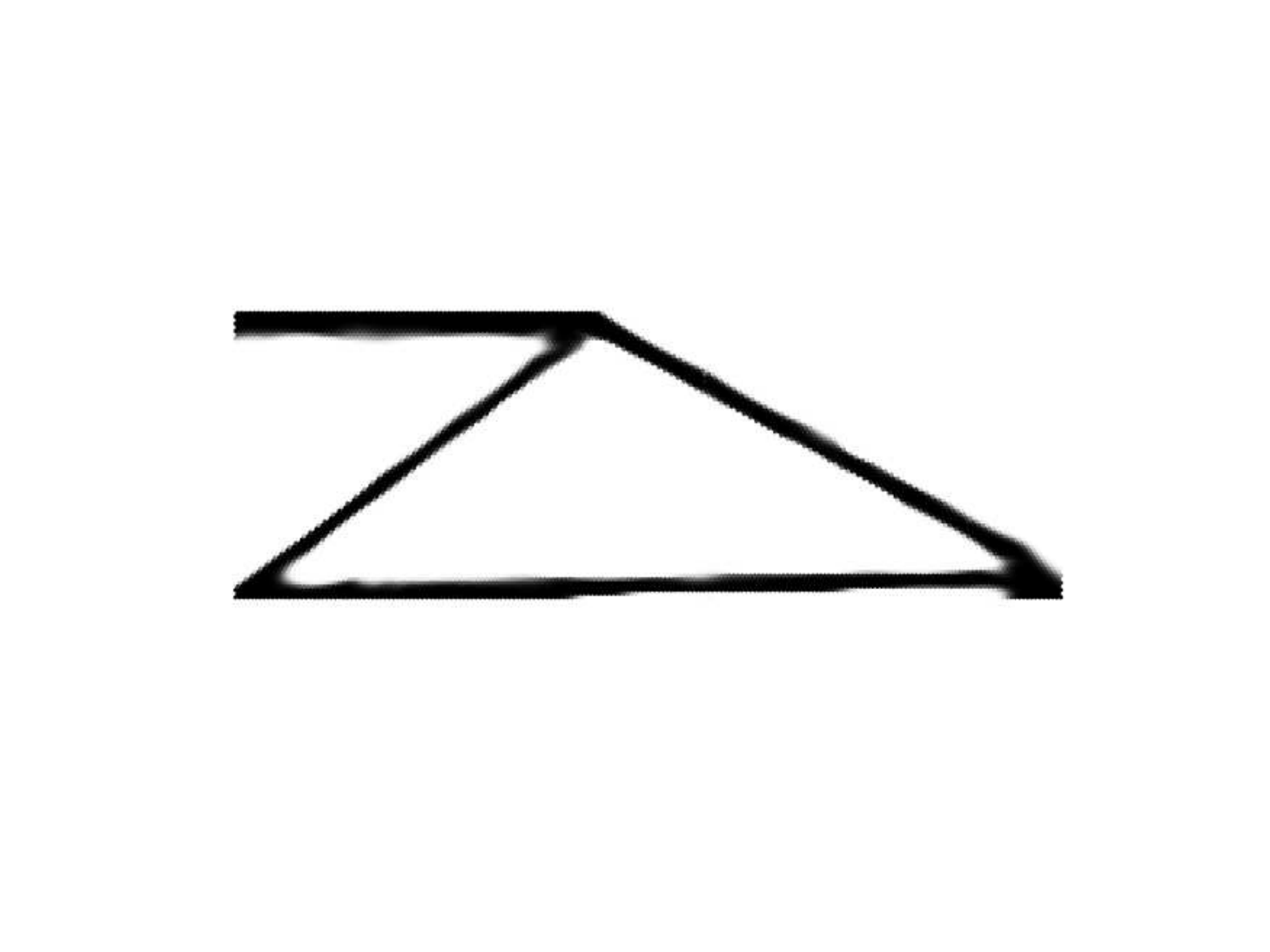}     
		\caption{ $V_{f} =0.20$, $BWI = 0.06$}
		\label{fig:Eg2_test_minimum_ls_a}
	\end{subfigure}~~~~
	\begin{subfigure}[b]{.3\textwidth}
		\centering
		\captionsetup{font=scriptsize}
\includegraphics[trim={2cm 2.5cm 2cm 2cm}, clip, scale = 0.5]{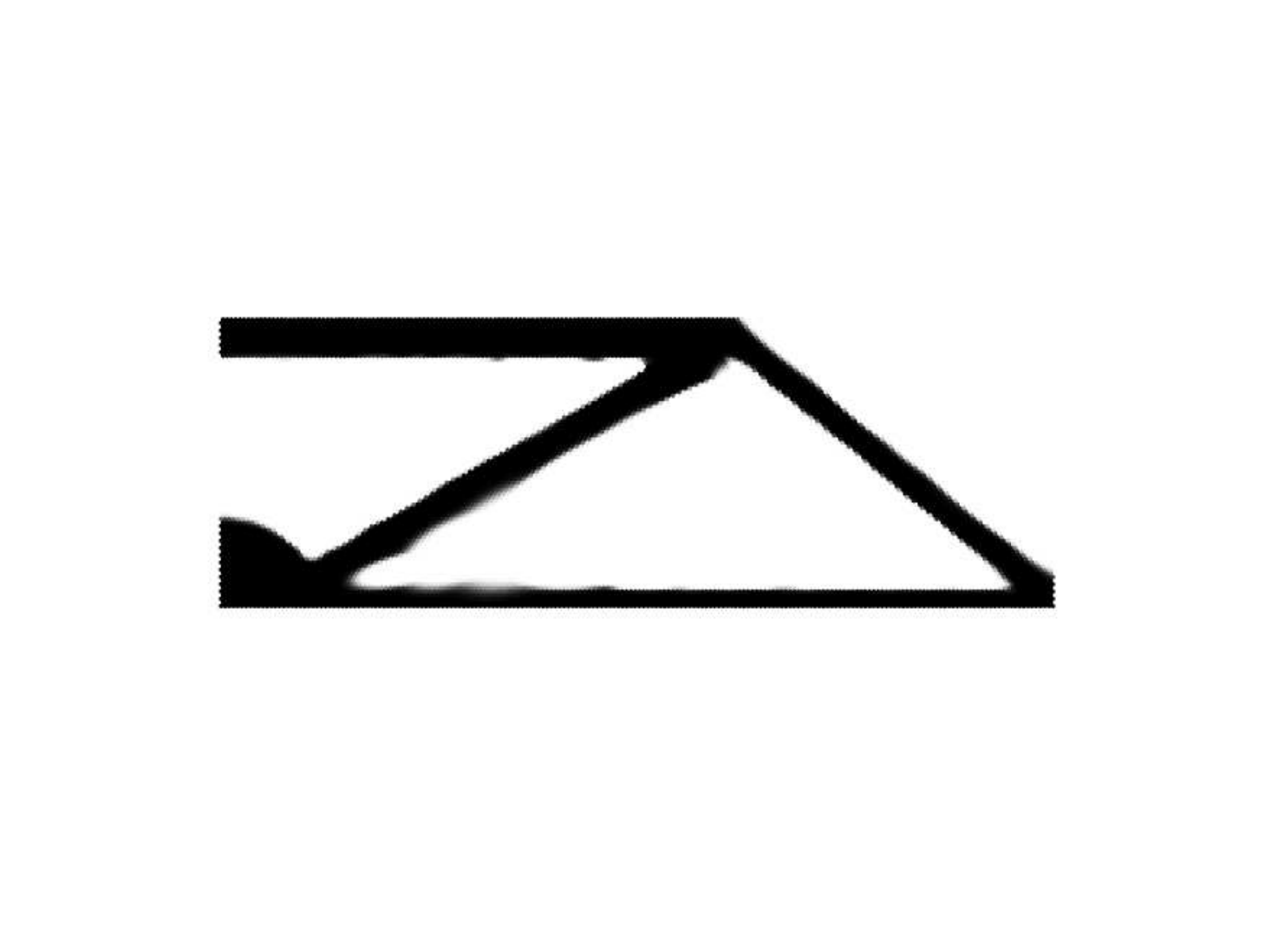}   
		\caption{ $V_{f} =0.30$, $BWI = 0.04$}
		\label{fig:Eg2_test_minimum_ls_b}
	\end{subfigure}~~~~
	\begin{subfigure}[b]{.3\textwidth}
		\centering
		\captionsetup{font=scriptsize}
\includegraphics[trim={2cm 2.5cm 2cm 2cm}, clip, scale = 0.5]{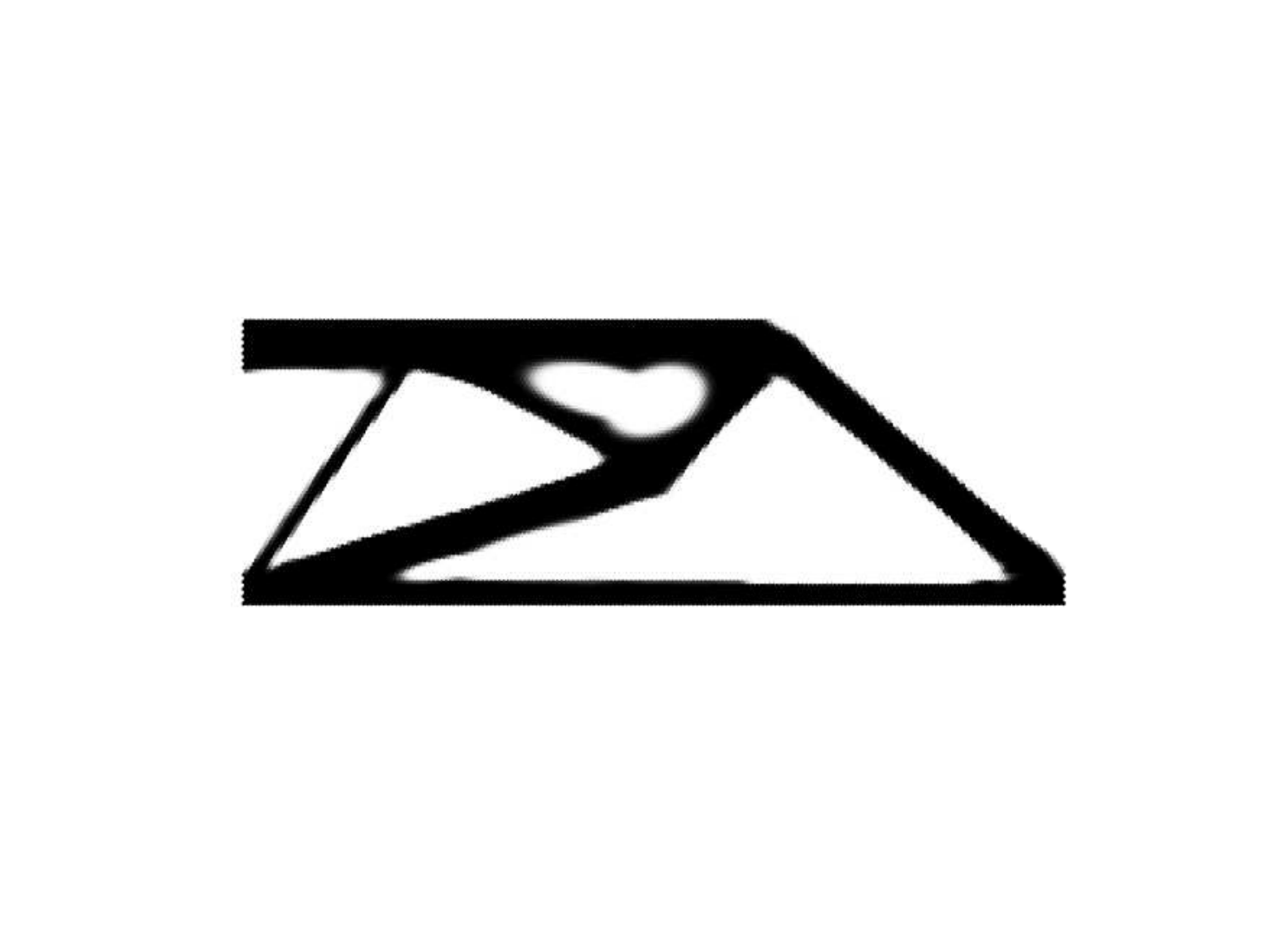}  
		\caption{ $V_{f} =0.40$ $BWI = 0.05$}
		\label{fig:Eg2_test_minimum_ls_c}
	\end{subfigure}
	\\
	\begin{subfigure}[b]{.3\textwidth}
		\centering
		\captionsetup{font=scriptsize}
		\includegraphics[trim={2cm 2.5cm 2cm 2cm}, clip, scale = 0.5]{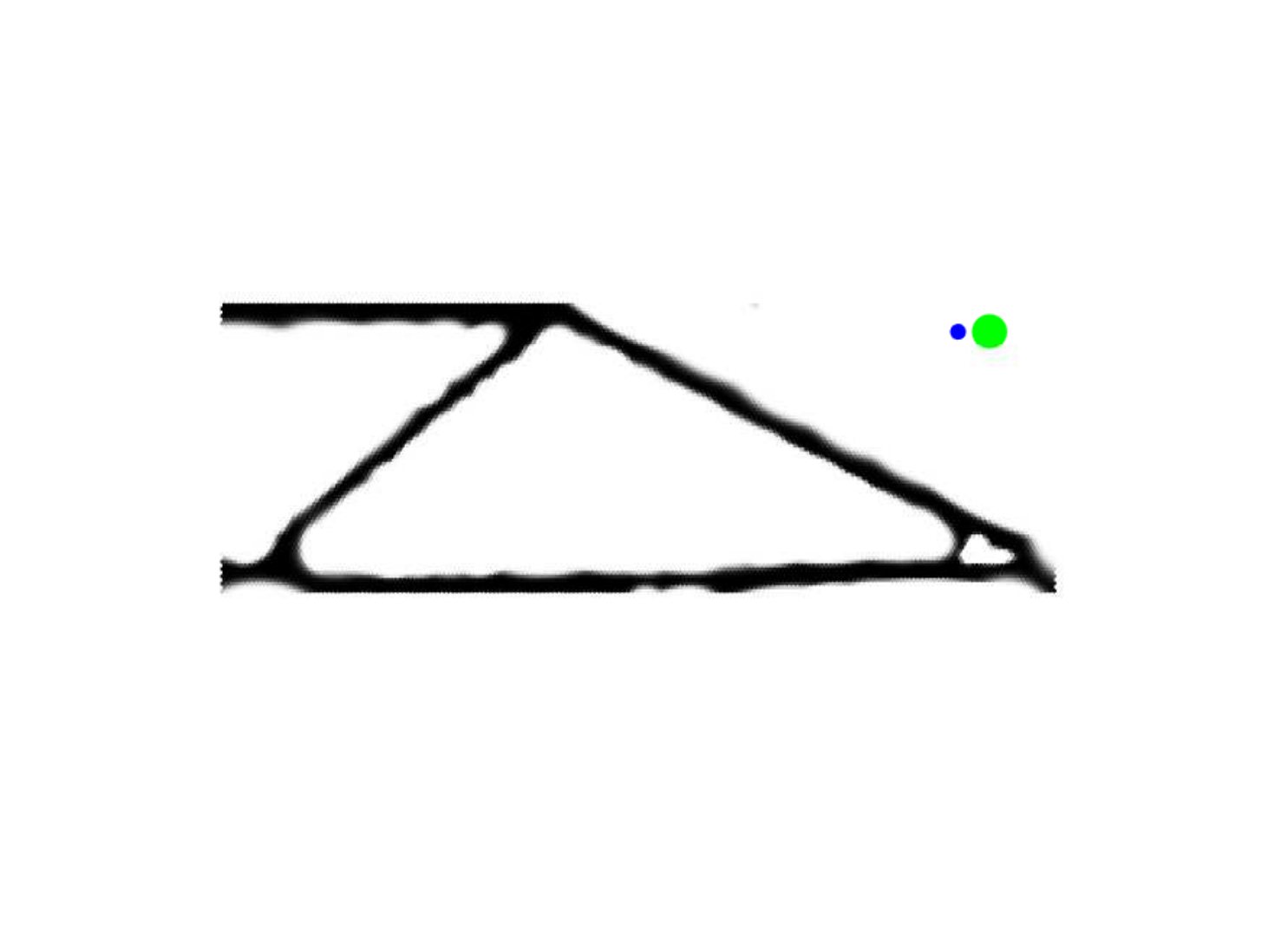}   
		\caption{ $V_{f} =0.20$, $min_{ls} = 3cs$, $max_{ls} = 7cs$,  \\ $BWI = 0.08$, $g_{min}(\bm{\rho})  = 18.2$, $g_{max}(\bm{\rho})  = 0.82$}
		\label{fig:Eg2_test_minimum_ls_d}
	\end{subfigure}~~~~
	\begin{subfigure}[b]{.3\textwidth}
		\centering
		\captionsetup{font=scriptsize}
		\includegraphics[trim={2cm 2.5cm 2cm 2cm}, clip, scale = 0.5]{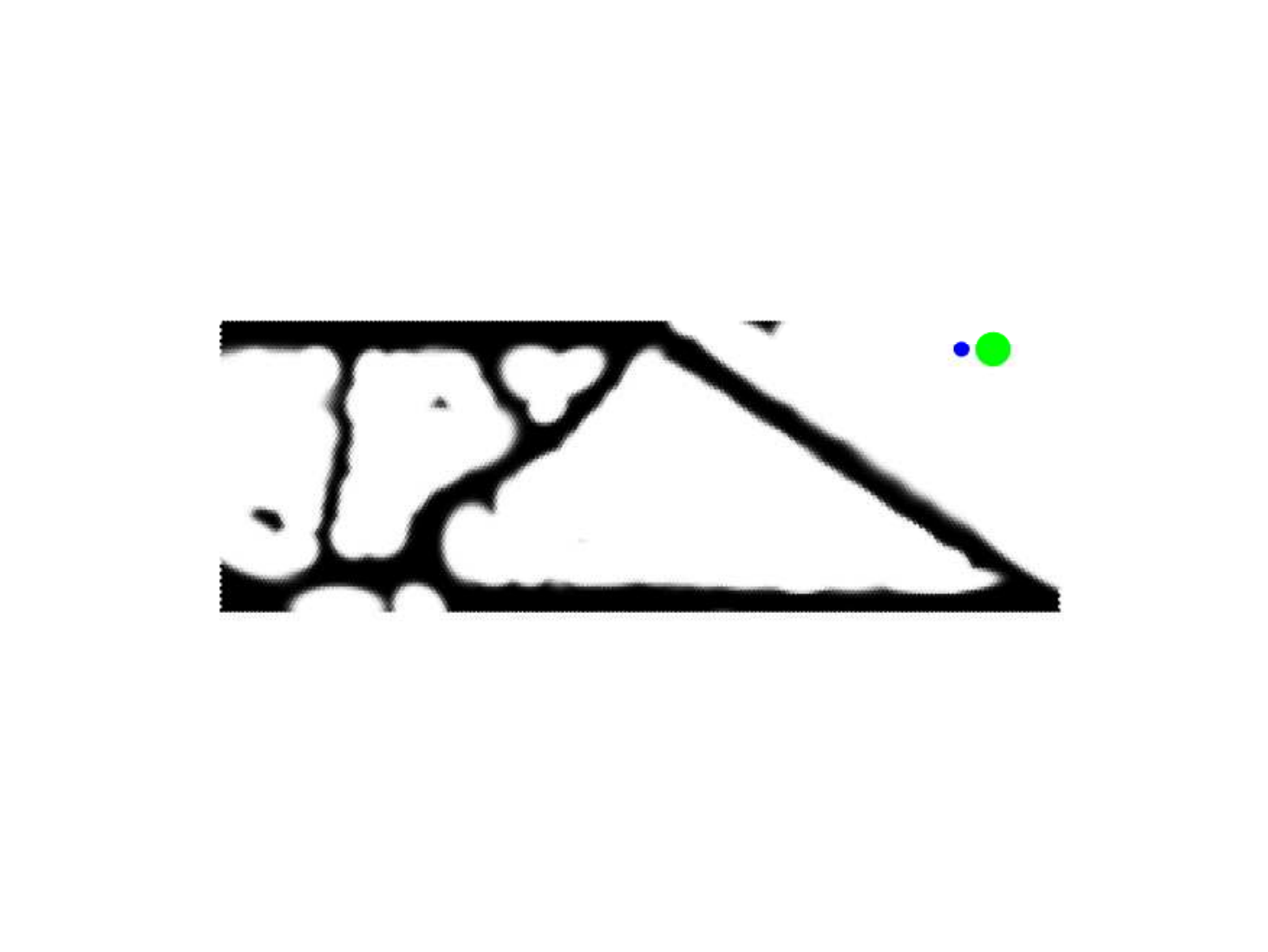}    
		\caption{ $V_{f} =0.30$, $min_{ls} = 3cs$, $max_{ls} = 7cs$,  \\ $BWI = 0.08$, $g_{min}(\bm{\rho})  = 45.5$, $g_{max}(\bm{\rho})  = 40.2$}
		\label{fig:Eg2_test_minimum_ls_e}
	\end{subfigure}~~~~
	\begin{subfigure}[b]{.3\textwidth}
		\centering
		\captionsetup{font=scriptsize}
		\includegraphics[trim={2cm 2.5cm 2cm 2cm}, clip, scale = 0.5]{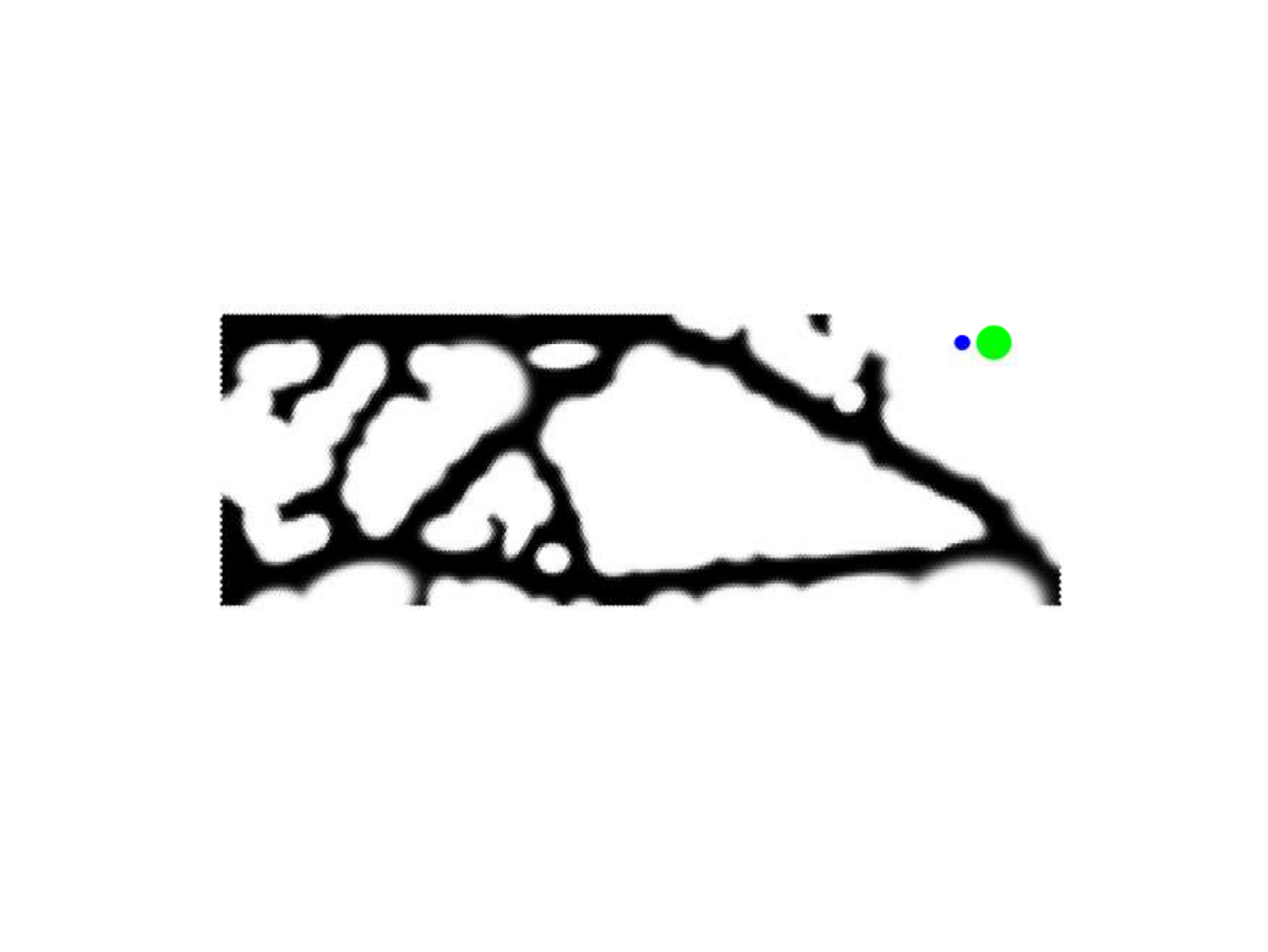}    
		\caption{ $V_{f} =0.40$, $min_{ls} = 3cs$, $max_{ls} = 7cs$, \\ $BWI = 0.12$, $g_{min}(\bm{\rho})  = 12.0$, $g_{max}(\bm{\rho})  = 33.71$}
		\label{fig:Eg2_test_minimum_ls_f}
	\end{subfigure}

	\begin{subfigure}[b]{.3\textwidth}
		\centering
		\captionsetup{font=scriptsize}
		\includegraphics[trim={2cm 2.5cm 2cm 2cm}, clip, scale = 0.5]{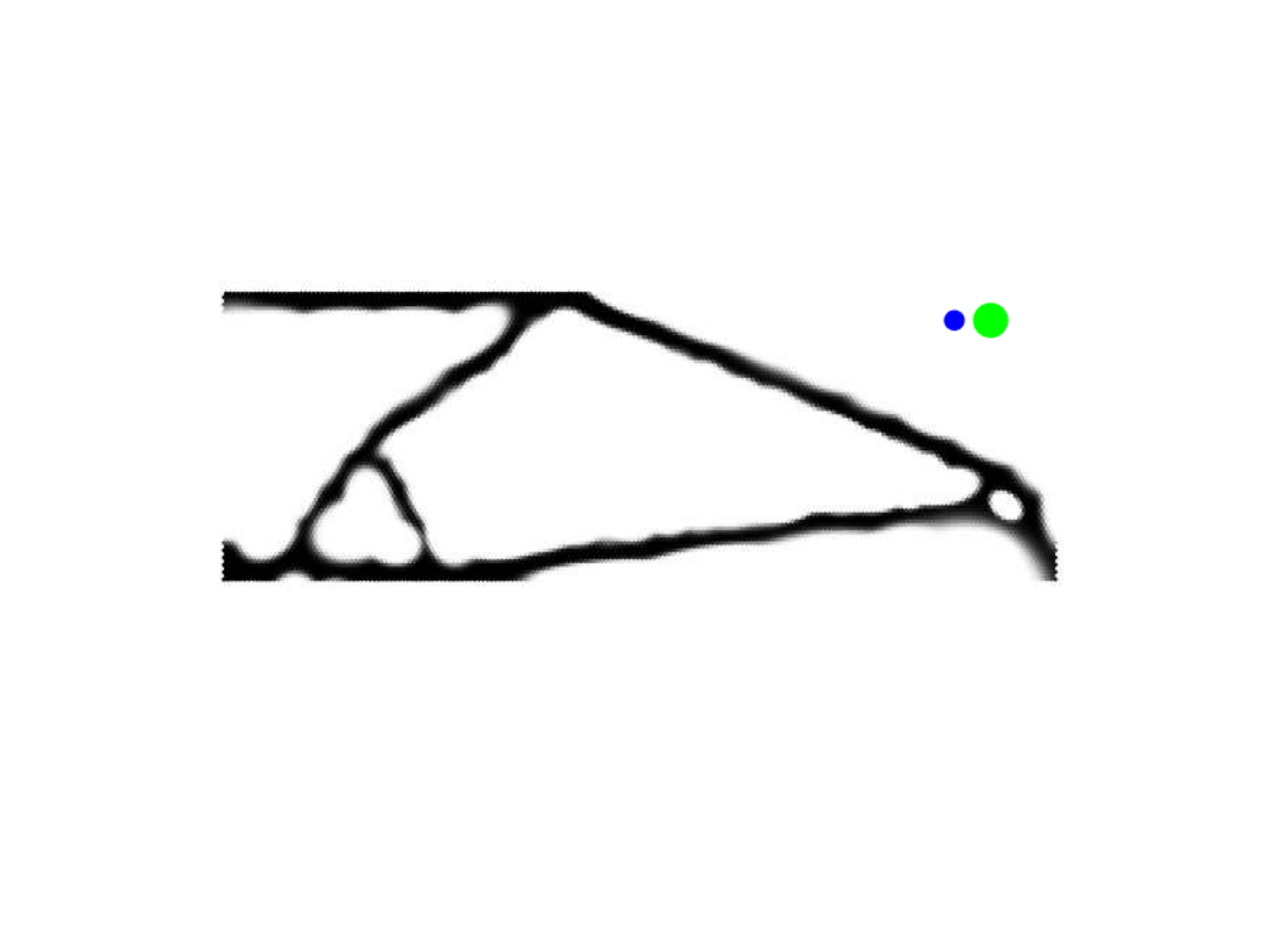}   
		\caption{ $V_{f} =0.20$, $min_{ls} = 4cs$, $max_{ls} = 7cs$,\\ $BWI = 0.10$, $g_{min}(\bm{\rho})  = 174.9$, $g_{max}(\bm{\rho})  = 4.00$}
		\label{fig:Eg2_test_minimum_ls_g}
	\end{subfigure}~~~~
	\begin{subfigure}[b]{.3\textwidth}
		\centering
		\captionsetup{font=scriptsize}
		\includegraphics[trim={2cm 2.5cm 2cm 2cm}, clip, scale = 0.5]{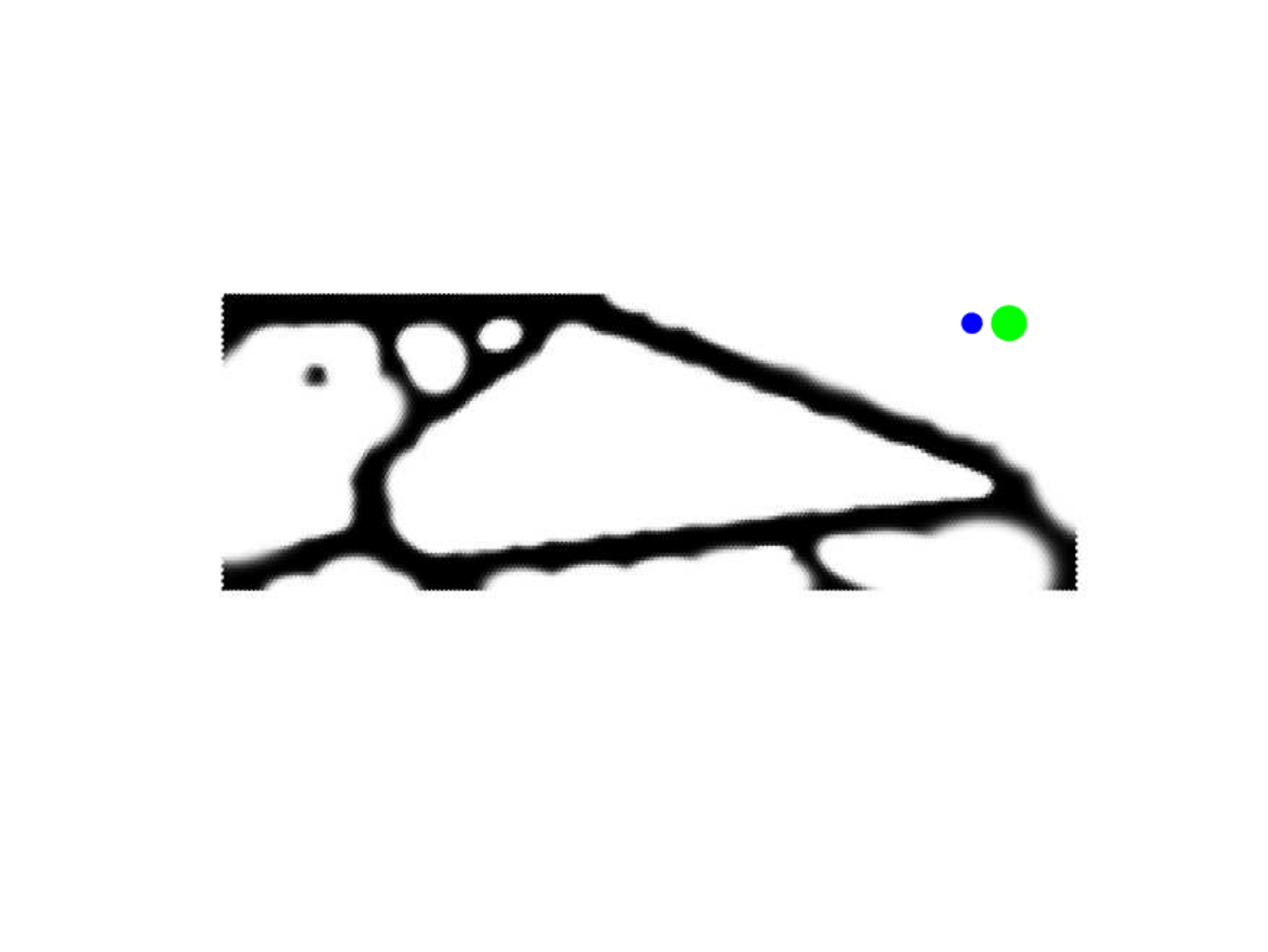}    
		\caption{ $V_{f} =0.30$, $min_{ls} = 4cs$, $max_{ls} = 7cs$,\\ $BWI = 0.1$, $g_{min}(\bm{\rho})  = 32.9$, $g_{max}(\bm{\rho})  = 38.7$}
		\label{fig:Eg2_test_minimum_ls_h}
	\end{subfigure}~~~~
	\begin{subfigure}[b]{.3\textwidth}
		\centering
		\captionsetup{font=scriptsize}
		\includegraphics[trim={2cm 2.5cm 2cm 2cm}, clip, scale = 0.5]{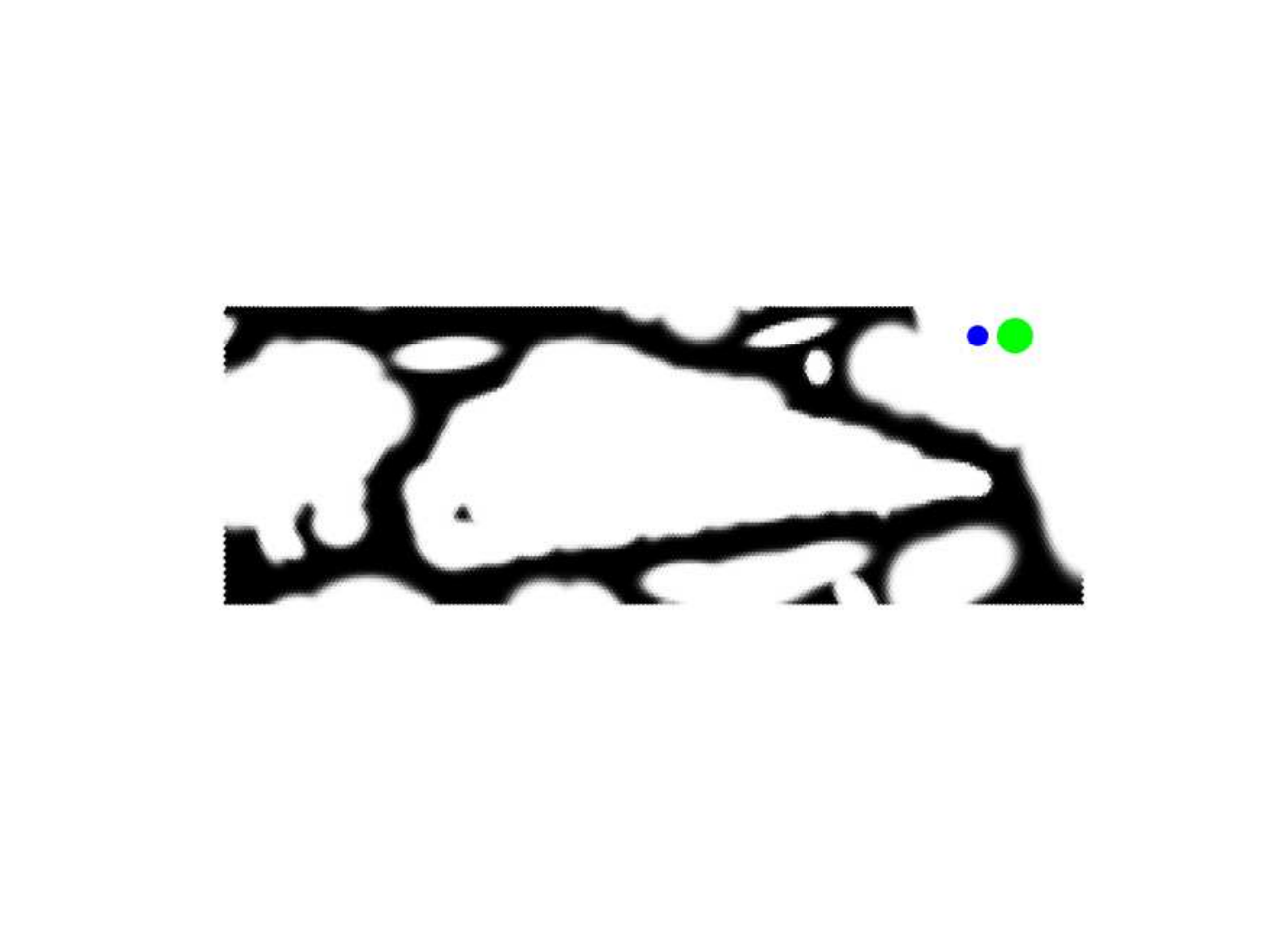}    
		\caption{ $V_{f} =0.40$, $min_{ls} = 4cs$, $max_{ls} = 7cs$,\\ $BWI = 0.11$, $g_{min}(\bm{\rho})  = 157.5$, $g_{max}(\bm{\rho})  = 150.5$}
		\label{fig:Eg2_test_minimum_ls_i}
	\end{subfigure}

	\begin{subfigure}[b]{.3\textwidth}
		\centering
		\captionsetup{font=scriptsize}
		\includegraphics[trim={2cm 2.5cm 2cm 2cm}, clip, scale = 0.5]{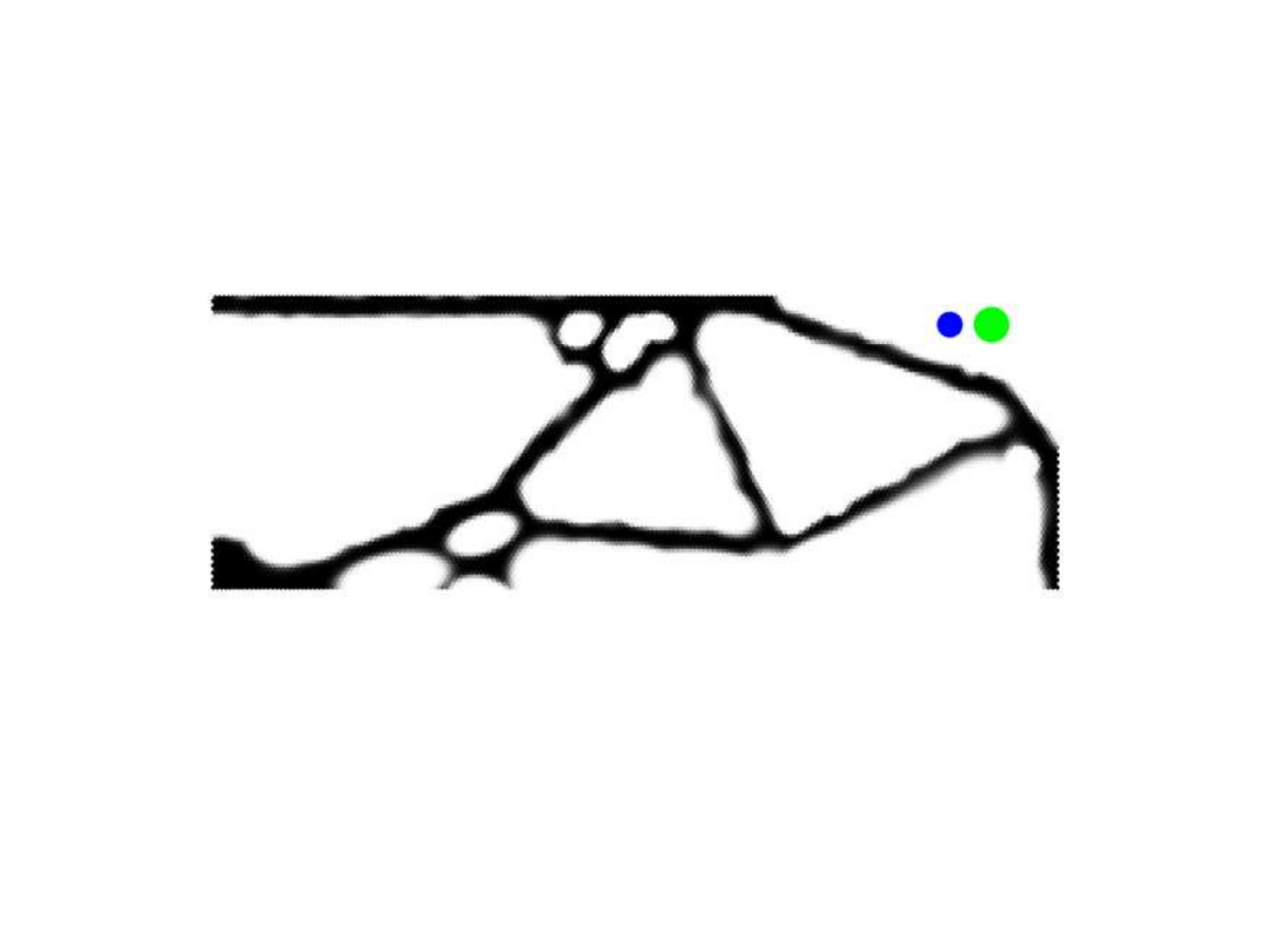}    
		\caption{ $V_{f} =0.20$, $min_{ls} = 5cs$, $max_{ls} = 7cs$,\\ $BWI = 0.10$, $g_{min}(\bm{\rho})  = 703.2$, $g_{max}(\bm{\rho})  = 27.2$}
		\label{fig:Eg2_test_minimum_ls_j}
	\end{subfigure}~~~~
	\begin{subfigure}[b]{.3\textwidth}
		\centering
		\captionsetup{font=scriptsize}
		\includegraphics[trim={2cm 2.5cm 2cm 2cm}, clip, scale = 0.5]{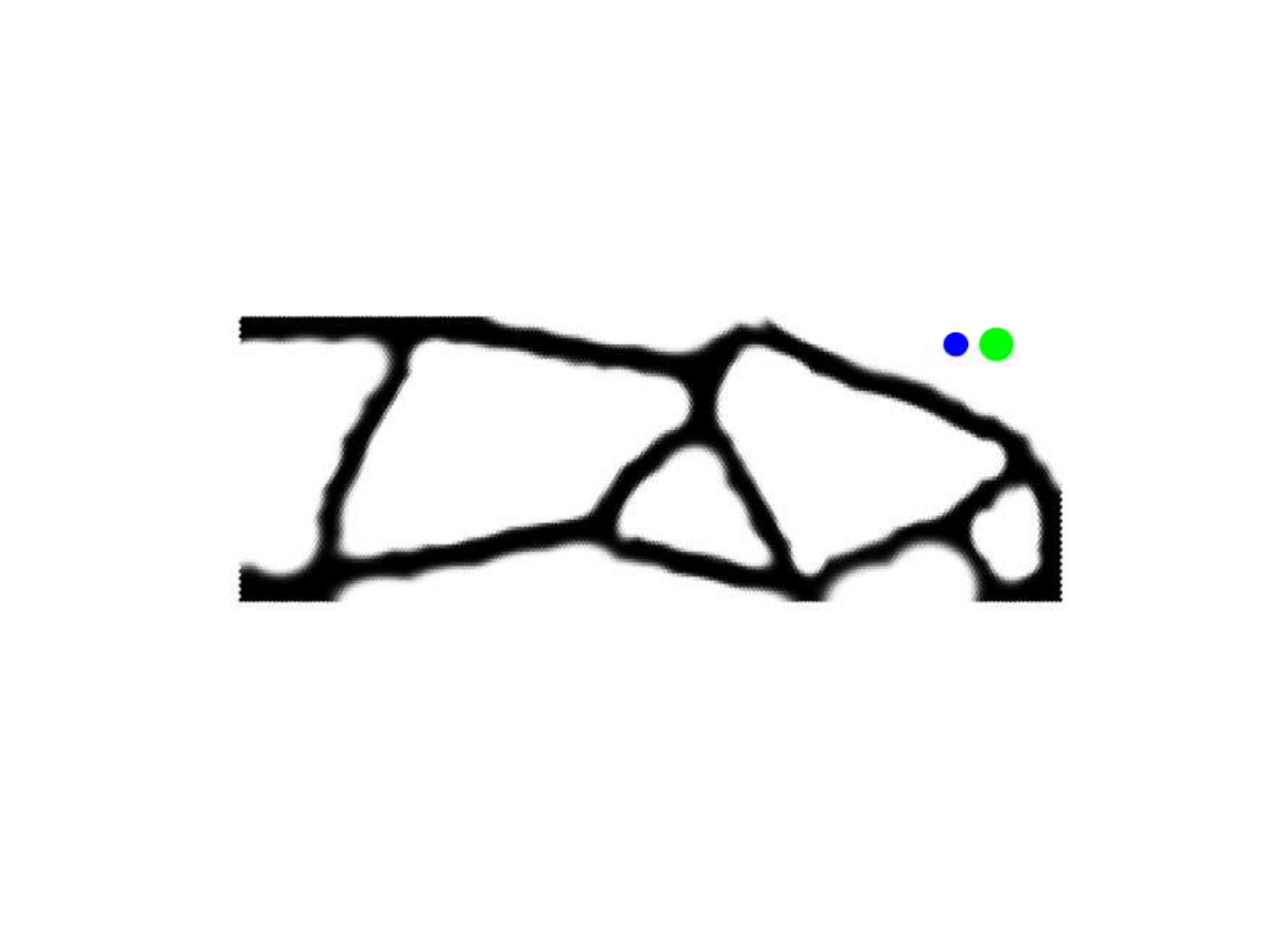}   
		\caption{ $V_{f} =0.30$, $min_{ls} = 5cs$, $max_{ls} = 7cs$, \\ $BWI = 0.11$, $g_{min}(\bm{\rho})  = 181.1$, $g_{max}(\bm{\rho})  = 29.3$}
		\label{fig:Eg2_test_minimum_ls_k}
	\end{subfigure}~~~~
	\begin{subfigure}[b]{.3\textwidth}
		\centering
		\captionsetup{font=scriptsize}
		\includegraphics[trim={2cm 2.5cm 2cm 2cm}, clip, scale = 0.5]{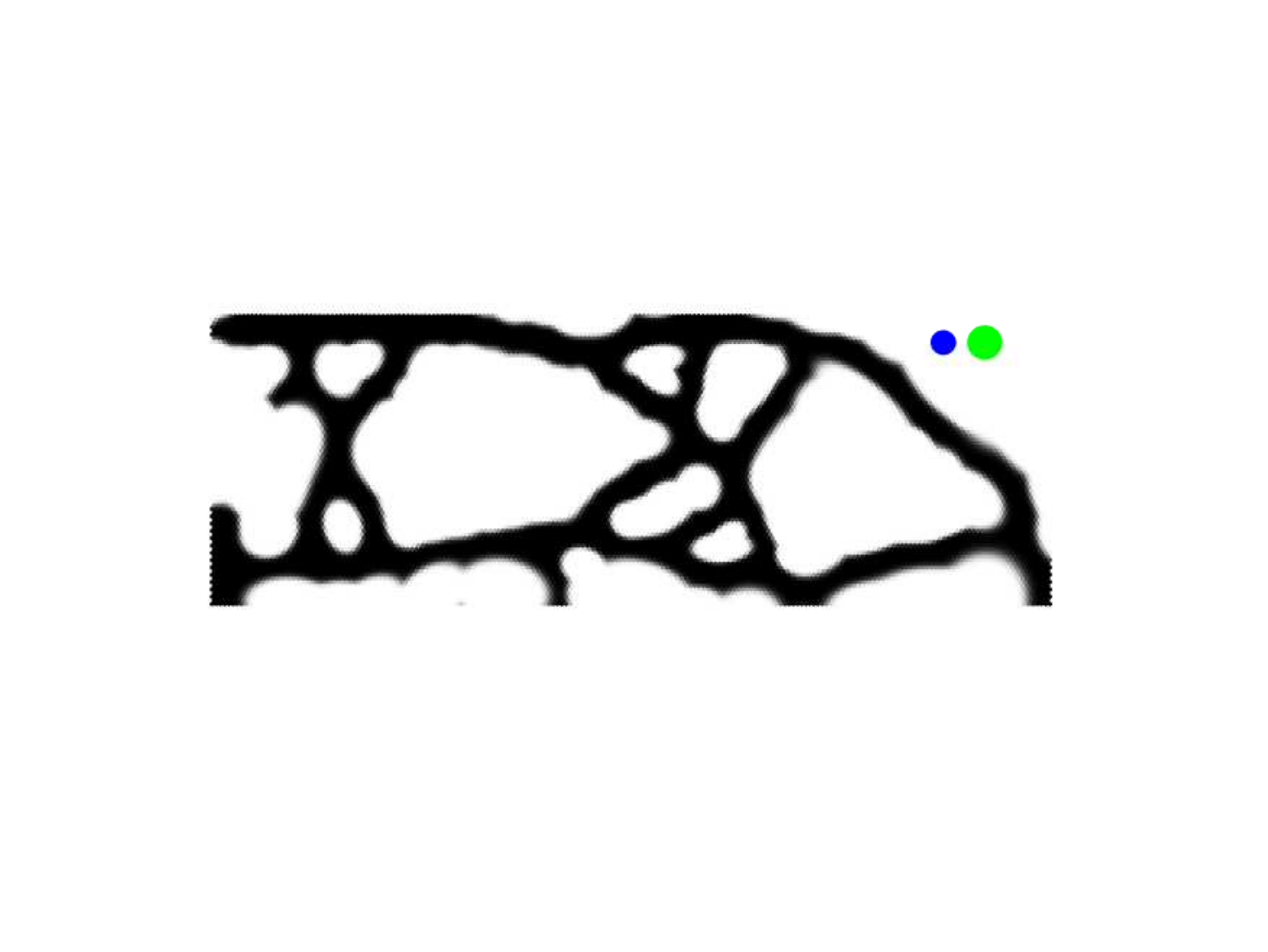}    
		\caption{ $V_{f} =0.40$, $min_{ls} = 5cs$, $max_{ls} = 7cs$, \\ $BWI = 0.08$, $g_{min}(\bm{\rho})  = 78.2$, $g_{max}(\bm{\rho})  = 51.3$}
		\label{fig:Eg2_test_minimum_ls_l}
	\end{subfigure}
	\caption{Topological Solutions for Example II: Solutions are obtained using negative circular masks as design variables. Domain of size $100 \times 46 \mbox{ unit}^2$ is discretized via 200 by 80 regular honeycomb mesh. Cell size (radius of the circumcircle) used is $cs = 0.288$ units. Both, minimum ($2 min_{ls}$) and maximum ($2 max_{ls}$)  length scales are imposed, as multiples of the cell size. For all solutions, $\alpha = 6$, $\eta = 3$, $p = 2$ (Eqs. \ref{min_max_ls}) and maximum number of function evaluations is 400. Values of the minimum ($g_{min}(\bm{\rho}) $) and maximum length scale constraints $g_{max}(\bm{\rho}) $ are depicted. The volume constraint ($g_1$) is $0$ or negative in all cases. Circles on top right represent the dimensions of the length scales imposed. }
	\label{fig:Eg2_test_min_max_ls}
\end{figure}

Fig. \ref{fig:Eg1_test_minimum_ls_g}-\ref{fig:Eg1_test_minimum_ls_i} represent solutions for the same specifications as Fig. \ref{fig:Eg1_test_minimum_ls_d}-\ref{fig:Eg1_test_minimum_ls_f} respectively, but for a lower volume fraction of $0.2$. High positive values of $g_1$ and $g_{min}(\bm{\rho}) $ imply that solutions do not satisfy the volume or minimum length scale constraints respectively. Members are not well formed, some having slight curvature and most having undulating contours with cells at the boundaries, mostly in gray states. Cells at the member boundaries, if attaining their filled states, will contribute to lowering of $g_{min}(\bm{\rho}) $ to a desirable value. But, this seems in direct conflict with the volume constraint $g_1$ as it will increase. Numerical investigations reveal for this example that optimization often converges to infeasible topologies even if maximum number of function evaluations is increased. Fig. \ref{fig:Eg1_test_minimum_ls_j}-\ref{fig:Eg1_test_minimum_ls_o} are solutions for the same specifications as Fig. \ref{fig:Eg1_test_minimum_ls_d}-\ref{fig:Eg1_test_minimum_ls_f} respectively but for higher values of the minimum length scale and volume fraction. While in the latter set (Fig. \ref{fig:Eg1_test_minimum_ls_d}-\ref{fig:Eg1_test_minimum_ls_f}), the volume and minimum length scale constraints are considered satisfied, in Figs.  \ref{fig:Eg1_test_minimum_ls_j}-\ref{fig:Eg1_test_minimum_ls_l}, values of $g_1$ and $g_{min}(\bm{\rho}) $ are high suggesting that increasing the minimum length scale for constant volume fraction may lead to infeasible solutions. One may expect that a higher volume fraction may help in achieving the minimum length scale, which is verified through the solution in Fig. \ref{fig:Eg1_test_minimum_ls_o}, especially, in comparison to that in Fig. \ref{fig:Eg1_test_minimum_ls_l}. Similar observation can be made by comparing the solutions in Figs. \ref{fig:Eg1_test_minimum_ls_j} and \ref{fig:Eg1_test_minimum_ls_k} to solutions in Figs. \ref{fig:Eg1_test_minimum_ls_m} and \ref{fig:Eg1_test_minimum_ls_n} respectively. The above suggests an implicit, conflicting dependence between constraints involving the minimum length scale and the volume fraction permitted.

Example II is a compliance minimization problem for a beam with roller supports allowing movement along the vertical axis (line of symmetry) at the left boundary, and a roller support allowing movement along the  horizontal axis at the right bottom corner. A force of 2 units along the negative vertical direction is applied at the left bottom corner node as illustrated in Fig. \ref{Fig:2:Eg2}.  Topologies for Example II in Fig. \ref{fig:Eg2_test_min_max_ls} are obtained over a domain of $100\times 46 \mbox{ unit}^2$ using a 200$\times$80 mesh with 20$\times$10 negative elliptical masks spread evenly over the domain for the initial guess, with $\alpha=6$ and $\eta=3$. A relatively high $\alpha$ is chosen to seek close to black and white solutions. Topologies for different volume fractions, minimum and maximum length scales are presented. The arrangement in Fig. \ref{fig:Eg2_test_min_max_ls} is such that all solutions in a row have the same length scales while those in the same column have the same volume fraction.  $g_{min}(\bm{\rho}) $ and $g_{max}(\bm{\rho}) $ represent the final minimum and maximum length scale values respectively. The maximum length scale is held constant at $2max_{ls} = 2 \times 7 cs$, where $cs$ is the \textit{cell size} given by radius of the circumscribing circle of the hexagonal cell, while the minimum length scale is increased as one moves down the column.

Fig. \ref{fig:Eg2_test_minimum_ls_a}-\ref{fig:Eg2_test_minimum_ls_c} are solutions obtained with only the volume fraction specified, and no length scale imposed. Increase in volume fraction leads to increase in member thickness and/or addition of new members (holes) to the solution. Members are straight and well formed though their thicknesses vary, as expected, since no explicit control  is imposed on them. Figs. \ref{fig:Eg2_test_minimum_ls_d}-\ref{fig:Eg2_test_minimum_ls_f} present solutions for the same specifications as for Figs. \ref{fig:Eg2_test_minimum_ls_a}-\ref{fig:Eg2_test_minimum_ls_c} respectively but with the imposition of minimum length scale constraint of $2\times 3cs$ and maximum length scale constraints of $2\times 7cs$ where $cs$, the cell size, is $0.288$ units. For the solution in Fig.  \ref{fig:Eg2_test_minimum_ls_d}, both minimum and maximum length scales are (close to) satisfied. For those in Figs. \ref{fig:Eg2_test_minimum_ls_e} and \ref{fig:Eg2_test_minimum_ls_f}, $g_{min}(\bm{\rho}) $ and/or $g_{max}(\bm{\rho}) $ is relatively high. Further, more members, undulating contours, dangling appendages, and local islands appear with increase in the specified volume fraction. These appendages and islands have negligible strain energy densities,  and do not contribute to the stiffness of the continua. Therefore, the obtained structures are sub-optimal.  Comparing solutions obtained with a volume fraction of $vf = 0.2$ (Figs. \ref{fig:Eg2_test_minimum_ls_d}, \ref{fig:Eg2_test_minimum_ls_g} and \ref{fig:Eg2_test_minimum_ls_j}), with increase in the minimum length scale, values of $g_{min}(\bm{\rho}) $ increase. All solutions are free from appendages and islands. An increase in $vf$ may not always result in a drop of $g_{min}(\bm{\rho}) $ as one would expect but may also lead to higher number of branches in the solution and a higher $g_{min}(\bm{\rho}) $. This can be observed by comparing solutions in Figs. \ref{fig:Eg2_test_minimum_ls_d} and \ref{fig:Eg2_test_minimum_ls_h} to solutions in Figs. \ref{fig:Eg2_test_minimum_ls_e} and \ref{fig:Eg2_test_minimum_ls_i} respectively. Both, the volume and explicit minimum length scale constraints seem to be in conflict in that a low volume fraction may not help in achieving the minimum length scale while a high fraction could lead to suboptimal solutions with connectivity degeneracies like the appendages and/or islands. 

Most solutions in Figures \ref{fig:Eg1_test_minimum_ls}-\ref{fig:Eg2_test_min_max_ls} are sub-optimal possibly because of imposition of the minimum/ maximum length scale constraints on illformed skeletons, which change continuously throughout optimization thereby hindering the removal of unnecessary branches/appendages from the solution. This, along with boundary undulations, makes it difficult for the optimization process to converge to  better solutions. One way to address the issue is to allow the development of a primitive, well formed skeleton before imposing length scale constraints. This notion is adopted in the construction of Stage I of the proposed methodology. \\

\subsection{Results with the SLS methodology}
\label{main_results} 

\begin{figure}[H]
\hspace{5mm}
	\begin{subfigure}[b]{.45\textwidth}
		\centering
		\captionsetup{font=scriptsize}
		\includegraphics[trim={2cm 3.5cm 2cm 3cm}, clip, scale = 0.4]{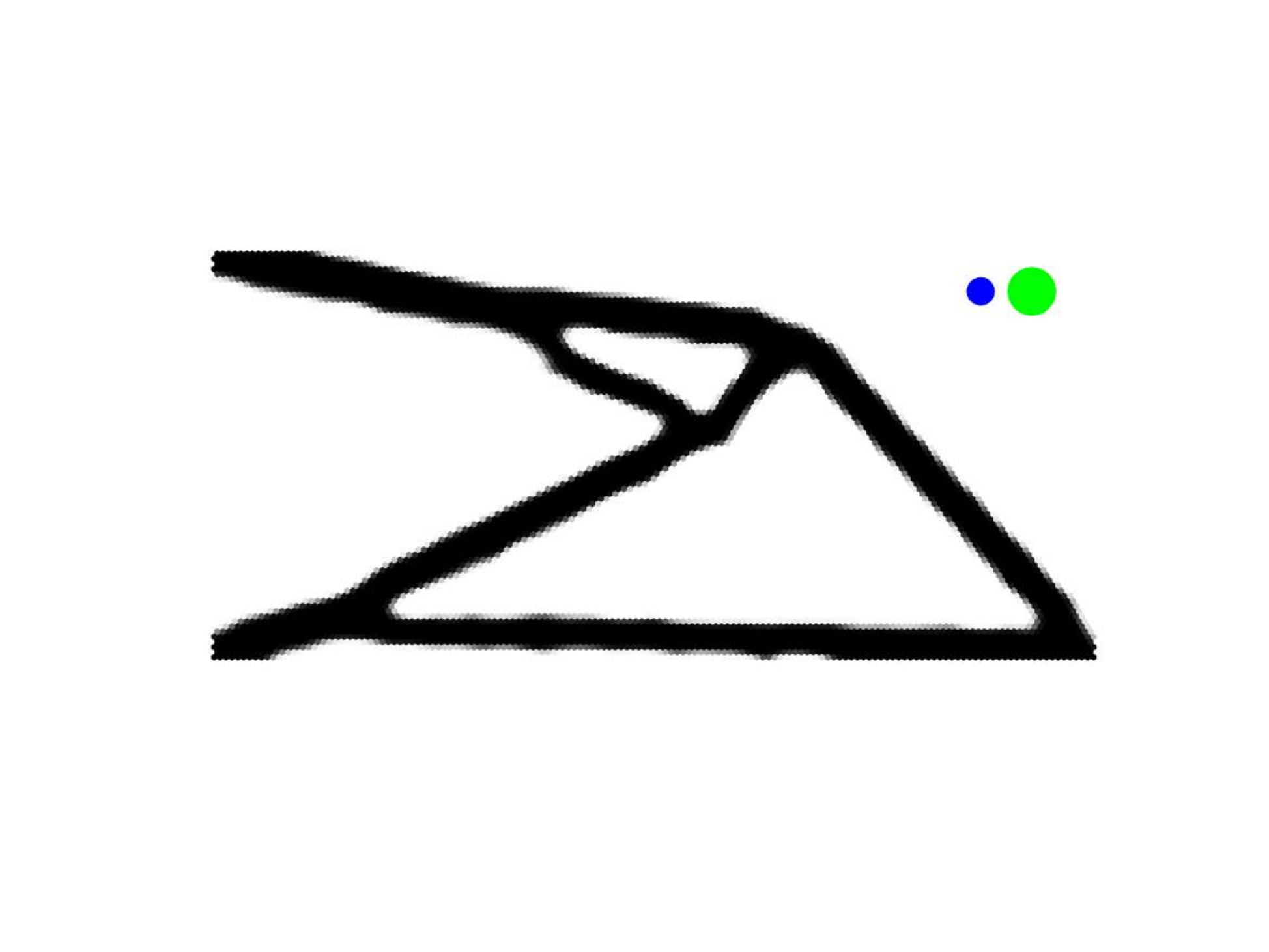}
		\caption{Example I:  $cs = 0.38$ units. $min_{ls} = 4cs$ units, $max_{ls} = 7cs$ units. Post optimization, $\Phi = 752.5$, $g_{min}(\bm{\rho})  = 38.9$, $g_{max}(\bm{\rho})  = 32.2$, 
	       $vf = 0.28$; $BWI = 0.05$.}
\label{fig:Eg1_Eg2_Eg3_Eg4_new_logic_NEMa}
	\end{subfigure}
	\hspace{10mm}		
\begin{subfigure}[b]{.45\textwidth}
		\centering
		\captionsetup{font=scriptsize}
		\includegraphics[trim={3cm 4.5cm 2cm 3cm}, clip, scale = 0.45]{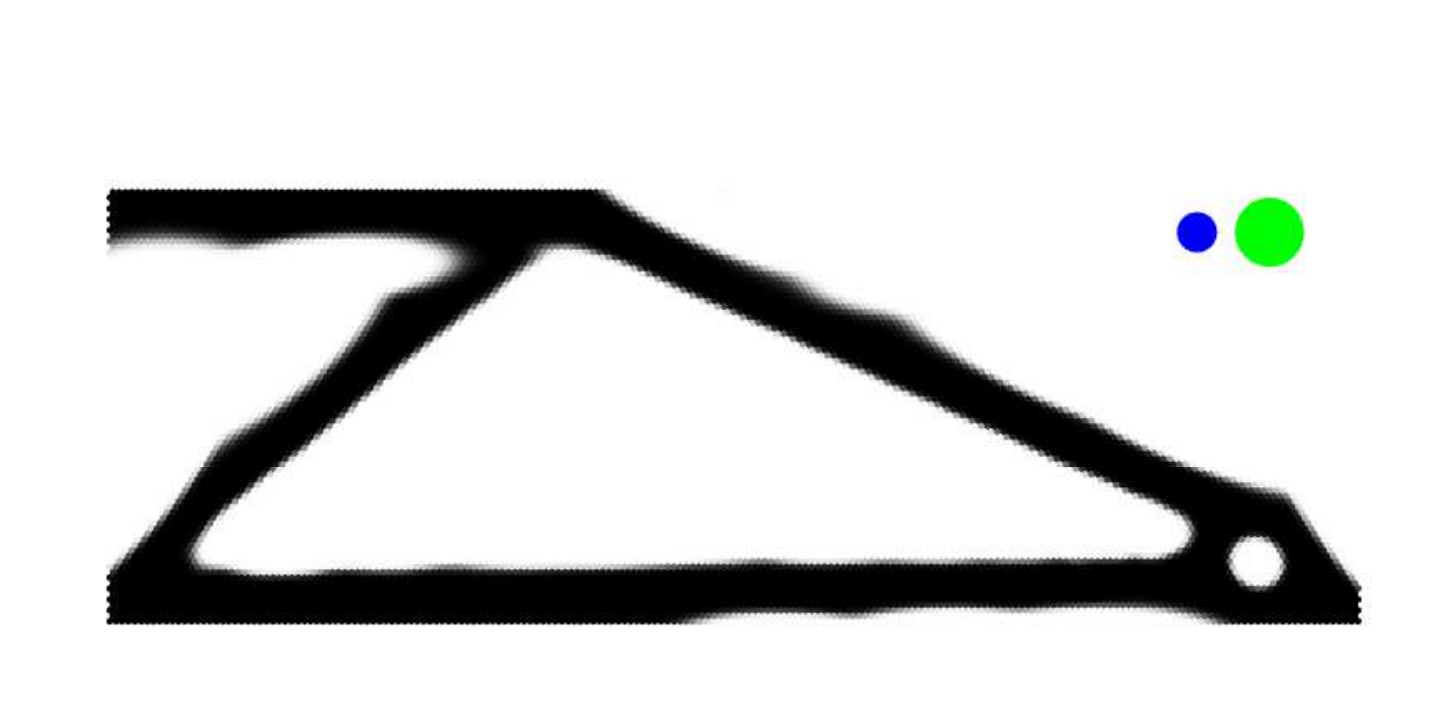}  
		\caption{Example II:  $cs = 0.28$ units. $min_{ls} = 4cs$ units, $max_{ls} = 7cs$ units. Post optimization,
	$\Phi = 1462.7$, $g_{min}(\bm{\rho})  =  81.8$, $g_{max}(\bm{\rho})  = 78.8$, 
	$vf = 0.33$; $BWI = 0.04$.}
\label{fig:Eg1_Eg2_Eg3_Eg4_new_logic_NEMb}
	\end{subfigure}

\hspace{5mm}		
	\begin{subfigure}[b]{.45\textwidth}
		\centering
		\captionsetup{font=scriptsize}
		\includegraphics[trim={3cm 4cm 2cm 3cm}, clip, scale = 0.45]{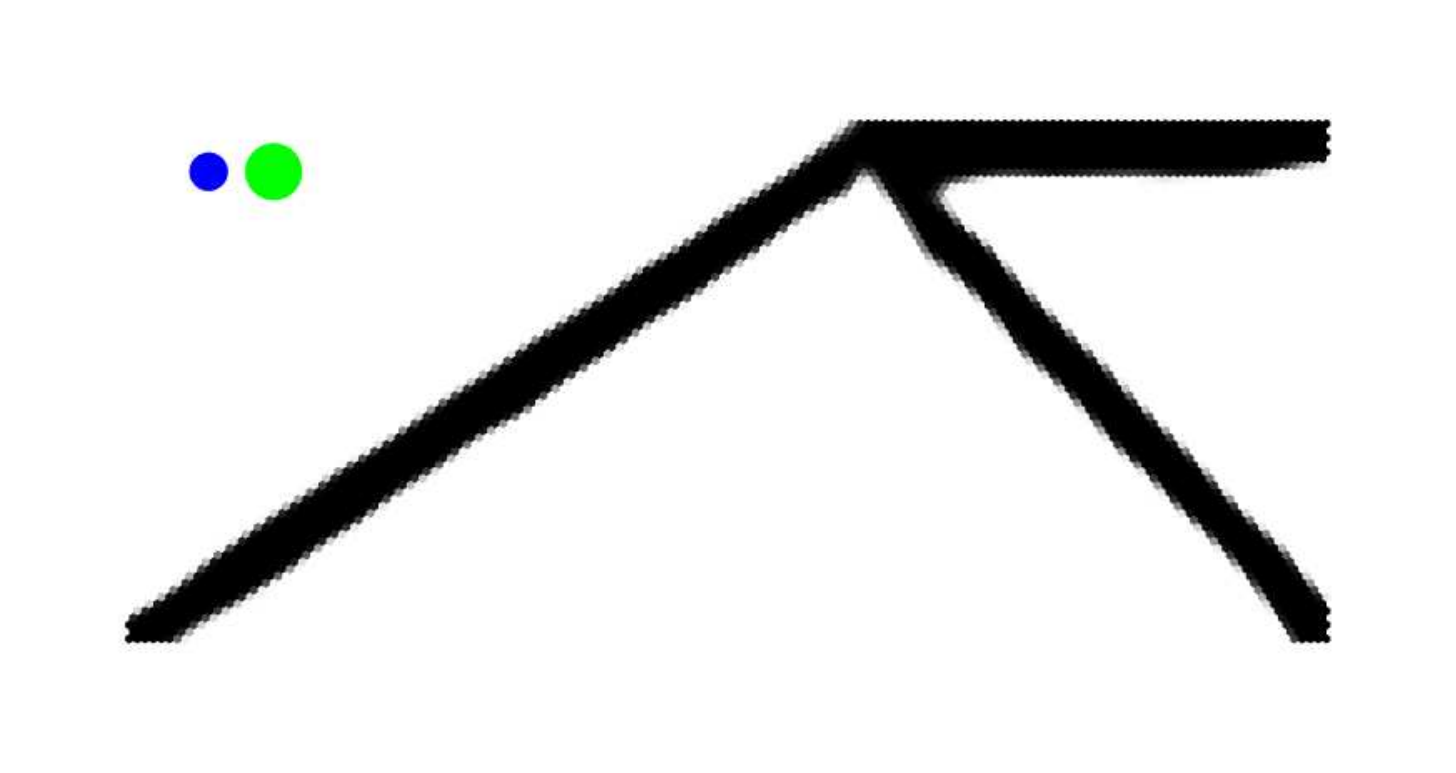}  
		\caption{Example III:  $cs = 0.38$ units. $min_{ls} = 4cs$ units, $max_{ls} = 6cs$ units. Post optimization,
	$\Phi = -0.174$, $g_{min}(\bm{\rho})  = 22.6$, $g_{max}(\bm{\rho})  = 13.4$, 
	$vf = 0.17$;  $BWI = 0.03$.}
\label{fig:Eg1_Eg2_Eg3_Eg4_new_logic_NEMc}
	\end{subfigure}	
	\hspace{10mm}			
\begin{subfigure}[b]{.45\textwidth}
		\centering
		\captionsetup{font=scriptsize}
		\includegraphics[trim={2cm 4cm 2cm 3cm}, clip, scale = 0.45]{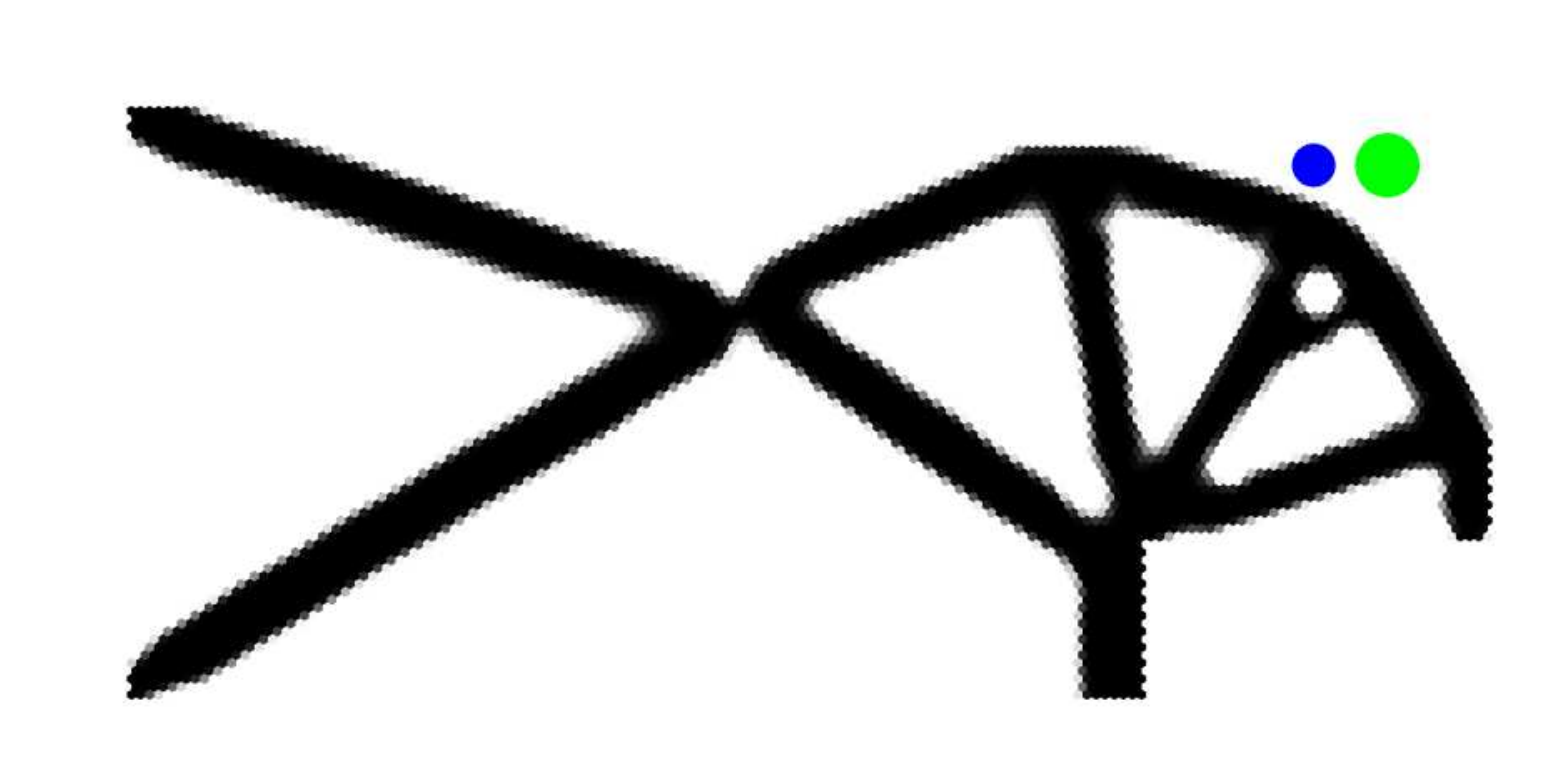}
		\caption{Example IV:  $cs = 0.38$ units. $min_{ls} = 4cs$ units, $max_{ls} = 6cs$ units. Post optimization,
	$\Phi = -0.077$, $g_{min}(\bm{\rho})  = 64.4$, $g_{max}(\bm{\rho})  = 35.6$, 
	$vf = 0.29$; $BWI = 0.05$.}
\label{fig:Eg1_Eg2_Eg3_Eg4_new_logic_NEMd}
	\end{subfigure}
\caption{Topologies generated with Negative Elliptical Masks with the methodology in Section \ref{method_TO}. Circles (blue/green) in the inset represent the (minimum/maximum) length scales }
	\label{fig:Eg1_Eg2_Eg3_Eg4_new_logic_NEM}
\end{figure}

\begin{figure}[H]
\hspace{15mm}
	\begin{subfigure}[b]{.3\textwidth}
		\centering
		\captionsetup{font=scriptsize} 
\includegraphics[trim={2cm 2.5cm 2cm 2cm}, clip, scale = 0.6]{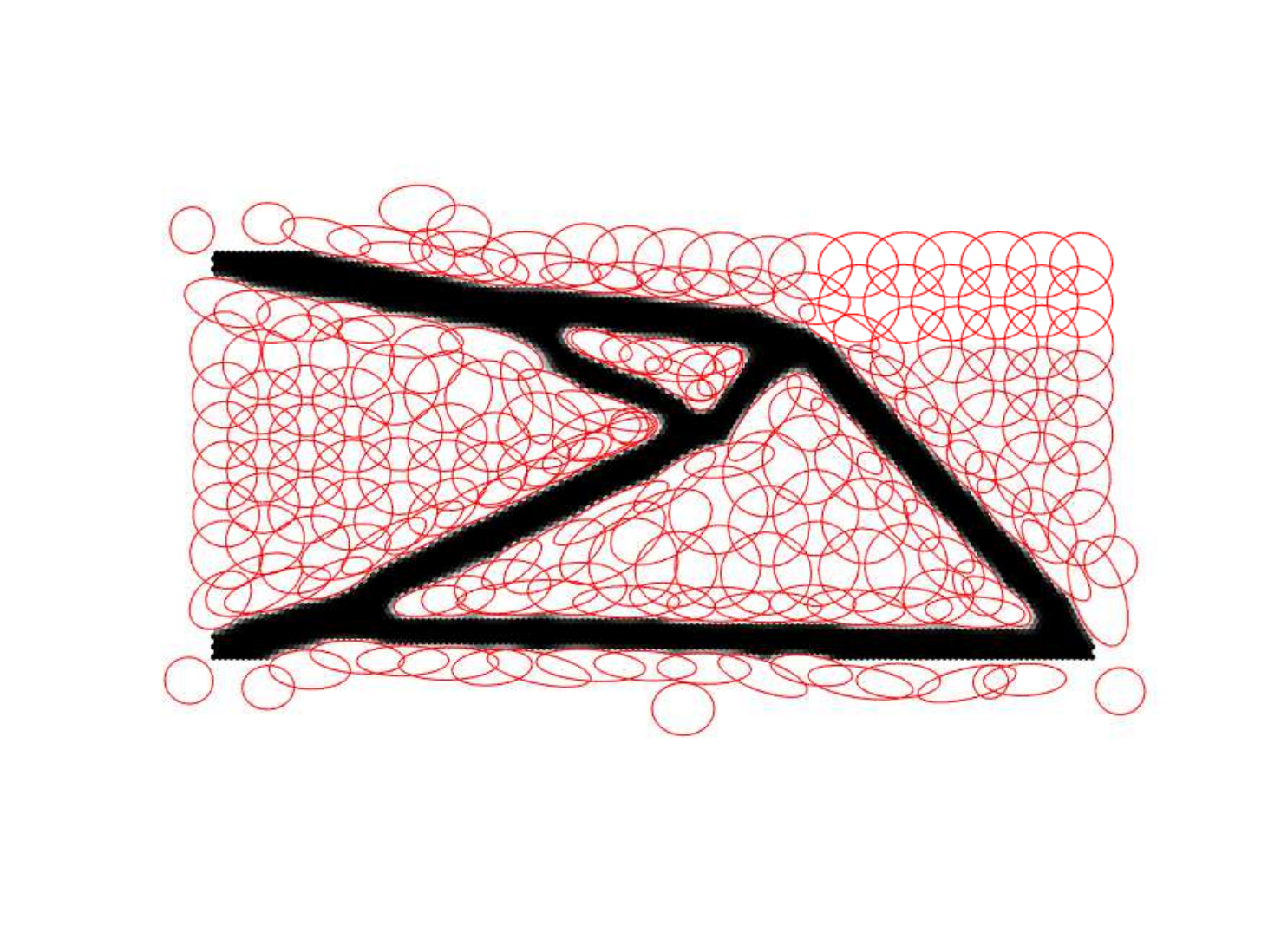} 
		\caption{}
\label{fig:Eg1_Eg2_Eg3_Eg4_new_logic_NEM_with_masksa}
	\end{subfigure}
	\hspace{30mm}		
\begin{subfigure}[b]{.3\textwidth}
		\centering
		\captionsetup{font=scriptsize}
\includegraphics[trim={2cm 2.5cm 2cm 2cm}, clip, scale = 0.6]{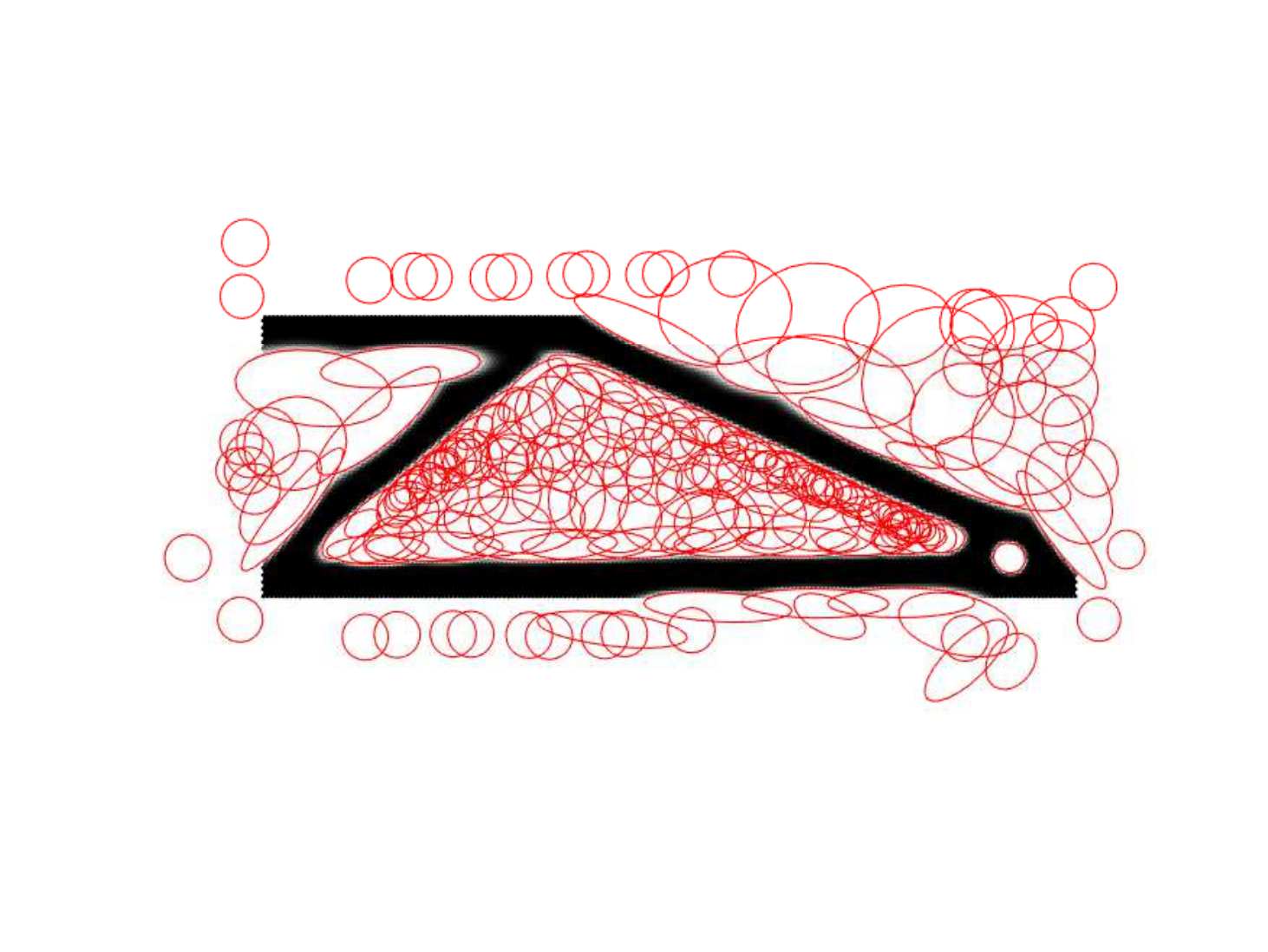} 
		\caption{ }
\label{fig:Eg1_Eg2_Eg3_Eg4_new_logic_NEM_with_masksb}
	\end{subfigure}

\hspace{15mm}		
	\begin{subfigure}[b]{.3\textwidth}
		\centering
		\captionsetup{font=scriptsize} 
\includegraphics[trim={2cm 2.5cm 2cm 2cm}, clip, scale = 0.6]{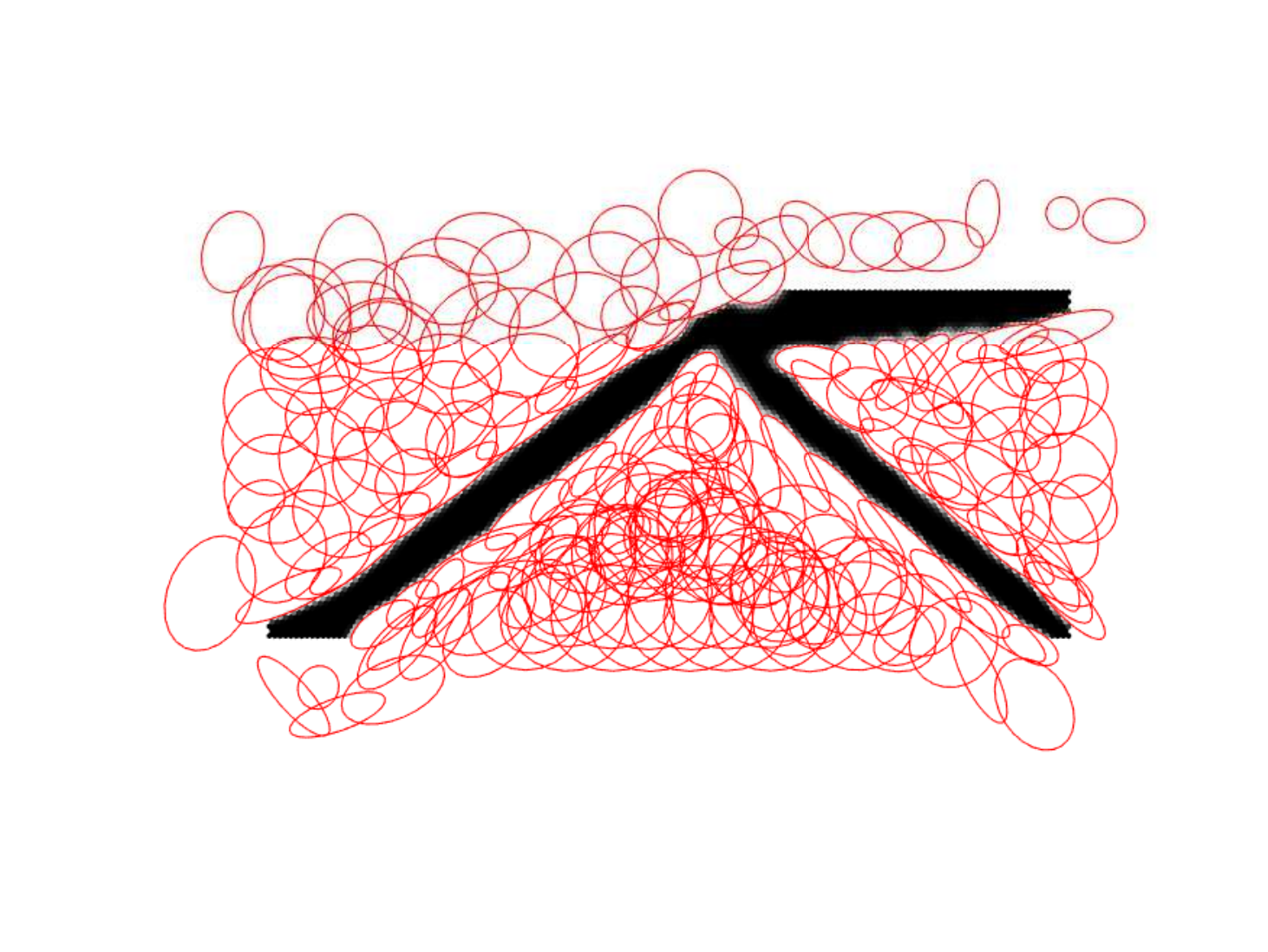} 
		\caption{ }
\label{fig:Eg1_Eg2_Eg3_Eg4_new_logic_NEM_with_masksc}
	\end{subfigure}	
	\hspace{30mm}			
\begin{subfigure}[b]{.3\textwidth}
		\centering
		\captionsetup{font=scriptsize}
\includegraphics[trim={2cm 2.5cm 2cm 2cm}, clip, scale = 0.6]{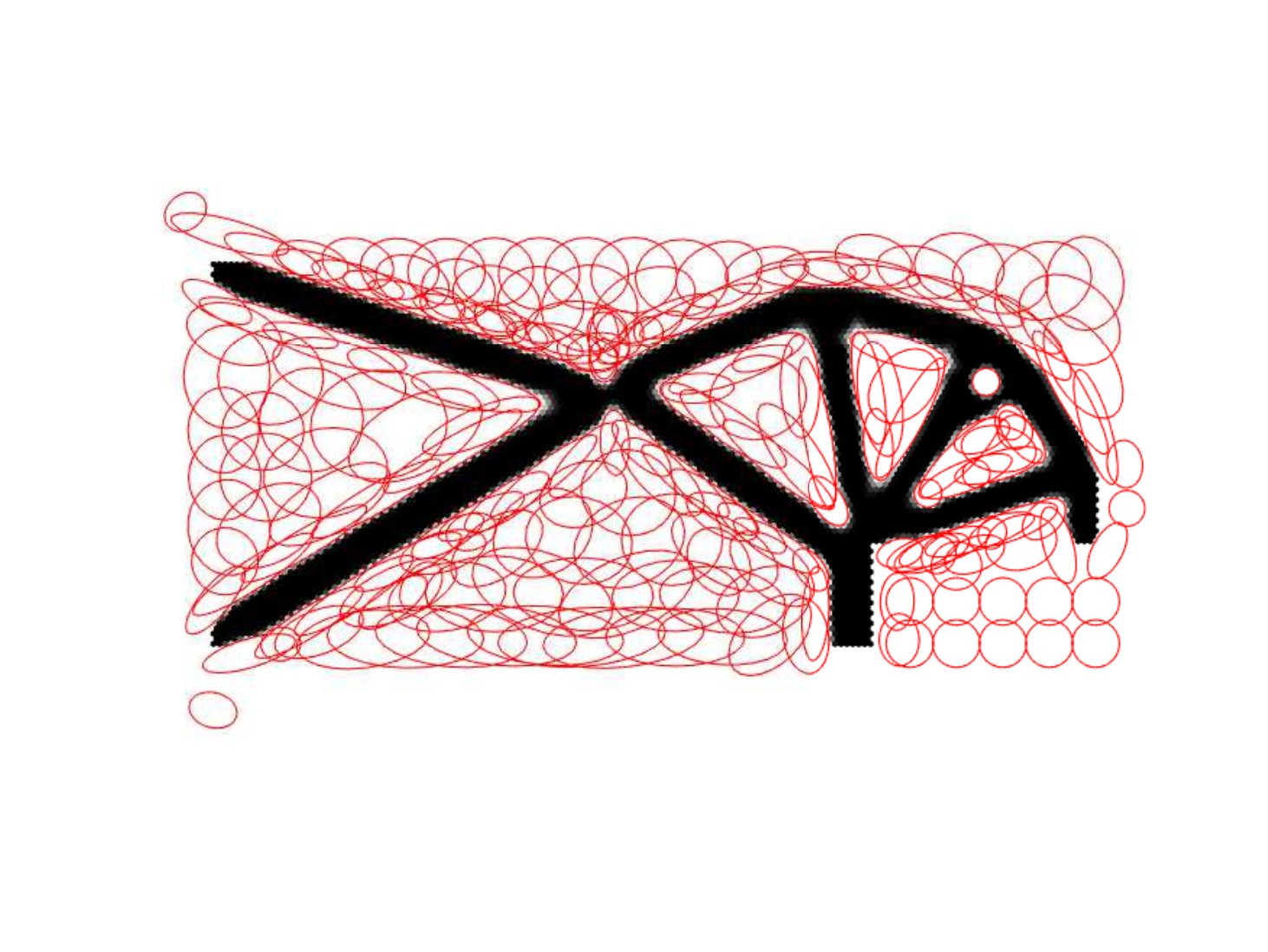} 
		\caption{ }
\label{fig:Eg1_Eg2_Eg3_Eg4_new_logic_NEM_with_masksd}
	\end{subfigure}
\caption{Respective solutions in Fig. \ref{fig:Eg1_Eg2_Eg3_Eg4_new_logic_NEM} depicted with Negative masks}
	\label{fig:Eg1_Eg2_Eg3_Eg4_new_logic_NEM_with_masks}
\end{figure}

\begin{figure}[H]
\hspace{5mm}
	\begin{subfigure}[b]{.4\textwidth}
		\centering
		\captionsetup{font=scriptsize}
		\includegraphics[trim={2cm 3.5cm 2cm 3cm}, clip, scale = 0.45]{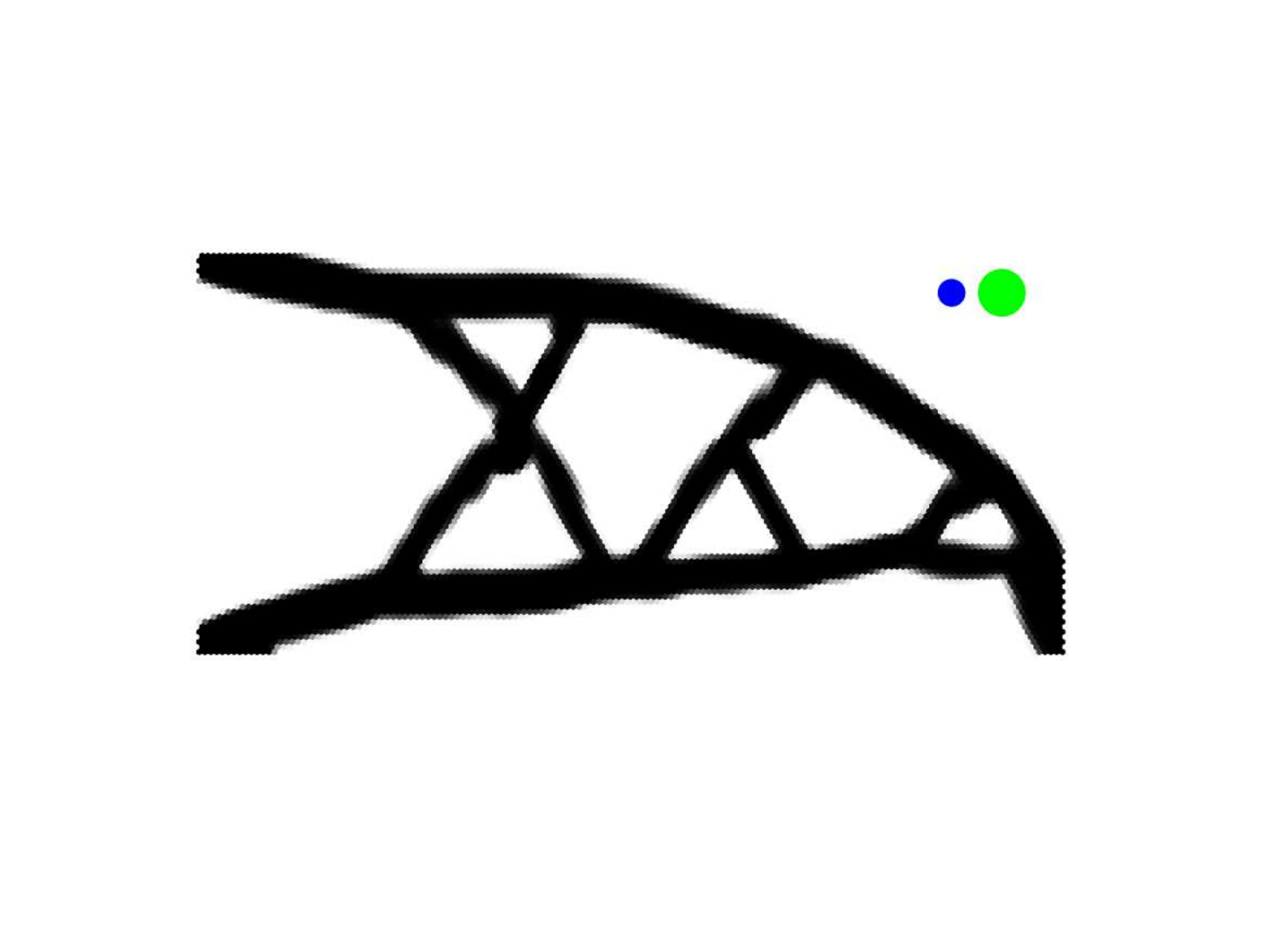} 
		\caption{Example I:  $cs = 0.38$ units. $min_{ls} = 4cs$ units, $max_{ls} = 7cs$ units. Post optimization,
	$\Phi = 654.1$, $g_{min}(\bm{\rho})  = 37.4$, $g_{max}(\bm{\rho})  = 35.3$, 
	$vf = 0.35$;  $BWI = 0.06$.}
\label{fig:Eg1_Eg2_Eg3_Eg4_new_logic_PEMa}
	\end{subfigure}
	\hspace{10mm}
	\begin{subfigure}[b]{.4\textwidth}
		\centering
		\captionsetup{font=scriptsize}
		\includegraphics[trim={2cm 3.5cm 2cm 3cm}, clip, scale = 0.45]{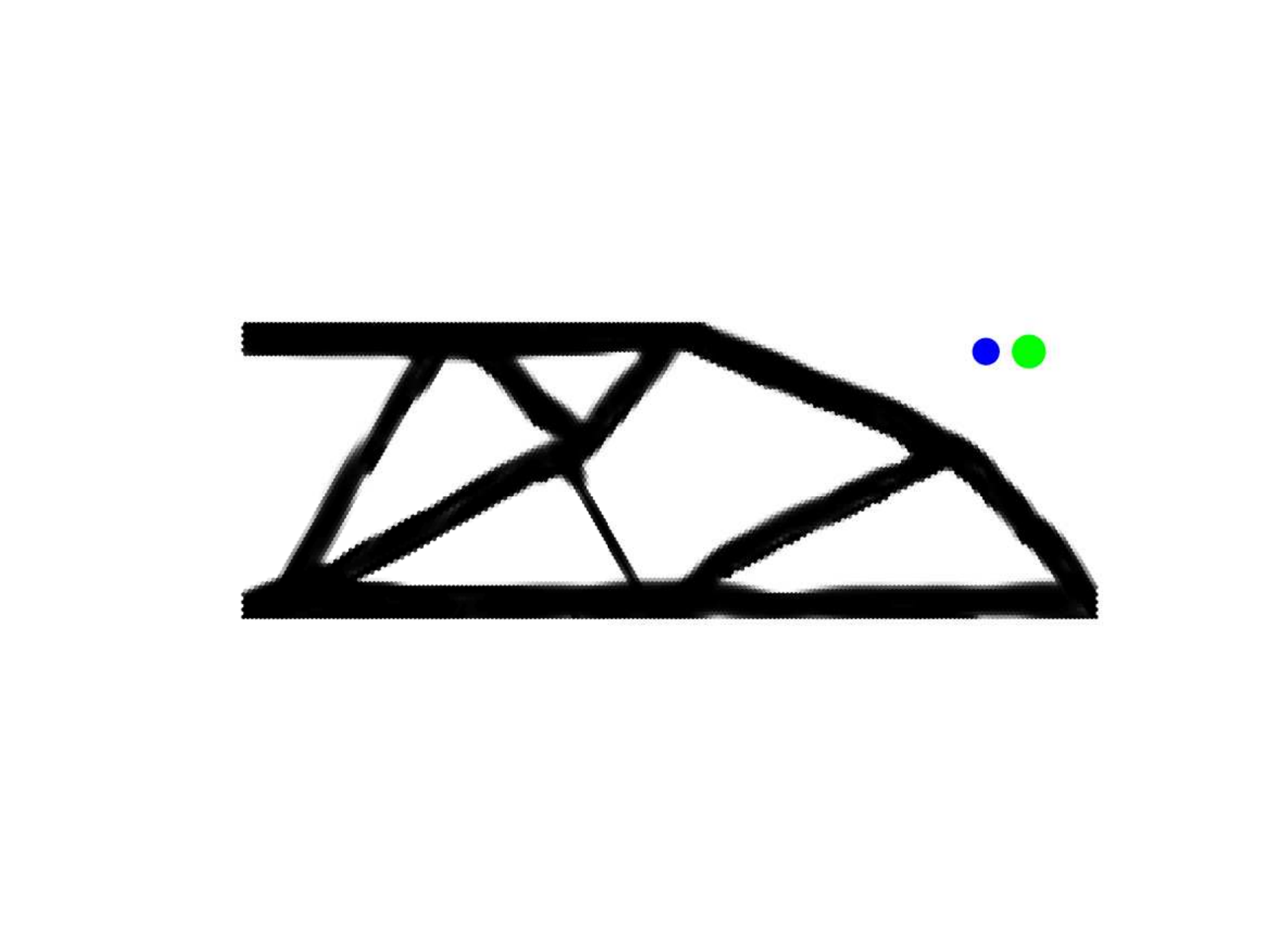} 
		\caption{Example II:  $cs = 0.28$ units. $min_{ls} = 4cs$ units, $max_{ls} = 5cs$ units. Post optimization,
	$\Phi = 1075.9$, $g_{min}(\bm{\rho})  = 296.4$, $g_{max}(\bm{\rho})  = 214.6$, 
	$vf = 0.4$;  $BWI = 0.06$.}
\label{fig:Eg1_Eg2_Eg3_Eg4_new_logic_PEMb}
	\end{subfigure}
	
\hspace{5mm}	
\begin{subfigure}[b]{.4\textwidth}
		\centering
		\captionsetup{font=scriptsize}
		\includegraphics[trim={2cm 4cm 2cm 3cm}, clip, scale = 0.45]{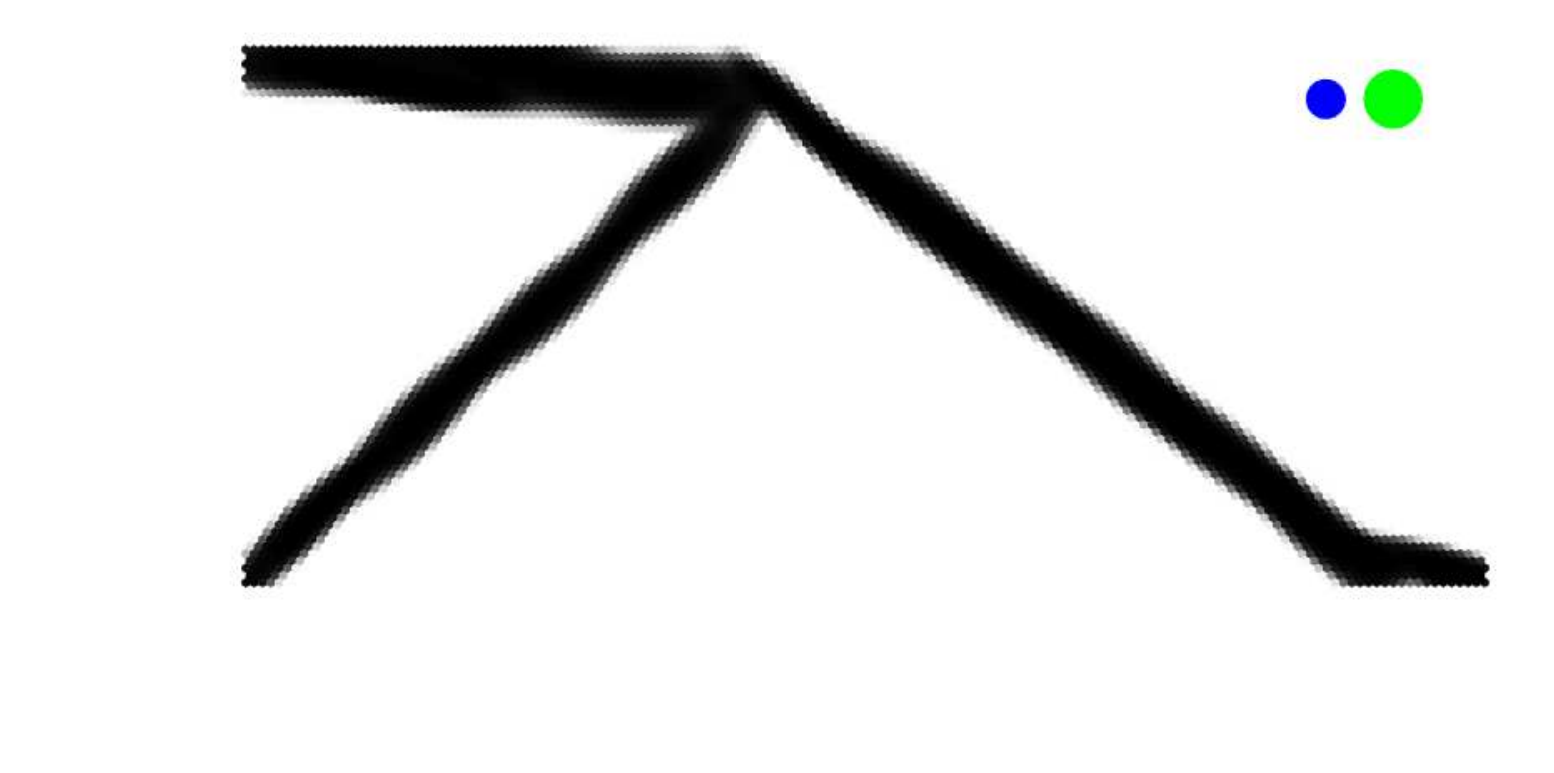}
		\caption{Example III:  $cs = 0.38$ units. $min_{ls} = 4cs$ units, $max_{ls} = 6cs$ units. Post optimization,
	$\Phi = -0.181$, $g_{min}(\bm{\rho})  = 55.9$, $g_{max}(\bm{\rho})  = 34.5$, 
	$vf = 0.18$;  $BWI = 0.05$.}
\label{fig:Eg1_Eg2_Eg3_Eg4_new_logic_PEMc}
	\end{subfigure}
	\hspace{10mm}
	\begin{subfigure}[b]{.4\textwidth}
		\centering
		\captionsetup{font=scriptsize}
		\includegraphics[trim={2cm 3.5cm 2cm 3cm}, clip, scale = 0.45]{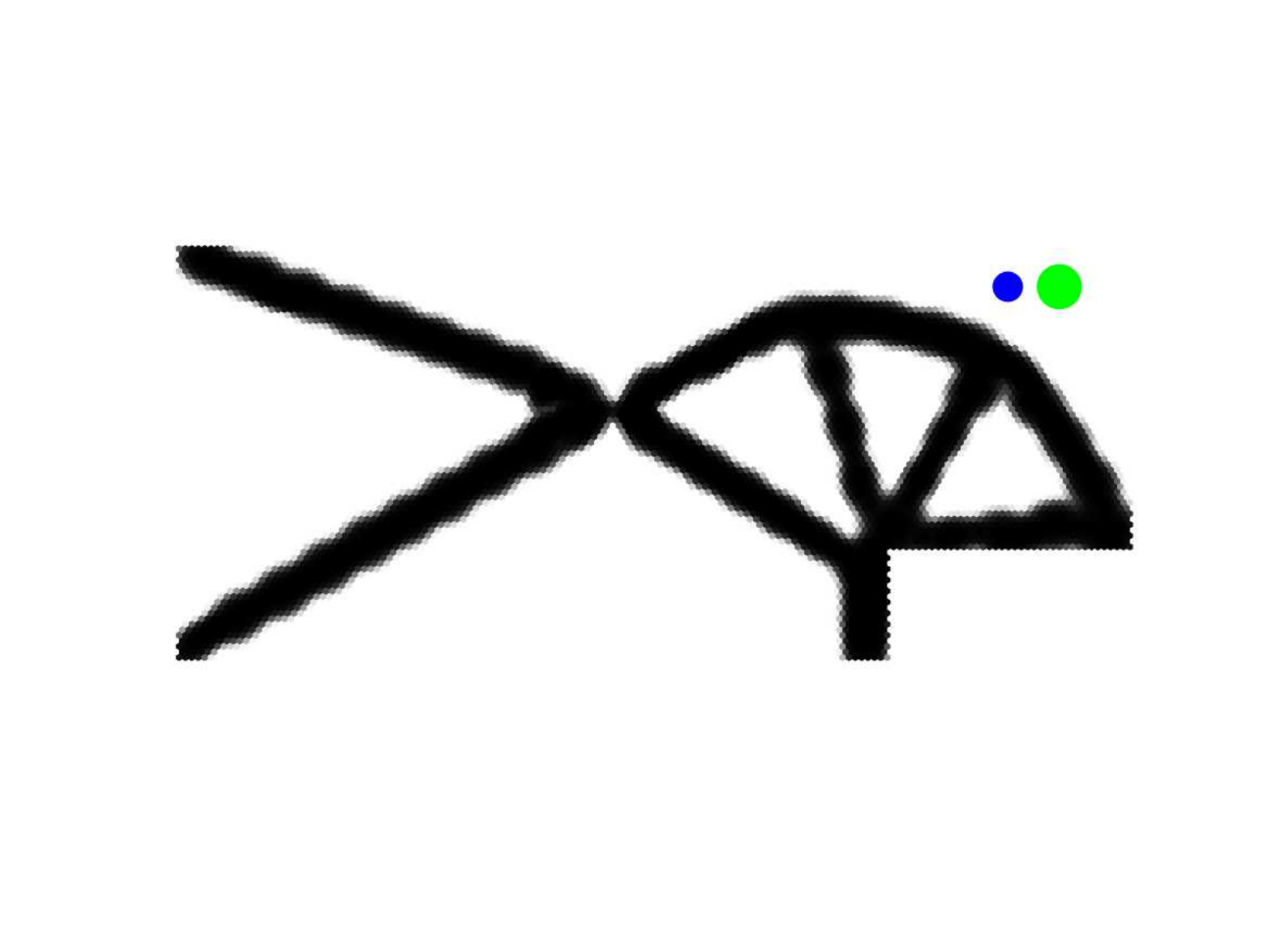}
		\caption{Example IV:  $cs = 0.38$ units. $min_{ls} = 4cs$ units, $max_{ls} = 6cs$ units. Post optimization,
	$\Phi = -0.079$, $g_{min}(\bm{\rho})  = 61.6$, $g_{max}(\bm{\rho})  = 161.7$, 
	$vf = 0.31$;  $BWI = 0.09$.}
\label{fig:Eg1_Eg2_Eg3_Eg4_new_logic_PEMd}
	\end{subfigure}
	\caption{Topologies generated with Positive Elliptical Masks with the methodology in Section \ref{method_TO}. Circles (blue/green) in the inset represent the (minimum/maximum) length scales }
	\label{fig:Eg1_Eg2_Eg3_Eg4_new_logic_PEM}
\end{figure}

\begin{figure}[H]
\hspace{15mm}
	\begin{subfigure}[b]{.3\textwidth}
		\centering
		\captionsetup{font=scriptsize}
		\includegraphics[trim={1.8cm 2.5cm 1cm 2cm}, clip, scale = 0.55]{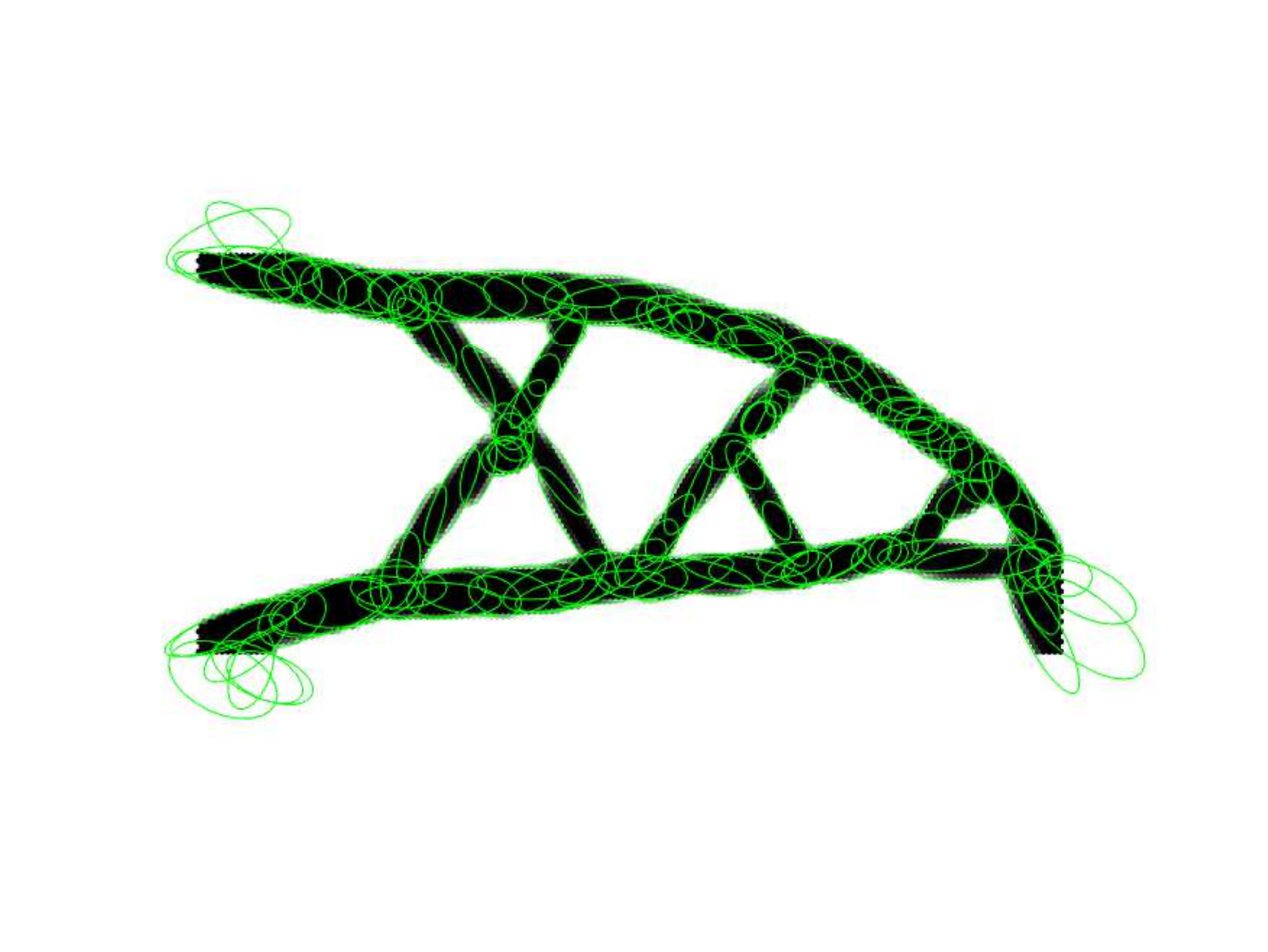} 
		\caption{ }
\label{fig:Eg1_Eg2_Eg3_Eg4_new_logic_PEM_with_masksa}
	\end{subfigure}
	\hspace{30mm}
	\begin{subfigure}[b]{.3\textwidth}
		\centering
		\captionsetup{font=scriptsize}
		\includegraphics[trim={2cm 2.5cm 1cm 2cm}, clip, scale = 0.55]{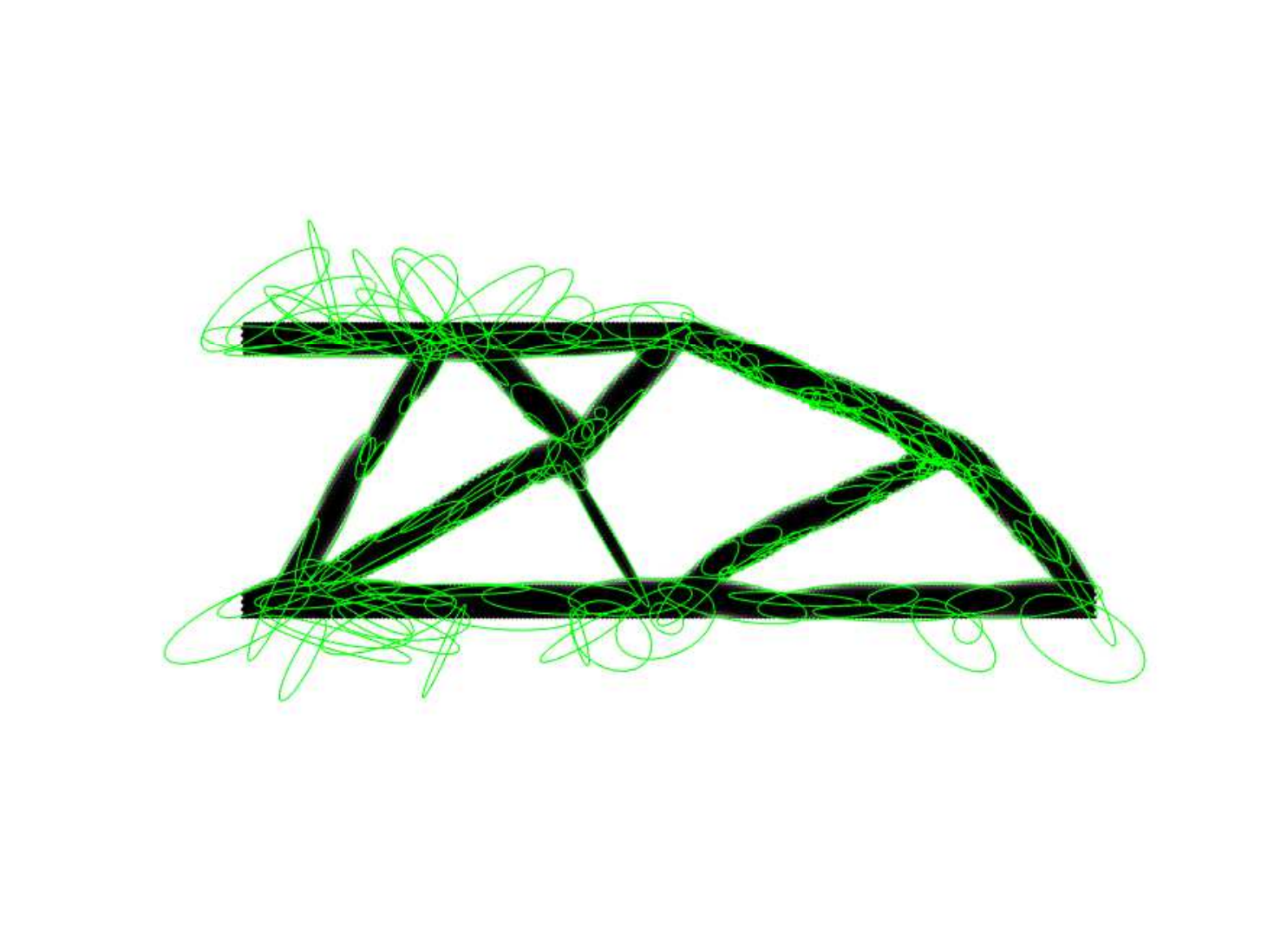} 
		\caption{ }
\label{fig:Eg1_Eg2_Eg3_Eg4_new_logic_PEM_with_masksb}
	\end{subfigure}

\hspace{15mm}	
\begin{subfigure}[b]{.3\textwidth}
		\centering
		\captionsetup{font=scriptsize}
		\includegraphics[trim={1.8cm 2.5cm 1cm 2cm}, clip, scale = 0.55]{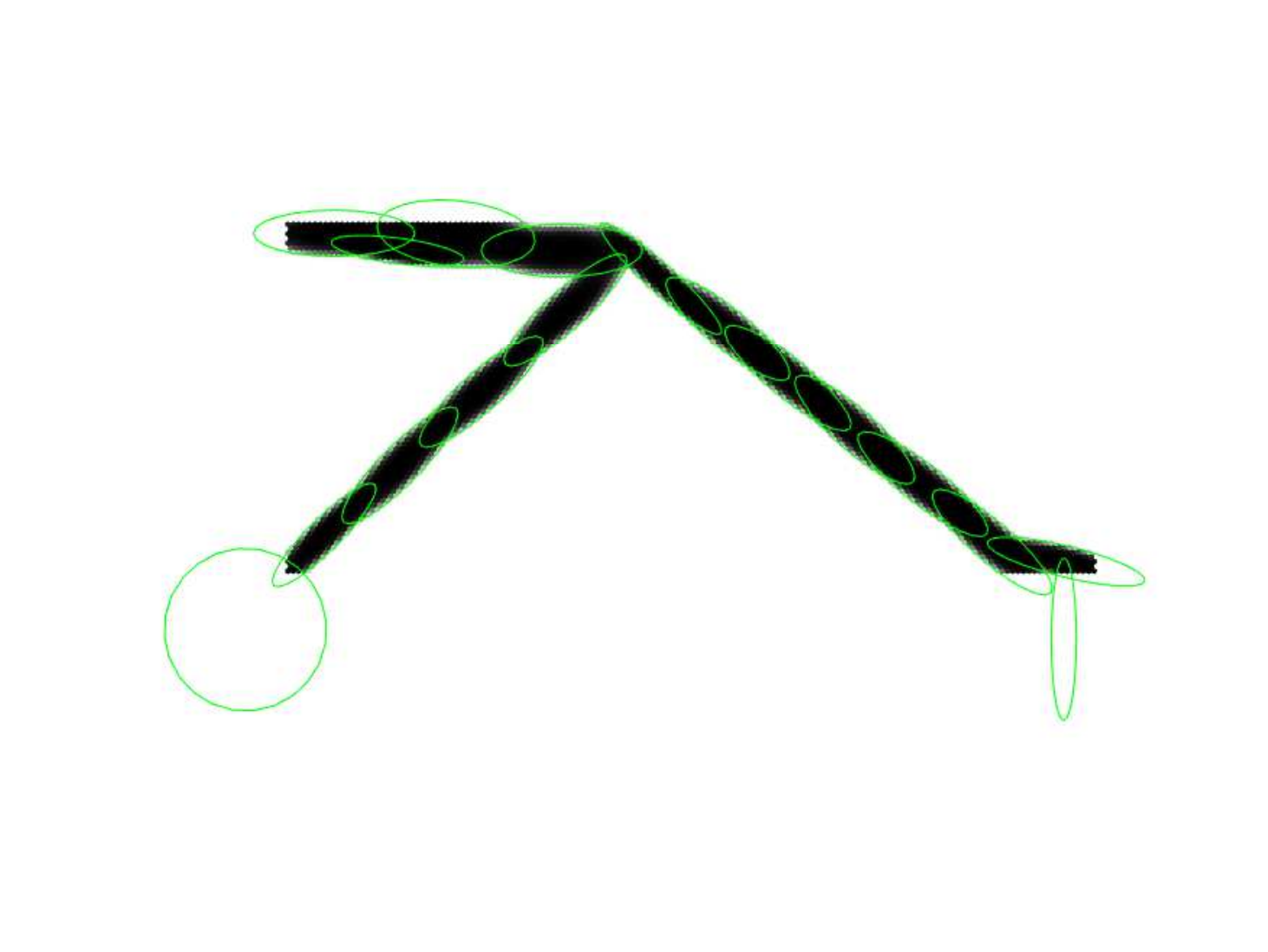} 
		\caption{ }
\label{fig:Eg1_Eg2_Eg3_Eg4_new_logic_PEM_with_masksc}
	\end{subfigure}
	\hspace{30mm}
	\begin{subfigure}[b]{.3\textwidth}
		\centering
		\captionsetup{font=scriptsize}
		\includegraphics[trim={2cm 2.5cm 1cm 2cm}, clip, scale = 0.55]{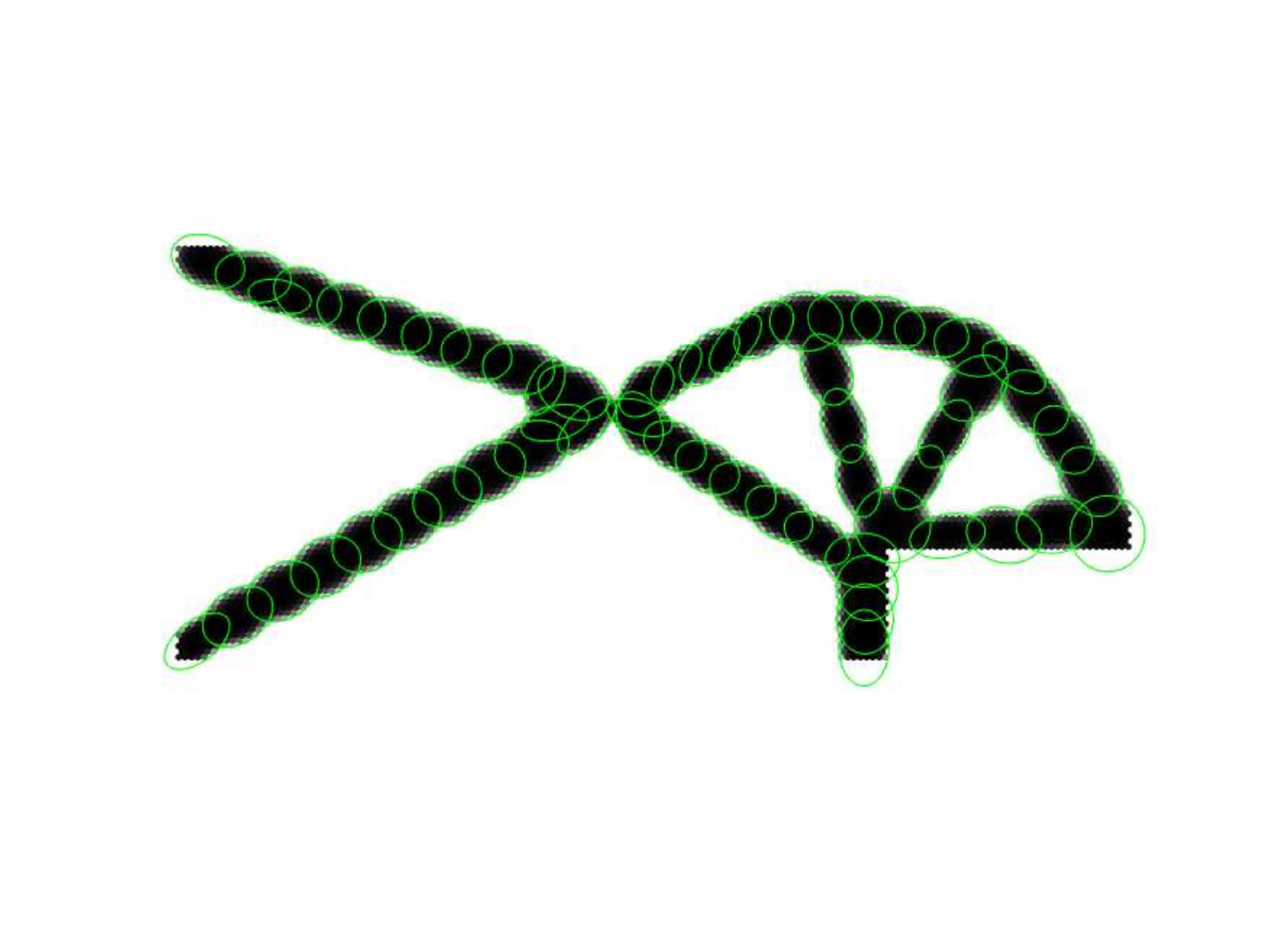} 
		\caption{ }
\label{fig:Eg1_Eg2_Eg3_Eg4_new_logic_PEM_with_masksd}
	\end{subfigure}
	\caption{Respective topologies in Fig. \ref{fig:Eg1_Eg2_Eg3_Eg4_new_logic_PEM} depicted with Positive Elliptical masks}
	\label{fig:Eg1_Eg2_Eg3_Eg4_new_logic_PEM_with_masks}
\end{figure}

The four examples in Fig. \ref{Fig:2} are solved with the methodology in section \ref{method_TO} for $\alpha = 6$ and $\eta = 3$. Mesh sizes are $150 \times 80$ for Example I, $200 \times 80$ for Example II, and $150 \times 75$ for both compliant mechanism problems (Examples III and IV). Except for Example II wherein the cell size(cs) is $0.28$ units, in all other examples, the cell size is $0.38$ units. In all examples, values of other variables are kept the same as those used for Examples I and II in section \ref{prelim_results}, however, specifications for $min_{ls}$ and $max_{ls}$ vary. Example I and II are solved with the starting volume fraction of $0.2$ while those on compliant mechanisms are solved with the starting $vf$ of 0.3. $S$ (Eq. \ref{formulation}) for Examples I and II is set as 1, and for Examples III and IV, it is set as $10^6$. Maximum number of function evaluations for each optimization step in Fig. \ref{Fig: methodology_flow} is set to 100. All examples are solved with both, negative (Fig. \ref{fig:Eg1_Eg2_Eg3_Eg4_new_logic_NEM}) and positive (Fig. \ref{fig:Eg1_Eg2_Eg3_Eg4_new_logic_PEM}) elliptical masks. Values of the objective, $g_{min}(\bm{\rho})$,  $g_{max}(\bm{\rho})$, final volume fraction, tolerance values, and the black and white measures $BWI$ are all indicated below each solution. Corresponding topologies are shown with negative masks in Fig. \ref{fig:Eg1_Eg2_Eg3_Eg4_new_logic_NEM_with_masks} and positive masks in Fig. \ref{fig:Eg1_Eg2_Eg3_Eg4_new_logic_PEM_with_masks}.

In Example I with negative masks (Fig. \ref{fig:Eg1_Eg2_Eg3_Eg4_new_logic_NEMa}), minimum and maximum length scales seem to be achieved within a  tolerance $\varepsilon_1 = \varepsilon_2 = \varepsilon = 39$. In Example II (Fig. \ref{fig:Eg1_Eg2_Eg3_Eg4_new_logic_NEMb}), the tolerance value suggested by the methodology is relatively high ($\varepsilon = 82$). In the compliant inverter problem (Example III) solved with negative masks (Fig. \ref{fig:Eg1_Eg2_Eg3_Eg4_new_logic_NEMc}), minimum and maximum length scales are quite close to each other, in an attempt to seek ‘volume-distributed’ solutions. Members seem  more or less of uniform thickness with length scales achieved within a tolerance of $\varepsilon = 23$. Similar is the case for the solution of Example II (Fig. \ref{fig:Eg1_Eg2_Eg3_Eg4_new_logic_NEMb}) although boundaries seem undulating suggesting that number of elliptical masks defining those boundaries are not adequate (see Fig. \ref{fig:Eg1_Eg2_Eg3_Eg4_new_logic_NEM_with_masksb}). In Example IV, one notices through visual inspection that the minimum length scale is not quite satisfied locally at two sites, one around the hinge and the second around the smaller void at the top right corner.  This could be attributed to $g_{min}(\boldsymbol{\rho})$ being a global measure and/or high relaxation parameter. Observing the solutions with negative masks (Fig. \ref{fig:Eg1_Eg2_Eg3_Eg4_new_logic_NEM_with_masks}), nearly all masks contribute in defining the respective topologies in that only a few masks are outside the specified domain. 

When observing topologies with positive masks,  in Example I (Fig \ref{fig:Eg1_Eg2_Eg3_Eg4_new_logic_PEMa}), minimum and maximum length scales are achieved within a tolerance of $\varepsilon = 38$. In Example II, (Fig \ref{fig:Eg1_Eg2_Eg3_Eg4_new_logic_PEMb}), however, this is not the case. A single member at bottom center of the topology, and oriented close to 120$^o$  with respect to the horizontal violates the minimum thickness specified. Values of $g_{min}(\bm{\rho})$ and $g_{max}(\bm{\rho})$ are quite high (see discussion). For the compliant inverter problem (Fig \ref{fig:Eg1_Eg2_Eg3_Eg4_new_logic_PEMc}), a higher tolerance ($\varepsilon = 56$) is required by the proposed methodology though members seem more or less to be of uniform thickness. In Example IV (Fig \ref{fig:Eg1_Eg2_Eg3_Eg4_new_logic_PEMd}), the minimum length scale is not achieved locally, near the hinge, just like the corresponding solution with negative masks (Fig \ref{fig:Eg1_Eg2_Eg3_Eg4_new_logic_NEMd}).  Boundary undulations are also observed, possibly due to the use of few positive masks (Fig. \ref{fig:Eg1_Eg2_Eg3_Eg4_new_logic_PEM_with_masksd}). One notes that in case of positive masks, \emph{mask deletion} is implemented in that final solutions for Examples I-IV require 130, 134, 17 and 58 masks respectively in comparison to 200 for all, specified initially.

\section{Discussion}
\label{discussion}

\indent Per \cite{Poulsen2003},  {\it mesh independence} is guaranteed if minimum length scale  is imposed when seeking optimal topologies. In this paper, minimum and maximum length scales are imposed explicitly, on  {\it skeletonized} (solid phase) intermediate topologies. It is shown via an analytical example (Section \ref{analytical_example}) that an arbitrary set of specified volume fraction ($vf$), and minimum length scale ($min_{ls}$) may not always yield a solution for a given skeletonized topology, or even if the skeleton changes. Rather, parameters $vf$, $min_{ls}$ and $max_{ls}$ tend to be interdependent. The SLS methodology suggested in Section \ref{method_TO} employs a two-stage heuristic approach to attain optimal solutions. In stage I, topologies are sought under only the volume constraint, and the volume fraction is lowered (if required) in sub-stages until the maximum length scale criterion is satisfied implicitly. With the stage I solution as the initial guess and skeleton well-formed, in stage II, all constraints are imposed, and solutions are sought by altering the volume fraction systematically in a manner that the explicit length scale constraints are satisfied within some tolerance which, is increased marginally within each sub-stage in optimization. Solutions are obtained with negative/positive elliptical masks that determine the material densities of hexagonal cells in groups over which they lay. Small deformation examples are used, and it is observed that minimum and maximum length scales are achieved by-and-large even if the length scale constraints are 'skeleton-based' and less restrictive in that they are not imposed more strictly and locally to keep the number of constraints to a minimum.   

\begin{figure}[H]
\hspace{5mm}	
\begin{subfigure}[b]{.3\textwidth}
		\centering
		\captionsetup{font=scriptsize}
		\includegraphics[trim={2cm 3.5cm 2cm 2cm}, clip, scale = 0.45]{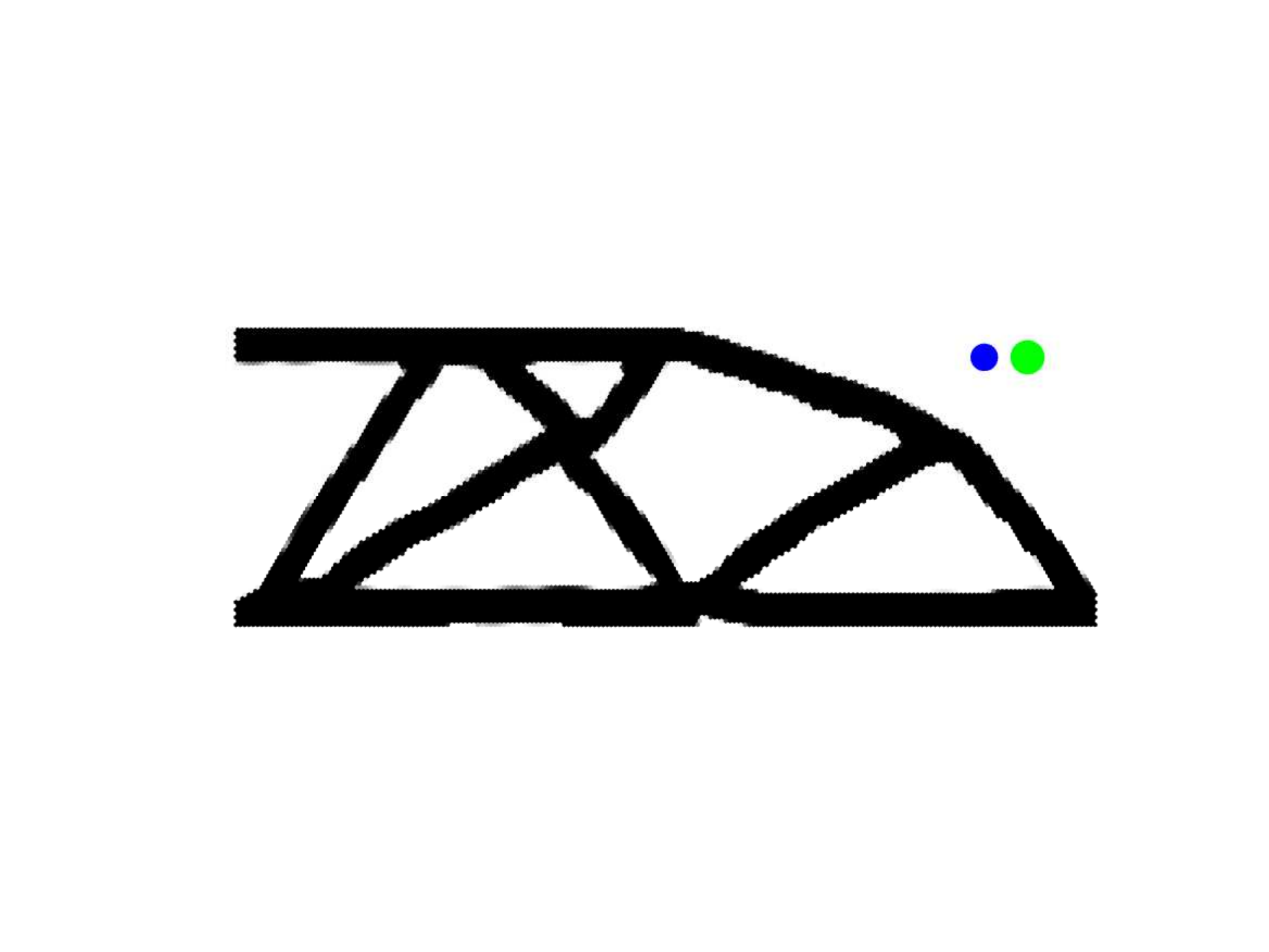}
		\caption{}
\label{fig:Eg2_Eg4_revisited_with_continuationa}
	\end{subfigure}
	\hspace{30mm}
\begin{subfigure}[b]{.3\textwidth}
		\centering
		\captionsetup{font=scriptsize}
		\includegraphics[trim={2cm 3.5cm 2cm 2cm}, clip, scale = 0.45]{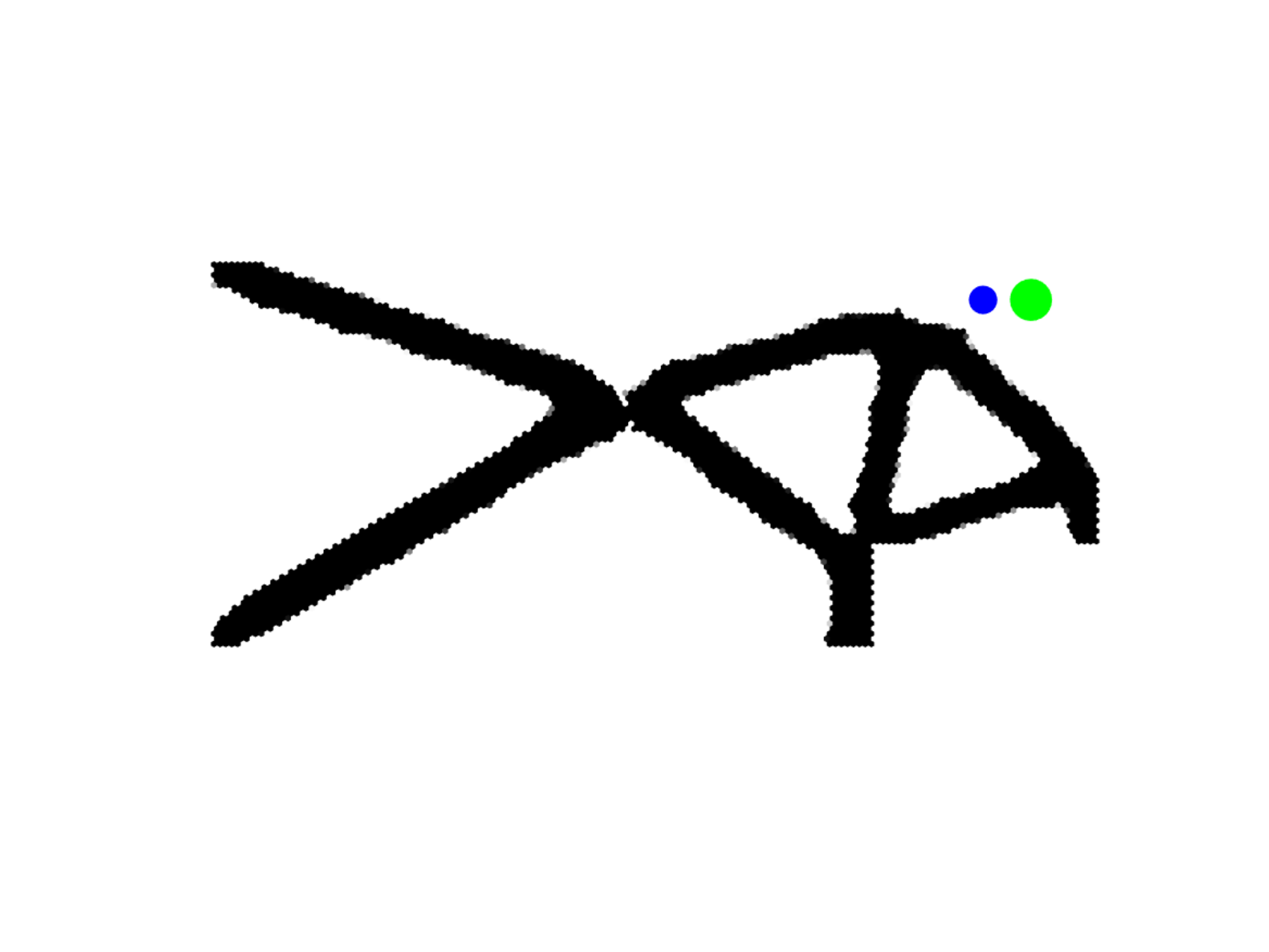}
		\caption{}
\label{fig:Eg2_Eg4_revisited_with_continuationb}
	\end{subfigure}
	\caption{(a) Example II in Figure \ref{fig:Eg1_Eg2_Eg3_Eg4_new_logic_PEMb}  revisited with continuation on $\alpha$ from 2 to 30. $g_{min}(\bm{\rho})  = 35.5$, $g_{max}(\bm{\rho})  = 276.7$. $\Phi = 1055.9$ (b) Example IV in Figure \ref{fig:Eg1_Eg2_Eg3_Eg4_new_logic_NEMd} revisited with continuation on $\alpha$ from 2 to 51. $g_{min}(\bm{\rho})  = 4.5$, $g_{max}(\bm{\rho})  = 73.5$. $\Phi = -0.071$.}	
	\label{fig:Eg2_Eg4_revisited_with_continuation}
\end{figure}

\indent 
In Example II generated using positive elliptical masks (Fig.  \ref{fig:Eg1_Eg2_Eg3_Eg4_new_logic_PEMb}), there exists a thin slender member that does not satisfy the minimum length scale. The member is defined by a single positive mask (Fig.  \ref{fig:Eg1_Eg2_Eg3_Eg4_new_logic_PEM_with_masksb}). It is reckoned that use of less number of masks, or high $\alpha$ could be the cause as sensitivities are close to zero and thus masks may not respond readily \cite{saxena2011topology}. With the solution in Figure \ref{fig:Eg1_Eg2_Eg3_Eg4_new_logic_PEMb} as the initial guess, optimization is performed again with continuation on $\alpha$, that is, $\alpha$ is  increased in steps from $2$ till $30$. The obtained solution is shown in Fig. \ref{fig:Eg2_Eg4_revisited_with_continuationa}. The minimum length scale $g_{min}(\bm{\rho}) $ reduces significantly, and the strain energy reduces from $\Phi = 1075.9$ to $\Phi = 1055.9$. 

\indent In Example IV, with both negative and positive masks (Fig.  \ref{fig:Eg1_Eg2_Eg3_Eg4_new_logic_NEMd} and \ref{fig:Eg1_Eg2_Eg3_Eg4_new_logic_PEMd} respectively), local hinges are observed. Also, in Figure \ref{fig:Eg1_Eg2_Eg3_Eg4_new_logic_NEMd}, minimum length scale is not satisfied at the top right corner, below the small void. Using the same rational as above, continuation is performed on $\alpha$, which is increased from 2 to 50 gradually, with the initial guess as that in Fig. \ref{fig:Eg1_Eg2_Eg3_Eg4_new_logic_NEMd}. The solution is shown in Fig. \ref{fig:Eg2_Eg4_revisited_with_continuationb} wherein hinges are more pronounced and are more in number. Value of the objective is marginally increased from $\Phi = -0.077$ to $\Phi = -0.071$. The above suggests that continuation with the setting in Fig. \ref{Fig: methodology_flow} may not always help in achieving the desired length scales. \textcolor{black}{A (set of) well-posed length scale constraint(s) should be effective in imposing the desired length scales, irrespective of the objective function used. Noting that $\varepsilon_1$ must be strictly positive (Section \ref{analytical_example}) and that $g_{min}(\bm{\rho}) = 4.5$ is quite low for this example, even though the minimum length-scale constraint in Eq. \ref{formulation} is satisfied mathematically, the optimization algorithm seems to exploit the loophole, that multi-criteria formulations in compliant mechanisms are prone to yielding local hinges \cite{Yin_Ananthasuresh_2003}. Another reason for appearance of hinge(s) is that the proposed approach controls length scales on only the solid and not the void states. Lazarov et al. \cite{lazarov2016length} state that length scale imposition on only one of the phases does not guarantee manufacturability in that it may not be possible to avoid hinges when designing small displacement compliant mechanisms. They recommend explicit length scale control on both phases. }

\begin{figure}[H]
\hspace{10mm}
\begin{subfigure}[b]{.45\textwidth}
		\centering
		\captionsetup{font=scriptsize}
		\includegraphics[trim={2cm 3.5cm 2cm 3cm}, clip, scale = 0.4]{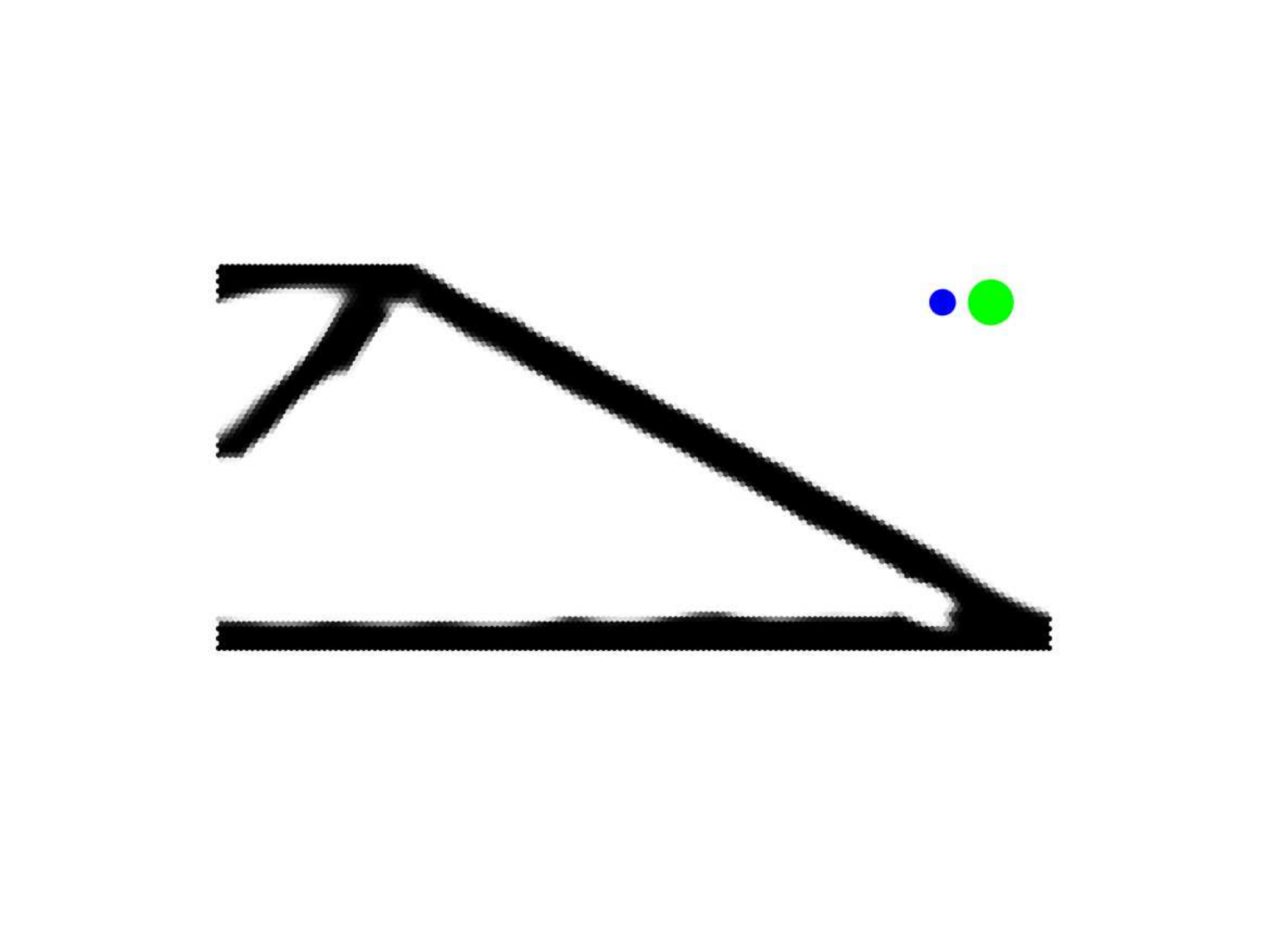}
		\caption{Example I:  $cs = 0.38$ units. $min_{ls} = 4cs$ units, $max_{ls} = 7cs$ units. Post optimization, $\Phi = 1073.6$, $g_{min}(\bm{\rho})  = 24.5$, $g_{max}(\bm{\rho})  = 7.4$, 
	       $vf = 0.22$;}
\label{fig:Eg1_post_review_NEM}
	\end{subfigure} \hspace{5mm}
\begin{subfigure}[b]{.45\textwidth}
		\centering
		\captionsetup{font=scriptsize}
		\includegraphics[trim={2cm 3.5cm 2cm 3cm}, clip, scale = 0.4]{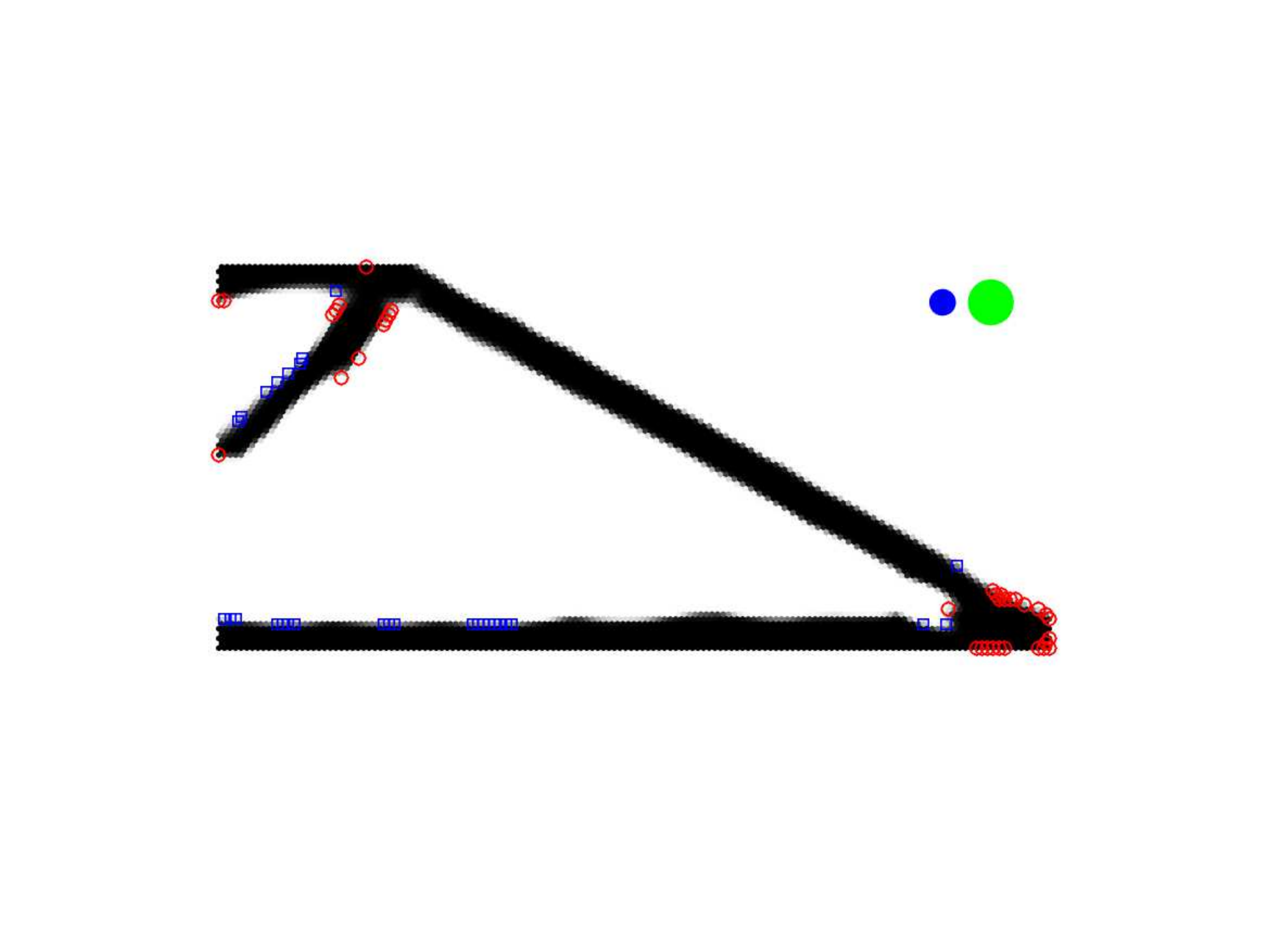}
		\caption{Number of (i) filled cells ($\rho > 0.99$): 1734, (ii) blue cells: 29, (iii) red cells:  36, skeletal cells: 360.}		
\end{subfigure}	

\hspace{10mm}	
\begin{subfigure}[b]{.45\textwidth}
		\centering
		\captionsetup{font=scriptsize}
		\includegraphics[trim={3cm 4.5cm 2cm 3cm}, clip, scale = 0.45]{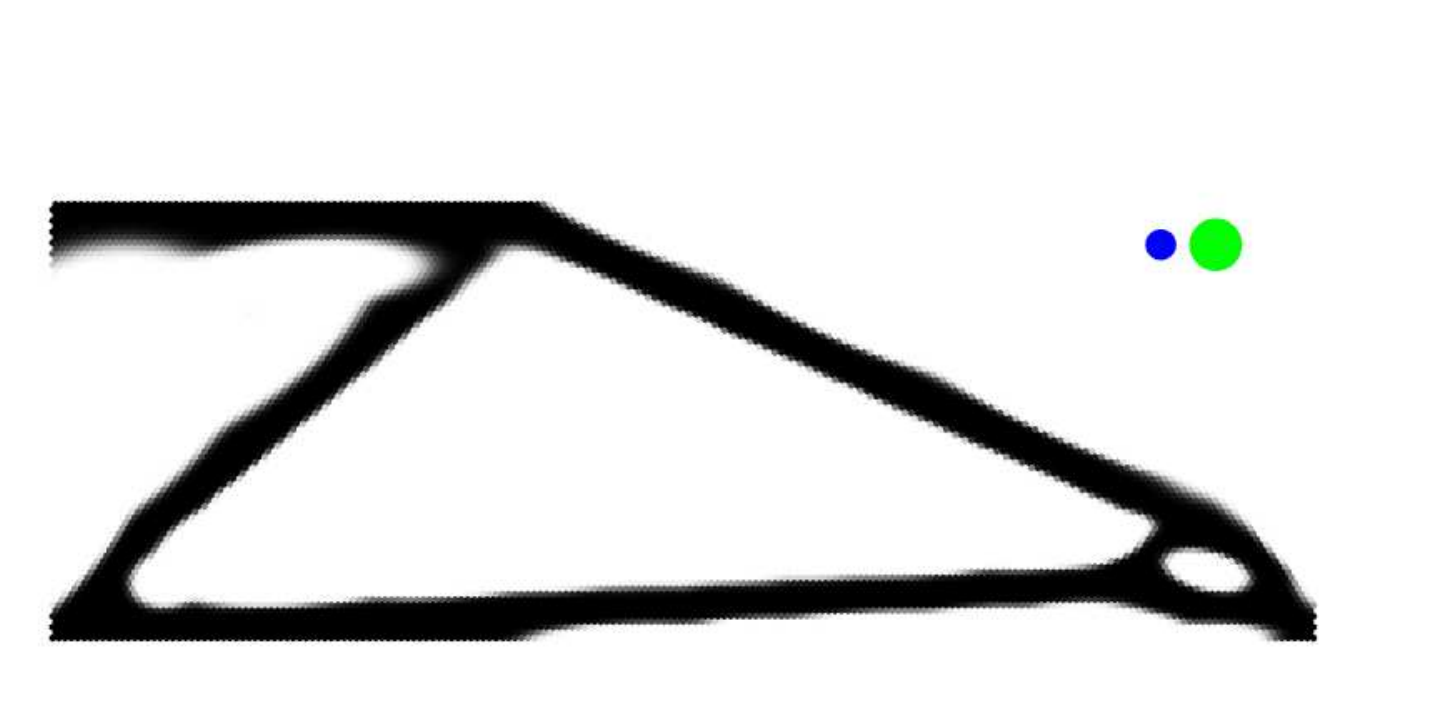}  
		\caption{Example II:  $cs = 0.28$ units. $min_{ls} = 4cs$ units, $max_{ls} = 7cs$ units. Post optimization,
	$\Phi = 1814.7$, $g_{min}(\bm{\rho})  =  49.79$, $g_{max}(\bm{\rho})  = 48.9$, 
	$vf = 0.26$;  }
\label{fig:Eg2_post_review_NEM}
	\end{subfigure}  \hspace{5mm}
\begin{subfigure}[b]{.45\textwidth}
		\centering
		\captionsetup{font=scriptsize}
		\includegraphics[trim={3cm 4.5cm 2cm 3cm}, clip, scale = 0.45]{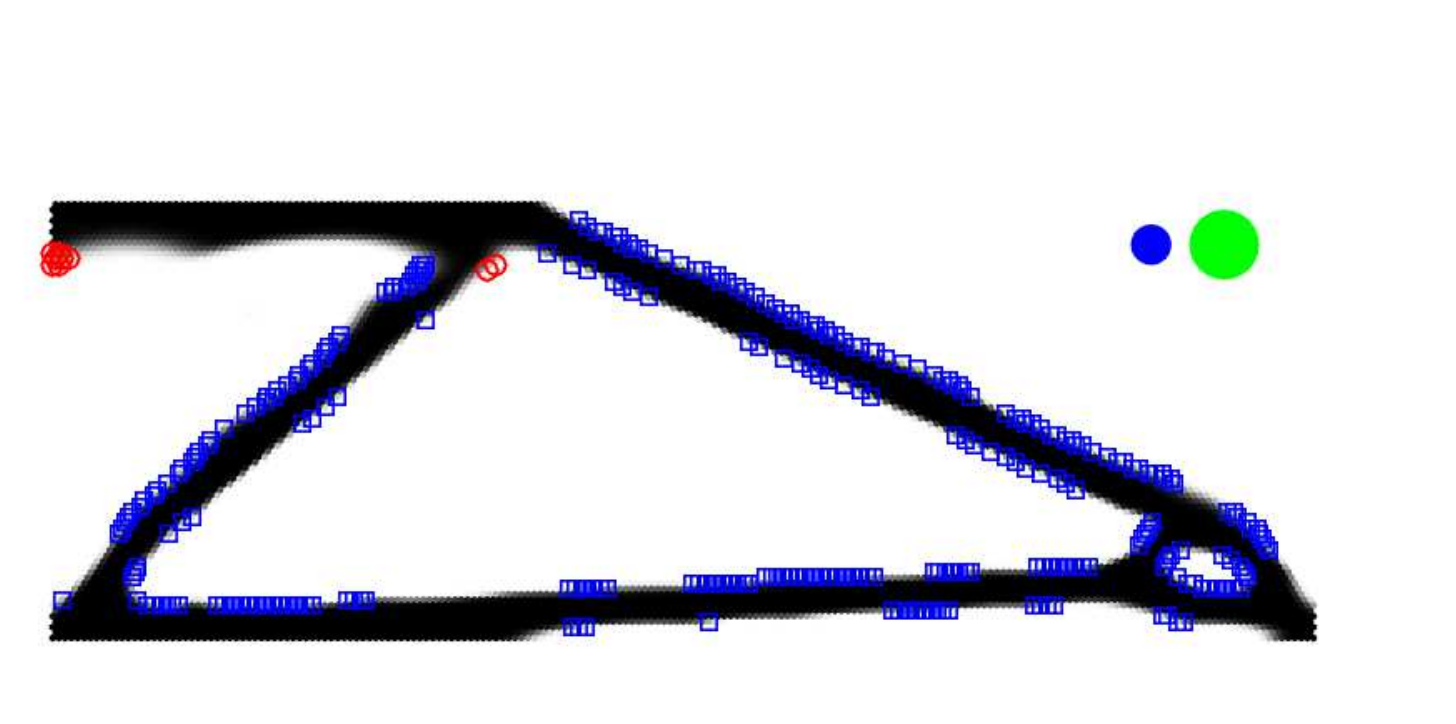}  
		\caption{Number of (i) filled cells ($\rho > 0.99$):  2589, (ii) blue cells: 294 (iii) red cells:  9, skeletal cells:  531.}
\end{subfigure}
		
\hspace{10mm}		
\begin{subfigure}[b]{.45\textwidth}
		\centering
		\captionsetup{font=scriptsize}
		\includegraphics[trim={3cm 4cm 2cm 3cm}, clip, scale = 0.45]{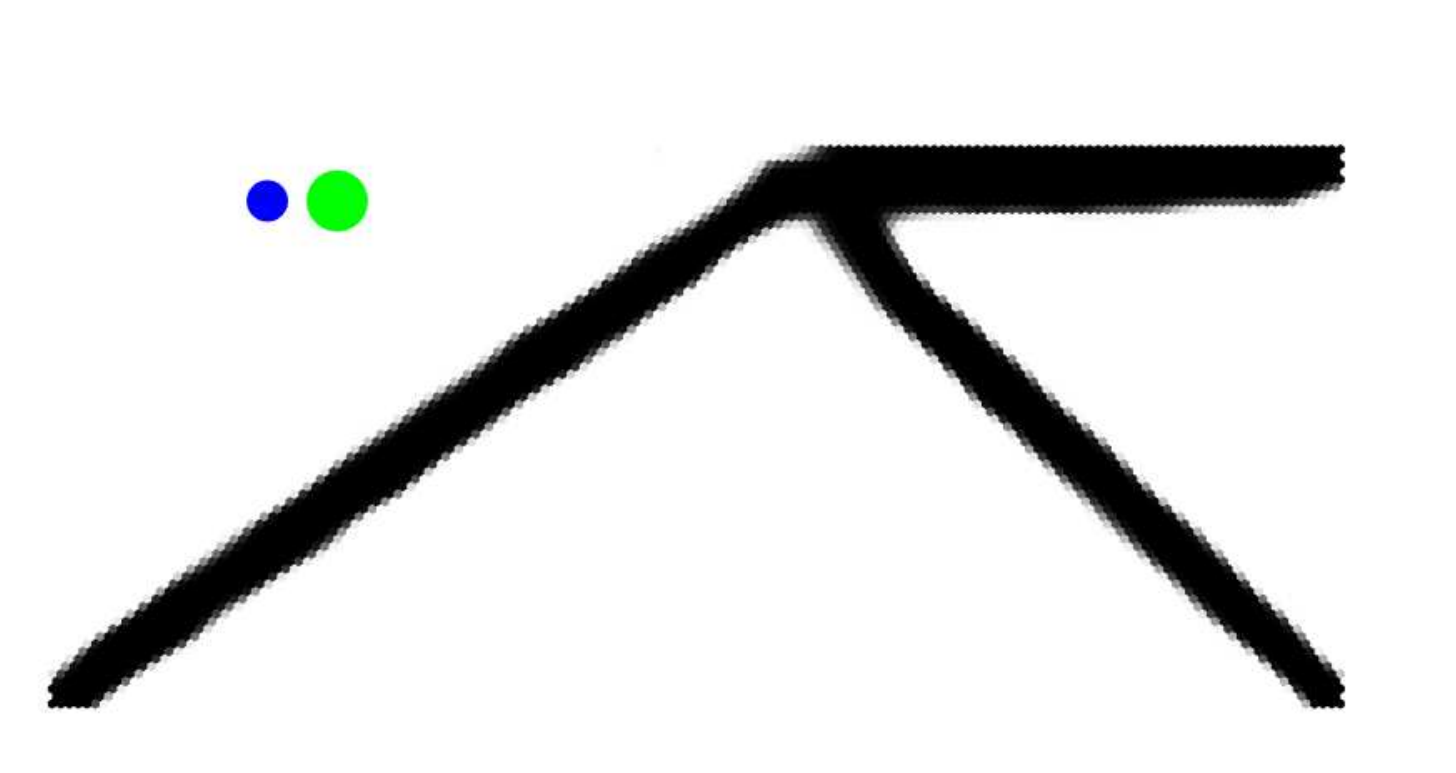}  
		\caption{Example III:  $cs = 0.38$ units. $min_{ls} = 4cs$ units, $max_{ls} = 6cs$ units. Post optimization,
	$\Phi = -0.185$, $g_{min}(\bm{\rho})  = 19.13$, $g_{max}(\bm{\rho})  = 30.7$, 
	$vf = 0.19$; }
\label{fig:Eg3_post_review_NEM}
	\end{subfigure}	\hspace{5mm}
\begin{subfigure}[b]{.45\textwidth}
		\centering
		\captionsetup{font=scriptsize}
		\includegraphics[trim={3cm 4cm 2cm 3cm}, clip, scale = 0.45]{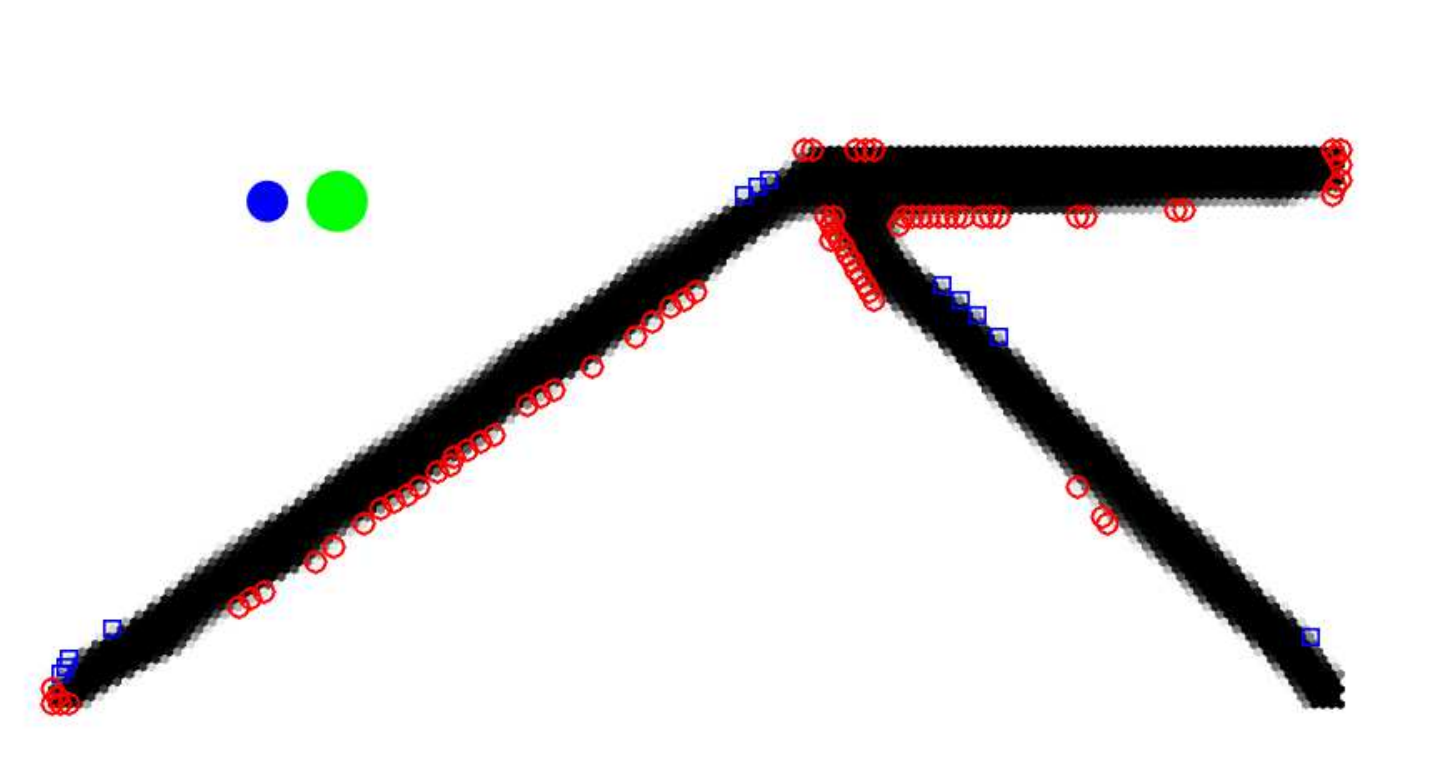}  
		\caption{ Number of (i) filled cells ($\rho > 0.99$):  1240, (ii) blue cells: 12, (iii) red cells:   75, skeletal cells:  267.}
\end{subfigure}	
\hspace{10mm}		
\begin{subfigure}[b]{.45\textwidth}
		\centering
		\captionsetup{font=scriptsize}
		\includegraphics[trim={2cm 4cm 2cm 3cm}, clip, scale = 0.45]{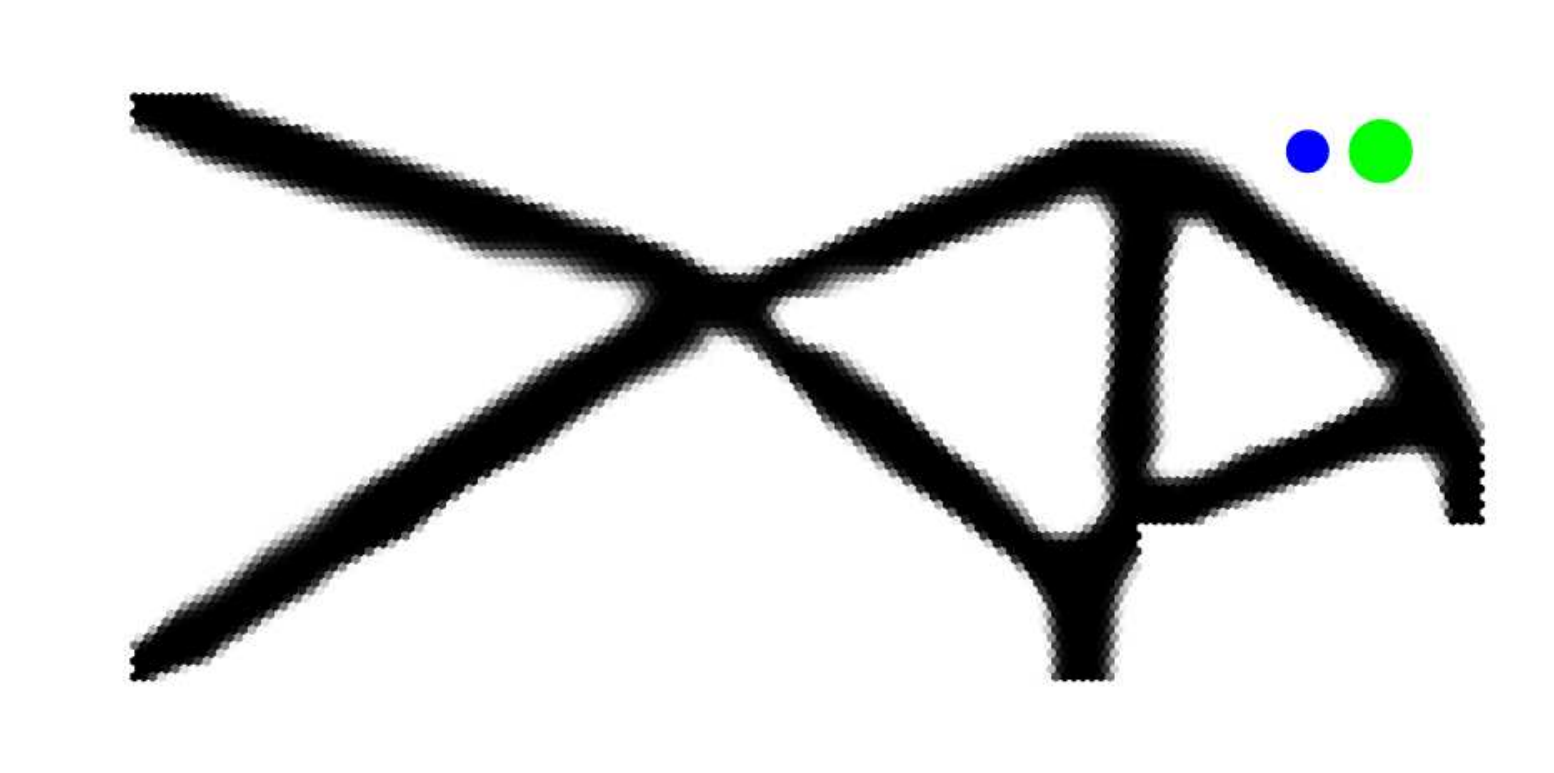}
		\caption{Example IV:  $cs = 0.38$ units. $min_{ls} = 4cs$ units, $max_{ls} = 6cs$ units. Post optimization,
	$\Phi = -0.068$, $g_{min}(\bm{\rho})  = 86.00$, $g_{max}(\bm{\rho})  = 70.44$, 
	$vf = 0.29$;  }
\label{fig:Eg4_post_review_NEM}
	\end{subfigure} \hspace{5mm}
	\begin{subfigure}[b]{.45\textwidth}
		\centering
		\captionsetup{font=scriptsize}
		\includegraphics[trim={2cm 4cm 2cm 3cm}, clip, scale = 0.45]{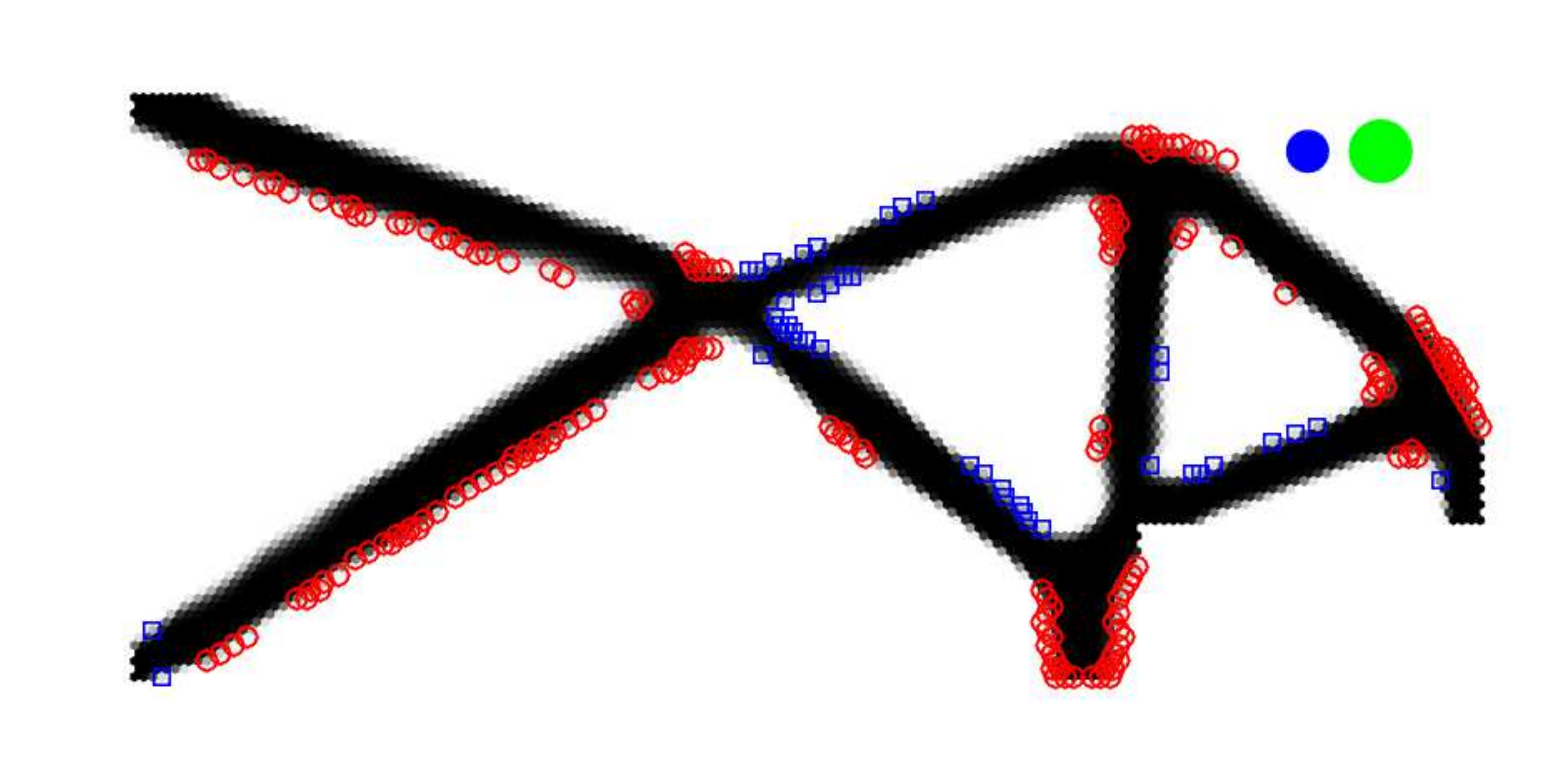}
		\caption{ Number of (i) filled cells ($\rho > 0.99$):  1497, (ii) blue cells: 42, (iii) red cells:  178, skeletal cells:  424.}
	\end{subfigure}	

\caption{Topologies generated with Negative Elliptical Masks with the SLS methodology. Circles (blue/green) in the inset represent the (minimum/maximum) length scales. Design specifications are identical to those for Figure \ref{fig:Eg1_Eg2_Eg3_Eg4_new_logic_NEM}. Commencing values $\varepsilon_1 = \varepsilon_2 = 1$. On the right column, cells in the respective topologies violating length scale constraints are shown with red circles and blue squares. Red circles represent cells violating the maximum length scale constraint. Blue squares enclose cells that should have density one, as required by the minimum length scale constraint.}
\label{fig:Eg1_Eg2_Eg3_Eg4_post_review_NEM}
\end{figure}

\begin{figure}[H]
\hspace{10mm}
\begin{subfigure}[b]{.45\textwidth}
		\centering
		\captionsetup{font=scriptsize}
		\includegraphics[trim={0cm 0cm 0cm 0cm}, clip, scale = 0.4]{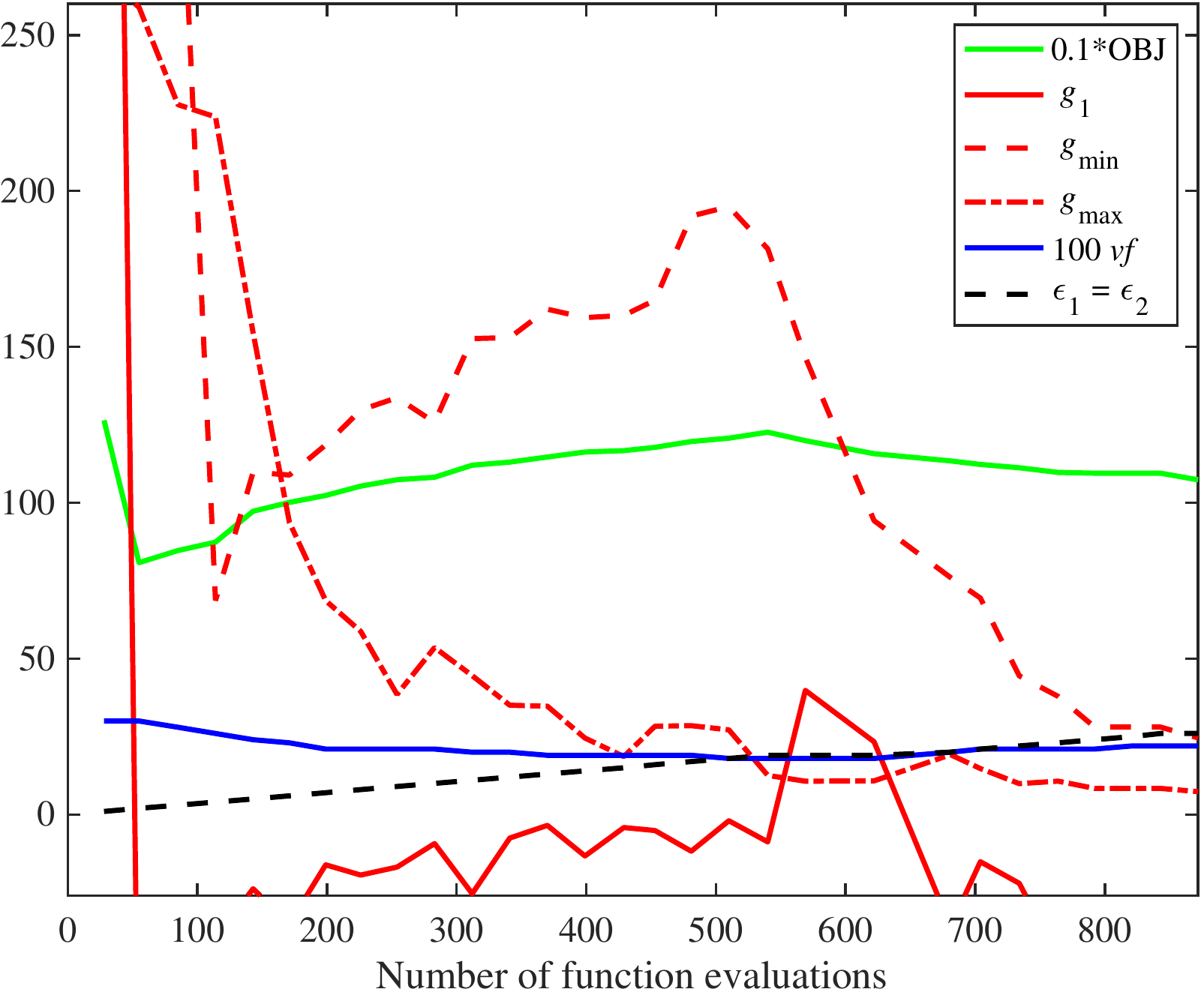}
		\caption{Example I:  Convergence history.}
\label{fig:Eg1_CH}
	\end{subfigure} \hspace{5mm}
\begin{subfigure}[b]{.45\textwidth}
		\centering
		\captionsetup{font=scriptsize}
		\includegraphics[trim={0cm 0cm 0cm 0cm}, clip, scale = 0.4]{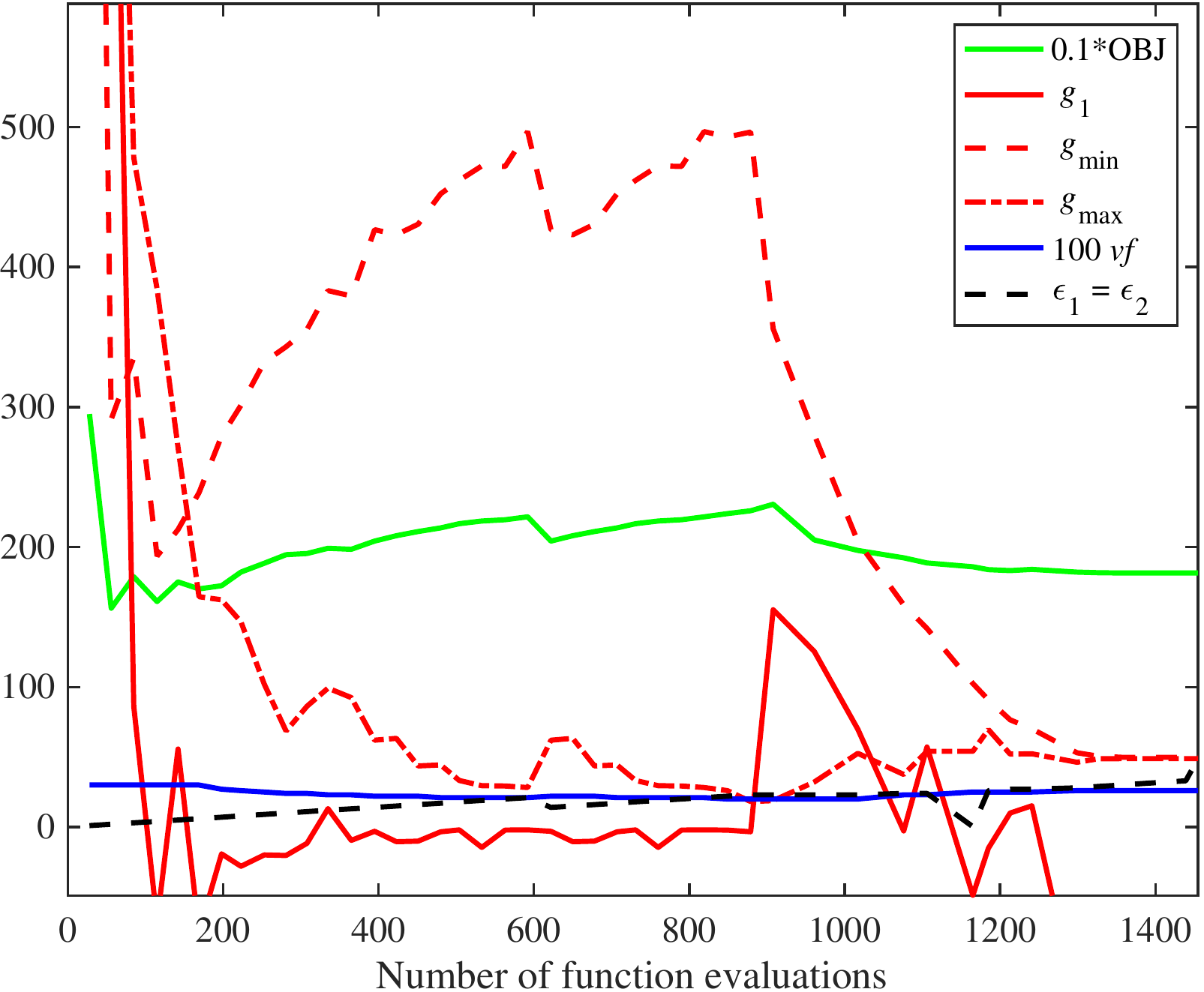}
		\caption{Example II:  Convergence history.}
\label{fig:Eg2_CH}
\end{subfigure}	

\vspace{10mm}
\hspace{10mm}	
\begin{subfigure}[b]{.45\textwidth}
		\centering
		\captionsetup{font=scriptsize}
		\includegraphics[trim={0cm 0cm 0cm 0cm}, clip, scale = 0.45]{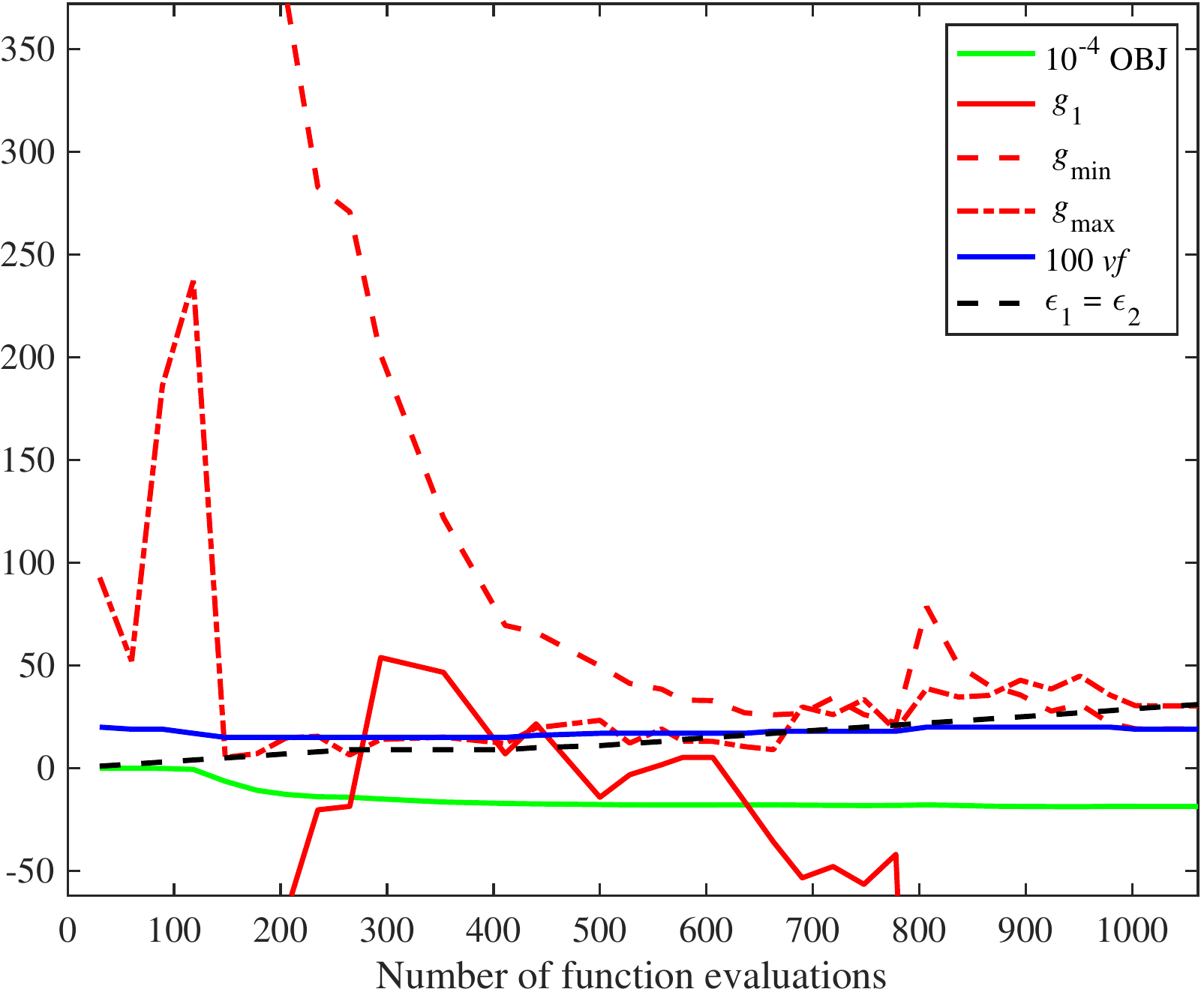}  
		\caption{Example III:  Convergence history.}
\label{fig:Eg3_CH}
	\end{subfigure}  \hspace{5mm}
\begin{subfigure}[b]{.45\textwidth}
		\centering
		\captionsetup{font=scriptsize}
		\includegraphics[trim={0cm 0cm 0cm 0cm}, clip, scale = 0.45]{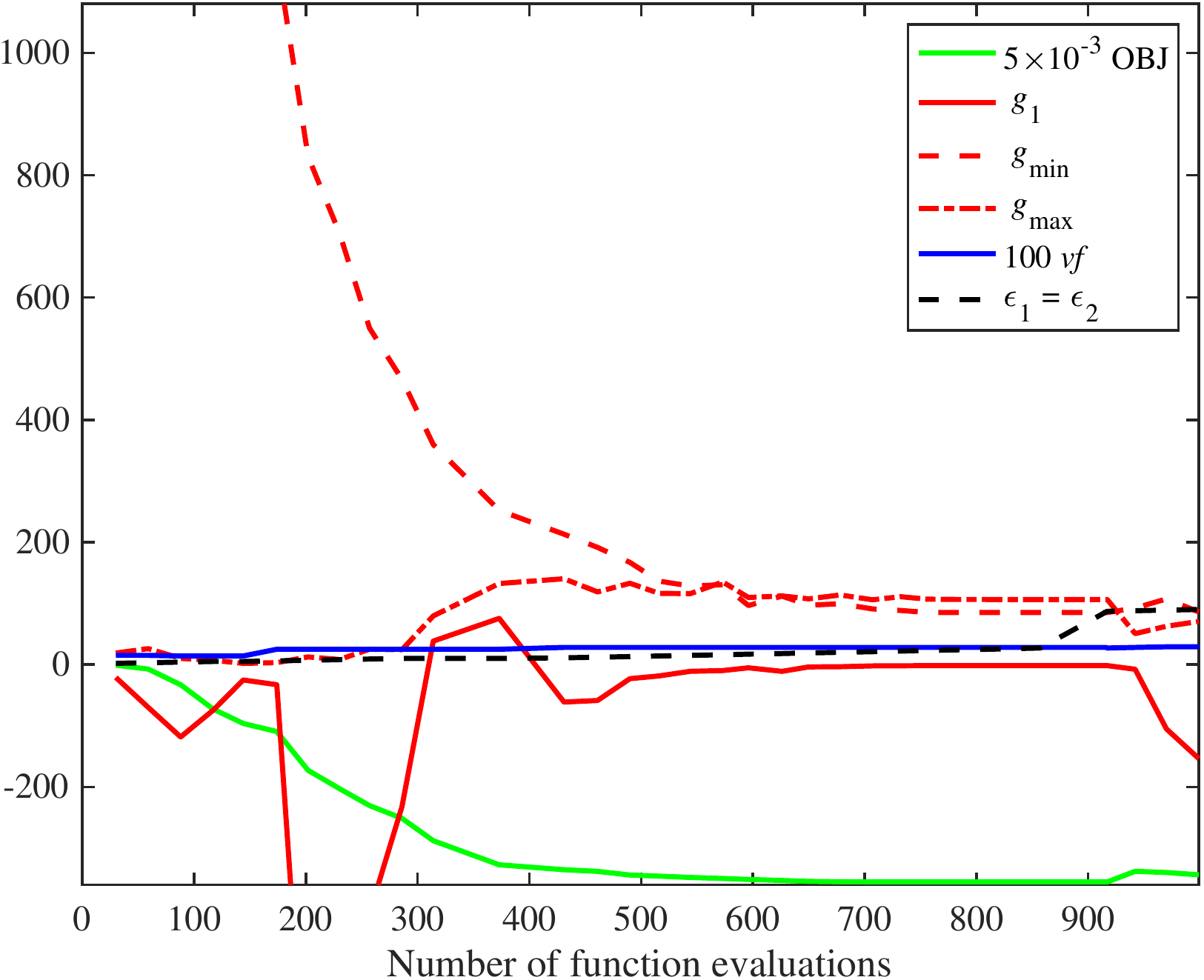}  
		\caption{Example IV:  Convergence history.}
\label{fig:Eg4_CH}
	\end{subfigure}
\caption{Convergence histories for the topologies in Figure. \ref{fig:Eg1_Eg2_Eg3_Eg4_post_review_NEM}.}
	\label{fig:EG1_to_4_conv_histories}
\end{figure}

To study the role of parameters $\varepsilon_{1}$ and $\varepsilon_{2}$, specifically their initial values and the way they are varied in intermediate stages, we solve the four benchmark problems again with negative elliptical masks. With all respective parameters identical, we commence these examples with $\varepsilon_{1} = \varepsilon_{2} = 1$. Maximum number of function evaluations permitted for each optimization stage (Figure \ref{Fig: methodology_flow}) is 30 (as opposed to 100 for examples in Figures \ref{fig:Eg1_Eg2_Eg3_Eg4_new_logic_NEM} and \ref{fig:Eg1_Eg2_Eg3_Eg4_new_logic_PEM}) after which, both,  $\varepsilon_{1}$ and $\varepsilon_{2}$ are incremented by 1. As mentioned earlier, the aim is to keep $\varepsilon_{1}$ and $\varepsilon_{2}$ as low as possible, though strictly positive. Final respective solutions are shown in Figure \ref{fig:Eg1_Eg2_Eg3_Eg4_post_review_NEM}. In the same figure, along the right column are shown the same solutions with cells marked with blue squares and red circles. Cells enclosed within the blue squares are those that are supposed to have the densities of the solid state. Those enclosed within the red circles are reckoned to stay void in accordance with the maximum length scale constraint. \\ \\
There is a topological change when comparing the solutions for Example I, in Figures \ref{fig:Eg1_Eg2_Eg3_Eg4_new_logic_NEMa} and \ref{fig:Eg1_post_review_NEM}. In the latter, the final volume fraction $vf$, $g_{min}(\boldsymbol{\rho})$ and $g_{max}(\boldsymbol{\rho})$ are all lower. Topologies for Example II remain the same, in Figures \ref{fig:Eg1_Eg2_Eg3_Eg4_new_logic_NEMb} and \ref{fig:Eg2_post_review_NEM}, even though $vf$, $g_{min}(\boldsymbol{\rho})$ and $g_{max}(\boldsymbol{\rho})$  are all lower. Solutions for Example III are identical (Figures \ref{fig:Eg1_Eg2_Eg3_Eg4_new_logic_NEMc} and \ref{fig:Eg3_post_review_NEM}) with $vf$, $g_{min}(\boldsymbol{\rho})$ and $g_{max}(\boldsymbol{\rho})$ comparable. Topologies for Example IV, in Figures \ref{fig:Eg1_Eg2_Eg3_Eg4_new_logic_NEMd} and \ref{fig:Eg4_post_review_NEM}, are significantly different. For the latter solution, even though $g_{min}(\boldsymbol{\rho})$ and $g_{max}(\boldsymbol{\rho})$ are significantly higher, local regions all seem to satisfy the minimum length scale constraint, not the case with the solution in Figure \ref{fig:Eg1_Eg2_Eg3_Eg4_new_logic_NEMd}. This seems to suggest that lower values of $\varepsilon_{1}$ and $\varepsilon_{2}$ are no guarantee for better solutions when one considers local length scales, especially with reference to the solution in Figure \ref{fig:Eg2_Eg4_revisited_with_continuationb}. \\ \\
$g_{min}(\boldsymbol{\rho})$ and $g_{max}(\boldsymbol{\rho})$, as modeled in Eq. \ref{min_max_ls}, are 'skeleton-dependent' global length scale measures on solid states, which by themselves, cannot guarantee local length scale control as expected. Nevertheless, the used length scale measures, along with the proposed 2-stage methodology by and large, do address length scale issues to a significant extent, if not comprehensively, as evident via the right column in Figure \ref{fig:Eg1_Eg2_Eg3_Eg4_post_review_NEM}. Very few cells (those whose centroids lie within blue squares or red circles), especially at continuum boundaries and mostly localized, violate the length scale constraints. \\ \\
Need for incrementing $\varepsilon_{1}$ and $\varepsilon_{2}$ is justified via the convergence histories in Figure. \ref{fig:EG1_to_4_conv_histories}, of the respective solutions in Figures \ref{fig:Eg1_post_review_NEM}-\ref{fig:Eg4_post_review_NEM}. In Figures \ref{fig:Eg2_CH}, \ref{fig:Eg3_CH} and \ref{fig:Eg4_CH}, $g_{min}(\boldsymbol{\rho})$, $g_{max}(\boldsymbol{\rho})$ and $vf$ stablize after 1300, 1000, and in between 600-900 function evaluations respectively. If  $\varepsilon_{1}$ and $\varepsilon_{2}$ are not incremented, the minimum and/or maximum length scales constraints may be violated. It is only after $\varepsilon_{1} = \varepsilon_{2} > \max(g_{min}(\boldsymbol{\rho}), g_{max}(\boldsymbol{\rho}))$ that constraints get satisfied. Especially, for Example IV (Figure \ref{fig:Eg4_CH}), after 900 evaluations, $g_{max}(\boldsymbol{\rho})$ gets lowered further. In all convergence histories, there is significant variation in $g_{min}(\boldsymbol{\rho})$, and it is only towards the end that the minimum length scale measure gets lowered than the relaxation parameter. $g_{max}(\boldsymbol{\rho})$ gets lowered relatively much earlier. Only for Example IV does $g_{max}(\boldsymbol{\rho})$ increase to almost the same value as $g_{min}(\boldsymbol{\rho})$ after 400 evaluations. Values of the objective stablize fairly early (600, 1200, 500 and 400 evaluations respectively).

\begin{figure}[H] \hspace{5mm}
\hspace{5mm}	
\begin{subfigure}[b]{.4\textwidth}
		\centering
		\captionsetup{font=scriptsize}
		\includegraphics[trim={2cm 3.5cm 2cm 2.5cm}, clip, scale = 0.45]{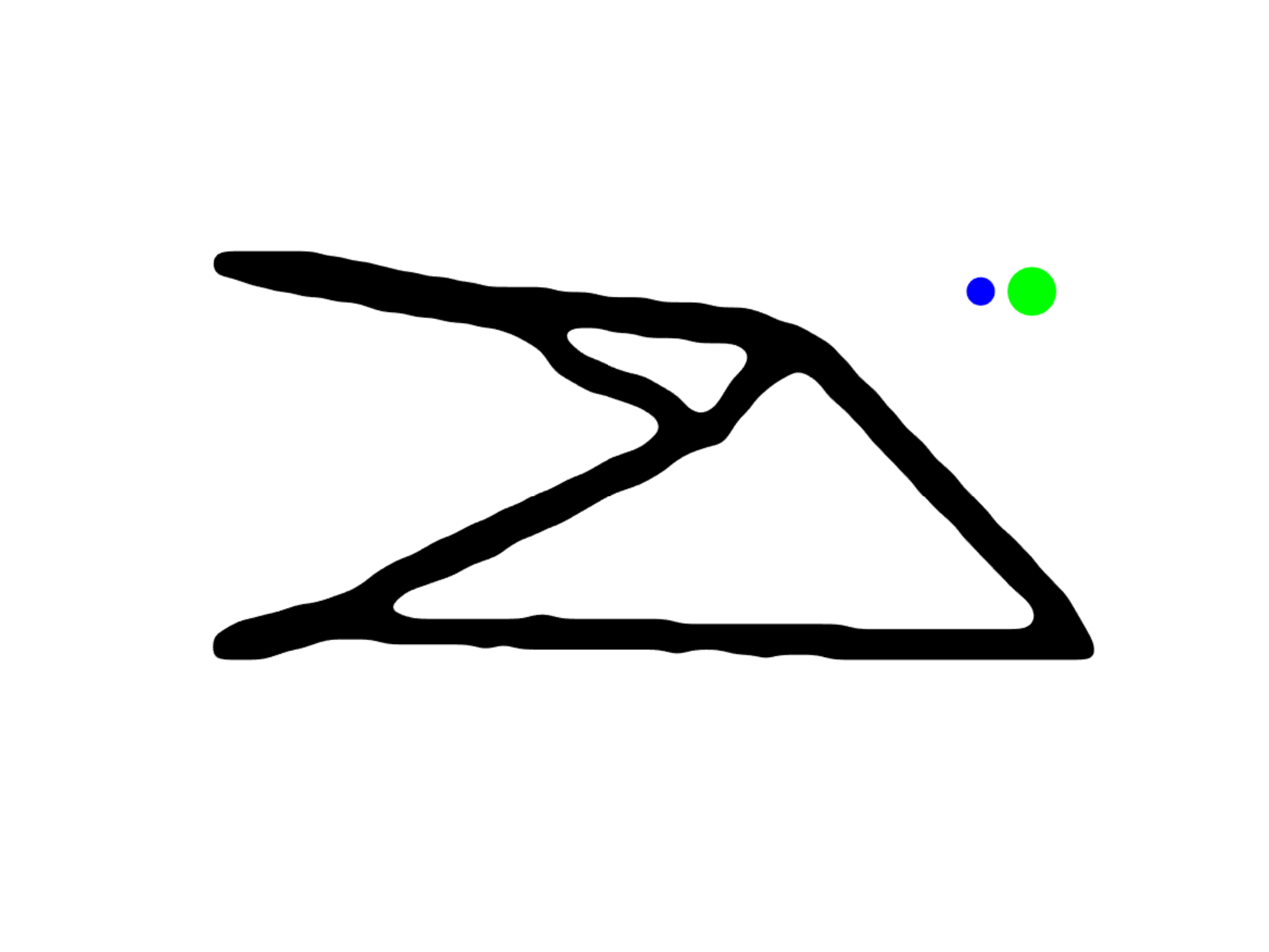}
		\caption{Smoothened solution for \ref{fig:Eg1_Eg2_Eg3_Eg4_new_logic_NEMa}. \\ $\Phi = 779.3$.} 
\label{solutions_with_imposed_min_ls_and_smoothened_boundariesa}
	\end{subfigure}
	\hspace{10mm}
\begin{subfigure}[b]{.4\textwidth}
		\centering
		\captionsetup{font=scriptsize}
		\includegraphics[trim={2cm 4.5cm 2cm 2.5cm}, clip, scale = 0.5]{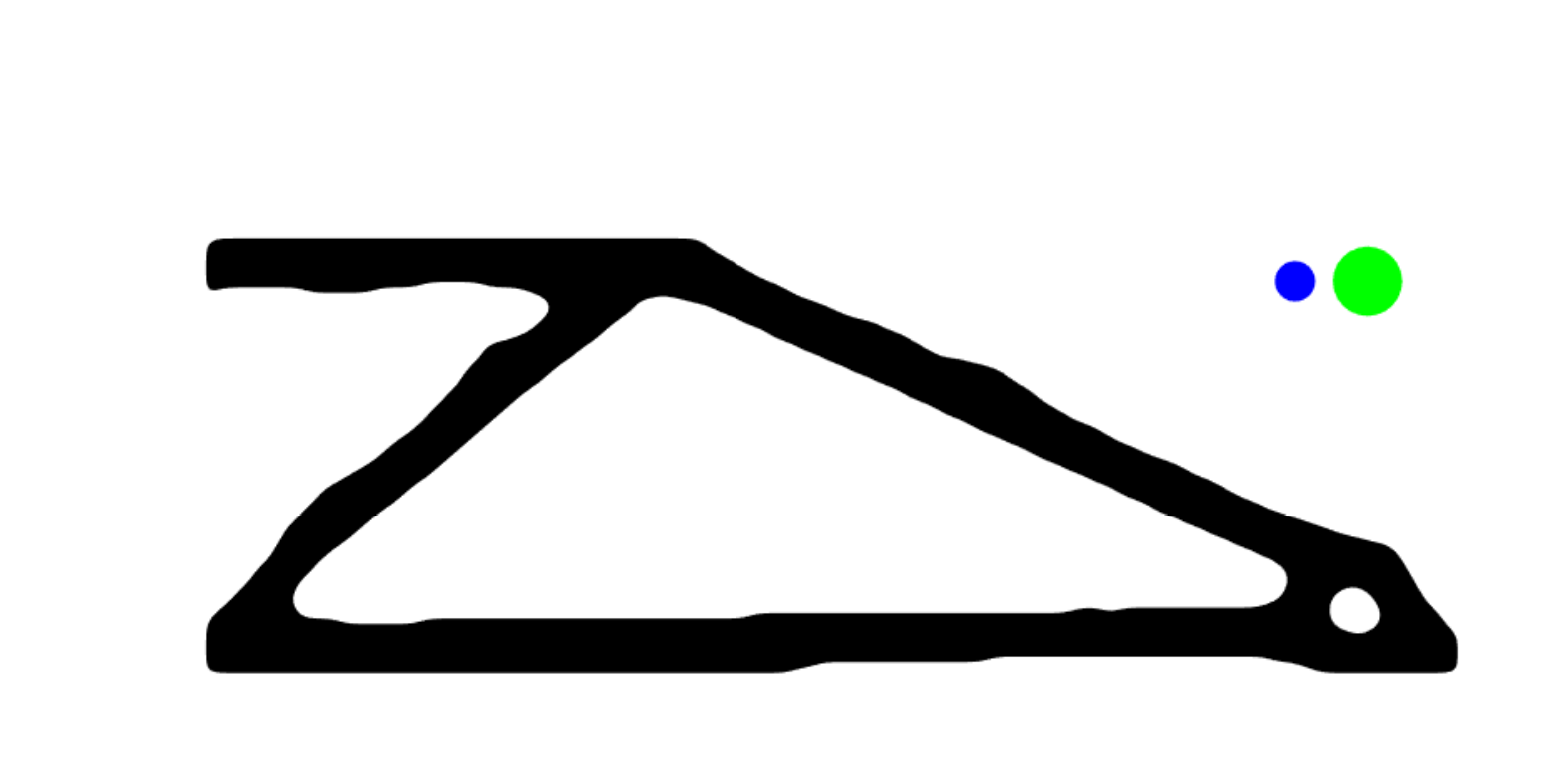}
		\caption{Smoothened solution for \ref{fig:Eg1_Eg2_Eg3_Eg4_new_logic_NEMb}. \\ $\Phi = 1495.2$.} 
\label{solutions_with_imposed_min_ls_and_smoothened_boundariesb}
	\end{subfigure}

 \hspace{5mm}	
\begin{subfigure}[b]{.4\textwidth}
		\centering
		\captionsetup{font=scriptsize}
		\includegraphics[trim={2cm 3.5cm 2cm 2.5cm}, clip, scale = 0.45]{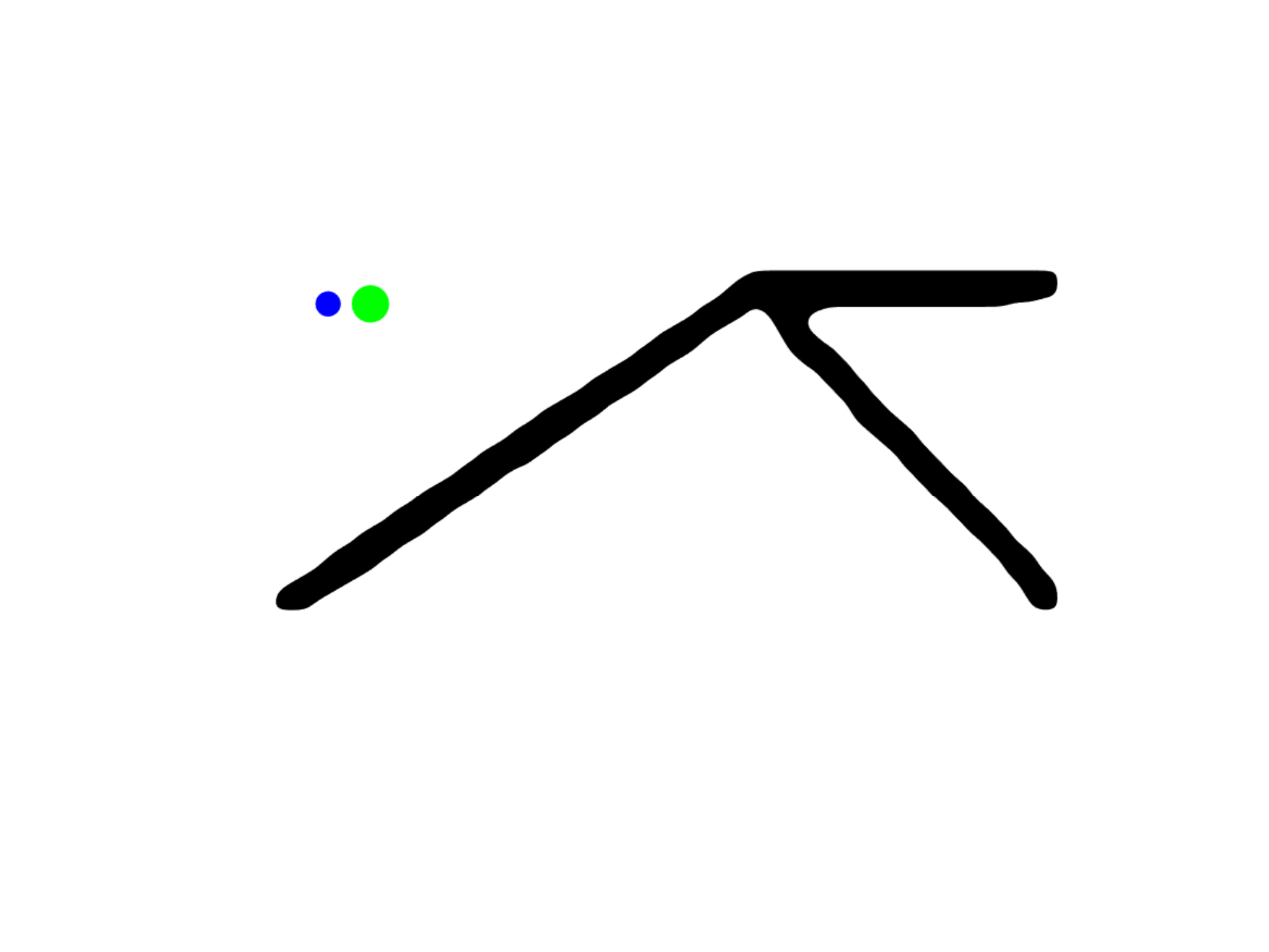}
		\caption{Smoothened solution for \ref{fig:Eg1_Eg2_Eg3_Eg4_new_logic_NEMc}. \\ $\Phi = -0.172$.} 
\label{solutions_with_imposed_min_ls_and_smoothened_boundariesc}
	\end{subfigure}
	\hspace{10mm}
\begin{subfigure}[b]{.4\textwidth}
		\centering
		\captionsetup{font=scriptsize}
		\includegraphics[trim={2cm 3.5cm 2cm 2.5cm}, clip, scale = 0.45]{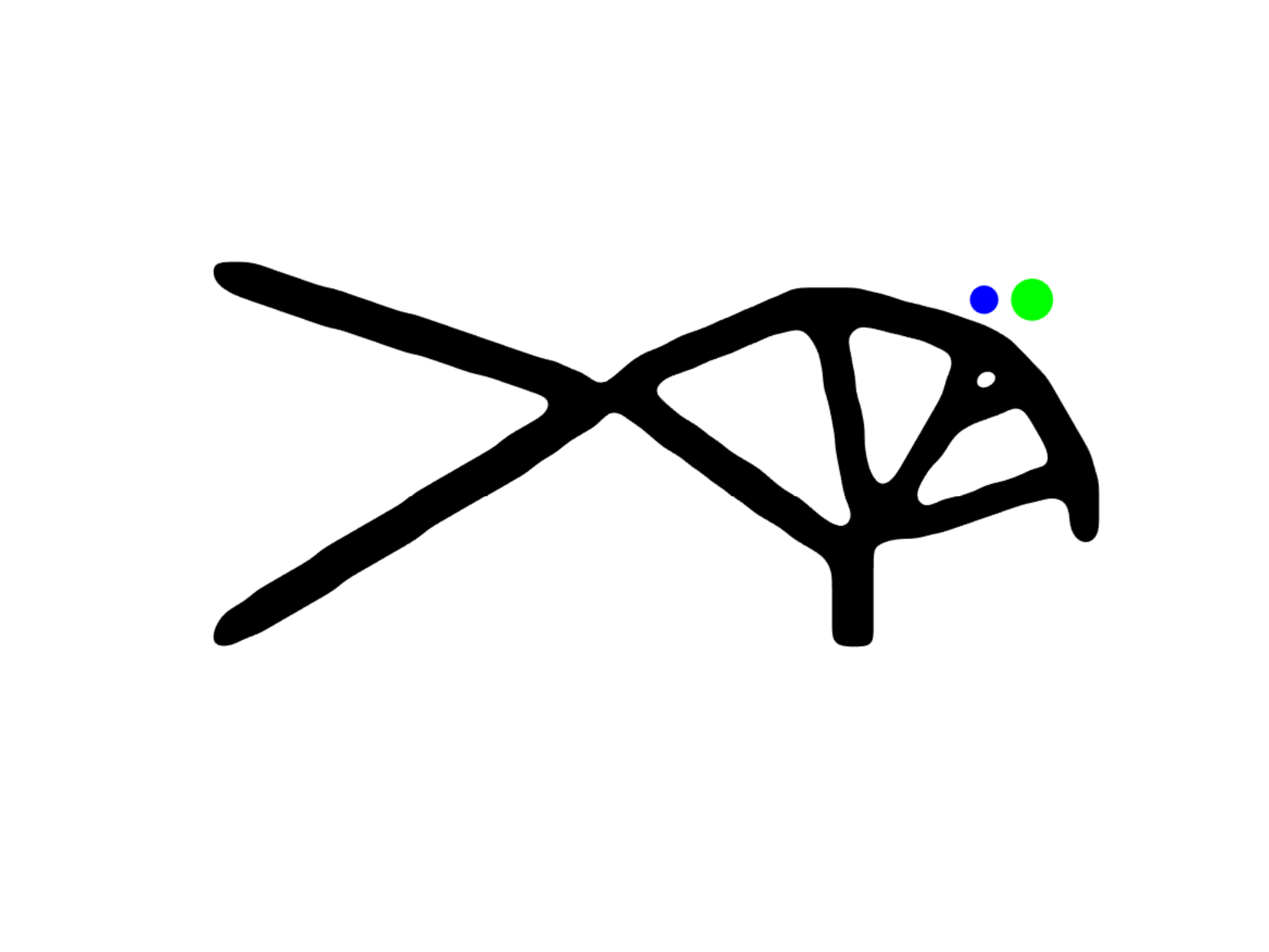}
		\caption{Smoothened solution for \ref{fig:Eg1_Eg2_Eg3_Eg4_new_logic_NEMd}. \\ $\Phi = -0.075$.} 
\label{solutions_with_imposed_min_ls_and_smoothened_boundariesd}
	\end{subfigure}
	
 \hspace{5mm}	
\begin{subfigure}[b]{.4\textwidth}
		\centering
		\captionsetup{font=scriptsize}
		\includegraphics[trim={2cm 3.5cm 2cm 2.5cm}, clip, scale = 0.45]{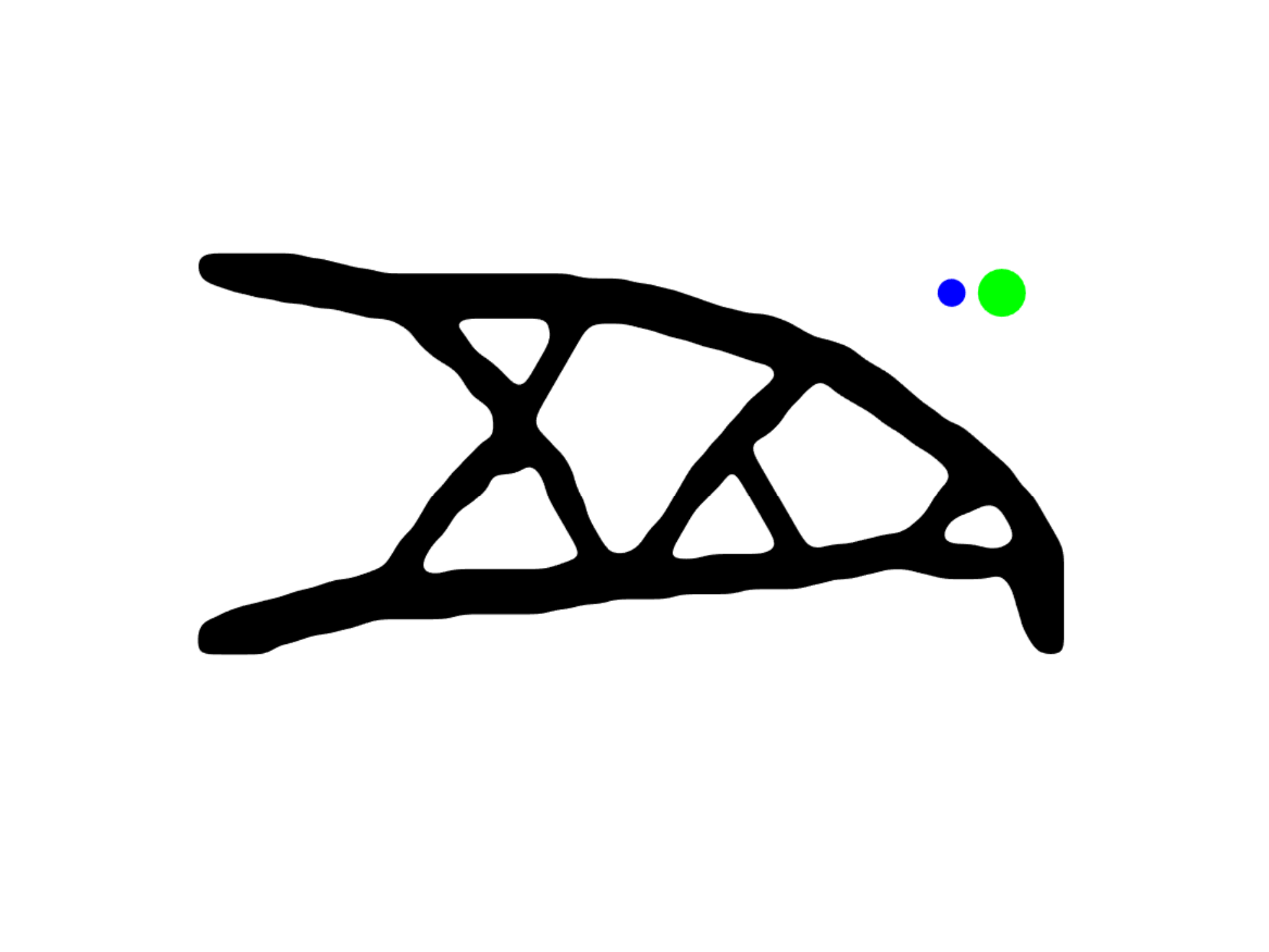}
		\caption{Smoothened solution for \ref{fig:Eg1_Eg2_Eg3_Eg4_new_logic_PEMa}. \\ $\Phi = 671.8$.} 
\label{solutions_with_imposed_min_ls_and_smoothened_boundariese}
	\end{subfigure}
	\hspace{10mm}
\begin{subfigure}[b]{.4\textwidth}
		\centering
		\captionsetup{font=scriptsize}
		\includegraphics[trim={2cm 3.5cm 2cm 2.5cm}, clip, scale = 0.45]{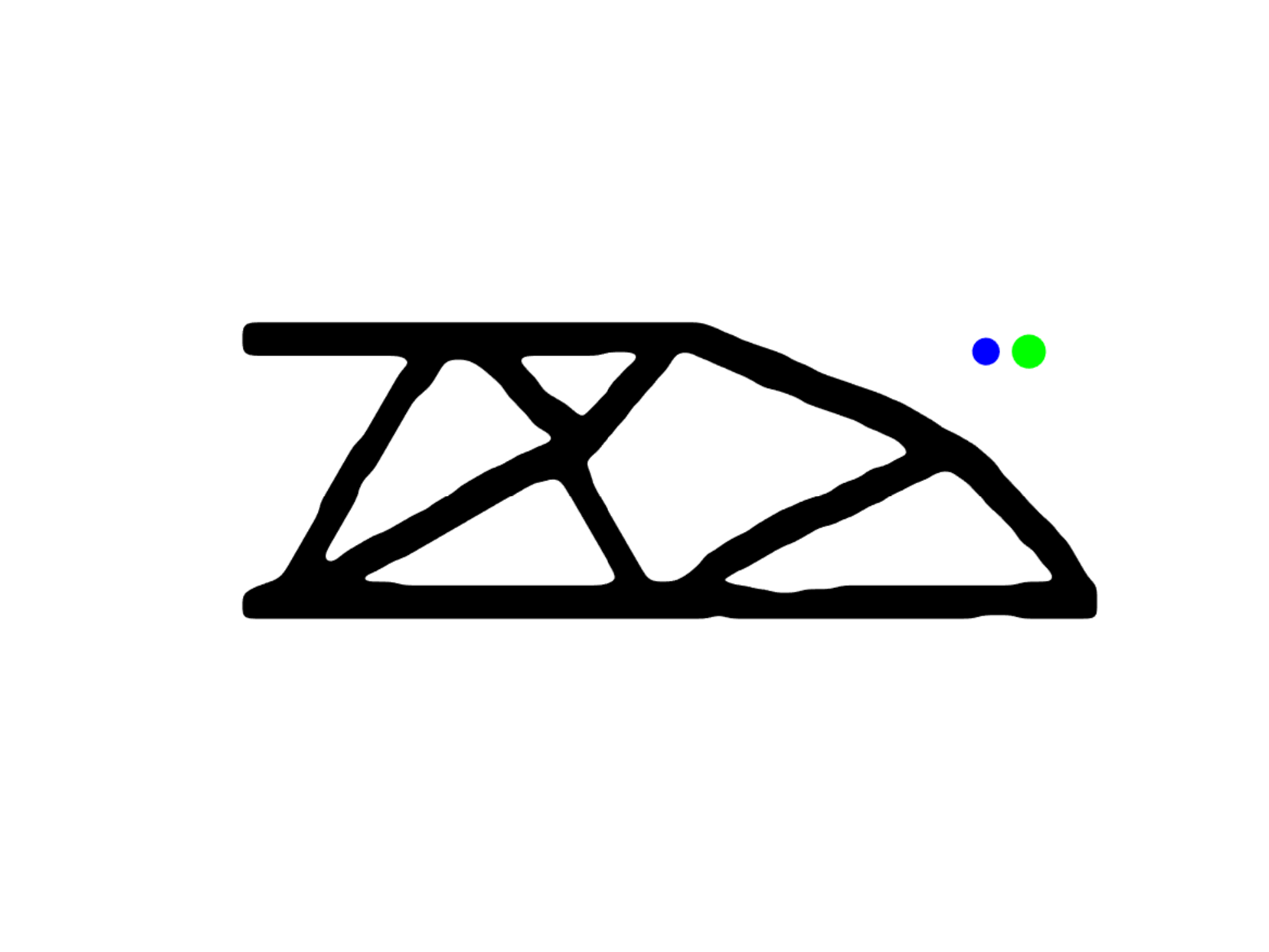}
		\caption{Smoothened solution for \ref{fig:Eg1_Eg2_Eg3_Eg4_new_logic_PEMb}. \\ $\Phi = 1092.9$.} 
\label{solutions_with_imposed_min_ls_and_smoothened_boundariesf}
	\end{subfigure}

 \hspace{5mm}	
\begin{subfigure}[b]{.4\textwidth}
		\centering
		\captionsetup{font=scriptsize}
		\includegraphics[trim={2cm 4.5cm 2cm 2.5cm}, clip, scale = 0.45]{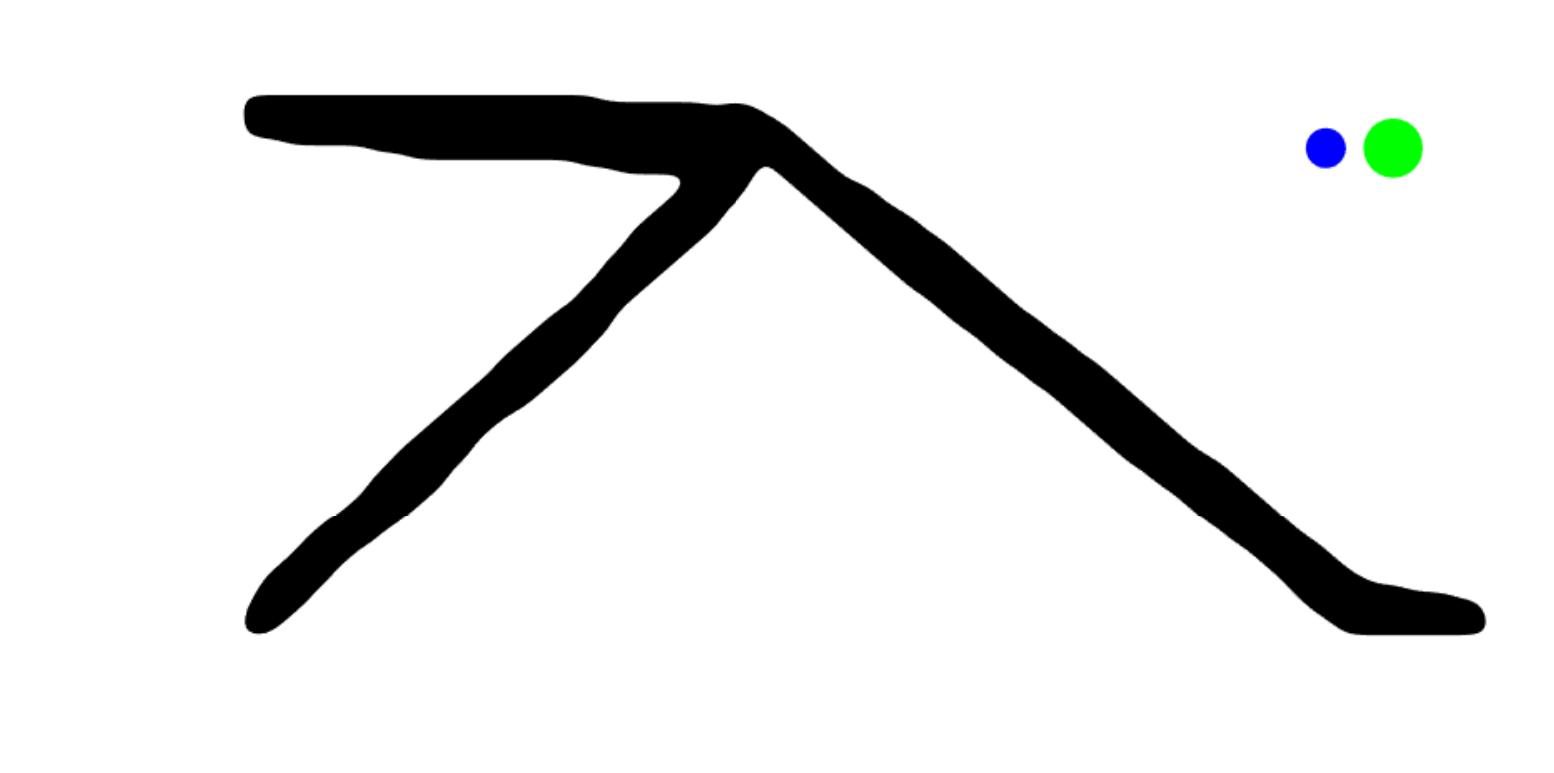}
		\caption{Smoothened solution for \ref{fig:Eg1_Eg2_Eg3_Eg4_new_logic_PEMc}. \\ $\Phi = -0.178$.} 
\label{solutions_with_imposed_min_ls_and_smoothened_boundariesg}
	\end{subfigure}
	\hspace{10mm}
\begin{subfigure}[b]{.4\textwidth}
		\centering
		\captionsetup{font=scriptsize}
		\includegraphics[trim={2cm 3.5cm 2cm 2.5cm}, clip, scale = 0.45]{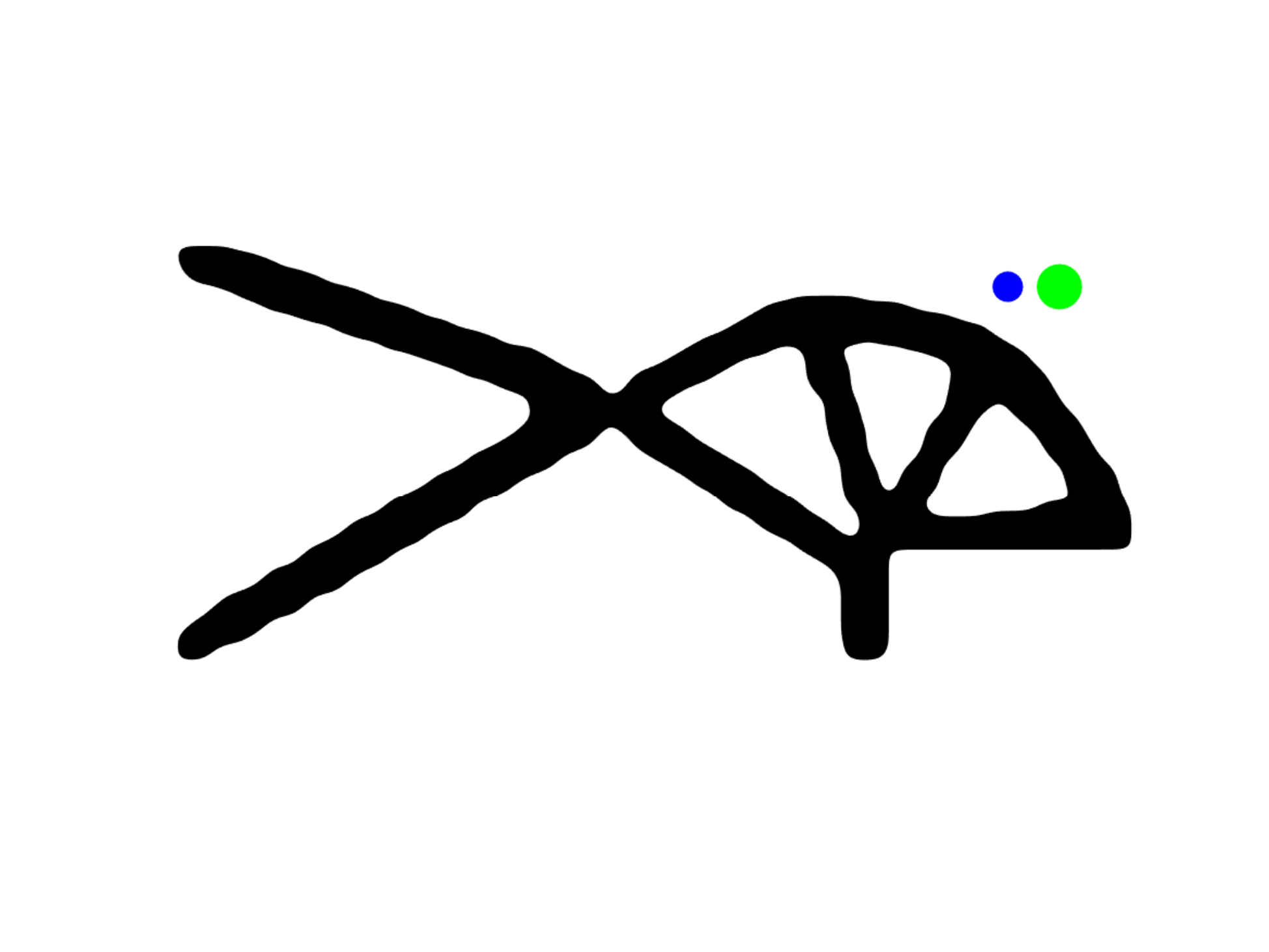}
		\caption{Smoothened solution for \ref{fig:Eg1_Eg2_Eg3_Eg4_new_logic_PEMd}. \\ $\Phi = -0.077$.} 
\label{solutions_with_imposed_min_ls_and_smoothened_boundariesh}
	\end{subfigure}	
		
	\caption{Perfectly {\it solid-void} solutions with smoothened boundaries (steps = 20) and imposed minimum length scales.}	
	\label{solutions_with_imposed_min_ls_and_smoothened_boundaries}
\end{figure}

\indent
While many solutions in Figs.  \ref{fig:Eg1_Eg2_Eg3_Eg4_new_logic_NEM}, \ref{fig:Eg1_Eg2_Eg3_Eg4_new_logic_PEM} and \ref{fig:Eg1_Eg2_Eg3_Eg4_post_review_NEM} have well defined boundaries, in some cases, e.g., in  Figs.  \ref{fig:Eg1_Eg2_Eg3_Eg4_new_logic_NEMb} and Fig.  \ref{fig:Eg1_Eg2_Eg3_Eg4_new_logic_PEMd} boundaries are undulated and have gray cells. Bare essential post processing which involves removal of void cells followed by smoothing of boundaries (meshes with any element type and howsoever fine will contain boundary notches once the void cells are removed) cannot be avoided if the obtained solutions are to be manufactured. Boundary smoothing \cite{kumar2015topology, saxena2011topology} is employed in the past with honeycomb meshes to address such undulations. Gray cells at mask boundaries are treated as filled. Such mask-based methods are capable of yielding crisp boundaries at any stage of topology optimization, an attribute that can be exploited to solve more involved problems, e.g., ones involving contact interactions \cite{kumar2016synthesis, kumar2019computational}. An added advantage with imposing length scales explicitly is that given a skeleton of any solution, cells which should be filled and void are known precisely, e.g, Fig. \ref{fig:Eg1_Eg2_Eg3_Eg4_post_review_NEM}. As an example, solutions in Figs. \ref{fig:Eg1_Eg2_Eg3_Eg4_new_logic_NEM} and Fig.  \ref{fig:Eg1_Eg2_Eg3_Eg4_new_logic_PEM} are shown in perfect 0-1 state, with imposed minimum length scales and smoothened boundaries in Fig. \ref{solutions_with_imposed_min_ls_and_smoothened_boundaries}. Undulations get reduced with the possibility of length scale definitions becoming better, e.g., in Figs. \ref{solutions_with_imposed_min_ls_and_smoothened_boundariesd} and \ref{solutions_with_imposed_min_ls_and_smoothened_boundariesh} wherein hinges seem to satisfy the minimum length scale. Respective change (increase) in the objective is marginal.

\section{Closure}
\label{closure}

\indent This paper investigates the role of elliptical masks, both negative and positive, in small deformation topology optimization. Explicit, skeleton-based,  length scales are imposed on solid states of the topologies defined by a group of hexagonal cells. To impose length scales explicitly, a novel skeletonization algorithm for hexagonal tessellation is presented and employed. Noting that there may exist an implicit interdependence between the volume fraction, minimum and maximum length scales, and that the length scale measures used work well only with well formed skeletons, a two-stage methodology that involves obtaining solutions by solving a sequence of optimization problems is proposed wherein length scales are specified as design parameters, and volume fraction and relaxation parameters are determined systematically. It is intended for the volume fraction to be as low as possible in order that length scales on the void states could be controlled indirectly. The procedure, though heuristic, yields solutions wherein length scales on solid states are satisfied by-and-large. However, certain sites may remain thinner or thicker than specified which is expected as the length scale constraints imposed are global in nature. While one demonstrates feasibility of attaining the desired length scales with the proposed methodology on bench mark problems, desirable solutions may not always be attainable. Investigations are planned in future for an improved approach to control  length scales on {\it solid} and {\it void} states more strictly/directly so that the obtained topologies with elliptical masks and honeycomb tessellation could be fabricated readily with concurrent, advanced manufacturing technologies. Better and effective, easy to implement, length scale measures are sought that are independent of topological skeletons. The proposed method is also computationally expensive and future effort will be geared towards making it more efficient.

\section{Replication of Results}
Results presented herein may be replicated by making modifications to the base MATLAB code provided in \cite{saxena2011topology}, and description on the SLS methodology and skeletonization in this paper. The corresponding author may be contacted in case there are additional queries. 

\section*{Acknowledgment}
The authors would like to acknowledge valuable comments from Prof. Ole Sigmund, Department of Mechanical Engineering, Section for Solid Mechanics, 
Technical University of Denmark, Building 404, Room 136, DK-2800, Lyngby, Denmark.

\section*{Contributions}
Nikhil Singh contributed in composing and coding the skeletonization process, working out some examples and preparation of the manuscript. Prabhat Kumar contributed in initial MMOS formulation with negative elliptical masks. Anupam Saxena contributed in coding the overall formulation with negative and positive elliptical masks, working on the analytical example, generation of examples and preparation of the manuscript. Significant part of this work was accomplished when the corresponding author was visiting AICES, RWTH Aachen University, Schinkelstrasse 2, 52062 Aachen, Germany.
\bibliography{My_Collection}
\bibliographystyle{hieeetr}

\begin{appendices}

\section{Skeletonization with honeycomb meshes}
\label{skeletonization}

To implement explicit length scale constraints on the solid states, a new algorithm for skeletonization of intermediate topologies obtained from honeycomb meshes is described. As cells enclosed within, on (the boundary), or outside the masks are precisely known, cell densities are known in their true 0-1 forms. A thresholding procedure to convert a gray scale solution into a binary one, as in \cite{expl_ls_simp_2014}, is not required.

The skeleton, or, \textit{medial topological contour} of a domain with void and filled regions is a (set of) curve(s) of unit cell thickness that captures topology of the domain such that each cell on the curve separates at least two void boundaries. Topology with the filled and void cells at any step of the algorithm is referred to as the \textit{configuration topology}.  Each iteration in the algorithm consists of three main steps, (a) \textit{contour detection}, (b) \textit{contour refinement} and (c) \textit{skeleton point retention}. There are certain cases under which the iterative process fails to give the desired result. Such cases, and the method to get the desired skeleton is also discussed. The overall notion is that one expands the voids continuously while retaining the path of collision between void boundaries. The stage just before the void boundaries merge into each other is one where the curve(s) thus generated form(s) the skeleton for the structure.

The proposed algorithm makes use of only the local information around a concerned cell, that is, information about its immediate neighbors. This makes the algorithm suitable for generic use. Also, the algorithm uses properties of neighboring cell arrangement which remain unchanged under rotation and reflection, taking care of multiple cases all at once, thus making it efficient.

\subsection{Contour cell detection}
A contour cell is a filled cell  present at the interface between filled regions and voids, or, between independent voids. In essence, the contour represents void boundaries. Cells are segregated into two categories, \textit{boundary} and \textit{interior} cells. Any cell surrounded by six neighboring cells is an interior cell, else, it is categorized as a boundary cell. An interior cell with density one, is part of the contour if at least one of its immediate neighbors is void, while a boundary cell, which is at the domain boundary, is part of the contour if its cell density is one.

Contour detection is achieved by summing the density values of neighboring cells and detecting the density of the cell itself. If the sum of values of neighboring cell densities is below six and the cell itself is filled, the latter is recognized as a contour cell. The topology thus created by the contour cells is referred as the \textit{contour topology}. The next step, contour refinement,  makes use of only the contour topology and does not require information about the configuration topology.

\subsection{Contour refinement}
\begin{figure}[h]
	\centering
	\includegraphics[width=.6\columnwidth]{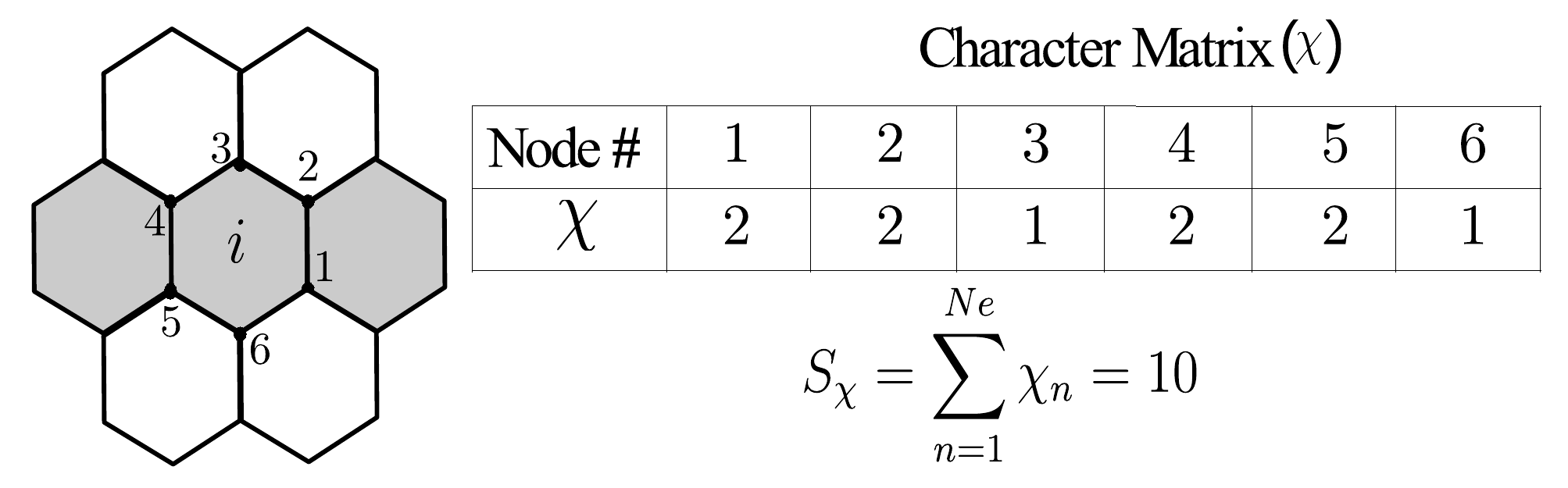}
	\caption{Character vector $\chi$ for a contour cell surrounded by 2 contour cells.}
	\label{character_matrix}
\end{figure}

\begin{figure}[h]
	\centering
	\begin{subfigure}[b]{0.3\textwidth}
		\centering
		\includegraphics[width=0.6\textwidth]{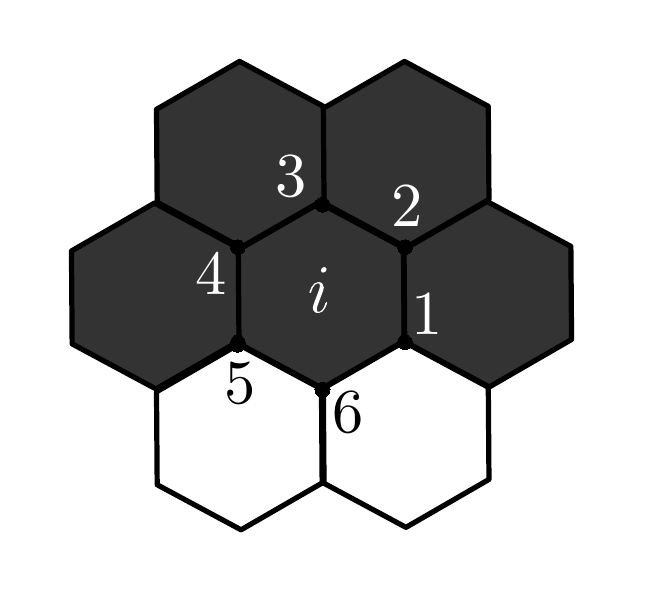}
		\label{fig:chi_config_1}
	\end{subfigure}
	\begin{subfigure}[b]{0.3\textwidth}
		\centering
		\includegraphics[width=0.6\textwidth]{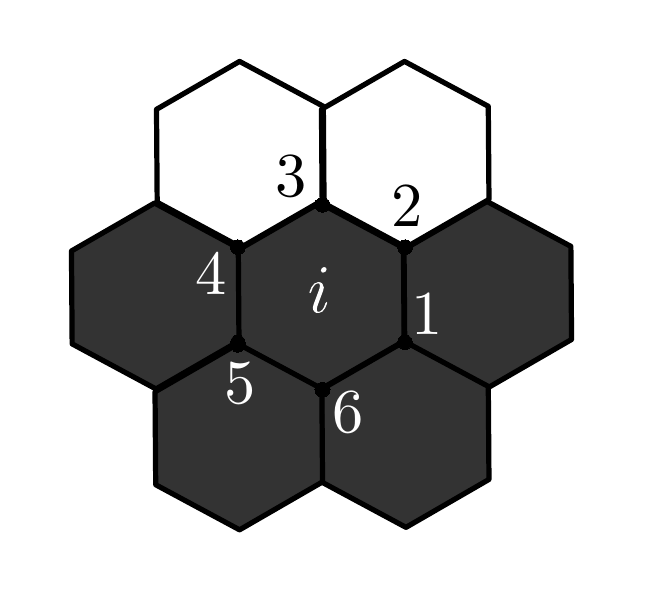}
		\label{fig:chi_config_2}
	\end{subfigure}
	\begin{subfigure}[b]{0.3\textwidth}
		\centering
		\includegraphics[width=0.6\textwidth]{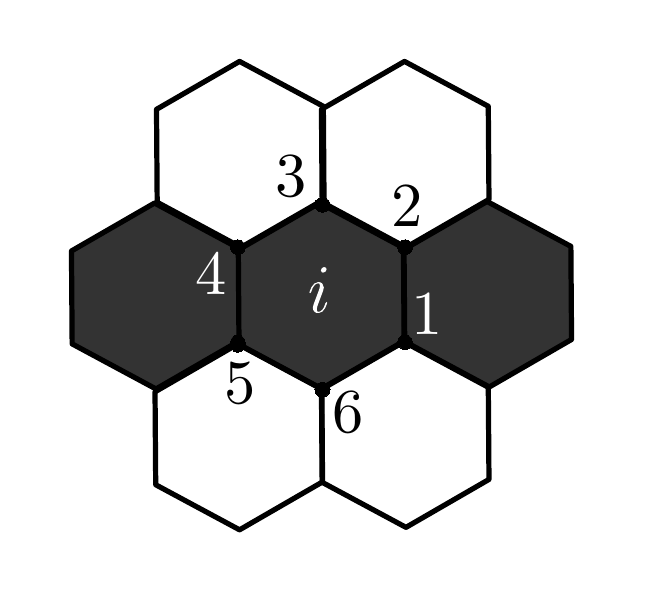}
		\label{fig:chi_config_3}
	\end{subfigure}
	\caption{Possible configuration topologies associated with contour topology in Fig. \ref{character_matrix}}
	\label{fig:config_topology_character_matrix}
\end{figure}

A contour cell is considered unnecessary if removing it from the contour does not alter the contour topology. Herein, we determine unnecessary cells on the contour and remove them. We explore immediate neighbors of a contour cell to determine the latter's importance on the contour. To distinguish between different configurations, we define a property of a contour cell called its character $\chi$. $\chi$ of a cell is a vector containing six entries displaying the number of surrounding contour cells around each node of the cell. Hence, entries of $\chi$ ranges between 1 and 3. An example is shown in Fig. \ref{character_matrix}, where highlighted (gray) cells are part of the contour. Also, all possible local configuration topologies which can lead to the local contour topology in Fig. \ref{character_matrix} are presented in Fig. \ref{fig:config_topology_character_matrix} where filled cells are highlighted in black and void cells in white. 


Vector $\chi$ is rotation variant but sum of its elements ($S_\chi$) for a contour cell is dependent only on the number of neighboring contour cells and independent of their specific arrangement. This property can be verified through the reasoning that when a contour cell is added adjacent to the reference cell, irrespective of the former's specific position, $S_\chi$ for the reference cell increases by 2. This is because, each neighboring cell shares two nodes with the reference cell and therefore, adding a contour cell in the neighborhood increases the value of elements of $\chi$ corresponding to the shared nodes by 1 each. Noting that a contour cell with zero surrounding contour cells has $S_\chi = 6$ and the observation above, the
 number of surrounding contour cells ($NSe$) for a contour cell can be given by:

\begin{eqnarray}
NSe = \left(S_\chi - 6\right)/2.
\end{eqnarray}

Contour cells are categorized based on the number of surrounding contour cells. For any contour cell, count of surrounding contour cells ranges from 0 to 5, therefore producing six possible cases. Amongst these, the case with 0 neighboring contour cells refers to a filled cell surrounded by void and hence has to be retained in the skeleton. The case of 1 neighboring contour cell can represent the end of a branch in the skeleton, hence, it also has to be retained. We now consider remaining 4 unique cases.

\subsubsection{Case I: Two surrounding contour cells} 
\label{case:1}
\begin{figure}[htb]
	\centering
	\begin{subfigure}[b]{0.3\textwidth}
		\centering
		\includegraphics[width=0.6\linewidth]{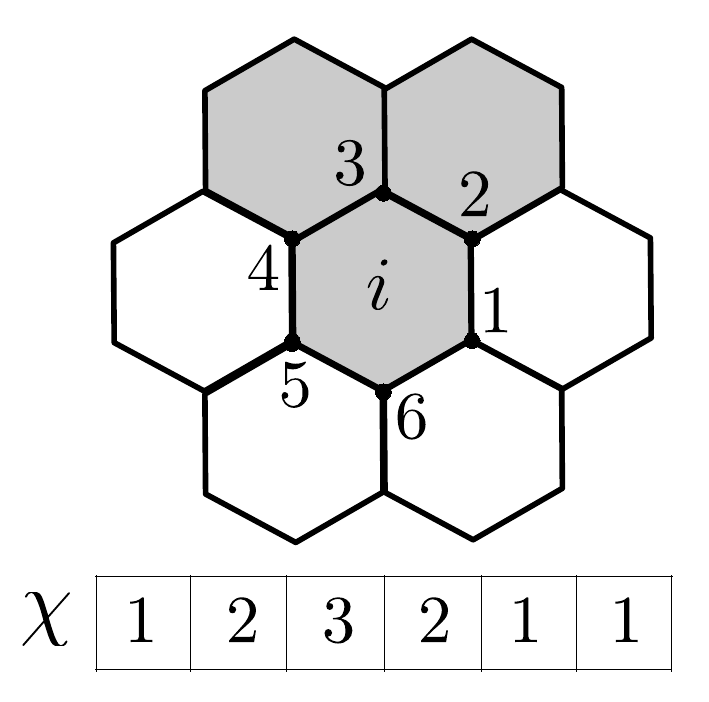}%
		\caption{Type (A) contour topology: (2+0)}
		\label{fig:2(a)}
	\end{subfigure}
	\begin{subfigure}[b]{0.54\textwidth}
		\centering
		\includegraphics[width=0.31\linewidth]{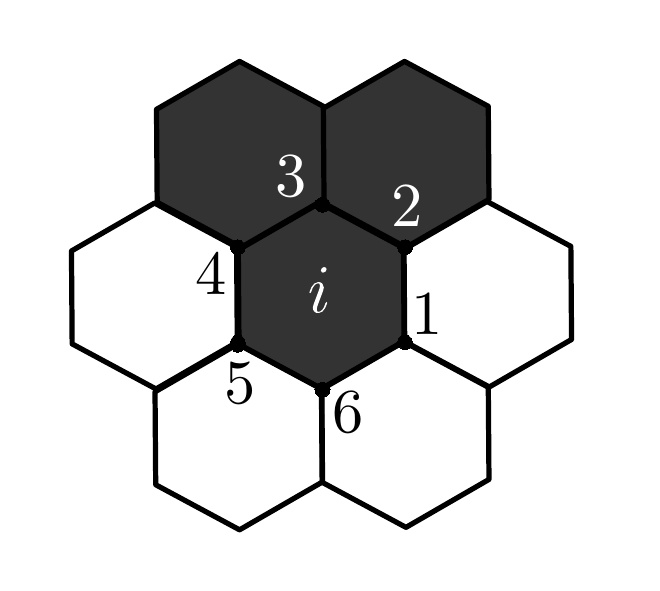}
		\vspace{6mm}
		\caption{Possible configuration topology for contour topology in Fig. \ref{fig:2(a)}}
		\label{fig:2(a)_config}
	\end{subfigure}
	\vskip\baselineskip
	\begin{subfigure}[b]{0.3\textwidth}
		\centering
		\includegraphics[width=0.6\linewidth]{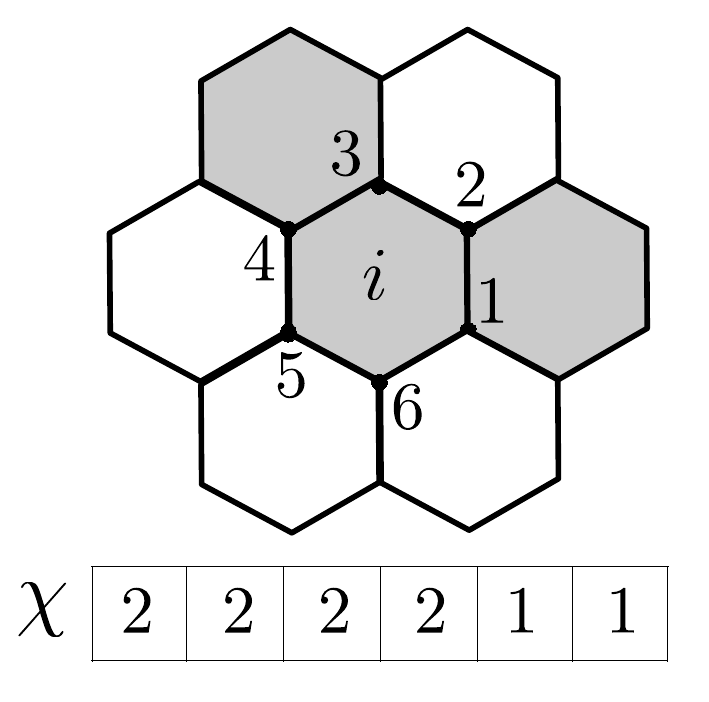}%
		\caption{Type (B) contour topology: (1+1)}
		\label{fig:2(b)}
	\end{subfigure}
	\begin{subfigure}[b]{0.54\textwidth}
		\centering
		\includegraphics[width=0.31\linewidth]{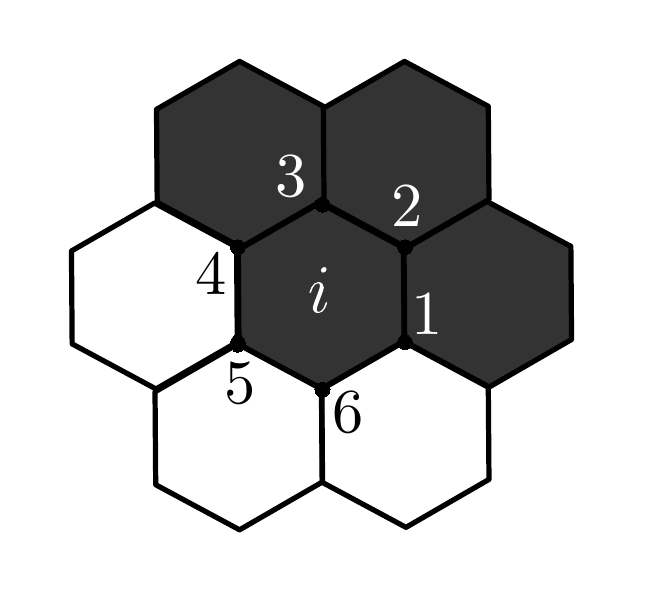}
		\includegraphics[width=0.31\linewidth]{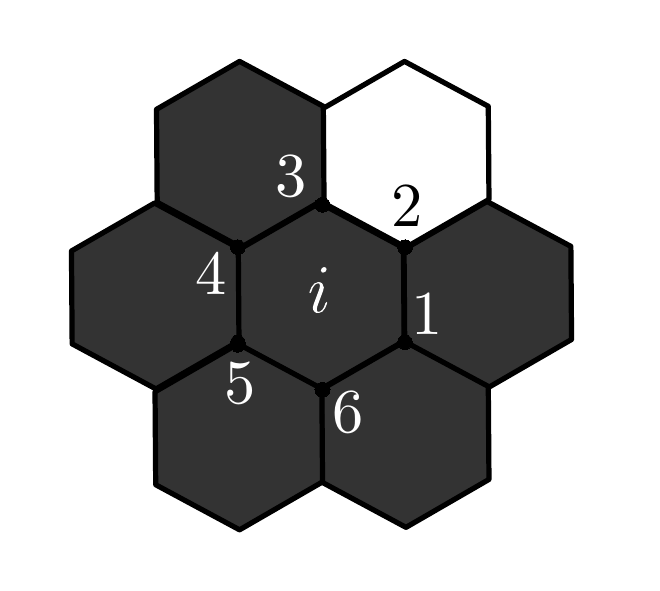}
		\includegraphics[width=0.31\linewidth]{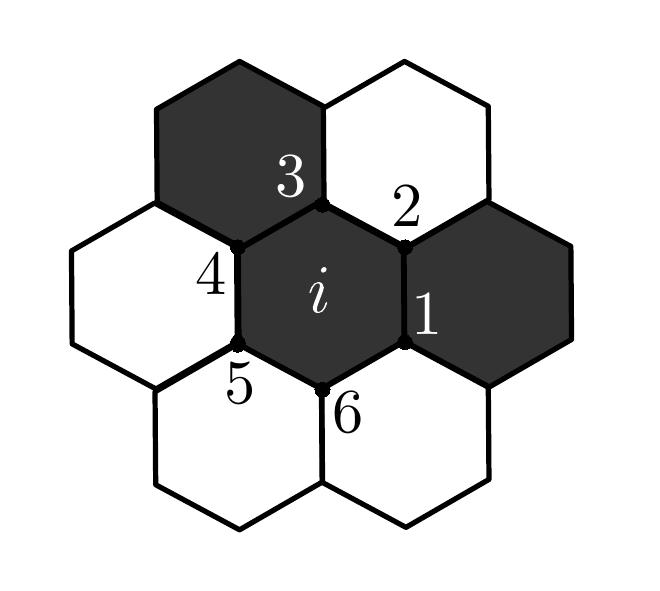}
		\vspace{6mm}
		\caption{Possible configuration topologies for contour topology in Fig. \ref{fig:2(b)}}
		\label{fig:2(b)_config}
	\end{subfigure}
	\caption{Case I: Possible contour topologies of a reference cell surrounded by 2 contour cells and corresponding configuration topologies.}
	\label{fig:2}
\end{figure}

All possible configurations of two contour cells around a reference contour cell can be distinguished into two types, type (A) and type (B). In type (A), the two contour cells are neighboring cells (Fig. \ref{fig:2(a)}) while in type (B), the two cells are placed separate from each other (Fig. \ref{fig:2(b)}). The possible local configuration topologies associated with contour topologies in Fig. \ref{fig:2(a)} and \ref{fig:2(b)}  are presented in Fig. \ref{fig:2(a)_config} and \ref{fig:2(b)_config} respectively. All possible contour topologies of type (A) are rotations of the configuration in Fig. \ref{fig:2(a)}. Similarly, all possible contour topologies of type (B) are rotations or reflections of the configurations in Fig. \ref{fig:2(b)} and Fig. \ref{character_matrix}.

\begin{figure}[htb]
	\centering
	\begin{subfigure}[b]{0.3\textwidth}
		\centering
		\includegraphics[width=0.6\linewidth]{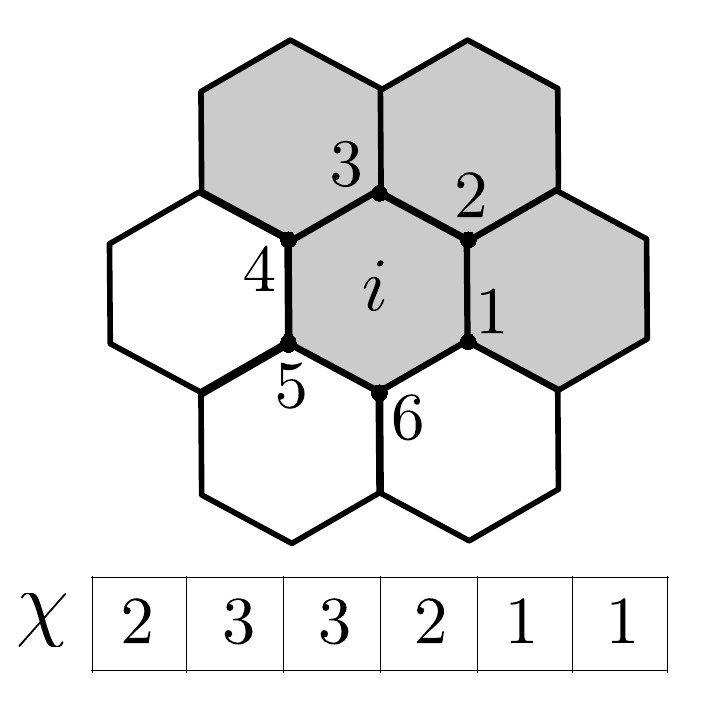}%
		\caption{Type (A) contour topology: (3+0)}
		\label{fig:3(a)}
	\end{subfigure}
	\begin{subfigure}[b]{0.54\textwidth}
		\centering
		\includegraphics[width=0.31\linewidth]{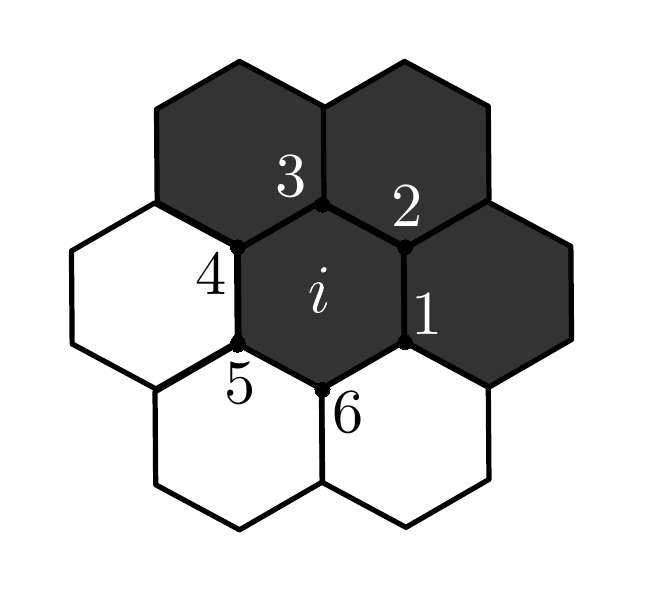}
		\vspace{6mm}
		\caption{Possible configuration topology for contour topology in Fig. \ref{fig:3(a)}}
		\label{fig:3(a)_config}
	\end{subfigure}
	\vskip\baselineskip
	\begin{subfigure}[b]{0.3\textwidth}
		\centering
		\includegraphics[width=0.6\linewidth]{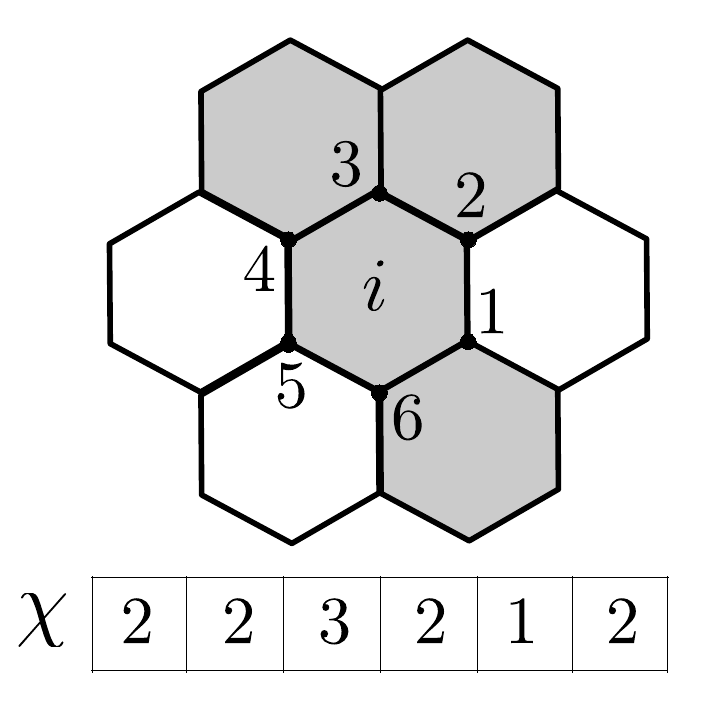}%
		\caption{Type (B) contour topology: (2+1)}
		\label{fig:3(b)}
	\end{subfigure}
	\begin{subfigure}[b]{0.54\textwidth}
		\centering
		\includegraphics[width=0.31\linewidth]{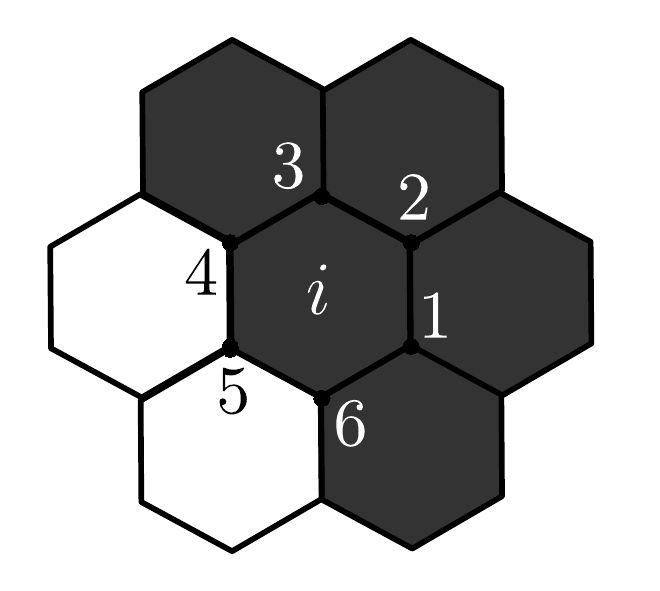}
		\includegraphics[width=0.31\linewidth]{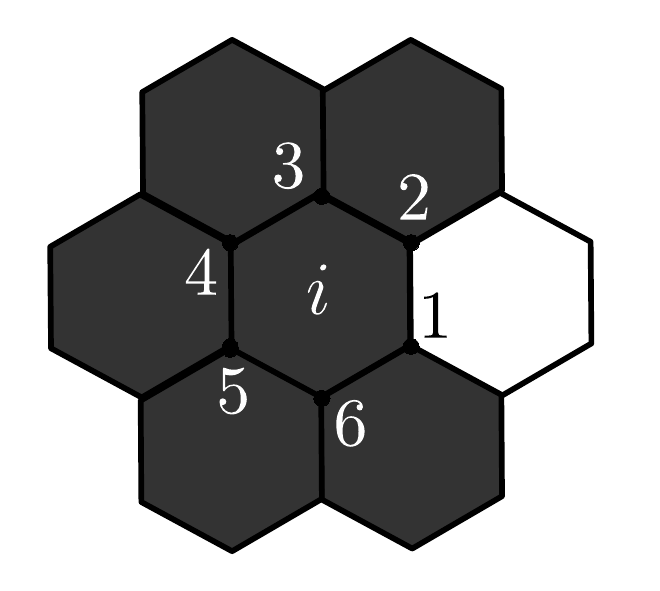}
		\includegraphics[width=0.31\linewidth]{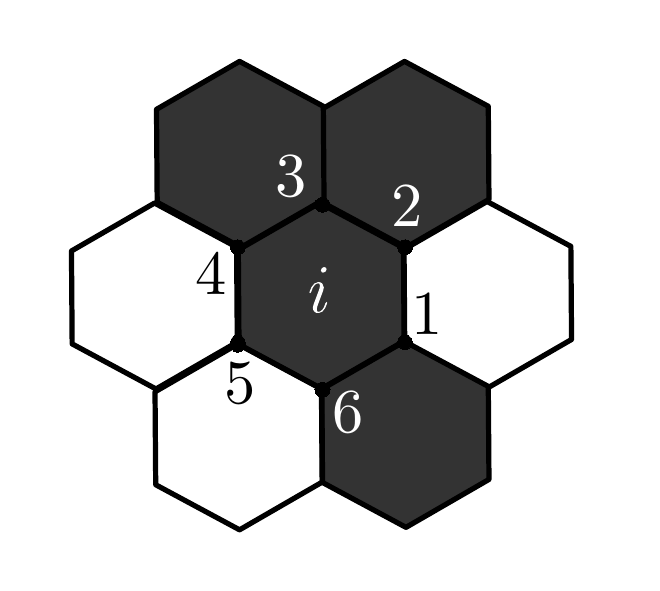}
		\vspace{6mm}
		\caption{Possible configuration topologies for contour topology in Fig. \ref{fig:3(b)}}
		\label{fig:3(b)_config}
	\end{subfigure}
	\vskip\baselineskip
	\begin{subfigure}[b]{0.3\textwidth}
		\centering
		\includegraphics[width=0.6\linewidth]{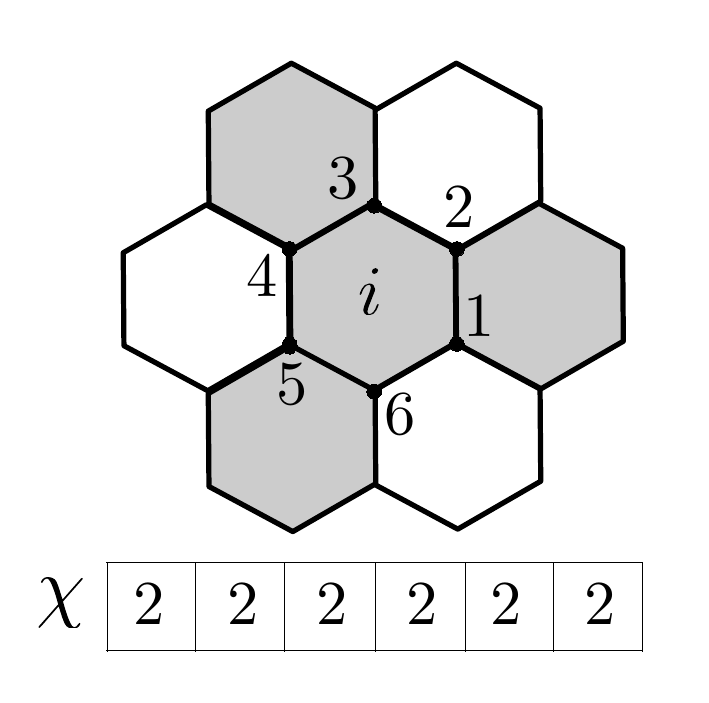}%
		\caption{Type (C) contour topology: (1+1+1)}
		\label{fig:3(c)}
	\end{subfigure}
	\begin{subfigure}[b]{0.54\textwidth}
		\centering
		\includegraphics[width=0.31\linewidth]{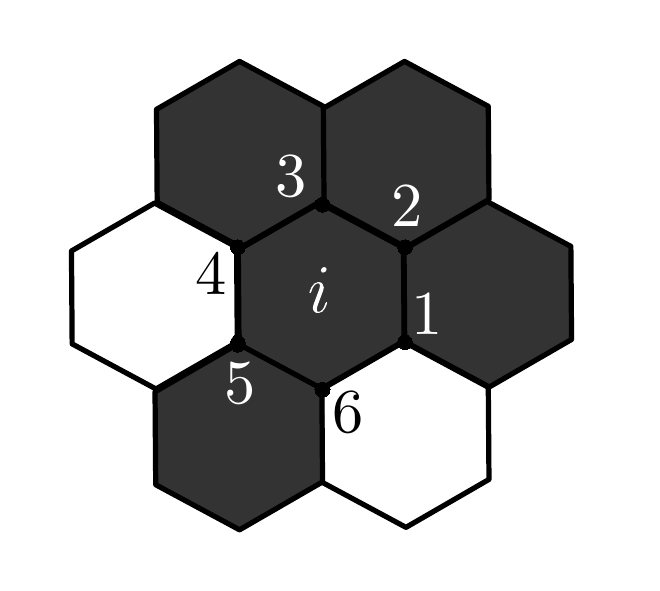}
		\includegraphics[width=0.31\linewidth]{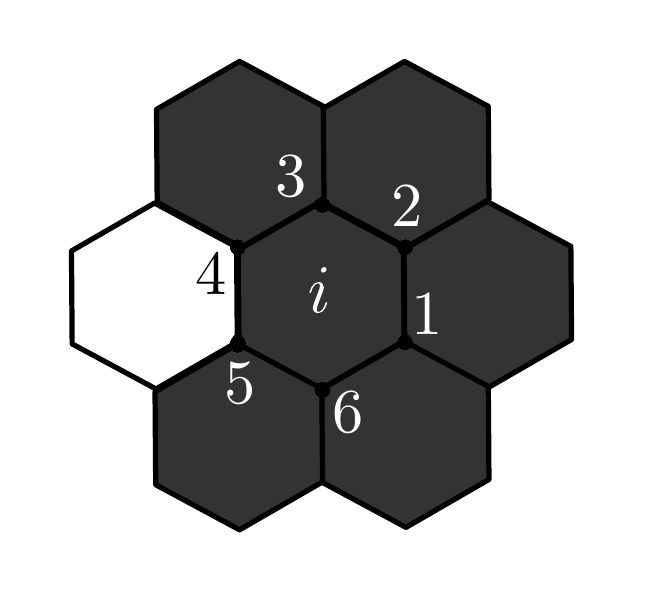}
		\includegraphics[width=0.31\linewidth]{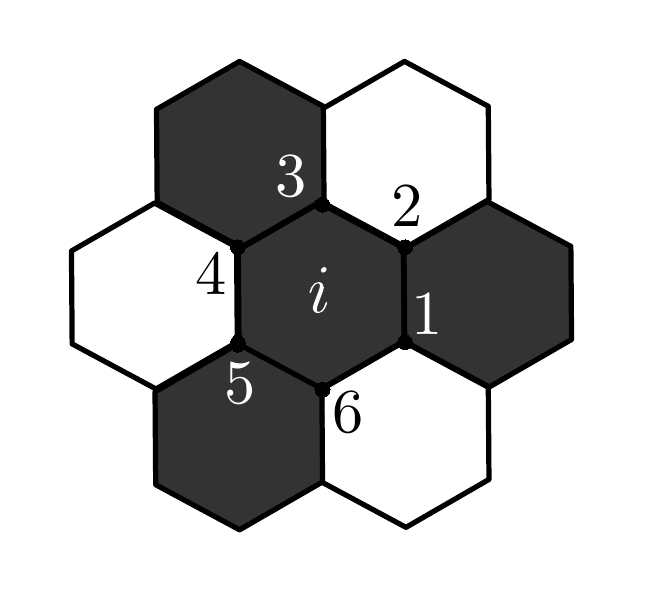}
		\vspace{6mm}
		\caption{Some possible configuration topologies for contour topology in Fig. \ref{fig:3(c)}}
		\label{fig:3(c)_config}
	\end{subfigure}
	\caption{Case II: Possible contour topologies of a reference cell surrounded by 3 contour cells and corresponding configuration topologies.}
	\label{fig:3}
\end{figure}

One observes that removing the reference cell $i$ from contour in type (A) configurations does not alter the topology of the contour. This is because removing the cell does not connect two regions which were initially seperated by the contour. Thus, the reference cell in type (A) is removed and its density is changed to 0 only during the skeletonization process. Reference cells in type (B) configurations are retained on the contour at this stage.

It is not necessary to distinguish between all possible configurations but to only categorise contour cells into type (A) or type (B) to determine their importance on the contour. This is achieved by counting the number of entries as 1 in $\chi$. For type (A) configurations, number of entries as 1 in $\chi$ is 3 while for type (B) it is 2. 

\subsubsection{Case II: Three surrounding contour cells}
All possible configurations of three contour cells around a reference contour cell can be distinguished into 3 types, type (A), type (B) and type (C). In type (A), three contour cells are consecutively placed as illustrated in Fig. \ref{fig:3(a)}. In type (B), two of the three cells are immediate neighbors while the third is placed separate from the two (Fig. \ref{fig:3(b)}). In type (C), all three cells are positioned seperate from each other (Fig. \ref{fig:3(c)}). All local configuration topologies associated with contour topologies in Fig. \ref{fig:3(a)} and \ref{fig:3(b)}  are presented in Fig. \ref{fig:3(a)_config} and \ref{fig:3(b)_config} respectively while Fig. \ref{fig:3(c)_config} only presents three of the seven possible configuration topologies associated with the contour topology in Fig. \ref{fig:3(c)}. Remaining four configuration topologies are rotations of the configuration topologies presented. All possible contour topologies of type (A) and type (C) are rotations of the topology in Fig. \ref{fig:3(a)} and Fig. \ref{fig:3(c)} respectively. Likewise, all possible contour topologies of type (B) are rotations or reflections of the one in Fig. \ref{fig:3(b)}.

Applying the same reasoning as in Case I, cell $i$ in type (A) is an unnecessary cell on the contour, and therefore is removed with its density changed to 0 only within the skeletonization process. Cells in configurations type (B) and type (C) are retained on the contour for this step. To identify reference cells in configuration type (A), one makes use of the same method as in Case I. Count of entries as 1 in $\chi$ for type (A), (B) and (C) configurations is 2, 1 and 0 respectively.

\subsubsection{Case III: Four surrounding contour cells} 
\begin{figure}[htb]
	\centering
	\begin{subfigure}[b]{0.3\textwidth}
		\centering
		\includegraphics[width=0.6\linewidth]{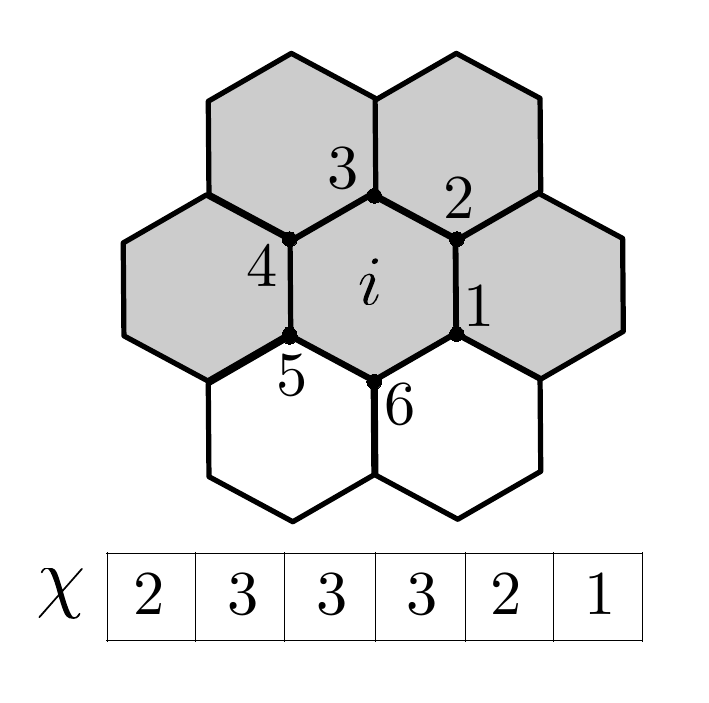}%
		\caption{Type (A) contour topology: (4+0)}
		\label{fig:4(a)}
	\end{subfigure}
	\begin{subfigure}[b]{0.54\textwidth}
		\centering
		\includegraphics[width=0.31\linewidth]{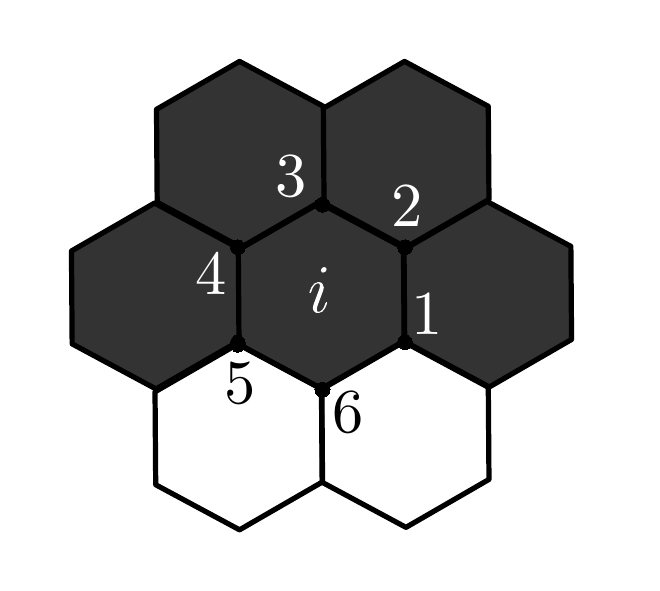}
		\vspace{6mm}
		\caption{Possible configuration topology for contour topology in Fig. \ref{fig:4(a)}}
		\label{fig:4(a)_config}
	\end{subfigure}
	\vskip\baselineskip
	\begin{subfigure}[b]{0.3\textwidth}
		\centering
		\includegraphics[width=0.6\linewidth]{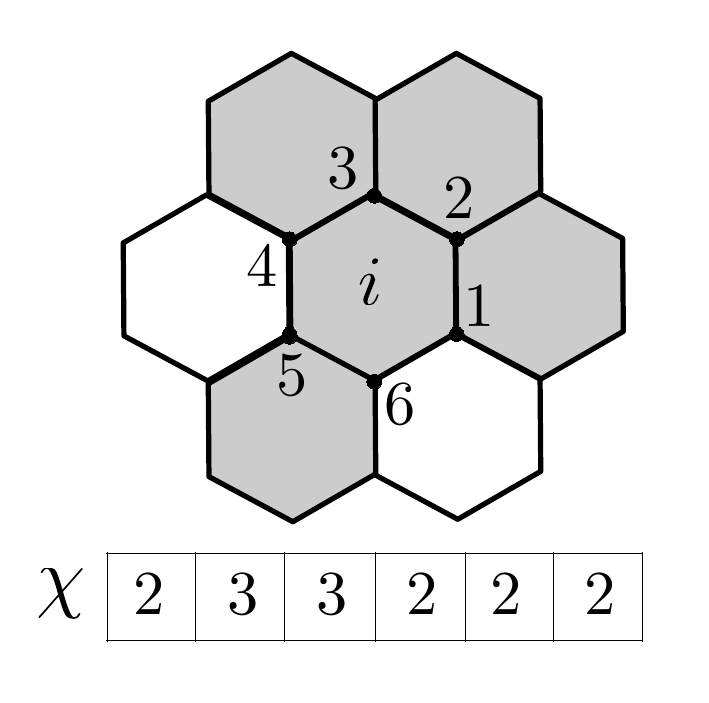}%
		\caption{Type (B) contour topology: (3+1)}
		\label{fig:4(b)}
	\end{subfigure}
	\begin{subfigure}[b]{0.54\textwidth}
		\centering
		\includegraphics[width=0.31\linewidth]{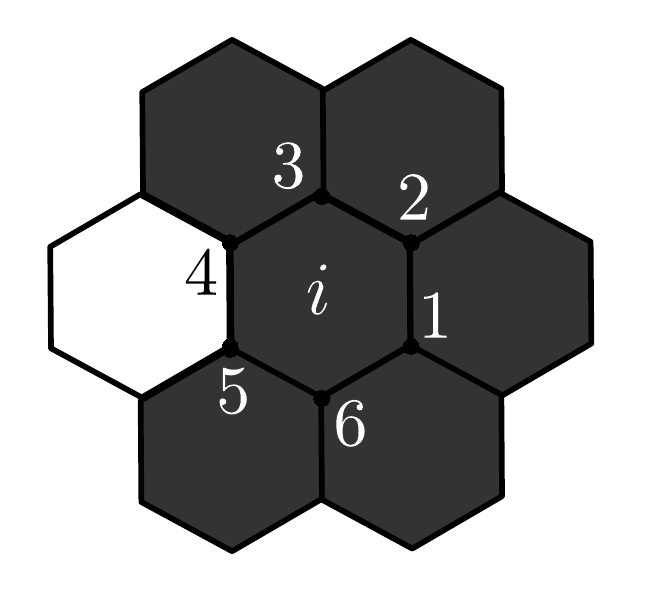}
		\includegraphics[width=0.31\linewidth]{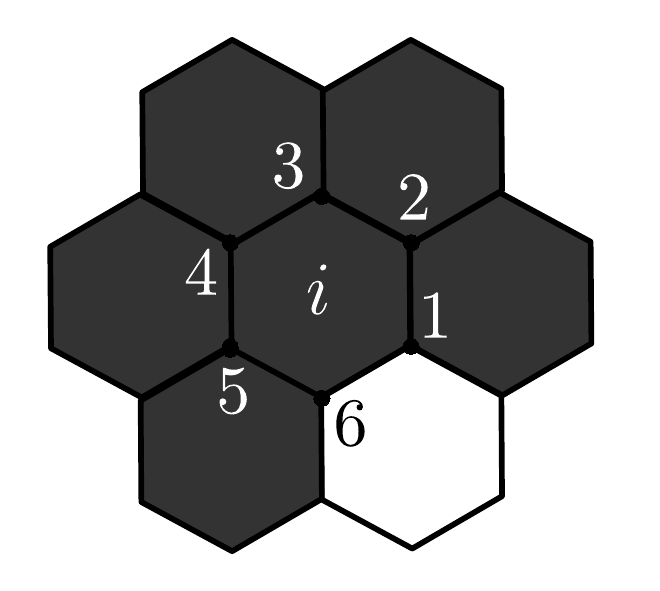}
		\includegraphics[width=0.31\linewidth]{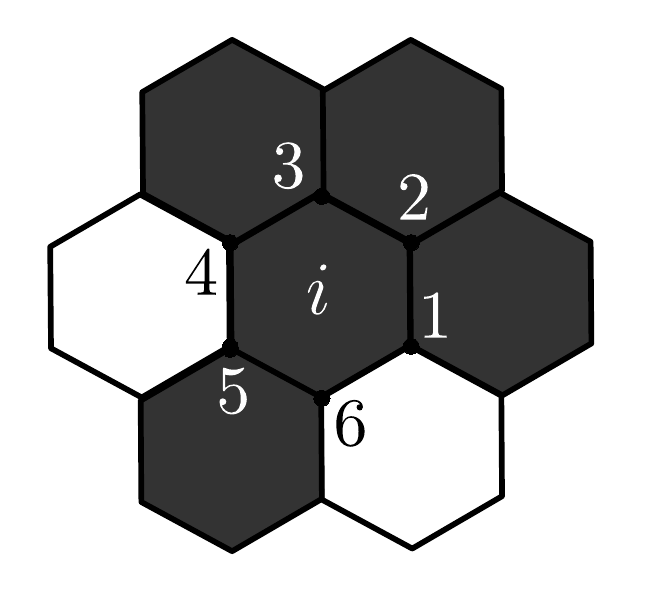}
		\vspace{6mm}
		\caption{Possible configuration topologies for contour topology in Fig. \ref{fig:4(b)}}
		\label{fig:4(b)_config}
	\end{subfigure}
	\vskip\baselineskip
	\begin{subfigure}[b]{0.3\textwidth}
		\centering
		\includegraphics[width=0.6\linewidth]{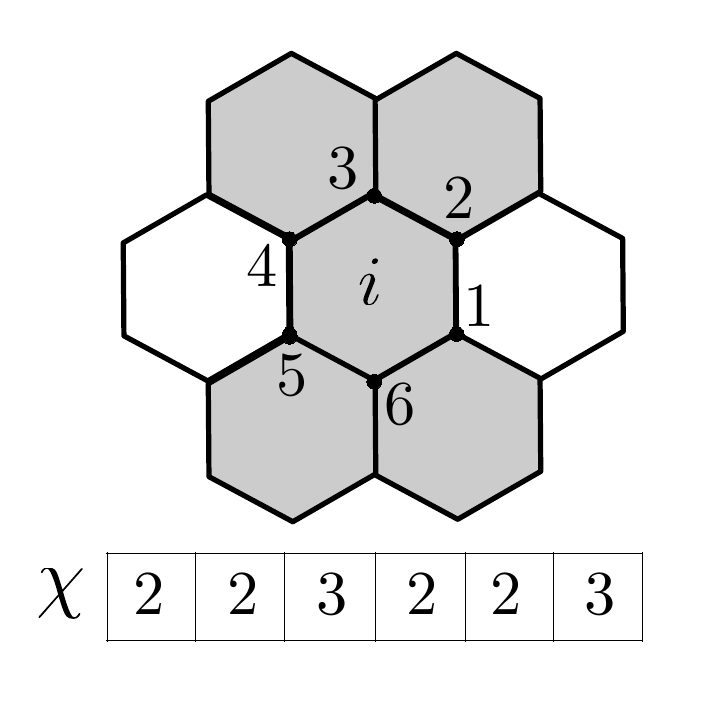}%
		\caption{Type (C) contour topology: (2+2)}
		\label{fig:4(c)}
	\end{subfigure}
	\begin{subfigure}[b]{0.54\textwidth}
		\centering
		\includegraphics[width=0.31\linewidth]{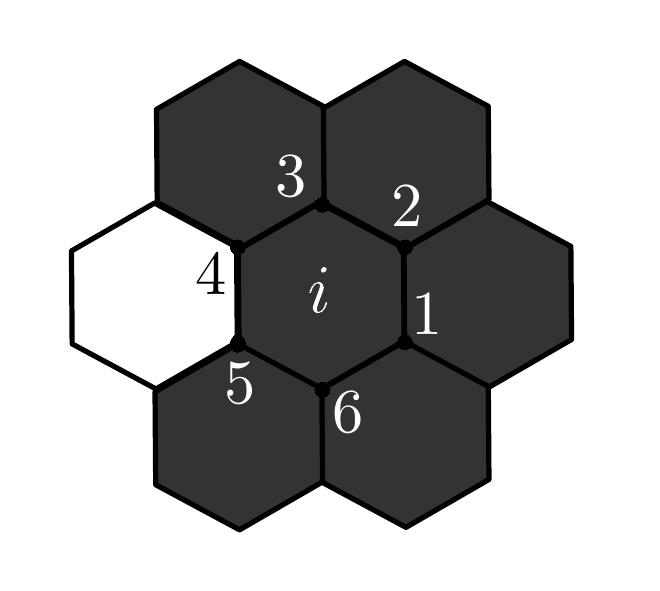}
		\includegraphics[width=0.31\linewidth]{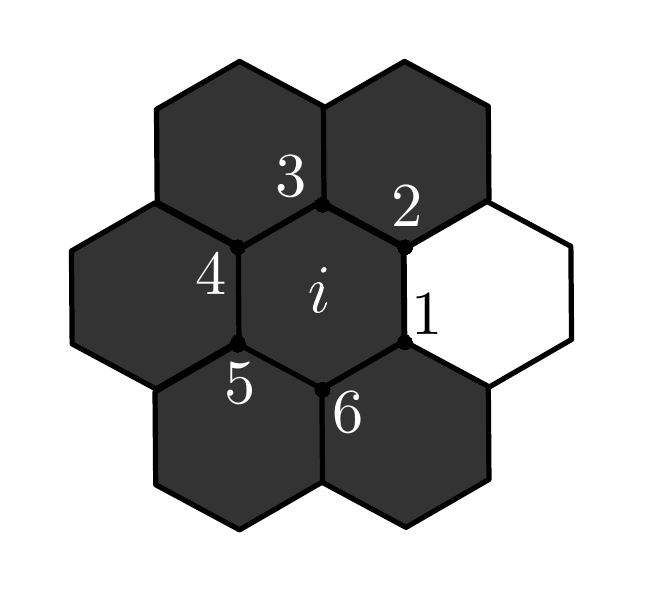}
		\includegraphics[width=0.31\linewidth]{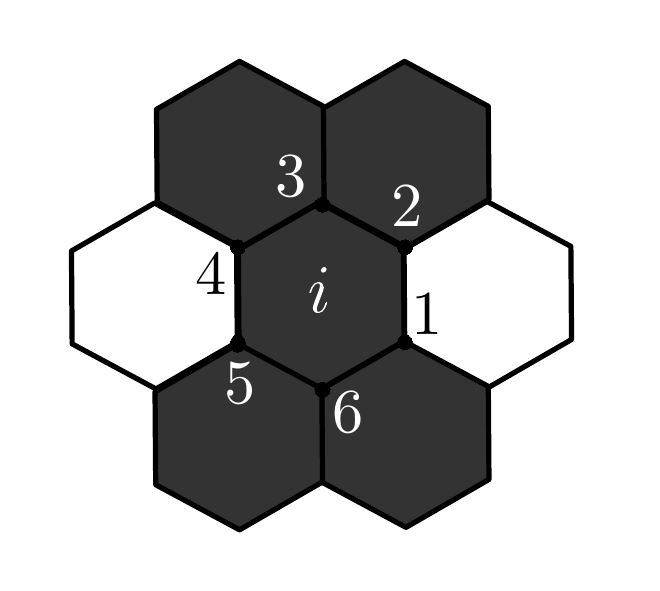}
		\vspace{6mm}
		\caption{Possible configuration topologies for contour topology in Fig. \ref{fig:4(c)}}
		\label{fig:4(c)_config}
	\end{subfigure}
	\caption{Case III: Possible contour topologies of a reference cell surrounded by 4 contour cells and corresponding configuration topologies.}
	\label{fig:4}
\end{figure}

All possible configurations of four contour cells around a reference contour cell can be distinguished into 3 types, type (A), type (B) and type (C). In type (A), the four contour cells are consecutive cells as illustrated in Fig. \ref{fig:4(a)}. In type (B), three of the four are consecutive cells while the fourth is placed separately (Fig. \ref{fig:4(b)}). In type (C), the four cells are divided into two pairs of neighboring cells and pairs are placed seperate from each other (Fig. \ref{fig:4(c)}). All local configuration topologies associated with contour topologies in Figs. \ref{fig:4(a)}, \ref{fig:4(b)} and \ref{fig:4(c)} are presented in Figs. \ref{fig:4(a)_config}, \ref{fig:4(b)_config} and \ref{fig:4(c)_config} respectively. All possible contour topologies of type (A) and type (C) are rotations of the topology in Fig. \ref{fig:4(a)} and Fig. \ref{fig:4(c)} respectively. Similarly, all possible contour topologies of type (B) are rotations or reflections of the topology in Fig. \ref{fig:4(b)}.

As in Case I, cell $i$ in configuration type (A) is unnecessary, and therefore is removed with its density changed to 0 within the skeletonization process. Reference cells in configuration type (B) and type (C) are retained. To identify the reference cells in configuration type (A), one counts entries as 1 in $\chi$ for type (A),  (B) and (C) configurations which are 1, 0 and 0 respectively.

\subsubsection{Case IV: Five surrounding contour cells} 
\begin{figure}[htb]
	\centering
	\begin{subfigure}[b]{0.3\textwidth}
		\centering
		\includegraphics[width=0.6\linewidth]{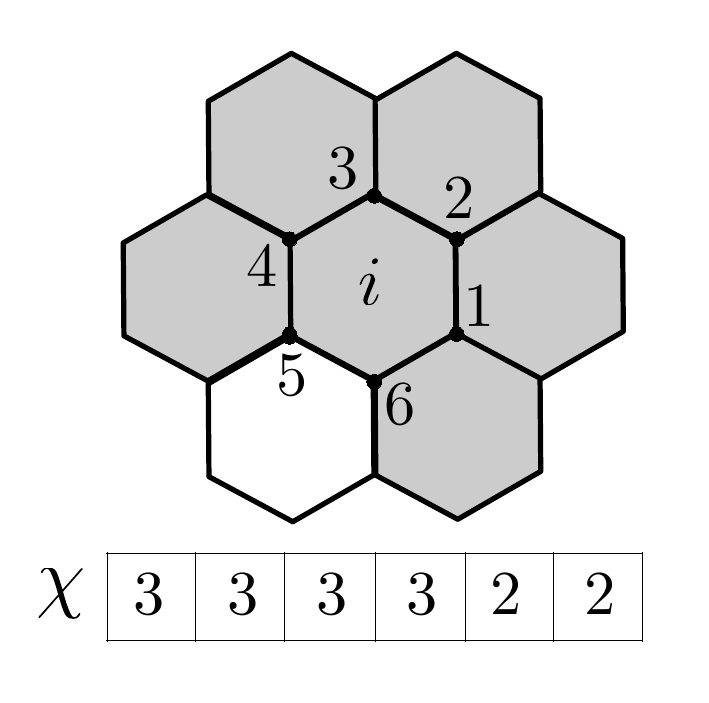}%
		\caption{Type (A) contour topology: (5+0)}
		\label{fig:5(a)}
	\end{subfigure}
	\begin{subfigure}[b]{0.54\textwidth}
		\centering
		\includegraphics[width=0.31\linewidth]{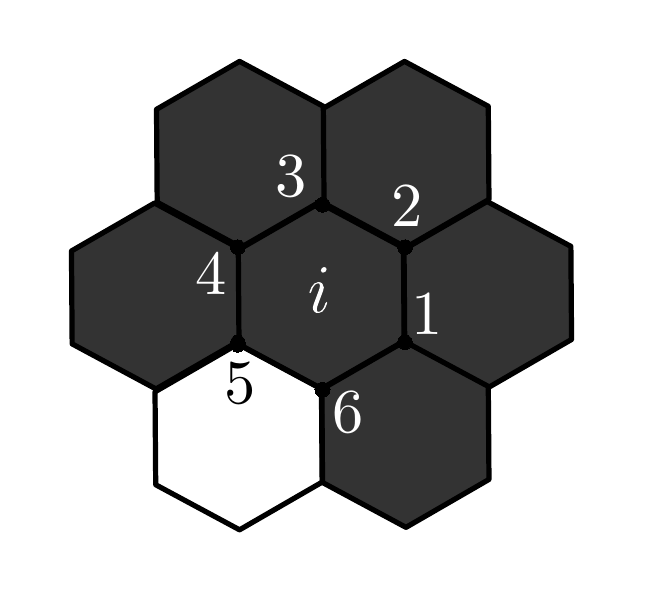}
		\vspace{6mm}
		\caption{Possible configuration topology for contour topology in Fig. \ref{fig:5(a)}}
		\label{fig:5(a)_config}
	\end{subfigure}
	\caption{Case IV: Possible contour topologies of a reference cell surrounded by 5 contour cells and corresponding configuration topologies.}
	\label{fig:5}
\end{figure}

Notwithstanding rotational symmetry, there is only one way to arrange five contour cells around a reference contour cell (Fig. \ref{fig:5}). Also, there is a unique configuration topology presented in Fig. \ref{fig:5(a)_config} associated with the configuration topology in Fig. \ref{fig:5(a)}.
As in Case I, cell $i$ in the given configuration is irrelevant to the contour, and therefore is removed with its density changed to 0 locally, within the skeletonization procedure.

\begin{figure}[h]	
	\centering	
	\includegraphics[width=.35\textwidth]{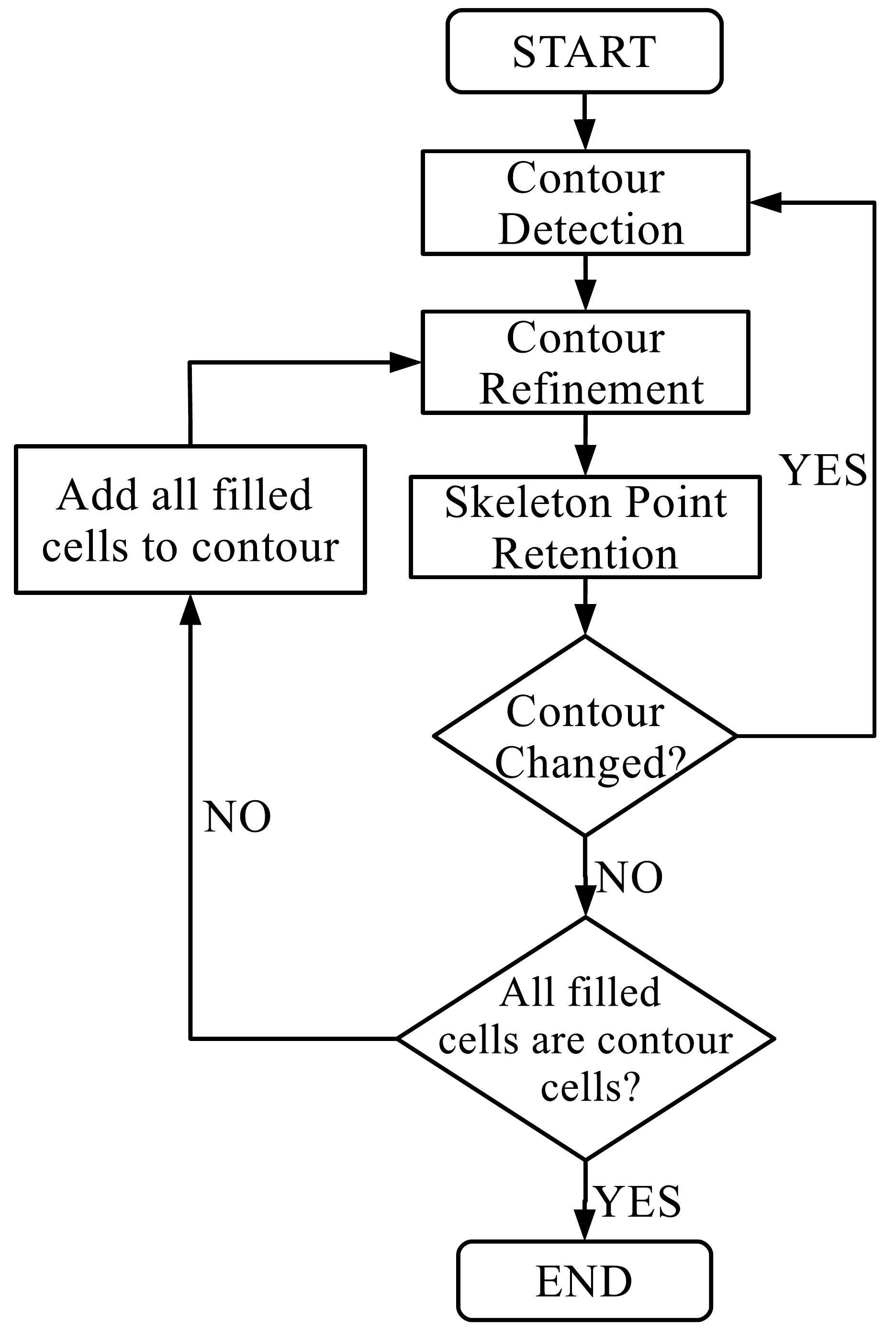}	
	\caption{Schematic of the skeletonization process}	
	\label{fig:6}	
\end{figure}

\begin{figure}[h]	
	\centering
	\includegraphics[width=.2\textwidth]{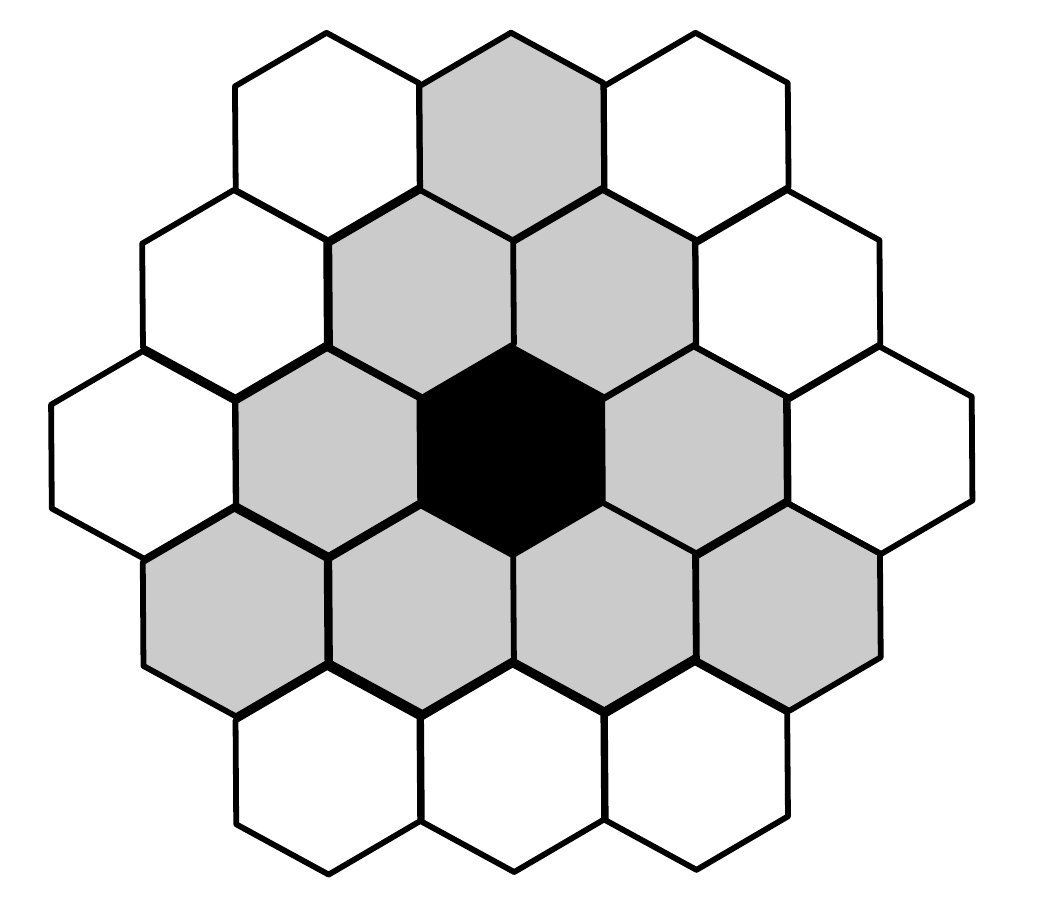}
	\caption{A special case in skeletonization}	
	\label{fig:7}	
\end{figure}

\subsection{Skeleton point retention}
After the contour is refined, the next step is to expand the void boundaries and retain necessary cells on the contour. To identify these, we use the fact that any closed contour topology of unit cell thickness homeomorphic to a circle has local contour topology of Case I, type (B). As contour refinement step 
eventually yields a contour topology of unit thickness, any contour cell with local contour topology, post contour refinement, other than the above signifies an intersection between two or more closed curves each of which is homeomorphic to a circle. Such cells need to be retained to preserve the original topology. Also, any contour cell with a local contour topology pertaining to Case I, type (B) and having physical voids on both sides has to be retained. This is because, removing such a point will connect two voids, and alter the parent topology. Contour cells with local contour topology of Case I, type B having atleast one neighboring cell with density 1 and not part of the contour are removed from the contour and their densities locally set to 0. The iterative process is continued until two consecutive iterations produce the same contour.

\subsection{Special cases}
At the end of the iterative process, one checks for cells with density 1 which are not part of the contour topology. If such cells exist, there are filled regions left and voids are yet to expand to produce the skeleton of the domain. Such conditions arise when there are multiple void boundaries collapsing at a single cell, leading to a structure in which the iterative process fails. Fig. \ref{fig:7} illustrates one such case where a filled cell, highlighted in black, is surrounded by contour cells, in gray, in a way that all contour cells have 3 neighboring contour cells and hence, are retained on the contour during the iterative process. All such cases that the iterative process fails to identify are treated as special cases.

\begin{figure}[H]	
\begin{subfigure}[b]{.24\textwidth} 
		\centering
		\captionsetup{font=scriptsize}
		\includegraphics[trim={3.5cm 2.5cm 2.5cm 2.5cm}, clip, scale = 0.3]{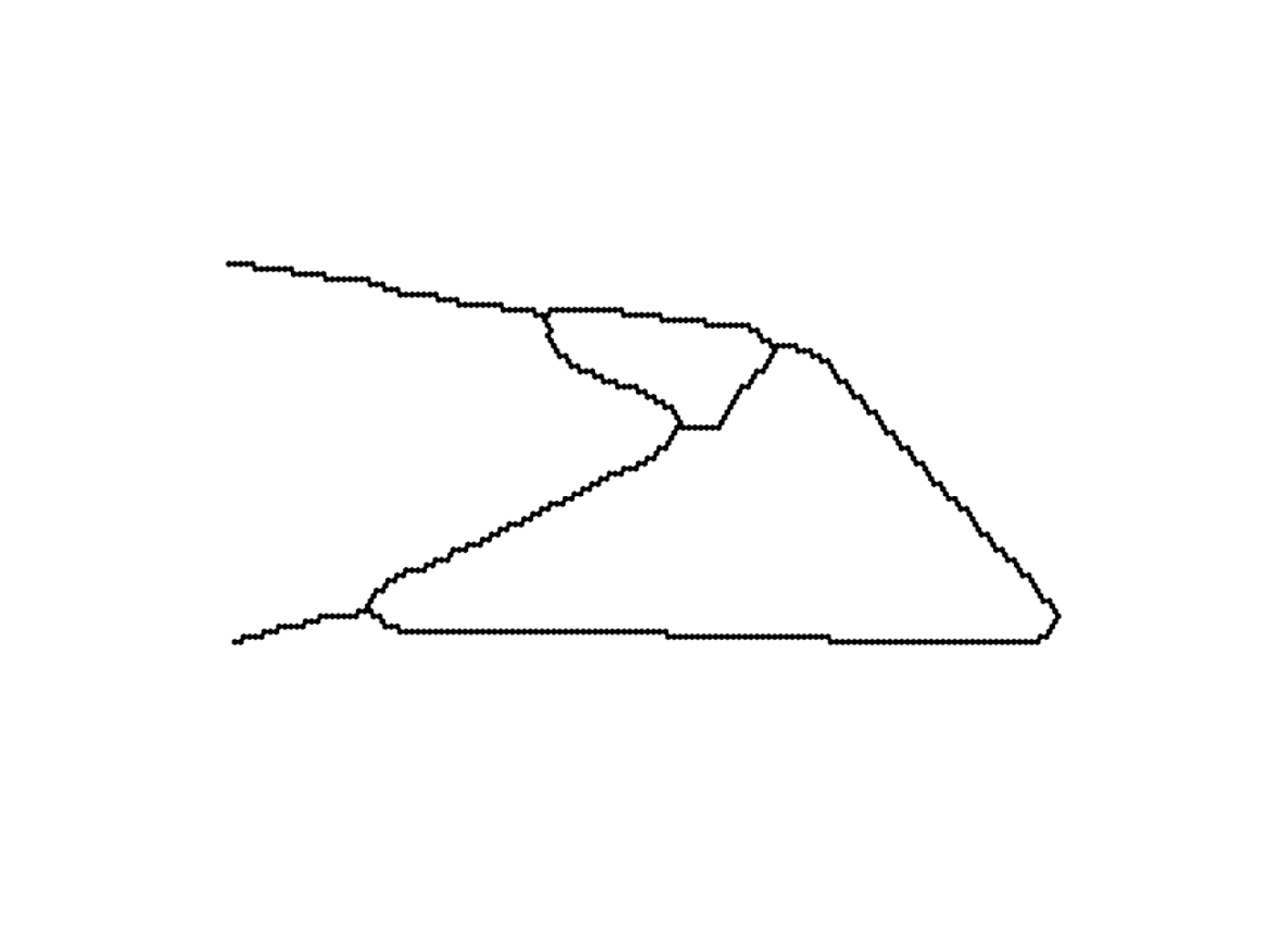}
		\caption{Skeleton (Solution Fig. \ref{fig:Eg1_Eg2_Eg3_Eg4_new_logic_NEM} a.)} 
	\end{subfigure}
\begin{subfigure}[b]{.24\textwidth}
		\centering
		\captionsetup{font=scriptsize}
		\includegraphics[trim={3.5cm 2.5cm 2.5cm 2.5cm}, clip, scale = 0.3]{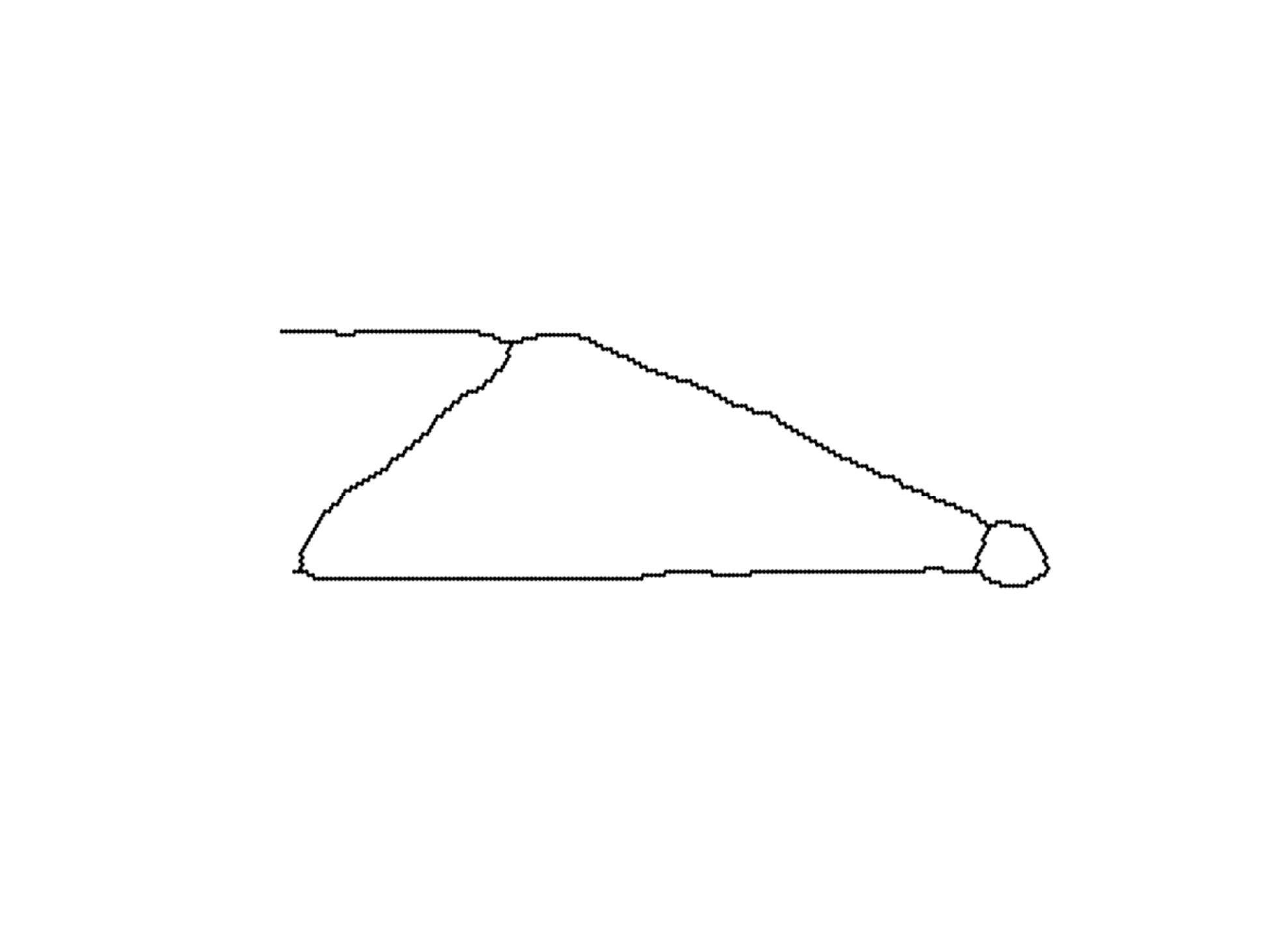}
		\caption{Skeleton (Solution Fig. \ref{fig:Eg1_Eg2_Eg3_Eg4_new_logic_NEM} b.)} 
	\end{subfigure}
\begin{subfigure}[b]{.24\textwidth}
		\centering
		\captionsetup{font=scriptsize}
		\includegraphics[trim={3.5cm 2.5cm 2.5cm 2.5cm}, clip, scale = 0.3]{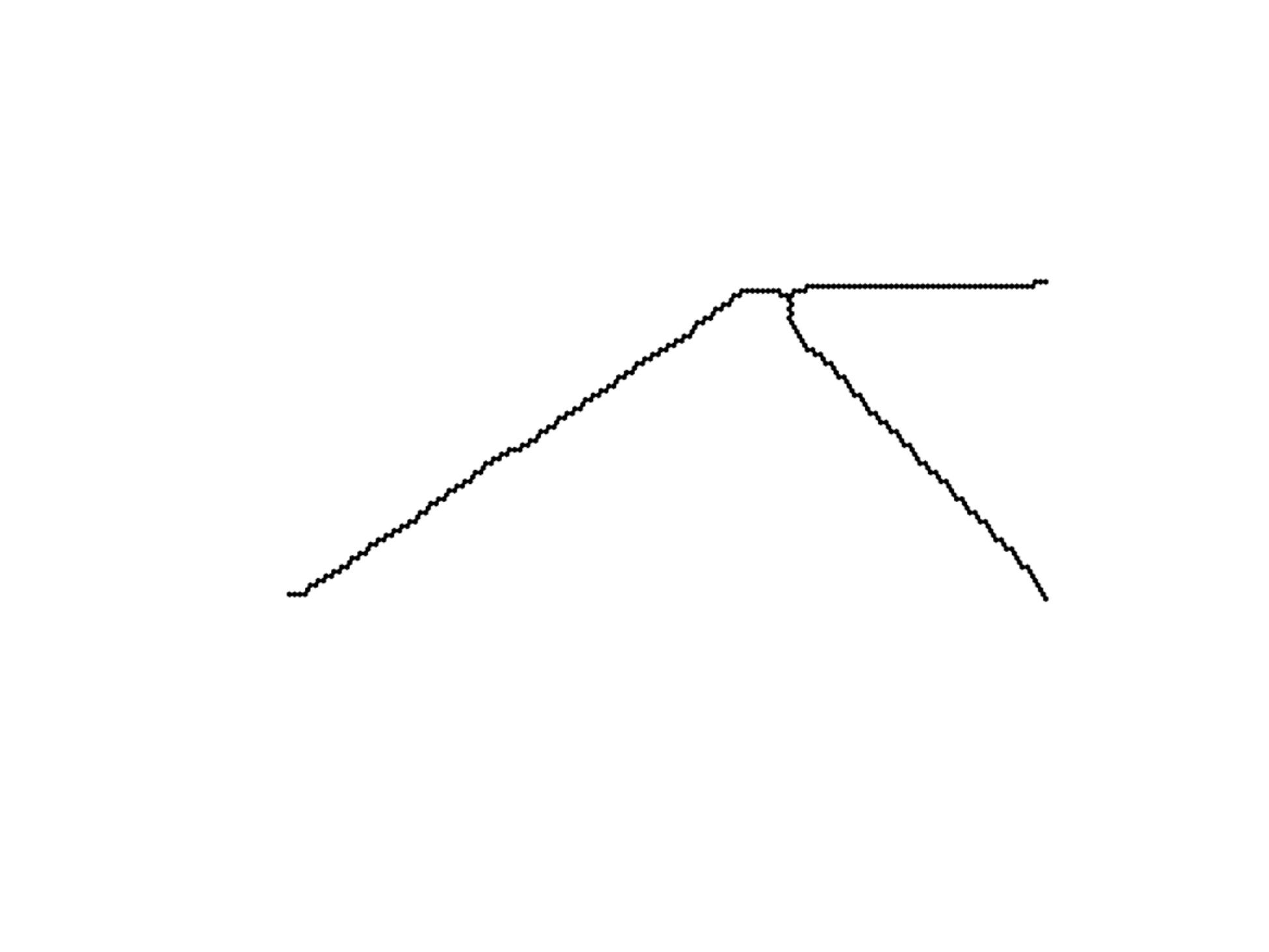}
		\caption{Skeleton (Solution Fig. \ref{fig:Eg1_Eg2_Eg3_Eg4_new_logic_NEM} c.)} 
	\end{subfigure}
\begin{subfigure}[b]{.24\textwidth}
		\centering
		\captionsetup{font=scriptsize}
		\includegraphics[trim={3.5cm 2.5cm 2.5cm 2.5cm}, clip, scale = 0.3]{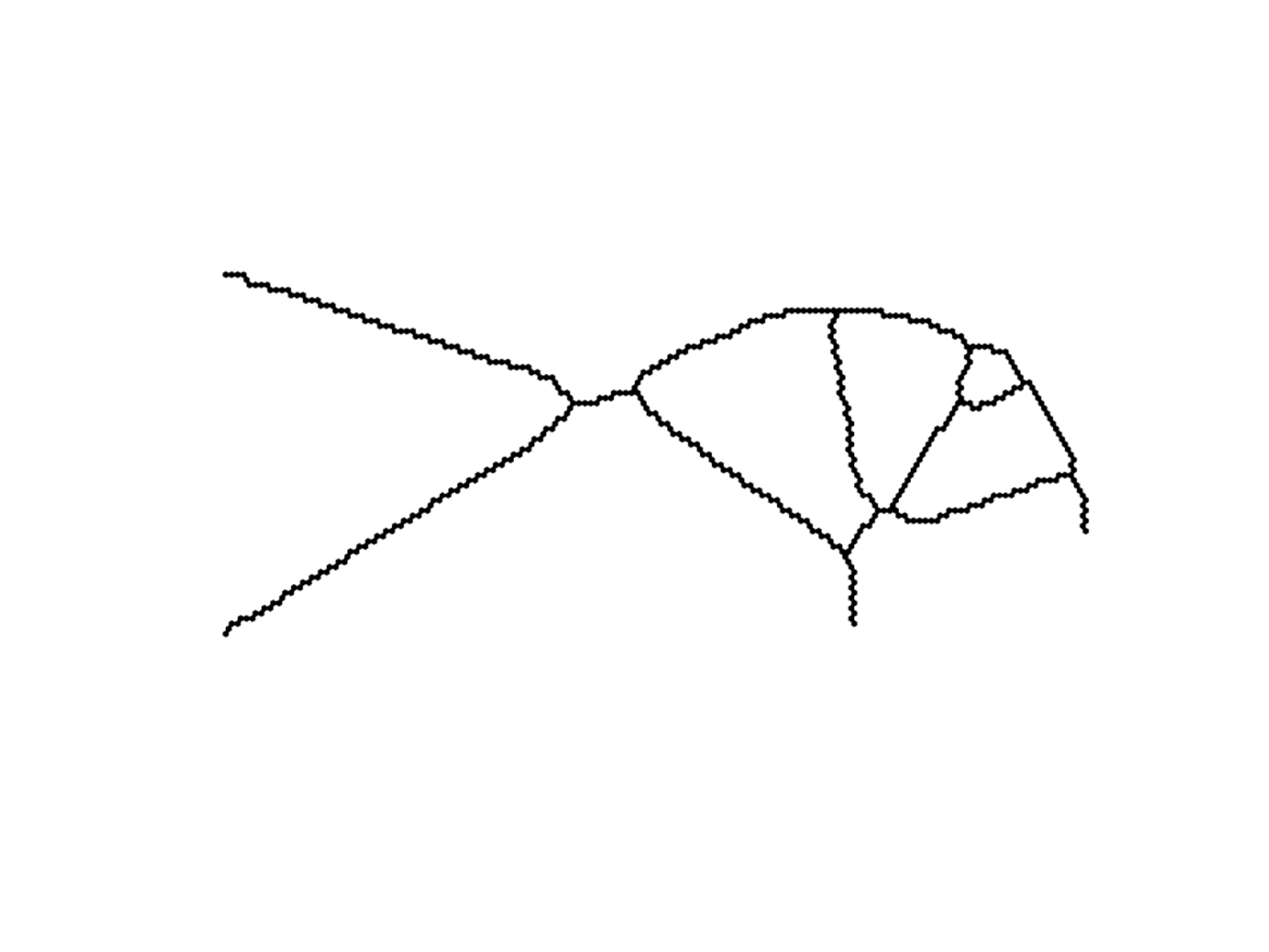}
		\caption{Skeleton (Solution Fig. \ref{fig:Eg1_Eg2_Eg3_Eg4_new_logic_NEM} d.)} 
	\end{subfigure}

\vspace{5mm}
\begin{subfigure}[b]{.24\textwidth}
		\centering
		\captionsetup{font=scriptsize}
		\includegraphics[trim={3.5cm 2.5cm 2.5cm 2.5cm}, clip, scale = 0.3]{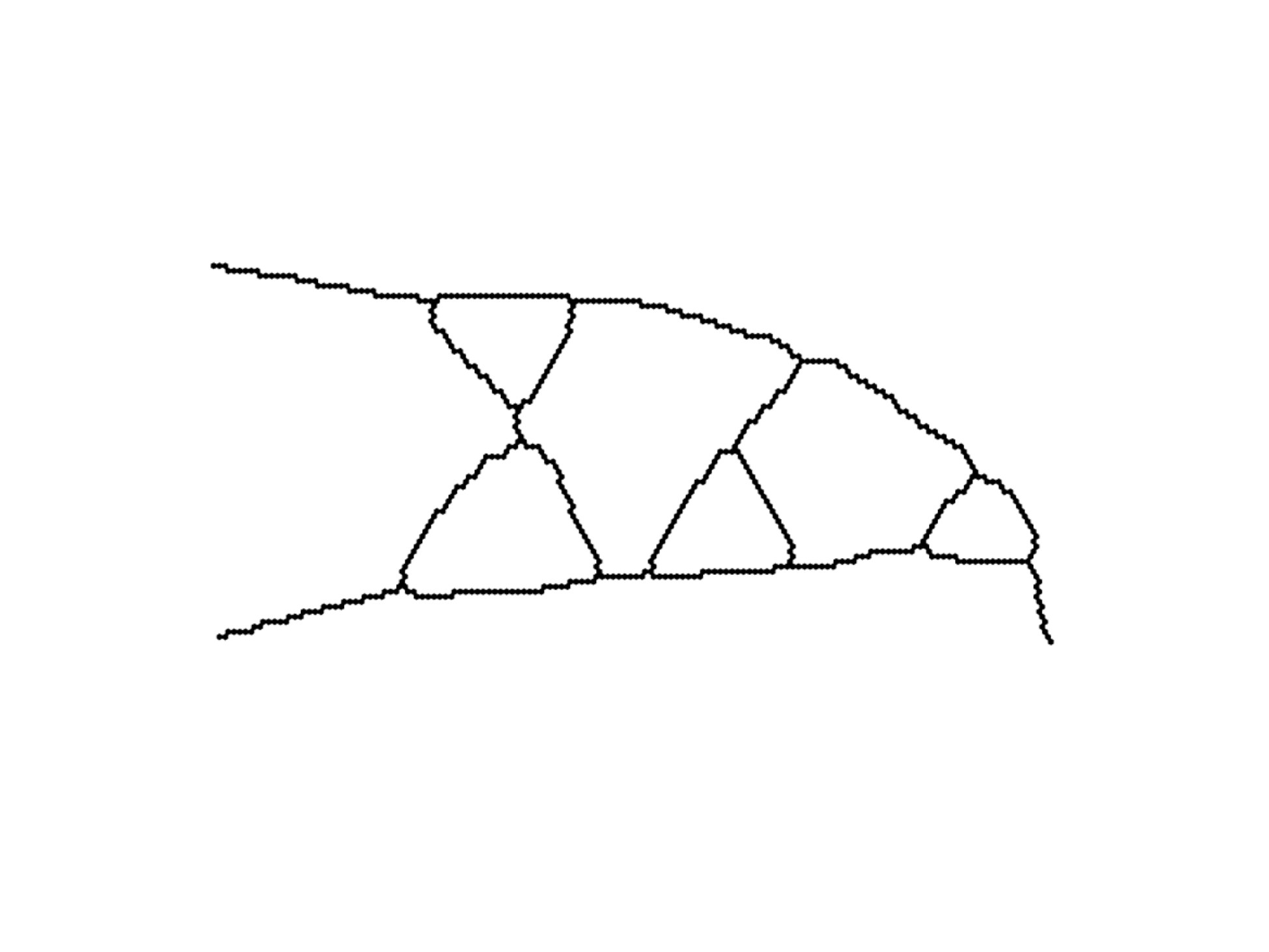} 
		\caption{Skeleton (Solution Fig. \ref{fig:Eg1_Eg2_Eg3_Eg4_new_logic_PEM} a.)} 
	\end{subfigure}
\begin{subfigure}[b]{.24\textwidth}
		\centering
		\captionsetup{font=scriptsize}
		\includegraphics[trim={3.5cm 2.5cm 2.5cm 2.5cm}, clip, scale = 0.3]{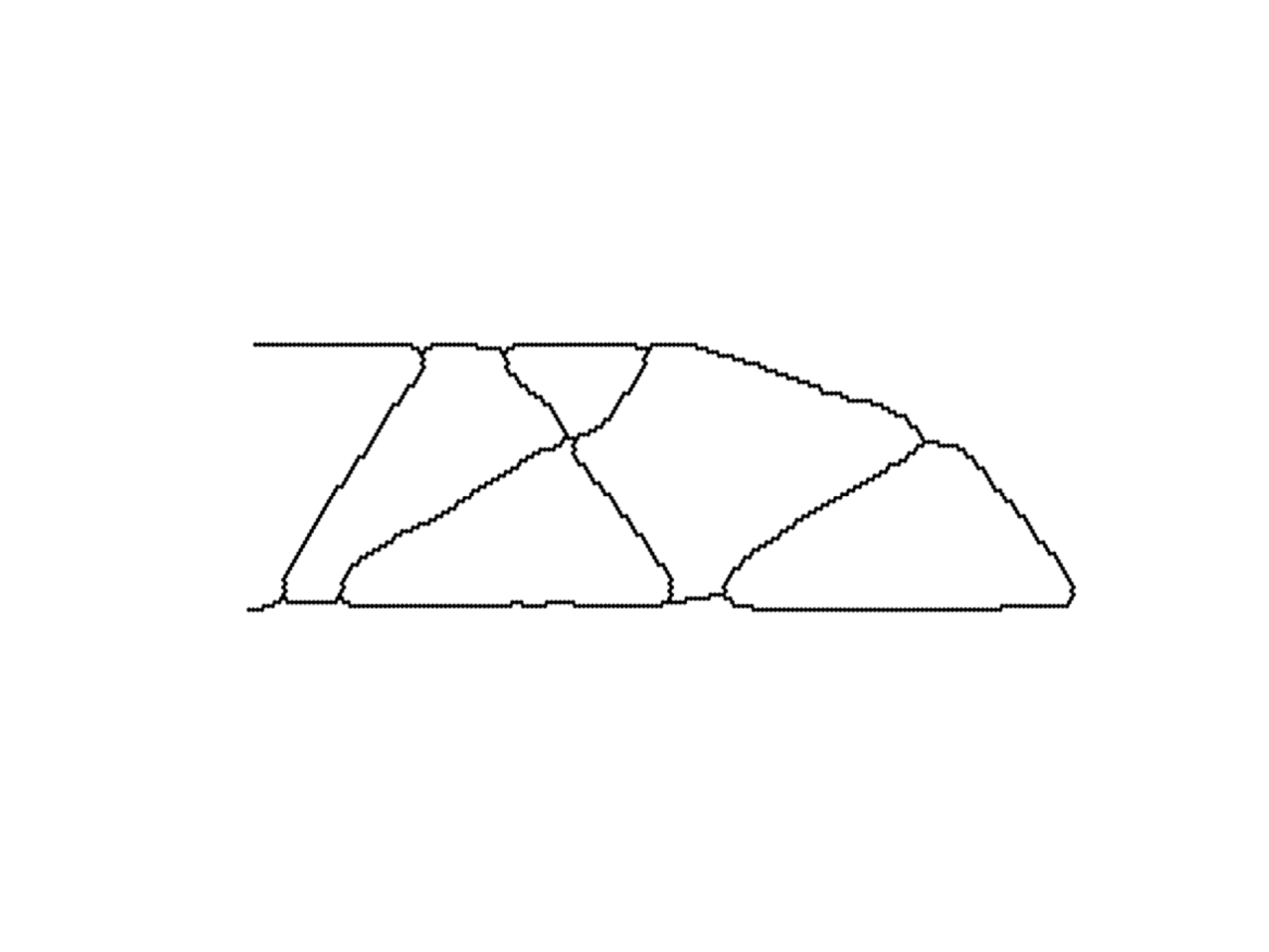}
		\caption{Skeleton (Solution Fig. \ref{fig:Eg1_Eg2_Eg3_Eg4_new_logic_PEM} b.)} 
	\end{subfigure}
\begin{subfigure}[b]{.24\textwidth}
		\centering
		\captionsetup{font=scriptsize}
		\includegraphics[trim={3.5cm 2.5cm 2.5cm 2.5cm}, clip, scale = 0.3]{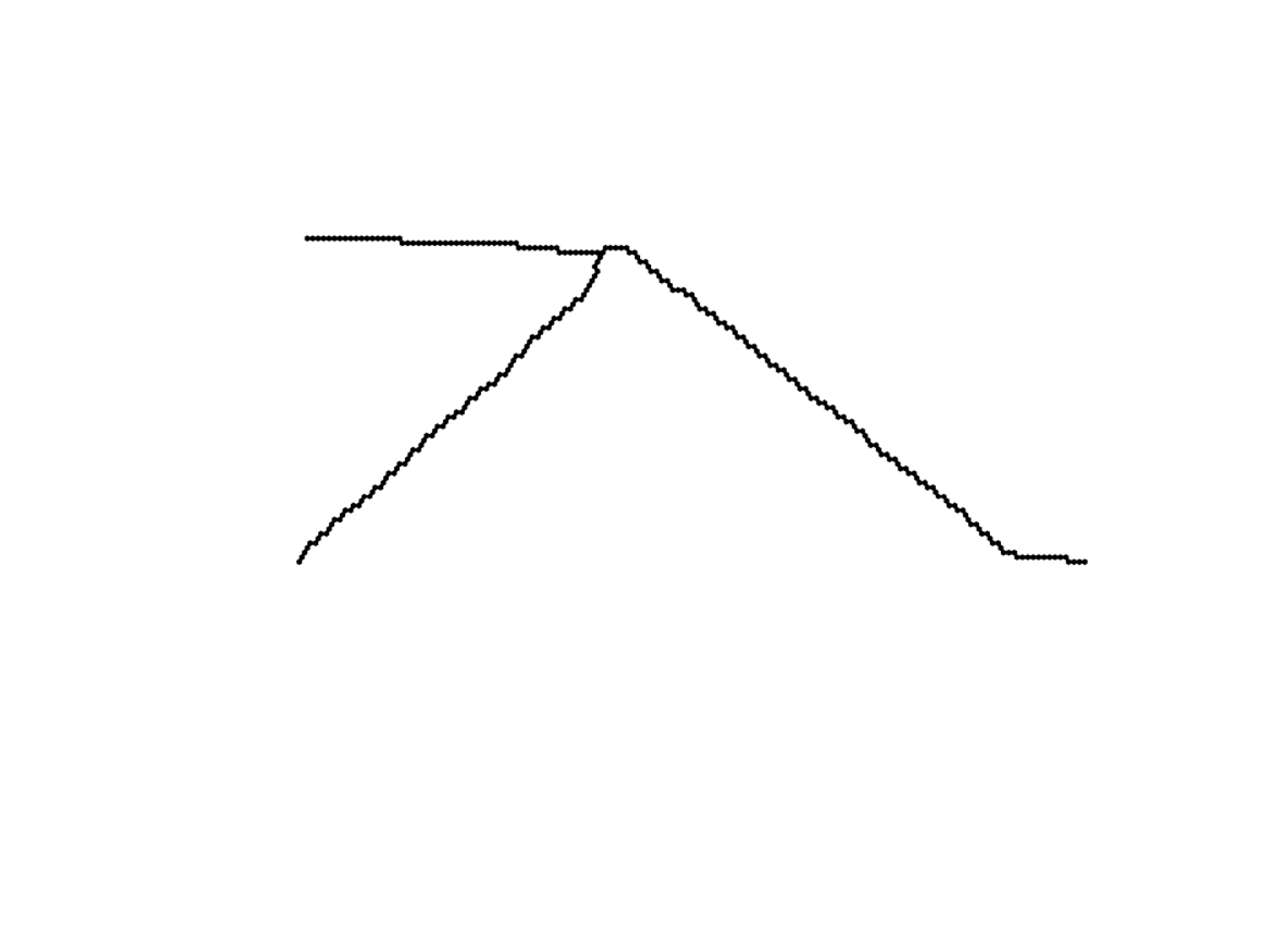}
		\caption{Skeleton (Solution Fig. \ref{fig:Eg1_Eg2_Eg3_Eg4_new_logic_PEM} c.)} 
	\end{subfigure}
\begin{subfigure}[b]{.24\textwidth}
		\centering
		\captionsetup{font=scriptsize}
		\includegraphics[trim={3.5cm 2.5cm 2.5cm 2.5cm}, clip, scale = 0.3]{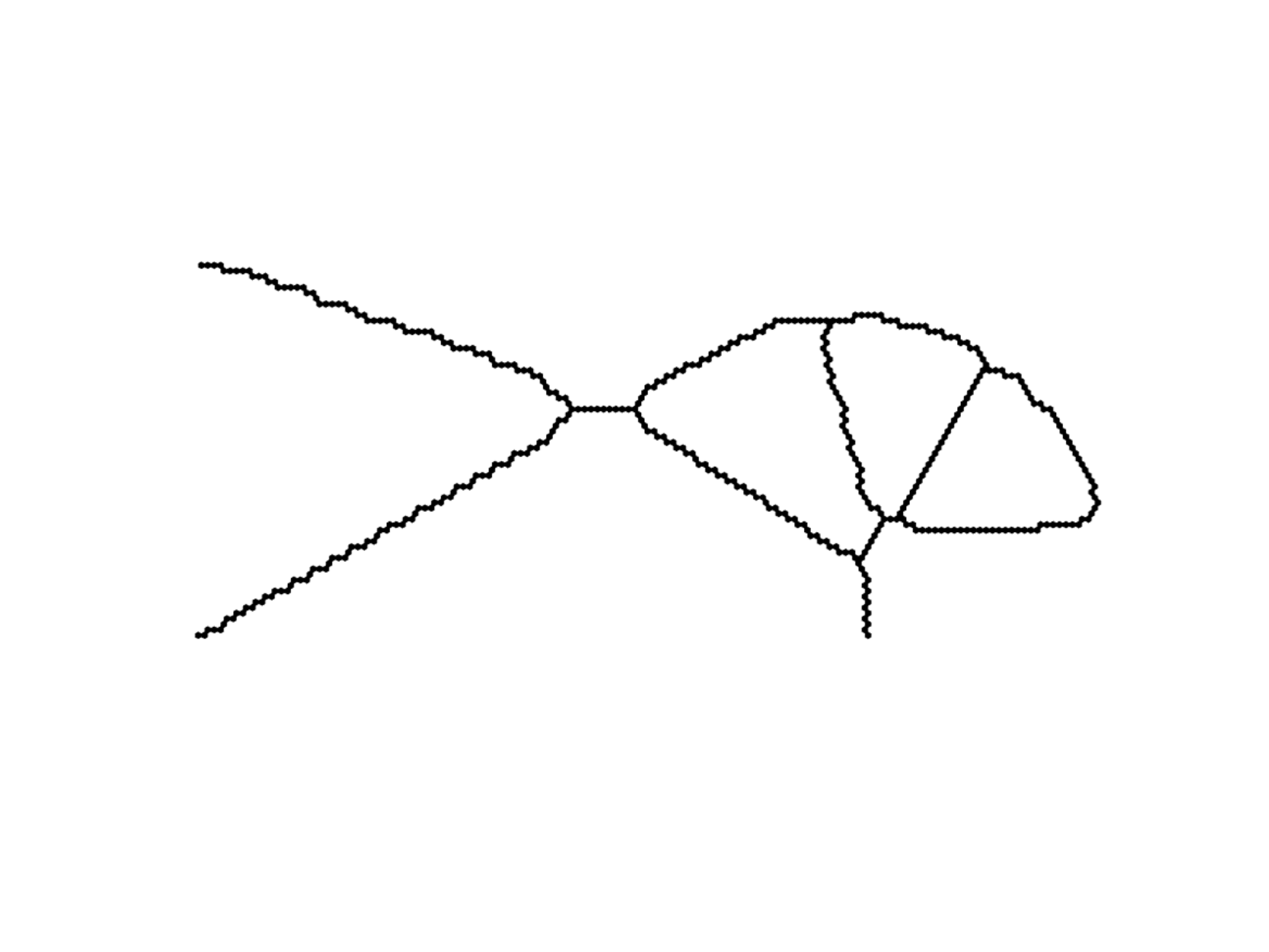}
		\caption{Skeleton (Solution Fig. \ref{fig:Eg1_Eg2_Eg3_Eg4_new_logic_PEM} d.)} 
	\end{subfigure}	
	\caption{Examples of skeletonization.}	
	\label{skeletonization_examples}
\end{figure}

A special case is generated when a filled region is enclosed by an even number of contour cells with precisely one branch of contour attached to every contour cell as in Fig. \ref{fig:7}. To address all such cases, filled cells are forcefully made part of the contour and then, the contour refinement process is implemented. The end result thereafter is the desired skeleton. A flow chart describing the complete skeletonization algorithm is shown in Fig. \ref{fig:6}. Examples of skeletonization are depicted in Fig. \ref{skeletonization_examples}. \\ \\

\vspace{-10mm}

\end{appendices}

 \end{document}